\documentclass[oneside,12pt]{Classes/CUEDthesisPSnPDF}

\ifpdf
    \pdfcatalog { /PageMode (/UseOutlines)
                  /OpenAction (fitbh)  }
\fi

\title{Investigation of astrophysical phenomena in short time scales with ,,Pi of the Sky'' apparatus}

\ifpdf
  \author{\href{mailto:msok@fuw.edu.pl}{Marcin Sokolowski}}
  \collegeordept{\href{http://hep.fuw.edu.pl/ipj/p6/index.php}{Department of High-Energy Physics}}
  \university{\href{http://www.ipj.gov.pl}{Soltan Institute for Nuclear Studies}}
  \crest{\includegraphics[width=30mm]{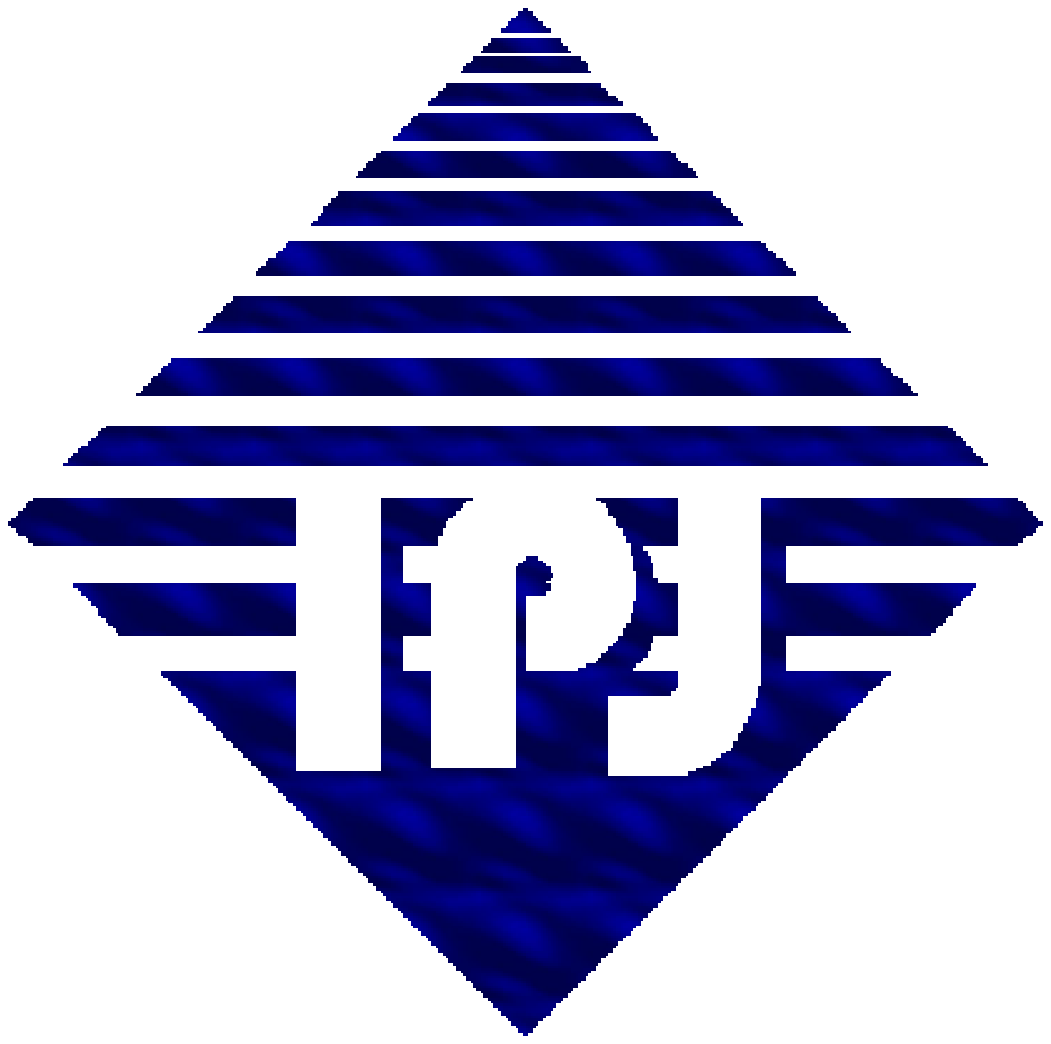}}
\else
  \author{Marcin Sokolowski}
  \collegeordept{Department of High-Energy Physics}
  \university{Soltan Institute for Nuclear Studies}
  \crest{\includegraphics[width=30mm]{ipj_bl.eps}}
\fi
\degree{Doctor of Philosophy in Physics}

\supervisor{dr hab. Grzegorz Wrochna} 

\degreedate{Warsaw, 2007}

\newcommand{\captionfonts}{\footnotesize}
\def\piname{ "Pi of the Sky" }

\makeatletter  
\long\def\@makecaption#1#2{%
  \vskip\abovecaptionskip
    \sbox\@tempboxa{{\captionfonts #1: #2}}%
      \ifdim \wd\@tempboxa >\hsize
          {\captionfonts #1: #2\par}
	    \else
	        \hbox to\hsize{\hfil\box\@tempboxa\hfil}%
		  \fi
		    \vskip\belowcaptionskip}
		    \makeatother   

\usepackage{StyleFiles/watermark}
\usepackage{amsmath}
\usepackage{lscape}

\bibpunct[,]{[}{]}{;}{n}{,}{,}

\begin{document}



\baselineskip=18pt plus1pt




\maketitle

\setcounter{secnumdepth}{3}
\setcounter{tocdepth}{3}

\begin{titlepage}
\newpage
\mbox{}
\end{titlepage}


\begin{acknowledgements}      

I am very grateful to my supervisor dr hab. Grzegorz Wrochna for giving me a
possibility of working in an interesting project and for many 
valuable suggestions and guidance throughout the course of preparing this
thesis.

I would like to thank dr hab. Lech Mankiewicz for many interesting and
fruitful discussions during the development of the system.

I would like to thank all my colleagues from the "Pi of the Sky" for
discussions, friendly atmosphere and their contribution to the project.

I would also like to thank dr hab. Tomasz Bulik and dr. Agnieszka Pollo for 
discussion and support in the final phase of writing this thesis.

Finally I would like to thank my wife Ela, family and friends for their support
and patience.

\end{acknowledgements}



\begin{titlepage}
\newpage
\mbox{}
\end{titlepage}


\begin{abstracts}        

In this thesis the data analysis designed by author for the "Pi of the Sky"
experiment is presented. The data analysis consists of data reduction and
specific algorithms for identification of short time scale astrophysical
processes. The algorithms have been tested and their efficiency has been
determined and described. The "Pi of the Sky" prototype is collecting data
since June 2004 and algorithms could be intensively studied and improved
during over 700 nights. A few events of confirmed astrophysical origin 
and above 100 events in 10s time scale of unknown nature have been 
discovered. During the data collection period 3 Gamma Ray Bursts ( out of
231 ) occurred in the field of view of the telescope, but no optical
counterpart has been found. The upper limits for brightness of the
 optical counterpart have been determined. 
The continuous monitoring of the sky and own trigger for optical flashes
allowed to determine limits on the number of GRBs without corresponding
$\gamma$-ray detection. This allowed determining limits on the ratio
 of emission collimation in optical and $\gamma$ bands, which is $R \leq 4.4$.
The perspectives of the full "Pi of the Sky" system has been studied and 
number of positive detections has been estimated on the level of $\approx$2.5 events per year.

\end{abstracts}



\begin{titlepage}
\newpage
\mbox{}
\end{titlepage}

\begin{romanpages}
\tableofcontents
\listoffigures
\printglossary  
\end{romanpages}


\begin{titlepage}
\newpage
\mbox{}
\end{titlepage}

\ifpdf
    \graphicspath{{Introduction/IntroductionFigs/PNG/}{Introduction/IntroductionFigs/PDF/}{Introduction/IntroductionFigs/}}
\else
    \graphicspath{{Introduction/IntroductionFigs/EPS/}{Introduction/IntroductionFigs/}}
\fi

\chapter*{Introduction}

For thousands of years people believed that the sky is a constant being. The
Universe was believed to be eternal and unchanged. This idea has been
questioned by Galileo Galilei who used supernova discovered by Johannes
Kepler in 1604 as an argument. Since that time big progress in astronomy 
was achieved. The Universe is not believed to be eternal and unchanged
anymore. Several observations prove that the Universe began in the Big Bang,
about 14 billion year ago and this is the well established theory now. 
The evolution of the Universe is studied as well as evolution of its
components. The evolution of stars is quite well understood and 
it is now well known fact that these objects evolve. The timescale of this 
evolution in comparison the human life is huge. However, there are many 
processes in the Universe which occur in much smaller timescales.
The first example can be already mentioned process, supernova explosion.
These events occur at the end of star life and timescale of the explosion 
itself starts on the level of seconds. 
Recent observations suggest that the most violent and spectacular processes in the Universe occur in
timescales of seconds. Among these processes the most energetic are 
Gamma Ray Bursts (hereafter GRB), which were main motivation of this thesis.
The GRBs are processes occurring in timescales ranging from milliseconds to
hundreds of seconds. 
In order to study such rapid processes in optical domain a new approach was
required. Such a need triggered development of the "Pi of the Sky" project.
Data analysis presented in this thesis was performed on the data collected by the 
"Pi of the Sky" detector, which was optimized for investigating short
timescale astrophysical processes with main focus on GRBs. \\
The thesis is divided in five chapters. In the first chapter short timescale
astrophysical processes are reviewed with main focus on GRB. In the second
chapter the "Pi of the Sky" experiment is presented, with several 
technical solutions described in details. The third chapter contains
description of data analysis, including description of flash recognition
algorithms created by author. In the fourth chapter results and 
perspectives for future are presented.

\chapter{Short timescale astrophysical phenomena}
\label{chapter_short_time_astro}

For centuries astronomical observations were performed in long timescales. 
It was mainly due to instrumental limitations. Evolution of stars
and the Universe is a very slow process, so for long time it was enough
for understanding many astrophysical processes in the Universe.
The main limitation was due to detectors, since the middle of XIX century
photographic films were used. They didn't allow for short exposures due 
to very poor quantum efficiency. The next generation of detectors,
photomultipliers allowed much better time resolution, however they were 
limited to observations of single object. Big progress in electronic\
technologies allowed to introduce new type of detectors Charged Coupled
Devices ( CCD ). This type of detector together with computers gives 
a powerful tool to the researchers. 
Using the CCD detectors it is possible to perform astrophysical observations
with time resolution of seconds. 
Except the optical domain big progress was also achieved in the area of 
X-ray, gamma and particle detectors, which allowed to begin studies of 
the Universe in other bands.
Progress in detector techniques allowed to monitor simultaneously many objects in short time scales.
We now know that there is a number of such fast astrophysical events 
occurring on very short timescales down to milliseconds.
List of interesting astrophysical processes acting in short timescales is
long. The most rapid, but still possible to observe are listed below :

\begin{itemize}
\item Gamma Ray Bursts (GRB)
\item Supernovae stars
\item Active Galactic Nuclei (AGN) in particular blasars
\item Nova stars
\item Flare and other variable stars
\end{itemize}

It seems that the most violent astrophysical processes occur on short
timescales. In most cases optical observations of these fast processes are
performed some time after the main explosion, when new object is observed in the
sky or signal from other bands is detected and distributed to the community.
Due to sudden character of these events it is very difficult to catch such
event when it is going on. This is a strong motivation for wide field
observations. Such observations give a big chance to discover many events
with flash-like signature, when suddenly new object appears in the sky.
Such signal may be used as trigger signal for other experiments to observe.
The "Pi of the Sky" experiment was designed to be a good tool for performing
this kind of observations.

\subsection{Gamma Ray Bursts}
\label{sec_grb}
The GRBs are the most violent and energetic events known in the Universe.
They were first observed in late 60s by American satellites Vela \cite{grb_vela}.
For many years they were the most mysterious astrophysical processes.
The main problems of early GRB researchers were the following :

\begin{itemize}
\item Origin - are the GRBs galactic or extra-galactic ? The extra-galactic origin 
   of GRBs would require huge energetics of these processes which was very 
   hard to accept by the researchers.
\item If extragalactic, how such amounts of energy can be produced ?
\item What is the central engine and progenitor responsible for such kind of explosion ?
\end{itemize}

\begin{figure}[!htbp]
  \begin{center}
    \leavevmode
    \ifpdf
      \includegraphics[width=4in,angle=90]{fig2_2704.gif}
    \else
      \includegraphics[width=4in,angle=90]{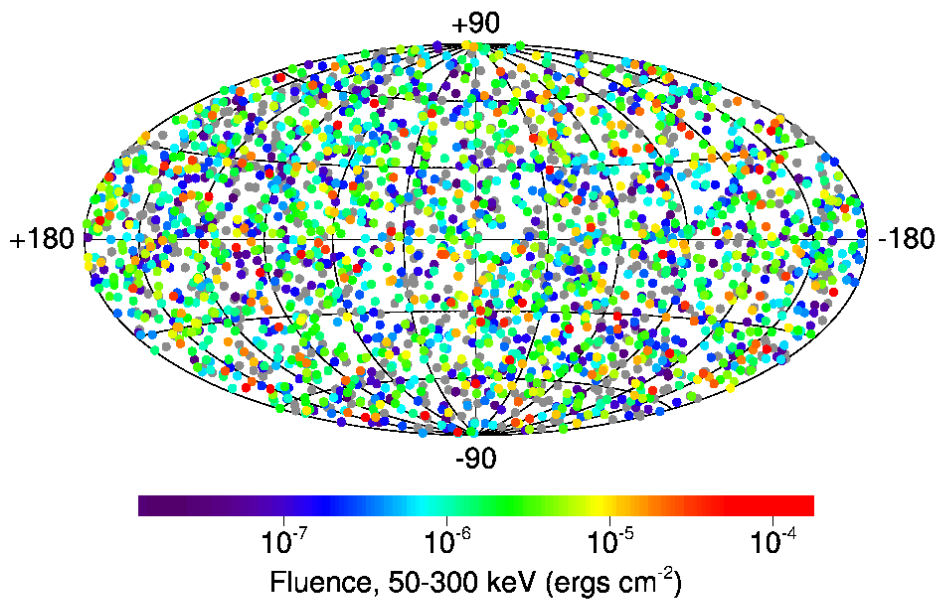}
    \fi
    \caption{Spatial distribution of 2704 GRBs detected by BATSE detector on board the CGRO satellite (\cite{batse_www})}
    \label{fig_batse_grbs}
  \end{center}
\end{figure}

\begin{figure}[!htbp]
  \begin{center}
    \leavevmode
    \ifpdf
      \includegraphics[width=4in]{4b_t90.gif}
    \else
      \includegraphics[width=4in]{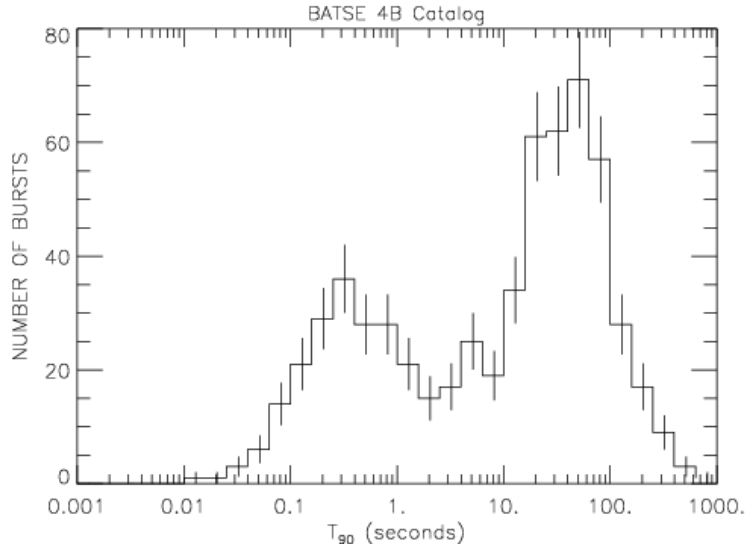}
    \fi
    \caption{Duration of GRBs detected by BATSE detector on board the CGRO satellite (\cite{batse_www})}
    \label{fig_batse_longshort}
  \end{center}
\end{figure}

A large progress was achieved in 90s due to BATSE instrument on CGRO
satellite which detected ~2700 GRBs. This allowed to determine the spatial
distribution of the GRBs (Fig. \ref{fig_batse_grbs}) giving a strong argument for cosmological
origin of the GRBs. Another success of the BATSE mission was discovery of 2
classes of short and long GRBs (Fig. \ref{fig_batse_longshort}). It seems that these
two kinds of GRBs are caused by different processes. 

\begin{figure}[!htbp]
  \begin{center}
    \leavevmode
    \ifpdf
		\includegraphics[width=2.8in,height=2.8in]{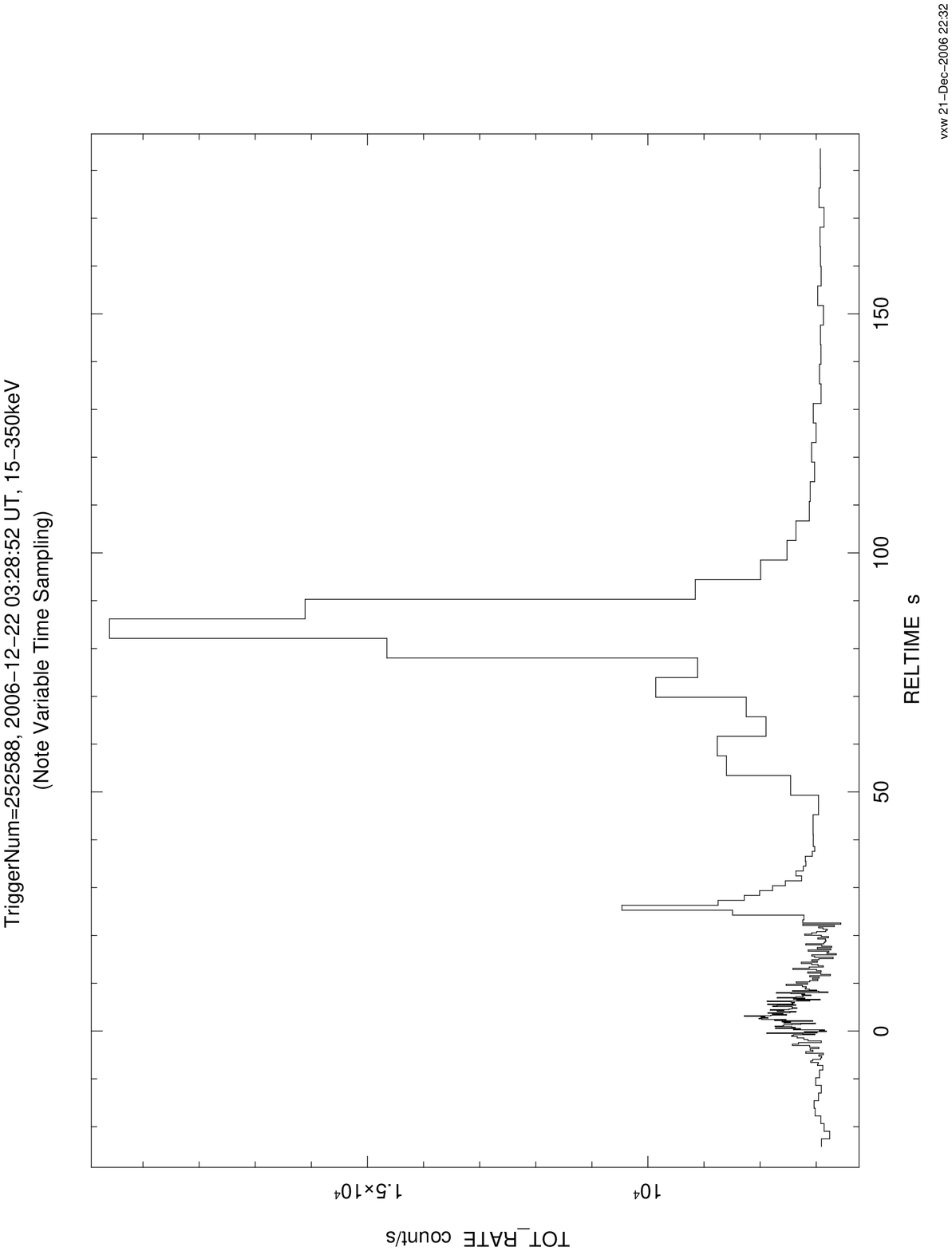}
      \includegraphics[width=2.8in,height=2.8in]{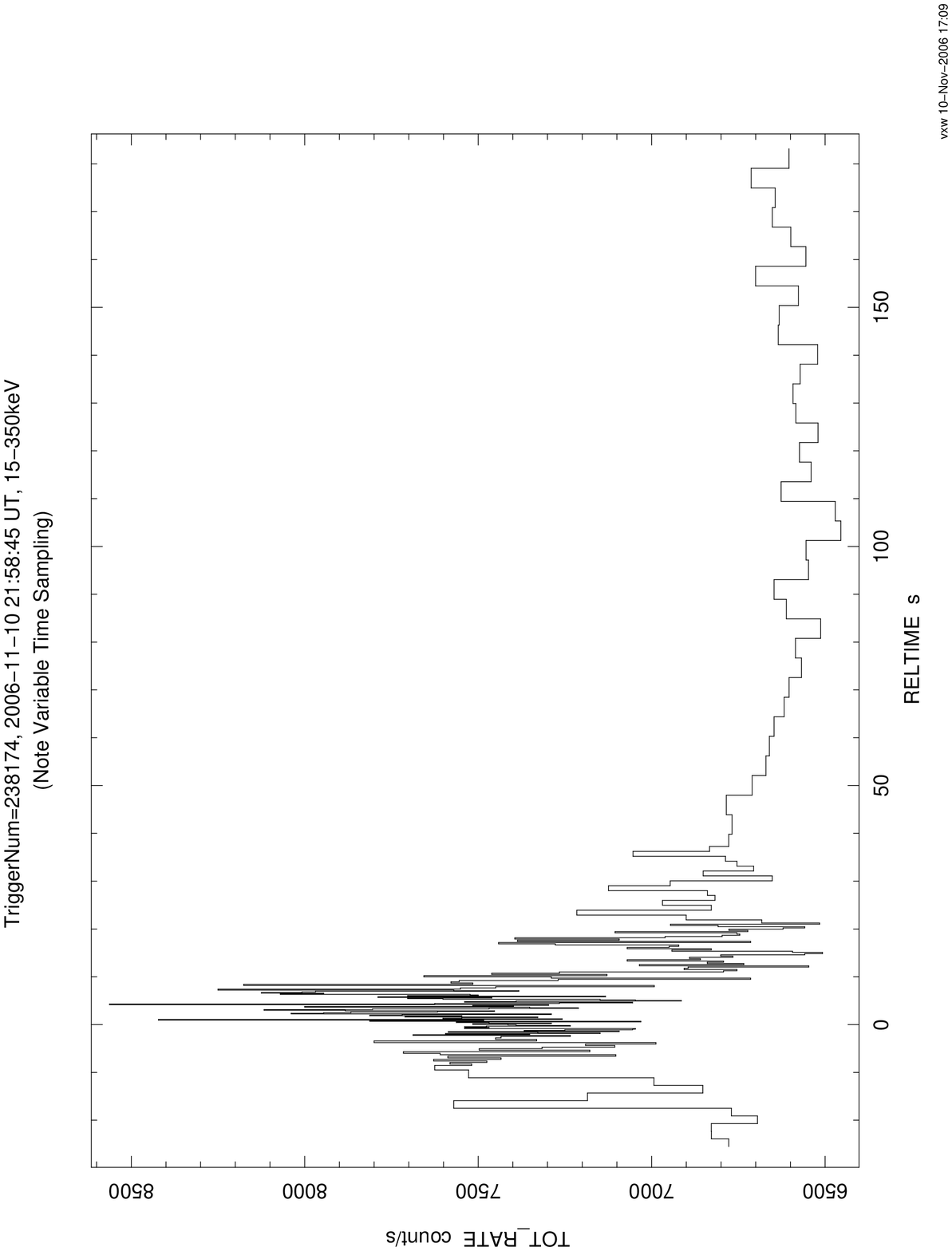}
      \includegraphics[width=2.8in,height=2.8in]{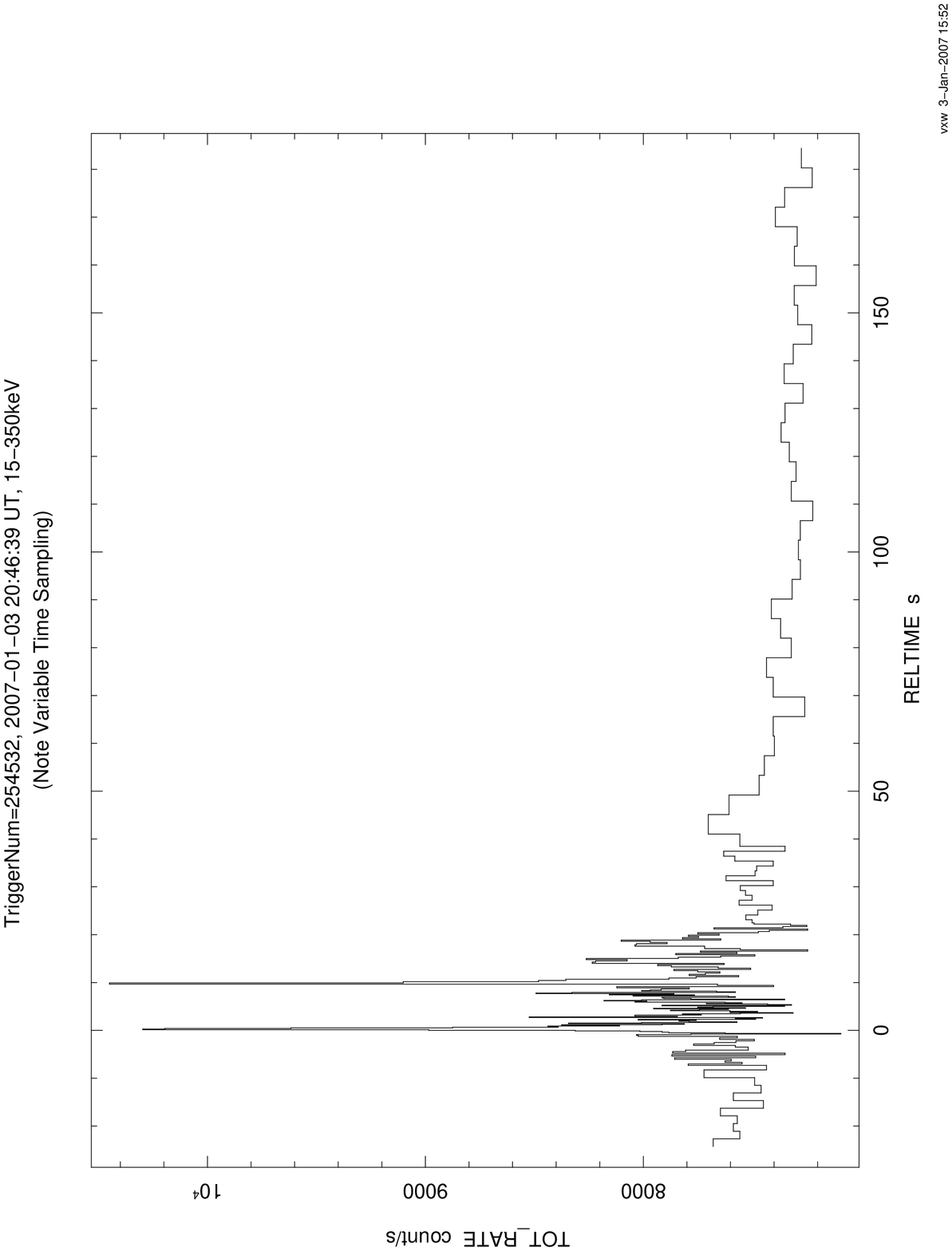}
      \includegraphics[width=2.8in,height=2.8in]{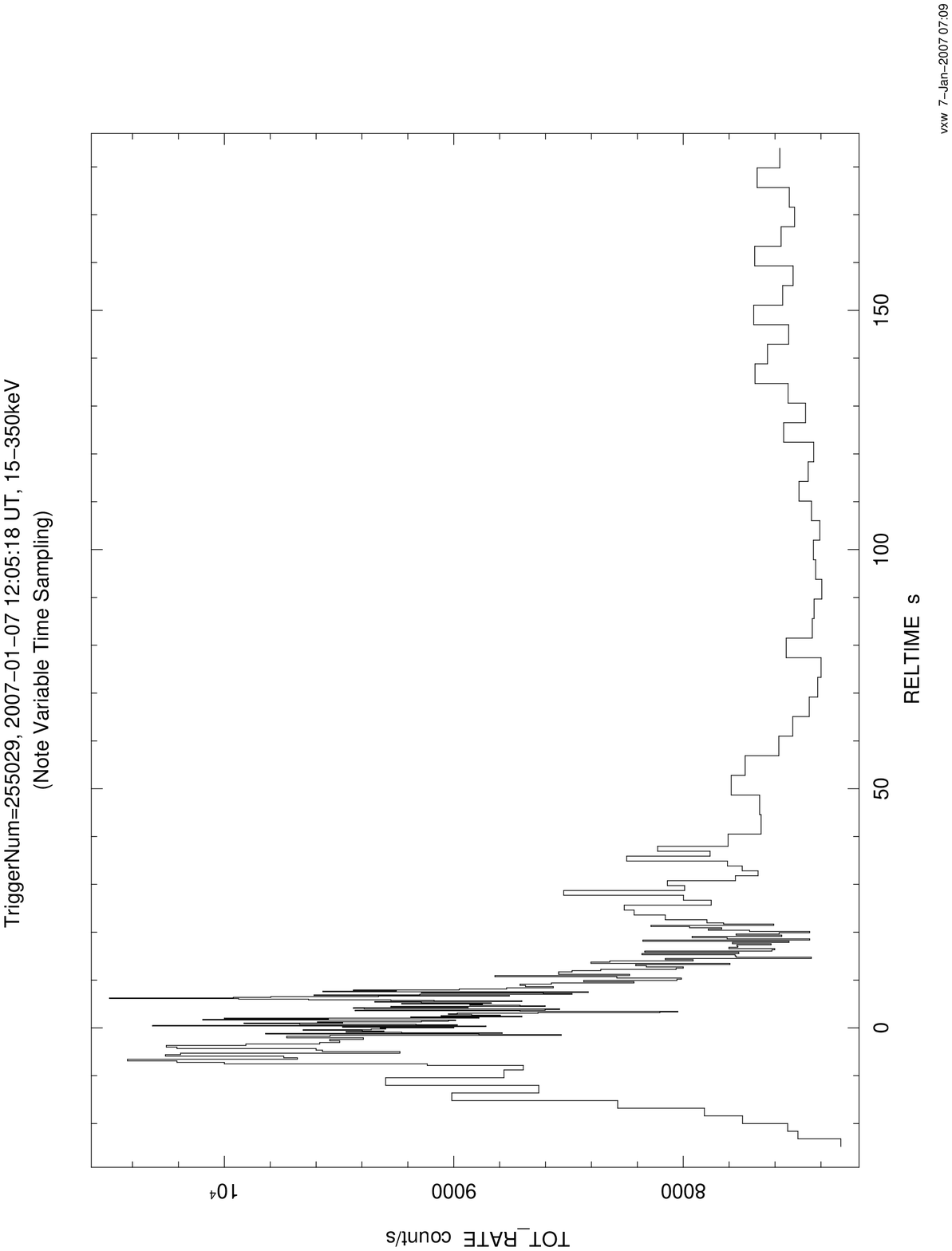}
    \else
      \includegraphics[width=2.8in,height=2.8in,angle=-90]{grb_lc/long/sw00252588000msb.eps}
      \includegraphics[width=2.8in,height=2.8in,angle=-90]{grb_lc/long/sw00238174000msb.eps}
      \includegraphics[width=2.8in,height=2.8in,angle=-90]{grb_lc/long/sw00254532000msb.eps}
      \includegraphics[width=2.8in,height=2.8in,angle=-90]{grb_lc/long/sw00255029000msb.eps}
    \fi
    \caption{Examples of long GRBs observed by SWIFT satellite in years 2006 and 2007 (\cite{grb_gcn_alerts})}
    \label{fig_grb_examples_long}
  \end{center}
\end{figure}

\newpage
\begin{figure}[!htbp]
  \begin{center}
    \leavevmode
    \ifpdf
		\includegraphics[width=2.8in,height=2.8in]{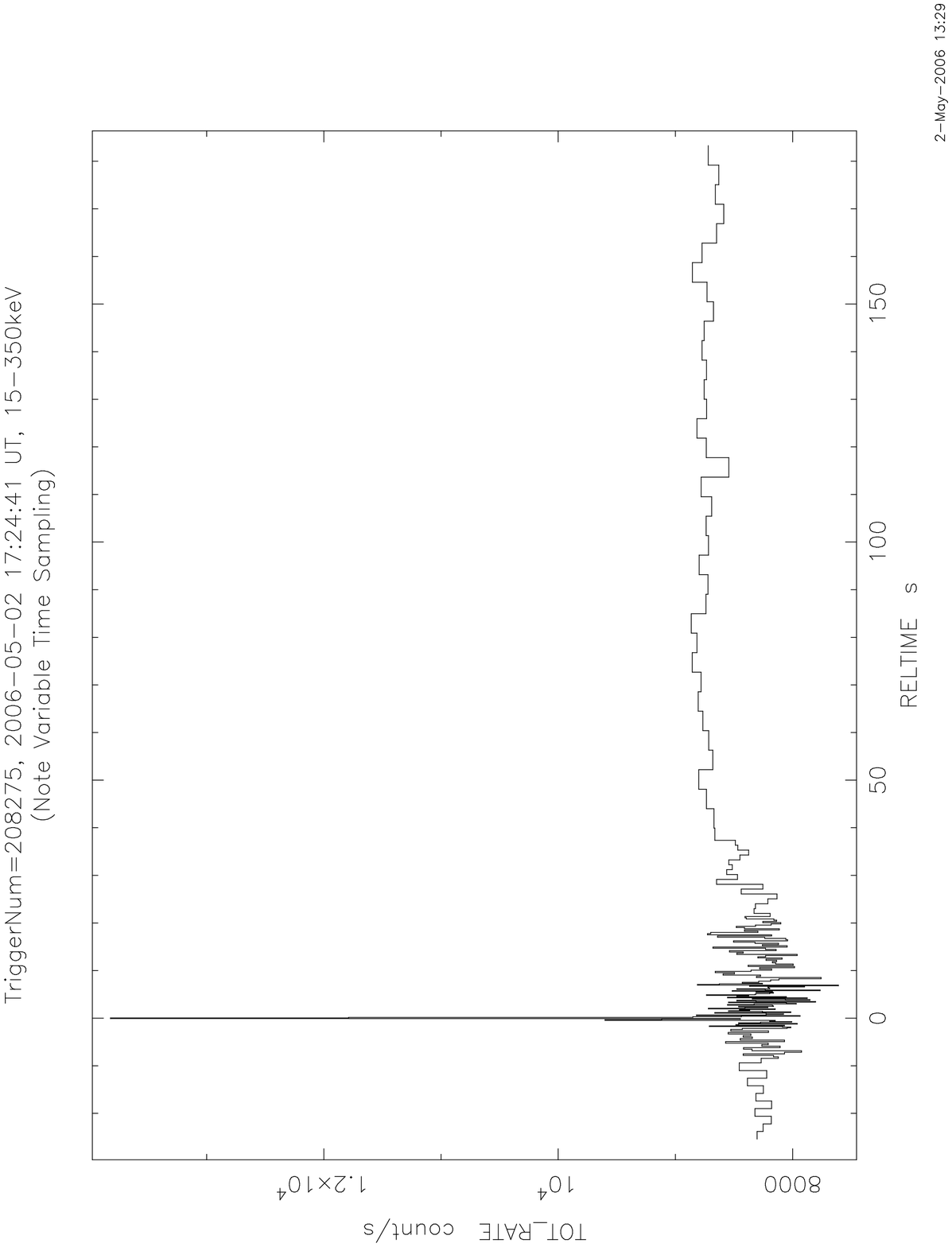}
		\includegraphics[width=2.8in,height=2.8in]{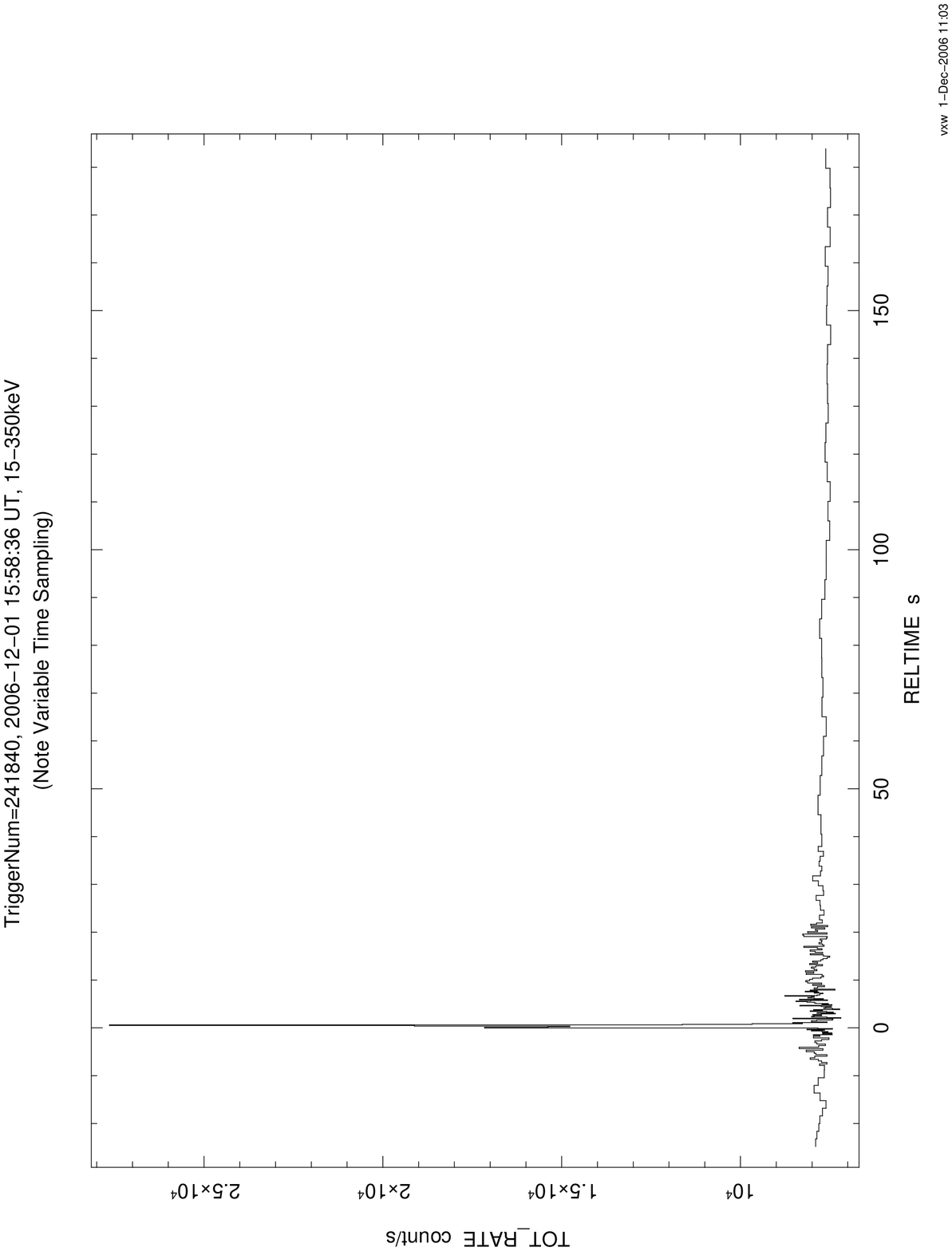}
		\includegraphics[width=2.8in,height=2.8in]{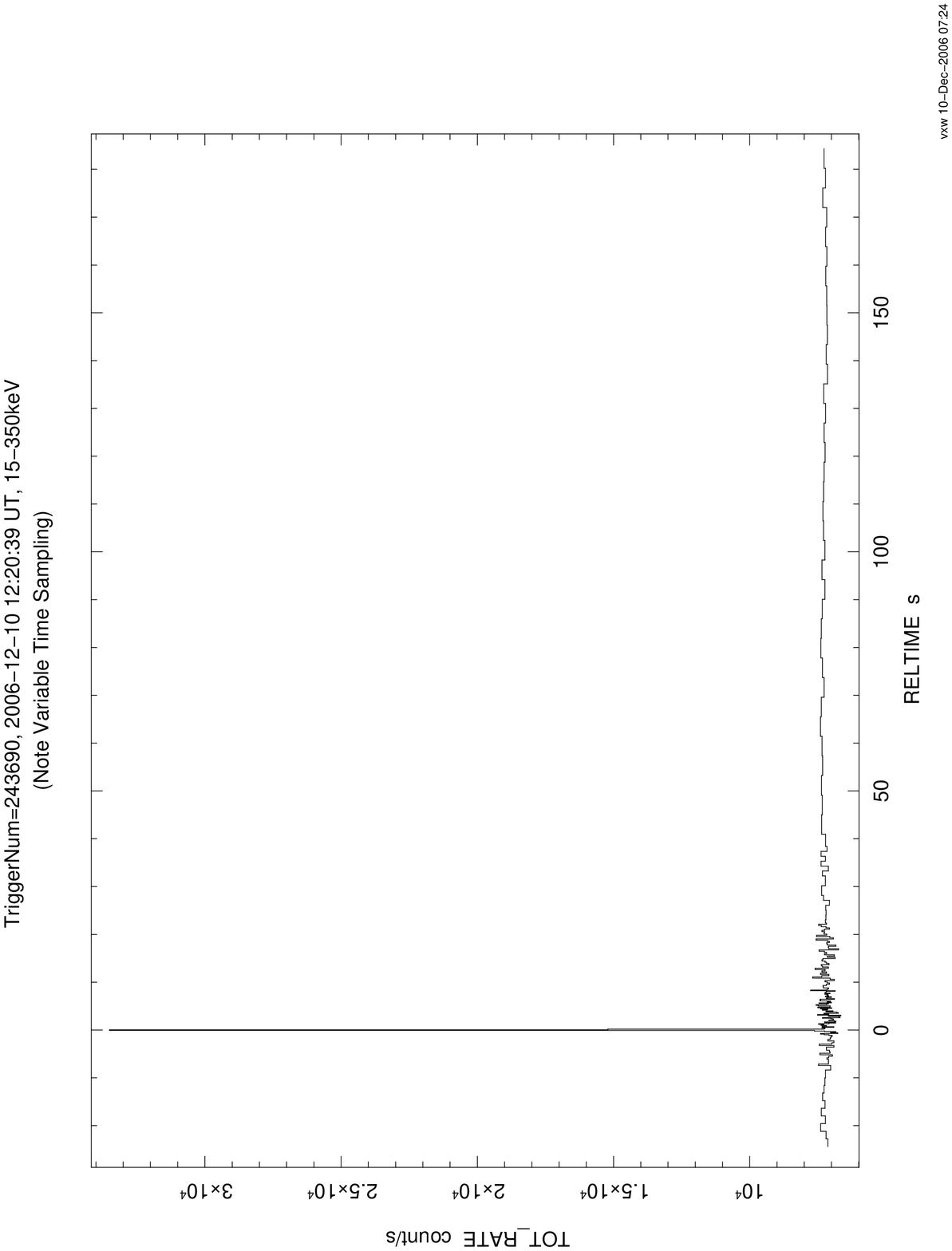}
		\includegraphics[width=2.8in,height=2.8in]{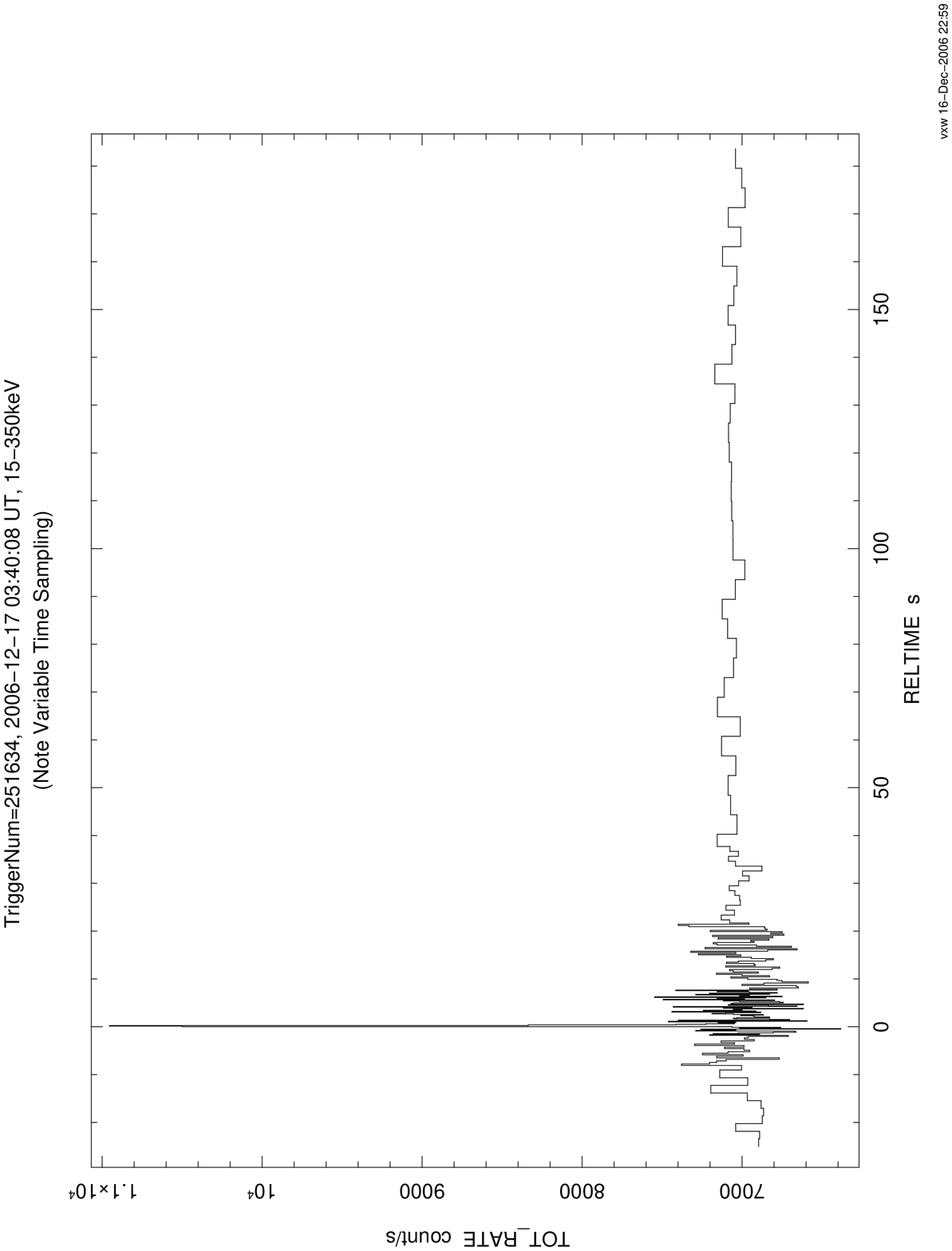}
    \else
		\includegraphics[width=2.8in,height=2.8in,angle=-90]{grb_lc/long/sw00208275000msb.eps}
		\includegraphics[width=2.8in,height=2.8in,angle=-90]{grb_lc/long/sw00241840000msb.eps}
		\includegraphics[width=2.8in,height=2.8in,angle=-90]{grb_lc/long/sw00243690000msb.eps}
		\includegraphics[width=2.8in,height=2.8in,angle=-90]{grb_lc/long/sw00251634000msb.eps}
    \fi
    \caption{Examples of short GRBs observed by SWIFT satellite in years 2006 and 2007 (\cite{grb_gcn_alerts})}
    \label{fig_grb_examples_short}
  \end{center}
\end{figure}

Figures \ref{fig_grb_examples_long} and \ref{fig_grb_examples_short} show examples of 
$\gamma$-ray light curves of long and short GRBs observed by the SWIFT satellite \cite{swift_www}.
Final prove of extragalactic origin of the GRBs was provided by the Beppo-SAX
satellite which on 1997-02-28 observed the first X-ray afterglow of the GRB970228.
The Beppo-SAX observation of GRB971214 allowed determination of its redshift z=3.1418
, which confirmed extragalactic origin of GRBs. 
The cosmological origin of GRBs was well established after redshifts were
measured for many more bursts. This fact implied huge energies produced 
in the explosion of the order of $10^{53}$ ergs when isotropic emission is
assumed. 

\begin{table}[htbp]
\begin{center}
\begin{tiny}
\begin{tabular}{|c|c|c|c| }
\hline
\textbf{Mission} & \textbf{Period} & \textbf{Properties of detectors} & \textbf{Main outcome} \\
\hline
VELA & 1969-1979 &  & Discovery of GRBs \\
\hline
 & & & \\
BATSE on CGRO & 1991-1999 & FOV=4$\pi$, $E_{\gamma}$=20keV-600keV  & \begin{minipage}{4cm}{Long and short, spatial distribution, detection of ~2700 GRBs}\end{minipage} \\
 & & & \\
\hline
Beppo-SAX & 1997-2003 &  & X and Optical afterglows \\
\hline
 & & & \\
Hete-2 & 2000-2006 & FOV$\approx$1.5 sr, $E_{\gamma}$=2keV-400keV & \begin{minipage}{4cm}{OT from short GRB, ~100 bursts detected and localized}\end{minipage} \\
 & & & \\
\hline
 & & & \\
\textbf{Swift} & 2004- & FOV$\approx$2 sr, $E_{\gamma}$=15-350keV & \begin{minipage}{4cm}{OT from short GRB, early optical and X-ray observations, X-ray flares. }\end{minipage} \\
 & & & \\
\hline
\textbf{Integral} & 2002-2010 & FOV$\approx 10^o$ x $10^o$, $E_{\gamma}$=3keV-10MeV & detection of many GRBs \\
\hline
Agile & April 2007- & 30MeV-50GeV & Explore unexplored regime \\
\hline
\textbf{GLAST} & 2008 (?) & FOV$\approx$2 sr, 5keV-100GeV & Explore unexplored regime \\
\hline
\end{tabular}
\end{tiny}
\caption{Past and future satellite missions most important for understanding the Gamma Ray Bursts}
\label{tab_grb_missions}
\end{center}
\end{table}

\begin{figure}[!htbp]
  \begin{center}
    \leavevmode
    \ifpdf
      \includegraphics[width=6in]{fireball.gif}
    \else
      \includegraphics[width=6in]{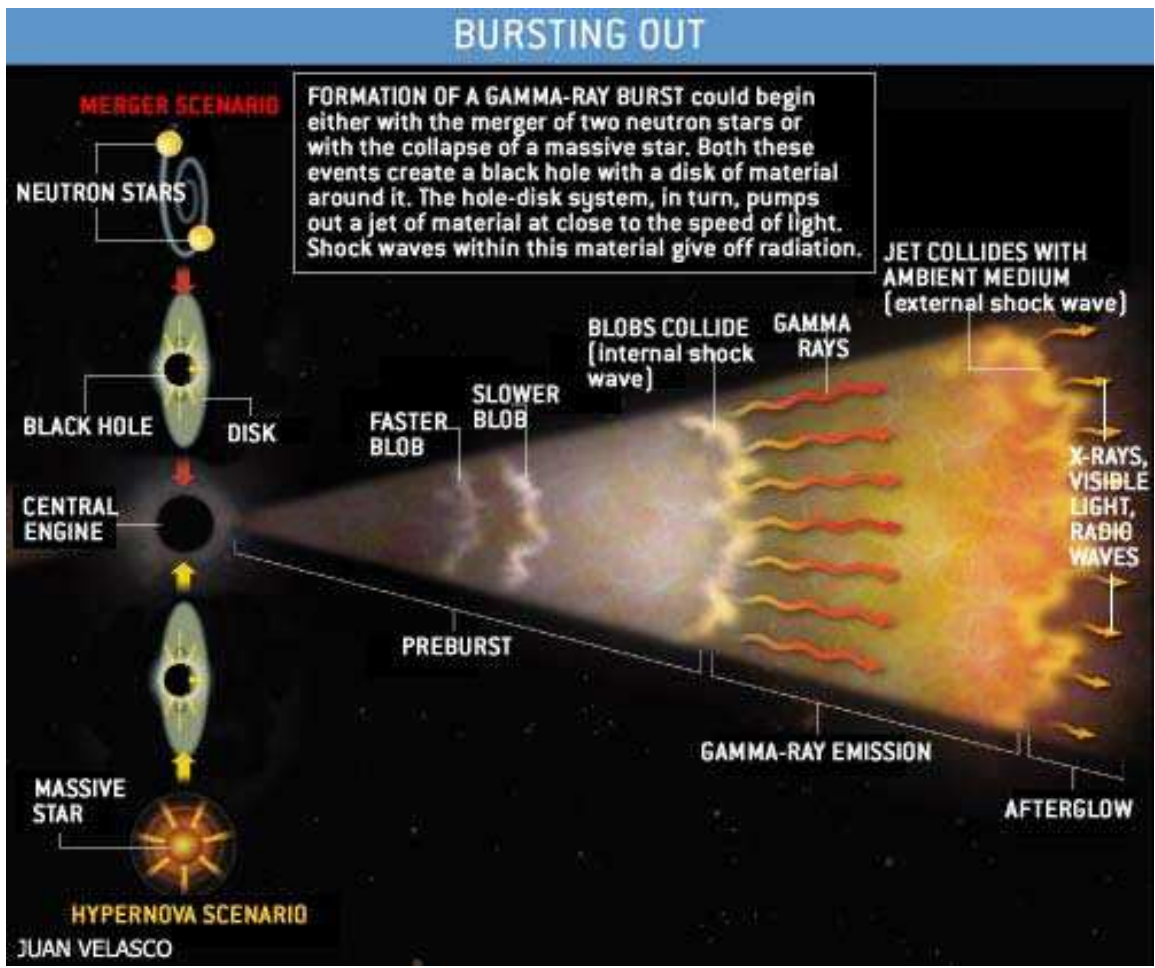}
    \fi
    \caption{Fireball model of the GRB ( image from \cite{grb_sciam_fireball_image} )}
    \label{fig_fireball_model}
  \end{center}
\end{figure}

The data collected by orbital missions (Tab. \ref{tab_grb_missions}) and ground base telescopes allowed to formulate 
the fireball model (Fig. \ref{fig_fireball_model}) which can describe most of the properties of GRBs \cite{e_waxman}.
The energy is produced by the central engine which in case of long bursts is
believed to be collapse of the massive star in the so called "collapsar"
scenario and it is a version of supernova explosion (\cite{collapsar_1999_macfadyen},\cite{collapsar_1998_paczynski},\cite{collapsar_2005_woosley},\cite{collapsar_1993_woosley}). In case of the short
bursts the central engine is believed to be a merging of 
two compact objects in the binary system (\cite{merger_1989_eichler_livio},\cite{merger_1992_meszaros},\cite{merger_1992_narayan_paczynski},\cite{merger_1986_paczynski}). Such compact objects can be 
two neutron stars or neutron star and a black hole. 

\begin{figure}[!htbp]
  \begin{center}
    \leavevmode
    \ifpdf
      \includegraphics[width=2.8in,height=2.8in]{grb990123_combined_lc.gif}
		\includegraphics[width=2.8in,height=2.8in]{raptor_041219.gif}
    \else
      \includegraphics[width=2.8in,height=2.8in]{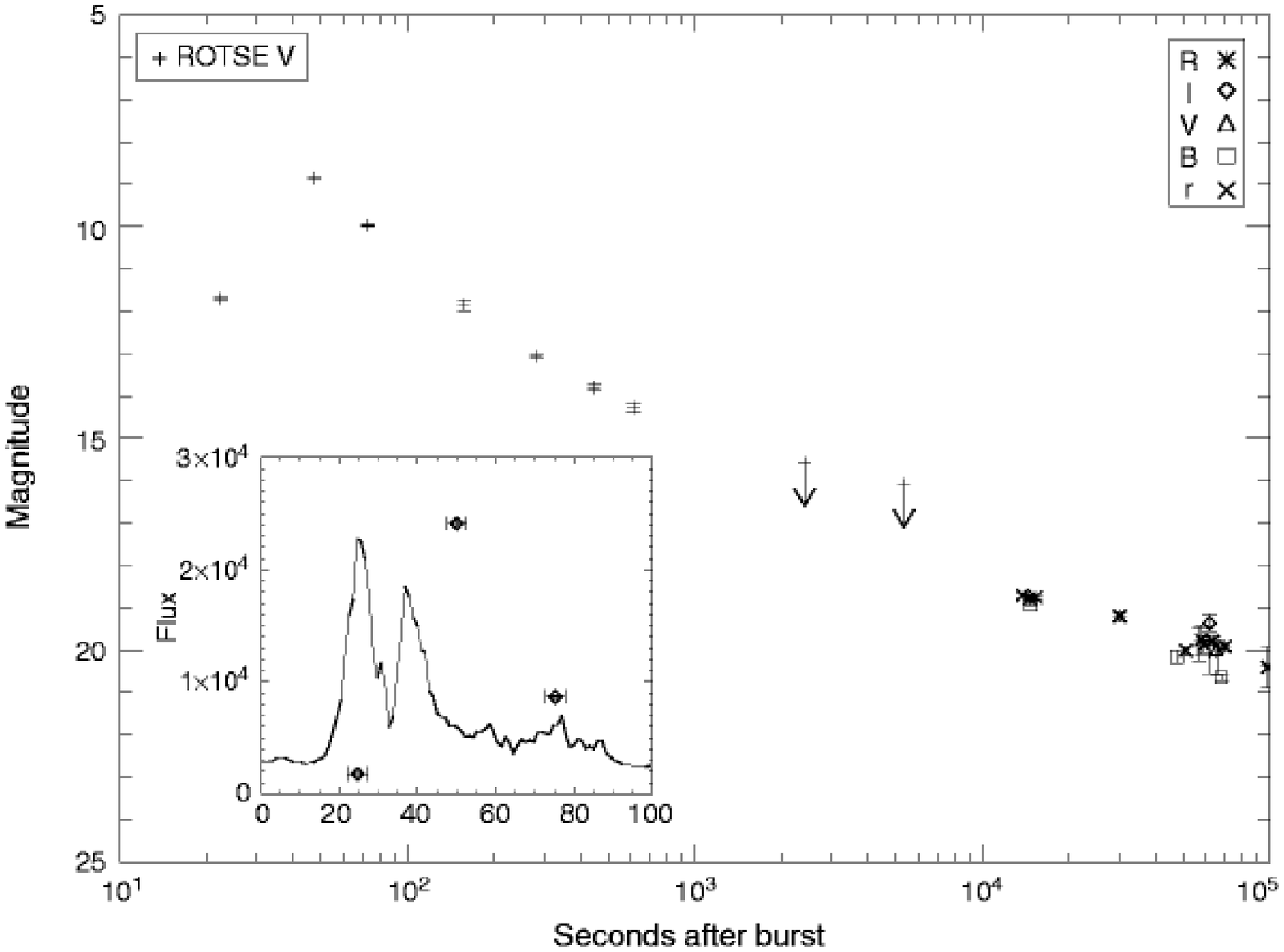}
		\includegraphics[width=2.8in,height=2.8in]{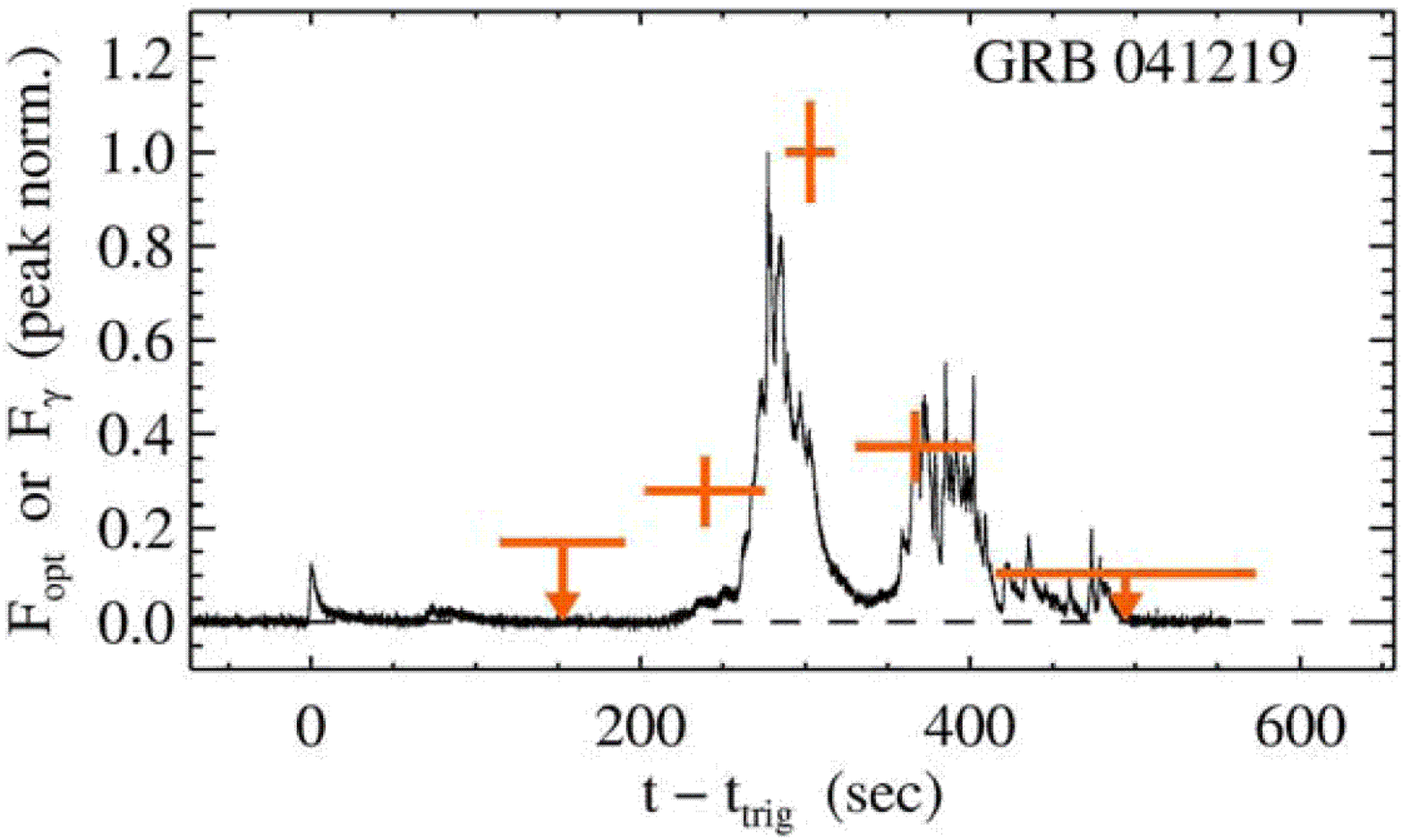}
    \fi
    \caption{Observation of prompt optical signal from GRB990123 by ROTSE (left image) and GRB041219 by RAPTOR (right image). In the case of GRB041219 the optical signal variability is correlated with $\gamma$ emission while in the case of GRB990123 it is not}
    \label{fig_rotse_99}
  \end{center}
\end{figure}

The central engine explosion causes ejection of matter in the collimated
jet. The matter is ejected in packets called "shells" with different Lorentz
factors $\Gamma \gg 100$ causing internal shocks when slower shells are crashed by
faster shells ejected later. This internal shocks mechanism is believed to 
produce the gamma rays observed as the GRB event. 
The ejected matter finally reaches the interstellar medium ( ISM ) and 
the collision with ISM causes the external shock. This processes is believed
to be responsible for the optical counterparts of the GRBs so called
"afterglows" which were observed for a number of bursts. Currently for about
50\% of GRBs optical counterparts are observed ( see Sec. \ref{sec_full_system_predict} ).
The jet opening angle was determined for number of bursts from the jet break time to be ~ few
degrees, which reduces the total GRB energy by factor of $\approx$100. 
Since the discovery of the GRBs a large progress in their understanding has
been achieved, however, the main question how the central engine works still
remains unanswered. There are also many other uncertainties and doubts concerning
mechanisms of jet production etc. 
More multiwavelength data is required to understand better these processes. 
Especially optical data is very important for understanding the puzzle of
the GRB. Due to technical limitations for many years optical counterparts
 of GRBs were observed many hours or even days after the gamma emission.
For long time, in fact until SWIFT satellite was launched in 2004, there was 
only one event 990123 \cite{rotse_99} for which the optical signal was observed 
when the GRB was still going on (Fig. \ref{fig_rotse_99}). For most of the bursts position, if known,
was determined long time after the GRB, causing optical observations to be delayed. 

\begin{figure}[!htbp]
  \begin{center}
    \leavevmode
    \ifpdf
      \includegraphics[width=6in]{GCN.gif}
    \else
      \includegraphics[width=6in]{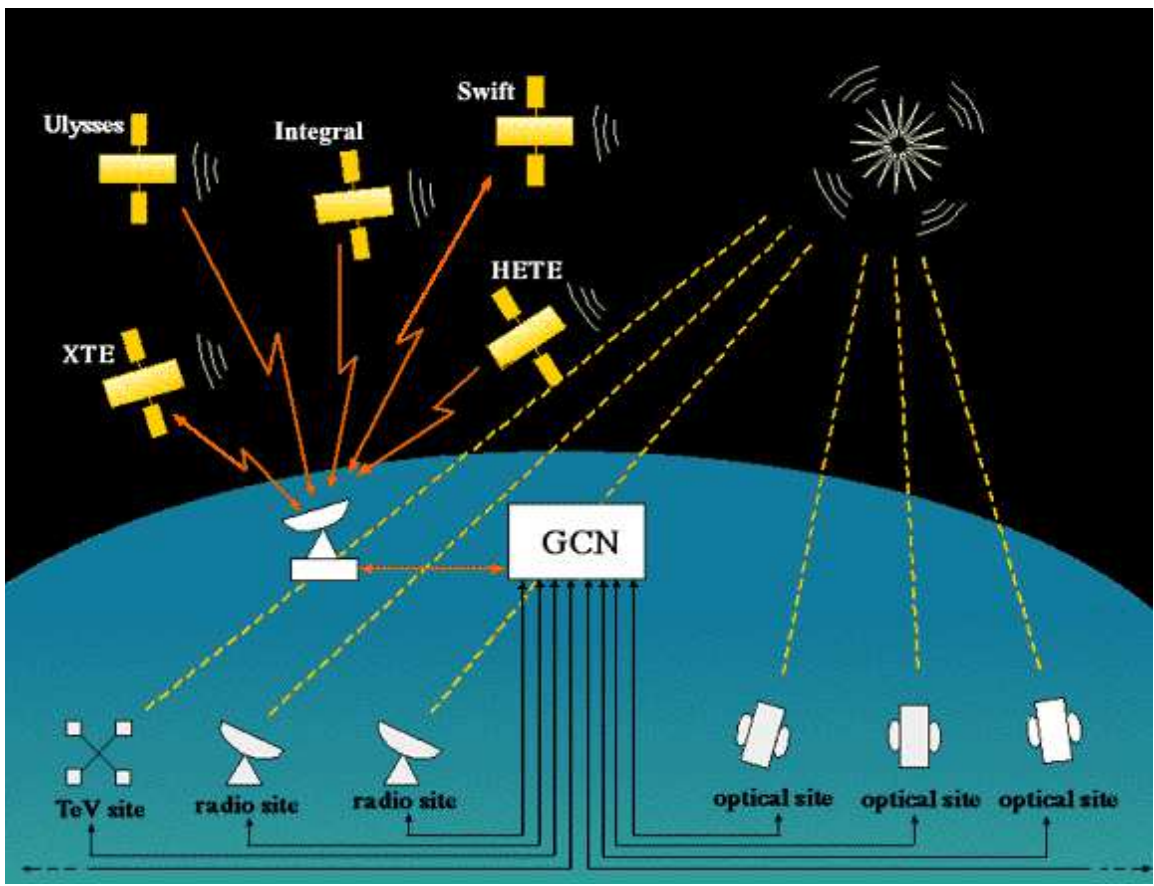}
    \fi
    \caption{The Gamma ray bursts Coordinates Network ( image from \cite{gcn_web_page} )}
    \label{fig_gcn}
  \end{center}
\end{figure}

Improvement of detection techniques allowed to measure the GRB position 
just after the gamma detection. In order to rapidly distribute alerts with 
burst position and time, the \textbf{G}amma ray bursts \textbf{C}oordinates \textbf{N}etwork ( GCN ) was developed
(Fig. \ref{fig_gcn}). The system works in such a way that in case GRB is detected
by any of the satellites and its position is determined, it is distributed among
the registered observatories. They point their telescopes to the burst
position and look for signal in other bands. Big immovable detectors of other messengers
like neutrinos or gravitational waves try to find signal correlated in time
and space with GRB, so far such correlation was not found.

New era of optical and X-ray GRB research was opened by the launch
of the SWIFT satellite on November 20, 2004 ( \cite{swift_paper} , \cite{swift_bat_paper2} ) 

\subsection{Optical counterparts in SWIFT era}
\begin{figure}[!htbp]
  \begin{center}
    \leavevmode
    \ifpdf
      \includegraphics[width=6in]{swift.gif}
    \else
      \includegraphics[width=6in]{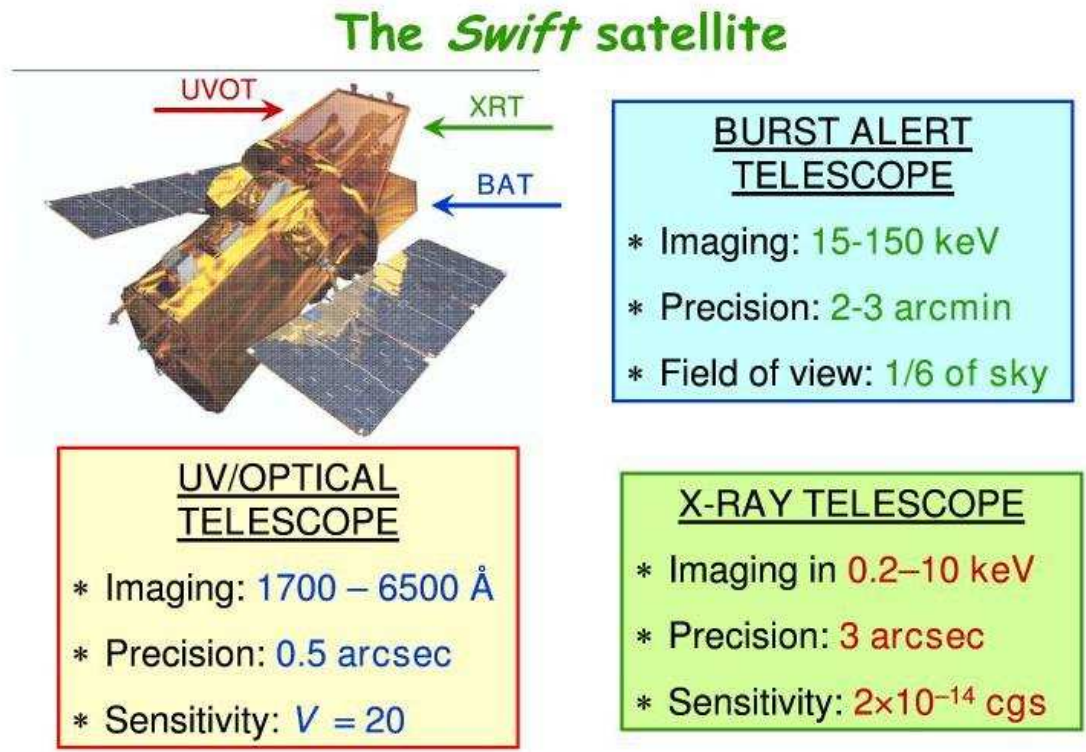}
    \fi
    \caption{Main scientific instruments on board the SWIFT satellite ( \cite{swift_www} )}
    \label{fig_swift}
  \end{center}
\end{figure}

The SWIFT satellite is dedicated to GRB research \cite{swift_www}. It has three scientific
instruments on board (Fig. \ref{fig_swift}), which are :\\

\begin{itemize}
\item Burst Alert Telescope ( BAT ) - gamma detector works in energy band 15-150 keV. 
This instrument detects GRBs and uses coded mask technique to determine their positions with
 the accuracy of $\approx$2-3 arcmin. It has large Field Of View (FOV) $\approx$2 steradians.
\item X-Ray Telescope ( XRT ) - X-ray detector works in energy band 0.2-10 keV.
After the BAT detects the GRB, XRT is able to determine the position of the X-ray counterpart with the 
precision of 3 arc sec. 
\item UV/Optical Telescope ( UVOT )- this is the optical telescope. After the position is determined by
the BAT detector, the satellite slews to the burst position and X-ray and optical
observations are performed. The limiting magnitude of the UVOT is 20$^m$
which allows to detect even very faint objects. The reaction time of UVOT is
limited by the slewing time and is on average 40-100 sec ( see Fig. \ref{fig_distr_mintime} ).
\end{itemize}

The SWIFT satellite allowed to access the "dark area" of GRB early optical and
X-ray data. It became possible to observe X-ray and optical counterparts seconds after the burst. 
Using XRT it became possible to discover previously not observed early X-ray flares. 
The fast position determination allowed UVOT, but also ground base
telescopes to observe optical signal only seconds after the gamma ray detection. 
It became clear that not all GRBs have bright optical counterpart. Since the
launch till this moment ( Oct 2007 ) SWIFT detected over 240 GRBs and for 
less then half of them optical counterpart was observed ( Tab. \ref{tab_swift_grb} on page \pageref{tab_swift_grb} ).
Early bright optical detections are listed in \mbox{the Table \ref{tab_swift_brightest_ot}} on page \pageref{tab_swift_brightest_ot}, it is clear that only small fraction of GRBs have 
very bright optical counterpart. 
Thanks to SWIFT ( and also HETE satellite ) it was also possible to observe the first
optical counterparts of the short GRBs which were previously undetected.
These counterparts were localized in the old elliptical galaxies which
strongly supports merger scenario expected to occur in old binary systems.
There is currently handful of long GRBs for which optical counterpart was observed 
during the GRB itself, the earliest and most bright are listed in Table \ref{tab_prompt_ot_grbs}. 

%
%
\begin{table}[htbp]
\begin{center}
\begin{tabular}{|c|c|c|c|c| }
\hline
\textbf{GRB} & \textbf{Telescope} & \textbf{\begin{minipage}{2.5cm}{Reaction Time [sec]}\end{minipage}} & \textbf{\begin{minipage}{2.7cm}{Max \hspace{0.8cm} Magnitude}\end{minipage}} & \textbf{Description} \\
\hline
990123 & ROTSE \cite{rotse_99} & 25 & 9 & \begin{minipage}{4cm}{Optical light curve uncorrelated with $\gamma$ light curve}\end{minipage}  \\ 
\hline
041219A & RAPTOR \cite{raptor_041219a} & 115 \footnotemark &  18.6  & \begin{minipage}{4cm}{Optical light curve correlated with $\gamma$ light curve, triggered by precursor}\end{minipage} \\
\hline
060111B & TAROT \cite{tarot_060111b} & 30 & 13.75  & - \\
\hline
061007A & ROTSE & 26.4 & 13.6 & - \\
\hline
060904B & TAROT & 23.1 & 15.8 & - \\
\hline
051111A & ROTSE & 26.9 & 13 & - \\
\hline
\end{tabular}
\caption{Observations of optical light curves of GRBs during the $\gamma$ emission}
\label{tab_prompt_ot_grbs}
\end{center}
\end{table}

\footnotetext{GRB 041219a was a very long GRB $T_{90}\approx$ 520 sec, $T_{90}$ is a time when 90\% of fluence was detected}

The light curves of these events have different properties. For
example the optical light curve of GRB990123 has no clear correlation with the 
gamma emission while light curve of GRB041219
is apparently correlated with the gamma emission ( Fig. \ref{fig_rotse_99} ).


This suggests that the optical emission mechanism can be different in these
two bursts. In the first case it is believed to be caused by 
back propagating reverse shock on the ISM ( external shock ). In the 
second case it is believed to be caused by the same
mechanism as $\gamma$-ray emission i.e. internal shocks. 
Moreover there are models suggesting that in some cases optical emission may
precede the $\gamma$-ray signal or can even be stronger \cite{paczynski_grb_ot}.
More early data is needed to resolve these doubts. Current
experiments have very limited chances to observe GRB in the optical band
during $\gamma$-ray emission, because time delay due to trigger propagation and pointing of the
telescope in best cases limits the reaction time to 30-40 seconds.
Another limitation of current experiments is the time resolution, only 
few ground base telescopes perform observations with a time resolution of
the order of seconds. An experiment observing full FoV of the satellite with
 a temporal resolution on the level of seconds would greatly contribute to resolving the GRB
puzzle \cite{paczynski_grb_ot}. The "Pi of the Sky" experiment is planned 
to realize exactly this type of strategy and observe GRBs in optical bands
during the $\gamma$ emission.

\subsubsection{The GLAST satellite}
\label{sec_glast_satellite}

The Gamma Ray Large Area Space Telescope ( GLAST ) mission is planned to be launched in the beginning of 2008. GLAST is
a next generation high-energy gamma-ray observatory designed for making
observations of celestial gamma-ray sources in the energy band extending
from 10 MeV to more than 100 GeV. It follows in the footsteps of the
CGRO-EGRET experiment, which was operational between 1991 and 1999.
The sensitivity of detectors have been greatly improved comparing to
EGRET detector on-board the CGRO satellite.
The main objective of the mission concerning the Gamma Ray Bursts is to determine
their behavior at high energy. The GLAST Burst Monitor (GBM) detector observing the 
FOV$\approx$10 steradians in energy range 5keV-30MeV will allow for fast identification of the burst and determination of its position
with accuracy of a few degrees within 1 second \cite{glast_gbm_www}. The Large Area Detector (LAT) allows to study
behavior of GRBs at higher energies in the wide energy range 20 MeV to 300 GeV, it has FOV$\approx$2.5 steradians.
It will also allow to determine the GRB position with the accuracy of $\approx$ 10 arcmin.
The strategy is very similar to the SWIFT's one, in case GRB is detected by GBM 
, but outside LAT's FOV, the GBM position is used to reorient the
spacecraft, so that the burst position is within the FOV of the LAT detector.
Bursts alerts will be sent to GCN network within 10 seconds  ( \cite{glast_stanford}, \cite{glast_www} ).
It is expected that the GLAST satellite will detect $\approx$ 200 bursts/year. 

\subsection{Orphan afterglows}
\label{sec_grb_orphaned_ag}

\begin{figure}[!htbp]
 \begin{center}
   \leavevmode
   \ifpdf
     \includegraphics[width=6in]{ox.gif}
   \else
     \includegraphics[width=6in]{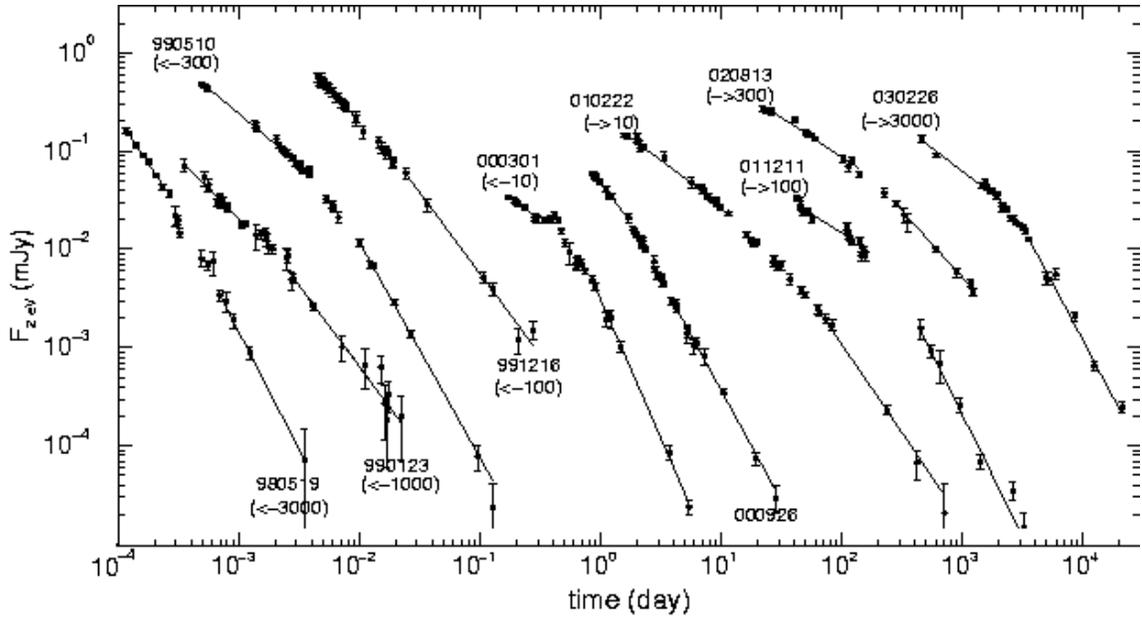}
   \fi
   \caption{Optical light-curves for 10 GRB afterglows with breaks and power-law fits to the pre- and post-break emission ( image taken from \cite{panaitescu_jet_breaks} ) }
   \label{fig_jet_break}
 \end{center}
\end{figure}

The jet structure of GRB explosions was originally proposed to resolve the
problem of huge energies ( $10^{54}$ erg ) resulting from cosmological
distances to GRBs. There is now extensive observational evidence for such
collimated emission from GRBs, provided by breaks in the optical/IR light
curves of their afterglows ( \cite{jetbreaks_1999_castro}, \cite{jetbreaks_1999_fruchter}, \cite{jetbreaks_1999_kulkarni} ).
Examples of light curves with observed "jet breaks" are shown in Fig. \ref{fig_jet_break}.
This effect is related to relativistic beaming effect, radiation isotropically emitted by 
relativistic matter becomes collimated.
Thus, only radiation emitted by the small part of the jet is visible from the Earth. When matter
reaches the Interstellar Medium ( ISM ) it is slowed down, $\Gamma$ factor
decreases and collimation becomes smaller, allowing observation of lower energy radiation emitted from other parts of the jet.
Finally matter is slowed down so much that collimation is of order 
of the jet opening angle $\theta_{jet}$. Since this moment observer 
 can detect radiation emitted from all parts of the cone ( \cite{oag_collim_totani} , \cite{oag_collim_huang} ).
This causes smaller energy fluxes observed from the Earth because
the radiation is no longer collimated and is emitted more isotropically.
Collimated structure of GRB explosions implies that only small fraction of
the total number of all GRBs in the Universe can be observed from the Earth.
Since radiation emitted in late times is observed in optical band, it is
expected that more optical afterglows should be observed then the GRBs.
Such kind of afterglows without GRB event are called "orphan afterglows".
They would appear as optical flashes without corresponding $\gamma$-ray detection. 
Observation of such events would allow for further tests of the GRB collimation
hypothesis ( \cite{rhoads_1997} , \cite{dalal_2002} ).
The typical orphan afterglow is rather related to late times, however the
similar situation may occur during prompt optical emission. This can be due
to early external shock ( GRB990123 ), reverse internal shocks or internal 
energetic structure of the jet (\cite{onaxis_ag_nakar_piran}). 
Such scenarios would lead to smaller collimation of optical
emission, which would lead to possibility of finding optical flashes related to
GRBs with similar timescale, but without corresponding $\gamma$ detection mbox{( \cite{onaxis_ag_nakar_piran} , \cite{oag_collim_huang})}.
Another possibility of finding optical flashes without corresponding
gamma detection is the failed GRB scenario ( \cite{oag_collim_huang} ). 
Observations of these kind of events would greatly improved understanding of
the GRB emission properties and geometry. The recent analysis give only
limitations on the number of orphan afterglow events and no confirmed event 
of this kind was observed ( \cite{klotz_orphans} , \cite{rau_orphans} ).
In order to observe such kind of events a wide field monitoring system is needed. 
One of the main results of this thesis is determination of limits on the
number of prompt orphan afterglow events on the whole celestial sphere and on the ratio of
optical and $\gamma$-ray emission collimation.

\begin{figure}[!htbp]
 \begin{center}
   \leavevmode
   \ifpdf
     \includegraphics[width=6in]{jet_break.gif}
   \else
     \includegraphics[width=6in]{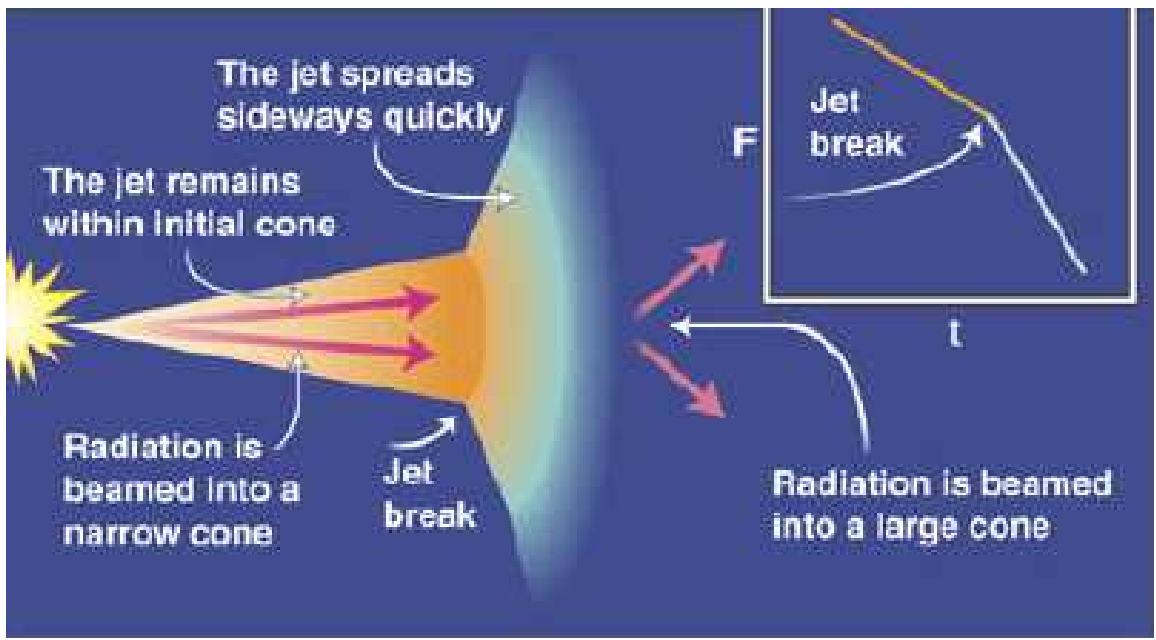}
   \fi
   \caption{The mechanism of the jet break caused by slow down of relativistic matter on the ISM (image from \cite{piran_jet_breaks_picture}). }
   \label{fig_jet_break_mechanism}
 \end{center}
\end{figure}

\subsection{Other fast astrophysical processes}
\label{sec_other_fast_proc}

Beside described in detail GRBs there is a variety of other astrophysical
processes manifesting themselves in time scales from seconds to days.
It is impossible to describe all of them in detail, however a short
description of those which can be observed with the "Pi of the
Sky " apparatus will be given : 

\begin{itemize}
\item[$\bullet$] \textbf{Nova explosions} - these processes occur in binary systems
with white dwarf accreting matter from a companion star. The matter
accumulates on the surface of the white dwarf and when certain critical
mass is reached a thermonuclear explosion occurs. This process
may repeat when enough matter is accreted again, these kind of novae are called \textit{recurrent novae}.
A very comprehensive introduction to subject of novae explosions is given in \cite{warner}.
Depending on the range of the telescope novae explosions signature is a sudden
increase of brightness of the system or appearance of a new object ( in case
normally it is below limiting magnitude of the telescope ).

\item[$\bullet$] \textbf{Variable stars} - about 1\% of all stars are variable stars. 
The time scale of brightness variations ranges from years and days down to milliseconds.
The variability mechanisms can be geometrical configuration ( e.g. eclipsing binaries ) or 
 internal properties of the star ( e.g. pulsations ).
A particular type of variable stars are flare stars ( also known as UV Ceti
variables ), which undergo sudden and unpredictable explosions related to release of magnetic energy. 
The mechanism is the same as in the case of Solar flares, however, flare 
stars explosions can be even thousand times brighter. The brightness increase
can be as large as 100-1000 times. The variability in the flare stars is
characterized as a rapid, irregular large-amplitude increase of brightness, followed by a much slower decay (
from minutes to hours ). 

\item[$\bullet$] \textbf{AGNs and particularly blasars} - these objects are 
active galactic nuclei which are powered by accretion of
matter on central super massive black hole. In many cases a jet of
relativistic matter is observed, which is most probably ejected in the direction of spin of the black hole.
If the jet is pointing towards the Earth, such AGN is called a blasar. 
AGNs manifest rapid ( time scales of days or even hours ) brightness variations in all wavelengths.
Monitoring of such objects and alerting about the increase of their activity is very
important for observations by big telescopes and can be realized by small wide field
telescopes.

\item[$\bullet$] \textbf{Supernovae} - these processes are related to death of
massive stars. When the thermonuclear reactions in the core are not able to 
balance gravitational pressure the star collapses and a huge explosion occurs ( $E_{SN} \approx 10^{51}$ erg ).
There are over 400 extragalactic supernovae discovered per year, which 
 means they are very dim and mostly below the range of the
telescopes like "Pi of the Sky". However, the brightest could also be
discovered by a system like this.

\end{itemize}

The above list is just an indication of what kind of processes are the aim of
analysis described in this thesis. In most cases the main goal of the analysis is  
a discovery of an object and sending alert to larger telescopes as the "Pi of the sky" system
is not able of performing spectroscopic measurements. However, in many cases 
good time resolution of the system gives a possibility of investigating 
brightness variations in time.



\newenvironment{mylisting}
{\begin{list}{}{\setlength{\leftmargin}{1em}}\item\scriptsize\bfseries}
{\end{list}}

\newenvironment{mytinylisting}
{\begin{list}{}{\setlength{\leftmargin}{1em}}\item\tiny\bfseries}
{\end{list}}

\chapter{The Pi of the Sky Experiment}
\ifpdf
    \graphicspath{{PiExperiment/figs/PNG/}{PiExperiment/figs/}{ThesisFigs/}}
\else
    \graphicspath{{PiExperiment/figs/EPS/}{PiExperiment/figs/}{ThesisFigs/}}
\fi


\section{General Idea}
In order to study rapidly varying astrophysical objects a telescope with time scale 
resolution at least of the order of seconds is needed. Exposures and dead time between images must be short.
This will allow to investigate light curve structure of rapidly varying objects.
However, this is not enough to study short and unpredictable processes like 
optical flashes related to GRB or any other kind of optical flashes.
In such cases it is impossible to predict where and when an event will
occur. Thus, it is not possible to point the telescope at certain star and 
wait for an event to occur. In order to be able to observe such class of
processes when they are going on a large field of view (FOV) is required.
If the telescope system will observe the whole sky continuously, then 
any outburst will already be in its FOV. There will be no delay, due to 
telescope movement and trigger information distribution. The area of interest will be 
observed continuously. The price for this is the range. The more sky 
a telescope observes the less sensitive it is to faint objects.

\begin{equation}
m_{MAX} = m_0 + 5 \cdot log\{ \frac{p \cdot N}{170.45 \cdot S \cdot FOV} \}
\label{eq_maglimit_vs_fov}
\end{equation}

Where $m_0$ is limiting magnitude of telescope with given detector and
aperture of 1cm, p is pixel size in $\mu$m, N is chip size in pixels, S is
light power and FOV is field of view in degrees.
For human eye detector $m_0=7.5^m$ and using FOV=21$^\circ$ $m_{MAX} \approx
11.8^m$ which is close to limiting magnitude of camera with Cannon f=85mm f/D=1.2.
However, for precise result $m_0$ should be determined from the specific CCD camera properties. 
It can be clearly seen that limiting magnitude decreases with field of view.
One needs to find a compromise between FOV of the single telescope and 
the maximum limiting magnitude it will be able to observe.
It is of course impossible to  build a single camera which would cover the 
whole celestial sphere at once with satisfying limiting magnitude.
The solution for this is to build a system of many telescopes, each covering
a fraction of the sky and pointing to different direction. 
The range of such a system will be limited by the range of a single telescope. 
In principle it would be possible to build a farm of large telescopes with 
high limiting magnitude, but the cost would be huge. One has to find 
another compromise between limiting magnitude ( number and size of
telescopes ) and the cost. 
The "Pi of the Sky" system was designed to cover significant fraction of the
sky with the satisfying limiting magnitude on the level of 14-15$^m$.
Data analysis presented in this thesis was performed on data collected by the prototype of the full system.
In the next section this prototype will be described in detail. The full detector
design will be presented in Section \ref{sec_full_pi}.

\section{The Prototype}
\label{sec_prototype}

\begin{figure}[!htbp]
  \begin{center}
    \leavevmode
    \ifpdf
      \includegraphics[width=6in,height=8.5in]{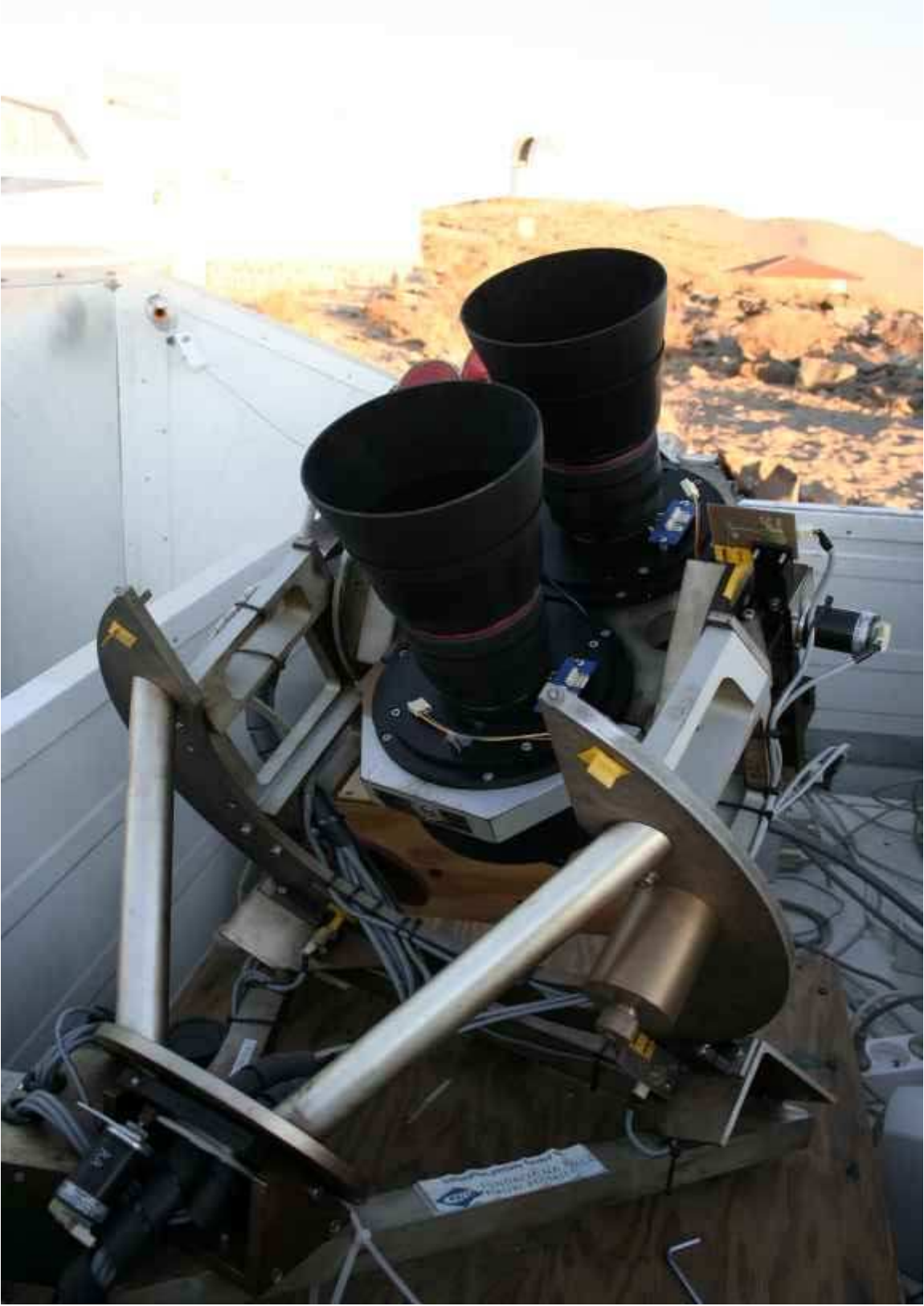}
    \else
      \includegraphics[width=6in,height=8.5in]{montaz559.eps}
    \fi
    \caption{The prototype in LCO, two cameras on a single mount in the ASAS dome}                                                                                
    \label{fig_prototype}
  \end{center}
\end{figure}

The prototype was build to test components of the final system including hardware and 
software solutions. One of the goals was to probe efficiency of optical flashes detection
and background rejection. The prototype was installed in June 2004 in the Las
Campanas Observatory (LCO) in Chile and it is working until now with a few months
break. The prototype after upgrade is shown in Figure \ref{fig_prototype}.
The system in LCO operates automatically and is fully controllable via the
Internet. There is no person in LCO who takes care of the system. 
It is very important requirement for both the prototype and the full system
that it must be remotely controlled and failure proof. 
The current setup of the prototype allows even some of the hardware failures 
to be handled remotely and continue operation of the detector.
The maintenance trips to LCO are expensive so the need for them must be
minimized. Current experience says that one maintenance trip a year should be
enough.

\subsection{Hardware}

The detector consists of two cameras on a single paralactic mount ( Fig. \ref{fig_prototype} ). The mount was adopted from the ASAS experiment. It is 
driven in two axis by step motors. The TMCM 300 microcontroller
driver by Trinamic was used to control mount from the PC \cite{trinamic}. The microcontroller box is
connected to the PC with the RS-232 interface. 
On the PC side the mount driver program controls settings and movements of the mount.
The position of the mount can be calculated from number of steps executed by
the step motors. The resolution of step motors is $\Delta \lambda \approx$ 3.5'' and $\Delta \delta
\approx$112.5'', where ($\lambda$,$\delta$) are equatorial coordinates right ascension and declination.
As a cross-check potentiometers were added, they give rough estimate of the
position with the accuracy of $\Delta \alpha \approx 0.2^\circ$ , this allows
to detect step motors position errors ( for example due to slip of belt drive ).
During observations mount performs tracking which is a rotation around the Earth axis compensating 
for the Earth daily rotation. \\
The cameras are custom designed. The development of own CCD cameras was 
motivated by several reasons. The first factor is the price of commercial
products available on the market. However, there are other reasons which
limit possibility of using commercial products in such kind of remotely
controlled experiment :

\begin{itemize}
\item reliability of internal mechanical shutter
\item remote control of camera settings which include lens heating and lens focus adjustments
\item temperature and humidity sensors
\item fast readout ( in most of the solutions limited to 1MHz )
\item possibility of changing firmware of the camera remotely
\end{itemize}

The cameras are based on Fairchild CCD 442A chip \cite{ccdchip} with
resolution 2048x2048 pixels ( 15 $\mu$ x15 $\mu$ each ). The CCD sensor is
placed in a separate chamber filled with a noble gas. The 16bit analog
digital converter (AD9826) has been used. Communication with the PC is realized by USB2.0
interface by the Cypress FX2 USB CYC68013 chip. Camera is controlled by the FPGA 
Altera chip. CCD chips are cooled with the thermoelectric Peltier junction,
down to 30$^\circ$ C  below ambient temperature. 
The following features have been implemented in cameras :

\begin{itemize}
\item Remote control of almost all functions : readout frequency, gain, CCD
 temperature, mechanical shutter, lens focus regulation, lens heating ( against water condensation )
\item Remote monitoring of atmospheric conditions ( temperature and humidity sensors )
\item Possibility of remote firmware upgrade ( Cypress microcontroller program and FPGA configuration )
\item Watchdog Timer which resets camera in case communication with PC is lost
\end{itemize}

The table below summarizes the parameters of the cameras  : \\

\begin{table}[htbp]
\begin{center}
\begin{tabular}{|c|c|}
\hline
\textbf{Parameter} & \textbf{Value} \\
\hline 
Readout Time & 1s - 1min \\
Readout noise & <16$e^-$ at 2MHz and <12$e^-$ at 1MHz \\
Shutter Durability & >$10^7$ cycles \\
USB2.0 max transfer speed & 52MB/s \\
Maximal cooling & 30$^\circ$ below ambient temperature \\
\hline
\end{tabular}
\caption{CCD cameras parameters}
\label{tab_ccdcam_params}
\end{center}
\end{table}

\begin{figure}[!htbp]
  \begin{center}
    \leavevmode
    \ifpdf
      \includegraphics[width=4in]{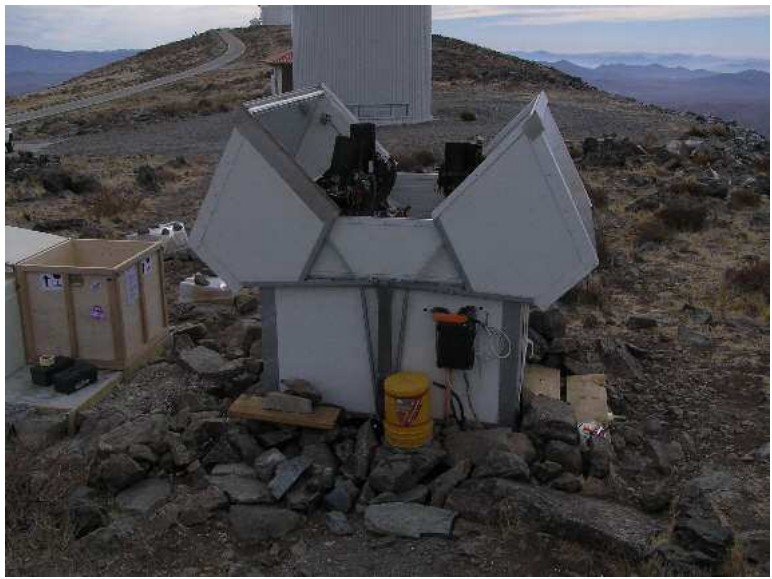}
    \else
      \includegraphics[width=4in]{dome.eps}
    \fi
    \caption{Dome of the ASAS experiment in LCO}
    \label{fig_asas_dome}
  \end{center}
\end{figure}

\begin{figure}[!htbp]
  \begin{center}
    \leavevmode
    \ifpdf
      \includegraphics[width=4in]{pi2_in_dome.gif}
    \else
      \includegraphics[width=4in]{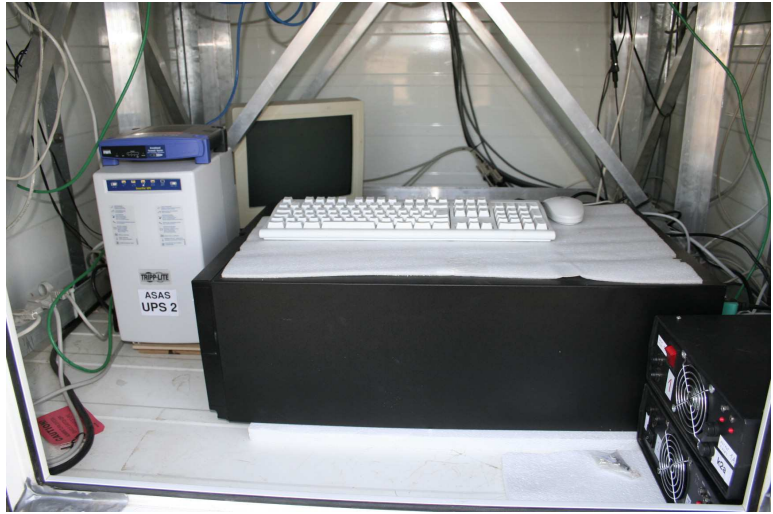}
    \fi
    \caption{Pi2 computer controlling nightly data acquisition is located in the lower part of the ASAS dome}
    \label{fig_foto_pi2}
  \end{center}
\end{figure}

More details on design of the cameras can be found in \cite{gk_k2a}.
The cameras are equipped with CANON EF f=85, f/d=1.2 lenses. 
Short time exposures (5-10 sec) imply many ($\approx 2000-4000$) exposures
during a single night. This results in $\approx 10^6$ images per year.
Huge number of collected images required special design of the shutter which can 
survive more then million cycles ( commercial shutters usually work up to $10^5$ cycles ). 
In order to save shutter it is possible to make exposures with shutter permanently opened.
The mount with cameras is installed in the ASAS \cite{ASAS} dome in the LCO ( Fig. \ref{fig_asas_dome} ). 
The dome is controlled by the OGLE telescope system (\cite{OGLE}). It opens when the OGLE
telescope dome opens, which depends on observer's decision motivated by the
weather conditions.
The mount and the cameras are controlled by the single PC computer which is
placed in the lower part of the dome (Fig. \ref{fig_foto_pi2}).
Custom driver was developed to control parameters of the
cameras and collect images, it will be described in the next section. 
The system contains two other PC computers which are used for off-line data analysis. 

\subsection{Software}
\subsubsection{Overview of the system components}

The main "Pi of the Sky" software components are running under Linux
(Fedora) operating system. Most of the software used in the experiment was custom developed. Some of
the procedures and formats were adopted from the ASAS experiment \cite{ASAS}.
Generally software can be divided into on-line part, which takes care of 
detector control during nightly data acquisition and off-line part which performs off-line
data analysis.

\begin{figure}[!htbp]
  \begin{center}
    \leavevmode
    \ifpdf
      \includegraphics[width=6in]{nightcontrol.gif}
    \else
      \includegraphics[width=6in]{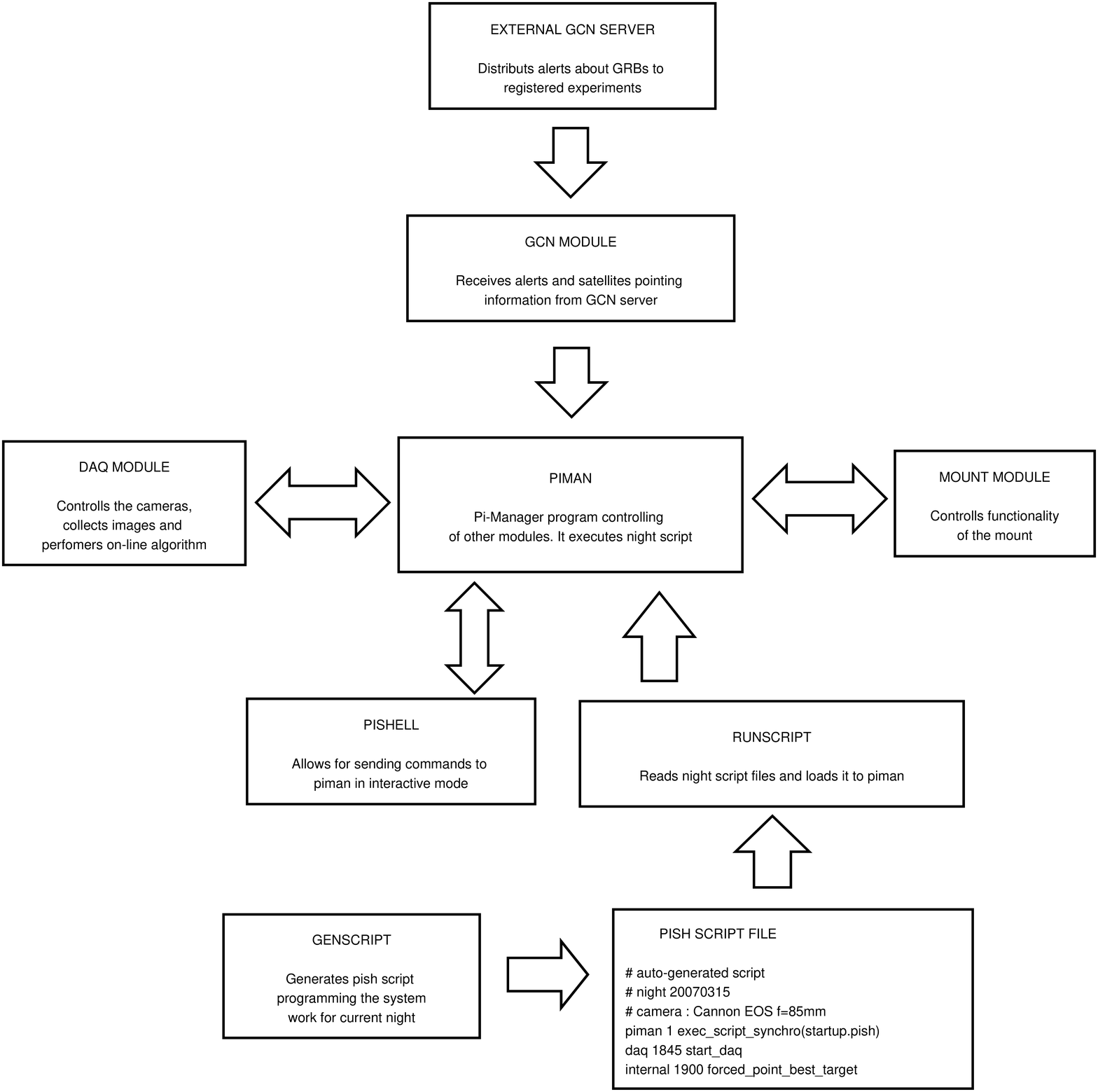}
    \fi
    \caption{General architecture of system controlling the detector during nightly data acquisition implemented in the prototype}
    \label{fig_prot_sys_arch}
  \end{center}
\end{figure}

General architecture of the night control system is presented in Figure \ref{fig_prot_sys_arch}. 
The system consists of several modules which communicate with each other by the CORBA interface \cite{CORBA}.
They were designed in the client server architecture.
The main components of the system are :\\
\begin{itemize}
\item \textbf{PIMAN} \\
	It is the main manager program which controls the whole system. 
	This program provides set of commands which can be executed by different modules. There
	are also complex commands, which use several different modules.
	In this case result of command executed by one module is used as an input to command for another module. 
	Commands can be executed manually from \texttt{pishell} program or loaded in form
   of the \texttt{pish} script by \texttt{runscript} program. Example of the night script is given in Figure \ref{lis_pish_old}.
	It is a set of commands with times at which they should be executed.
	Script is generated every evening by the \texttt{genscript} program. 
	After system is started the \texttt{runscript} program is executed to read the night script and sent it to
	the \texttt{piman} program. The \texttt{piman} program  executes commands at specified times. 
   In case GCN alert is received by the GCN module the alert information is passed to \texttt{piman} and handled ( See. \ref{sec_obsstrategy} ).
\item \textbf{PISHELL} \\
	This \texttt{pishell} program is a client program which has interactive command
	line interface. User can launch this program to sent commands to \texttt{piman} in
	interactive mode. It allows modification of night scheduled program
	loaded from the \texttt{pish} script. All necessary actions like checking stack of
	commands, deleting commands and adding new commands can be executed 
	by using this program.
\item \textbf{RUNSCRIPT} \\
	This is a simple client program for reading given \texttt{pish} script and sending 
	it to \texttt{piman} module.
\item \textbf{DAQ} \\
	This program is responsible for controlling cameras settings, collecting images
   and on-line data analysis. On-line data analysis consists of two
	main actions : astrometry and flash recognition algorithm.
	In general astrometry is a transformation of chip (x,y) to celestial ($\lambda$,$\delta$) coordinates. 	
	This transformation allows to calculate coordinates of image center. They
	can be compared with the expected coordinates and mount movement
	corrections can be determined.
   On-line flash recognition algorithm looks for optical flashes in single 10 sec exposures.
	More detailed description of the \texttt{DAQ} module will be given in the next section.		
\item \textbf{MOUNT} \\
	This module is responsible for controlling the mount. All functionality of 
   the mount hardware can be controlled by this program. This includes
	calibration, moves to 
   the desired position, enabling/disabling of tracking, changing the tracking speeds and 
   calibration of the pointing using exact information from astrometry ( \texttt{DAQ} module ) or 
   encoders ( potentiometers in the prototype ). 
	The core of this module is compiled in form of library \texttt{libmount.so}. There 
	are two programs using this library. Program \texttt{monit} allows to control the 
	mount in the interactive mode, while \texttt{mount\_server} is a server which 
	provides a set of functions which can be executed by \texttt{piman} when
	performing commands from a night script.
\item \textbf{GCN}
	This module is waiting for GCN alert messages sent through software socket\footnote{socket - term for type of interprocess communication} from the remote GCN server (Fig. \ref{fig_gcn}).
   During the night every package obtained from the GCN server is passed to \texttt{piman} module
   which decides weather it is an interesting event which is possible to be 
   observed by the "Pi of the Sky" system.
	In such case alert procedure is executed (Sec. \ref{sec_obsstrategy}).
	The \texttt{gcn}  module is also responsible for writing satellite pointing
	information to log files ( \texttt{swift\_pointdir.log} and \texttt{integral\_pointdir.log} ).
	This information is used by pointing procedure to follow FOV of the GRB
	detecting satellites (Sec. \ref{sec_obsstrategy}).
\item \textbf{GENSCRIPT}
	This program is responsible for preparing a plan of night observations. It is 
	generated in form of \texttt{pish} script (Appendix \ref{lis_pish_old}) which is loaded to the \texttt{piman} program memory
	and executed during the night. The syntax of the single command is the following :\\
		\hspace{2cm}\textbf{      MODULE TIME COMMAND( command parameters )} \\
   The module name \texttt{internal} is used for complex commands ( see above ).
	More detailed description of the script generator module will be given in 
	separate subsection \ref{sec_obsstrategy}.
\end{itemize} 

Every module writes its current status to the specified status file. They
also produce separate log files and most important information is saved 
to global \piname system log file (Tab. \ref{tab_log_files}).
The relations between "Pi" modules and external systems are
presented in Figure \ref{fig_prot_sys_arch}. Communication between modules is 
client-server architecture and is implemented in CORBA technology \cite{CORBA} 
which is efficient and reliable method of object oriented interprocess communication. 
Most of modules were written in C$++$ and C.

\subsubsection{Data Acquisition System}
The data acquisition system (\texttt{DAQ}) is a program which is responsible for
collecting images from the cameras and for on-line data analysis.
The program uses several custom developed libraries which provide
different functionality. The most important libraries are listed in Table
\ref{tab_daqlibs} in Appendix \ref{sec_appendix1}.

\begin{figure}[!htbp]
  \begin{center}
    \leavevmode
    \ifpdf
      \includegraphics[width=6in,height=8in]{daq.gif}
    \else
      \includegraphics[width=6in,height=8in]{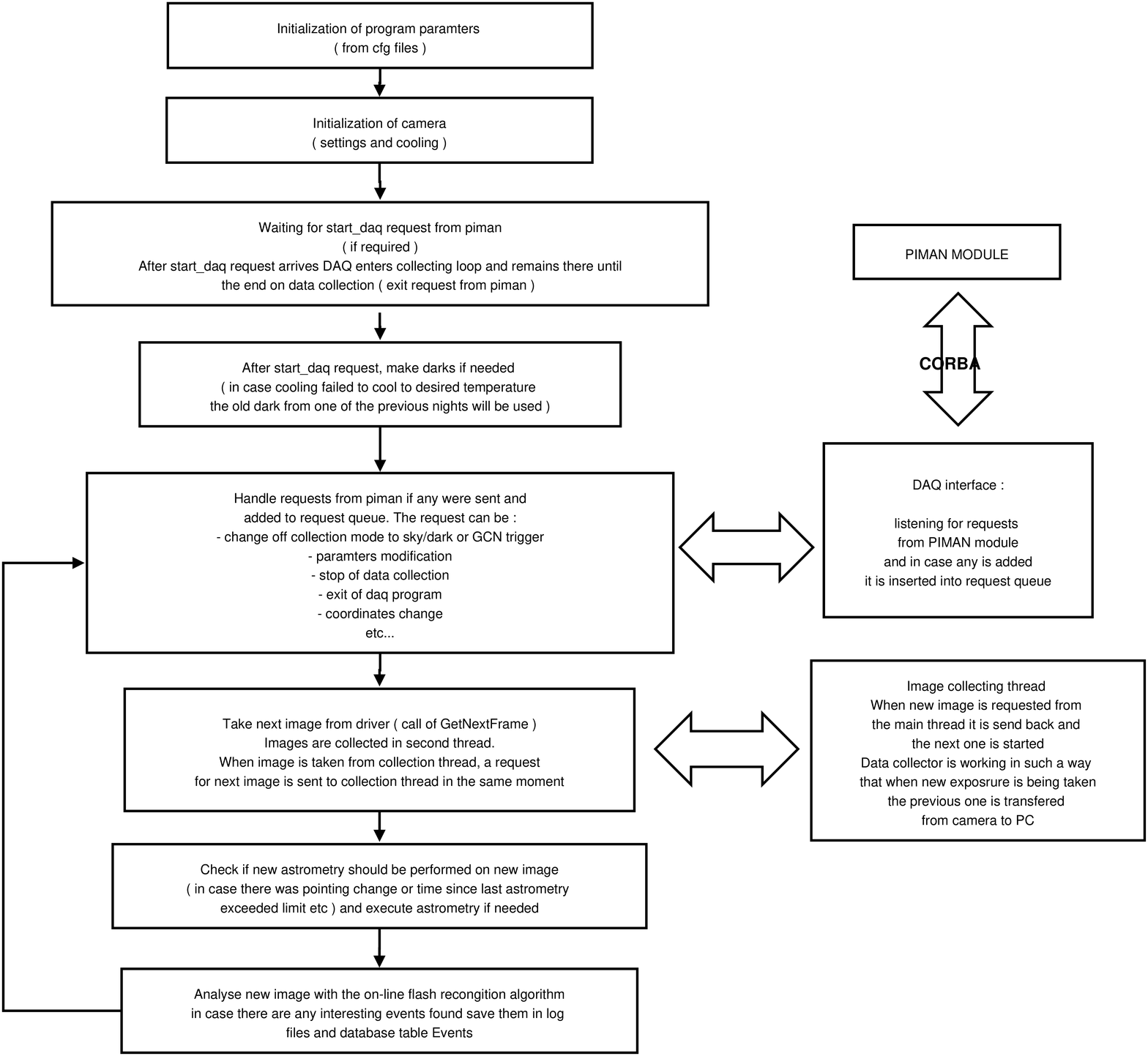}
    \fi
    \caption{Block diagram of the Data AcQuisition program}
    \label{daq_diagram}
  \end{center}
\end{figure}

This libraries are linked by several programs, the most important programs for
data collection and analysis are listed in Table \ref{tab_daqprograms}.
Programs which are used every night for data acquisition are
\texttt{ccdsingle} and \texttt{ccddouble}. The first one is used when \texttt{DAQ} uses single camera to collect data.
The program \texttt{ccddouble} is used to collect images from 2 cameras on the same mount working
in coincidence. The main parts of the collection program are shown in Figure \ref{daq_diagram}. 
The only difference in \texttt{ccdsingle} program is 
that images are collected from single camera and the algorithm for flash
recognition is different.\\

The most important task of these programs is to collect images from the
cameras and save them to a disk. 
There are also actions important for other modules in the system, the most 
important one is astrometry (Sec. \ref{sec_image_reduction}).
Generally astrometry is a procedure for transforming (x,y) coordinates of objects on
CCD chip to equatorial coordinates ($\lambda$,$\delta$). More details on
astrometry procedure can be found in Section \ref{sec_image_reduction}.
Astrometry is very important for finding real coordinates of image center
which can be slightly different then values calculated from mount step motors.
The position of the image center found by astrometry can be verified against
position calculated by the \texttt{mount} module. In case they differ, position
determined from astrometry can be used to correct mount position and 
also mount tracking speeds in both axis.
The third main task of \texttt{DAQ} program is to analyze images in search for
optical flashes (Sec. \ref{sec_online_flashes_algo}). Interesting flashes found by the
algorithm are saved on disk and are almost immediately published on the WWW
page in order to be reviewed by human.
\texttt{DAQ} program exports several functions and acts as CORBA server (Appendix \ref{sec_daq_controll_commands}). 
This functions are executed by \texttt{piman} in order to control the whole system.
Information from \texttt{DAQ} program is stored in night log file ( see Appendix \ref{app_log_files} )
Most important information is stored in the database allowing for fast
and easy access for reporting purposes (Sec. \ref{online_db_structure}).

\subsubsection{System nightly performance}
\label{system_nightly_performance}

The system is started every evening from \texttt{crontab}\footnote{crontab is 
a system tool for starting programs at specified dates and times}  by starting script \texttt{run\_pisys!}.
This script waits until the dome is opened and then it launches
all system modules like \texttt{piman}, \texttt{DAQ}, \texttt{mount}, \texttt{gcn} etc. After all modules are started the night 
script is loaded to piman's memory and subsequent commands are executed 
at specified times. 
Since this moment the system can be controlled from the \texttt{pishell}. After \texttt{DAQ} is
started communication with cameras is initialized and program waits until
cameras are cooled to temperature specified in the configuration file.
After cameras reach the desired temperature, program waits for command
\texttt{start\_daq}. This command is executed when dark images can be collected. 
Dark images are images collected in the same conditions as the sky images but
with closed shutter. This allows to obtain average values of noise and dark
current in every pixel. Before analysis of a the sky image the dark image is subtracted from it.
Typically 20 dark images are collected and median image is
calculated to be used as a dark image in further analysis.
After the dark image is ready, \texttt{DAQ} program is ready to collect sky images and
waits for a command to start analysis. The \texttt{piman} module chooses 
the best object to be observed by launching \texttt{point\_best\_target} command
and sends request to \texttt{mount} module to move to the position of this target.
The \texttt{start\_analysis} command passes sky position
obtained from \texttt{mount} module to \texttt{DAQ}, it is required to perform astrometry.
Before any change of the position the \texttt{DAQ} stops collection of images, it is
restarted after next \texttt{start\_analysis} command is obtained from \texttt{piman} after the desired position if reached.
In specified time intervals \texttt{piman} asks \texttt{DAQ} for current position
($\lambda,\delta$) of image center resulting from astrometry and sends this
information to mount in order to correct tracking speeds in both axis.
Twice a night the whole sky scan is performed. During the scan \texttt{DAQ} takes images
in single image mode without performing astrometry in order to complete the scan
as fast as possible. Astrometry for scan images is performed off-line.

\subsubsection{Camera Driver}
\label{sec_cam_driver}

\begin{figure}[!htbp]
  \begin{center}
    \leavevmode
    \ifpdf
      \includegraphics[width=6in]{driver_classes.gif}
    \else
      \includegraphics[width=6in]{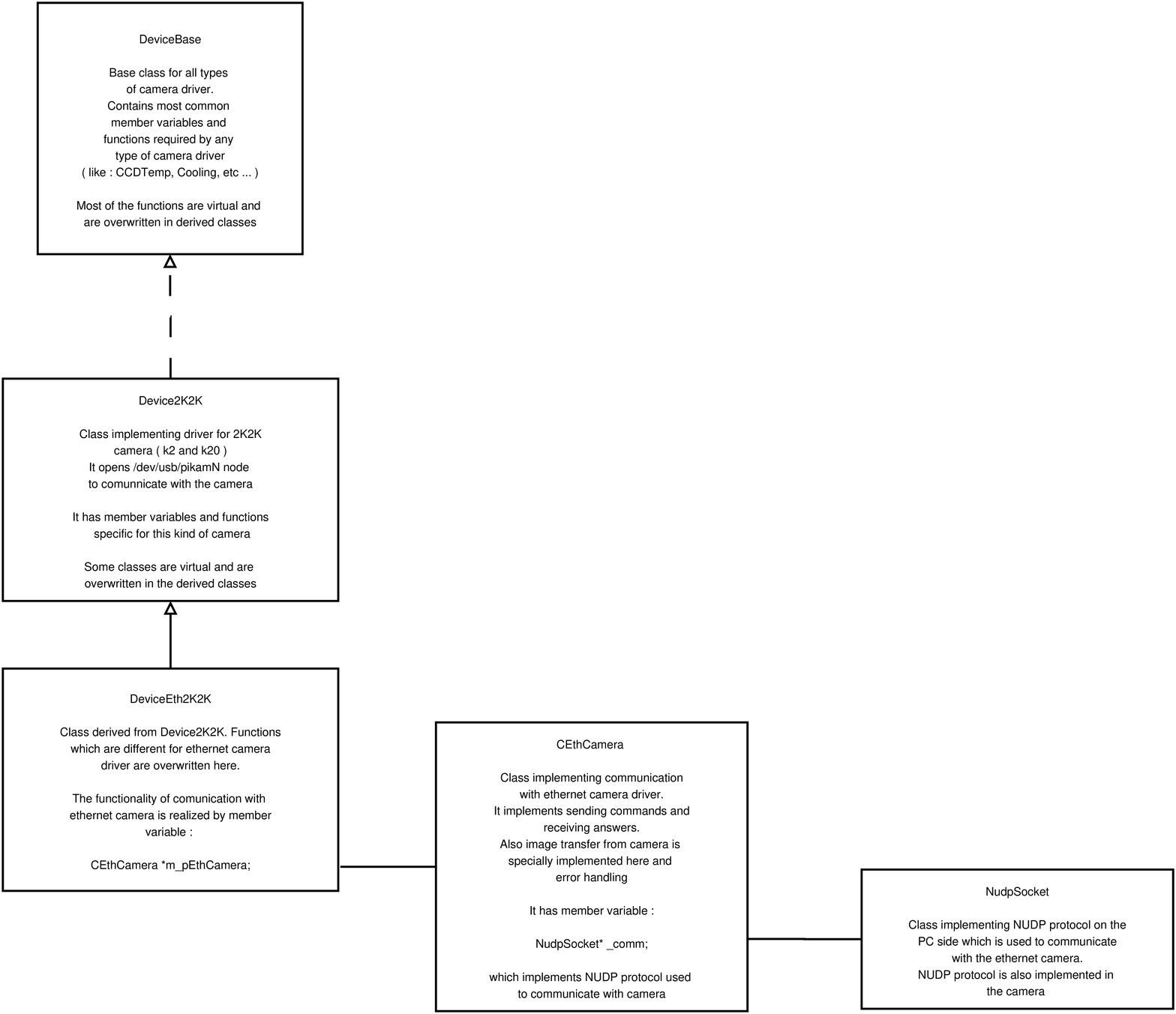}
    \fi
    \caption{Dependencies of camera driver C$++$ classes}
    \label{fig_driver_classes}
  \end{center}
\end{figure}

The CCD cameras used in the project are custom designed. Due to this fact it
was necessary to develop the software to control them. The camera driver was developed in C/C$++$.
The original cameras were equipped with USB2.0 interface only. The driver
for USB2.0 cameras consists of two parts. The first part is a kernel module pikam.ko
which can be compiled under Linux kernel $\geq$ 2.6.5. The second part is the C$++$
class \texttt{Device2K2K} which opens the device file ( /dev/usb/pikamN ) in order to communicate with the camera.
The final design of the camera was enriched in gigabyte Ethernet interface.
The Ethernet camera exports the same set of functionality, so only the low level
communication part of the driver has to be optionally replaced.
This required development of communication protocol NUDP \cite{ju_wilga2006}, it is implemented
in camera firmware and in PC driver. Dedicated kernel module is not needed
in this case. The C$++$ class \texttt{CEthCamera} using external library Sockets
\cite{ext_socktes} implements communication with camera via Ethernet
interface. The class \texttt{DeviceEth2K2K} derived from \texttt{Device2K2K}
overwrites several low level communication functions using member class \texttt{CEthCamera} ( see Fig. \ref{fig_driver_classes}).
New Ethernet camera gives a possibility of using USB2.0 or Ethernet interface,
depending on current user requirements. Change of communication can be easily done 
by editing configuration files or by command line option to the data acquisition program ( see Table \ref{tab_daqprograms} in Appendix \ref{sec_appendix1} ).
The driver is compiled in form of library \texttt{libpimandrv.a}.

Images are saved to \texttt{fits} files (\cite{fits_format},\cite{fits_format2}). 
Writing of \texttt{fits} files is realized by library \texttt{libmyfitslib.a}, but
this is the camera driver library which is responsible for preparing camera settings information to be saved in
\texttt{fits} file header. The keywords related to camera settings are listed in
Table \ref{tab_cam_fits_header} in Appendix \ref{sec_appendix1}. Most of this
information is also written to the database. 
The driver library is linked by \texttt{DAQ} programs (\texttt{ccdsingle} and \texttt{ccddouble}). The best way to easily test 
cameras is to use program \texttt{test2K2K} (Tab. \ref{tab_daqprograms}).
It is a simple text interface interactive program which allows to change settings of
the camera, to take image and to realize any kind of camera functionality. It can be
used to operate camera by the USB2.0 interface or by the Ethernet depending on options
passed in the command line. It is also possible to use this program in a batch
mode where actions to be performed are passed in the command line.

\subsubsection{Database for on-line data}
\label{online_db_structure}

\begin{figure}[!htbp]
  \begin{center}
    \leavevmode
    \ifpdf
      \includegraphics[width=6in,height=8.5in]{on_line_tables.gif}
    \else
      \includegraphics[width=6in,height=8.5in]{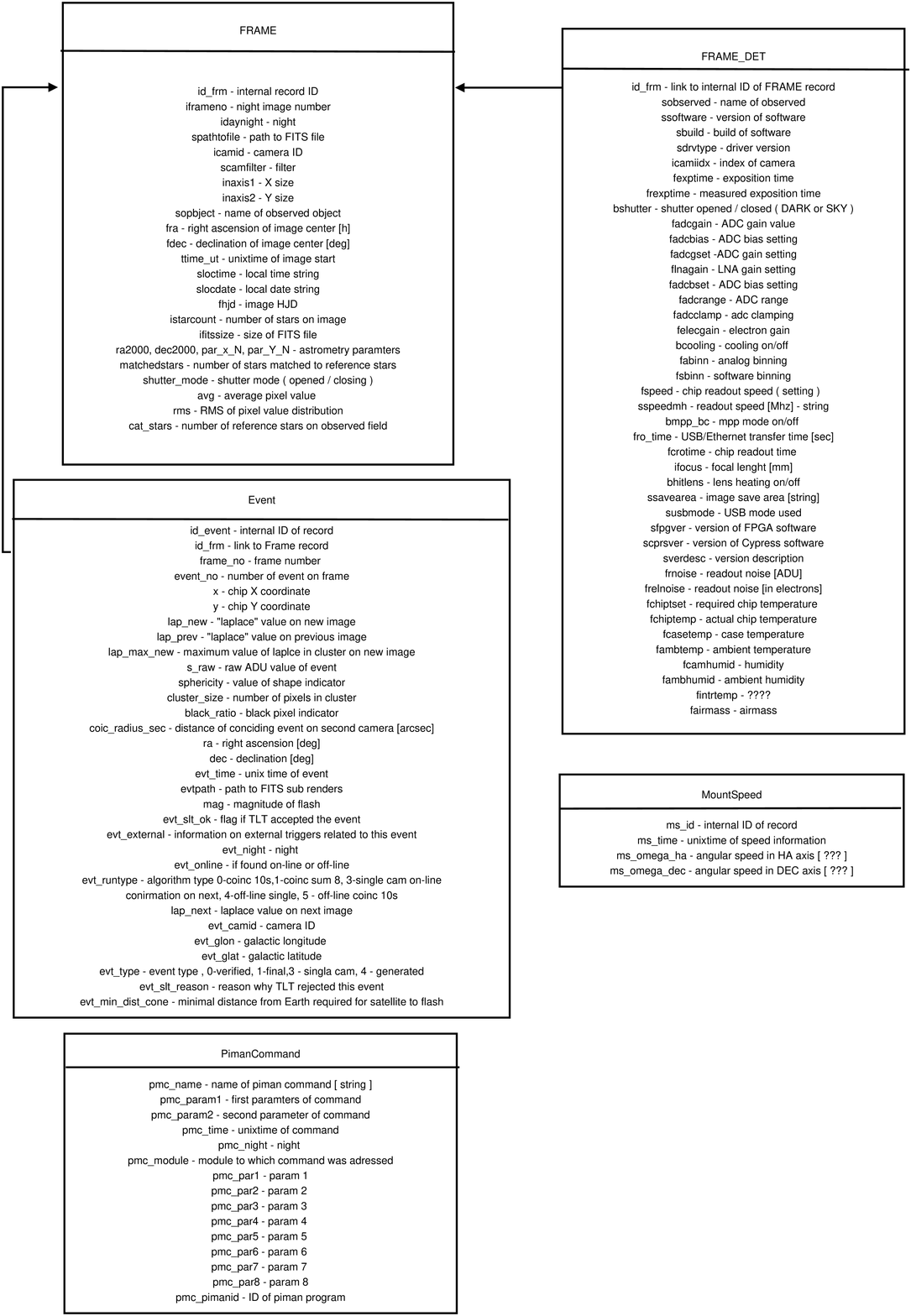}
    \fi
    \caption{Structure of database for storing on-line information from the system}
    \label{fig_online_db}
  \end{center}
\end{figure}

In order to store the most important information about system on-line performance
a database structure was designed and developed. The PostgreSQL database system is used for this purpose \cite{POSTGRES}.
The structure of the database used for storing on-line system information is shown in Figure \ref{fig_online_db}.
\texttt{DAQ} program stores on-line information in 3 tables : \\
\begin{itemize}
\item \textbf{FRAME} and \textbf{FRAME\_DET} tables contain information about every image
collected by the system. Generally all information from \texttt{fits} header is saved
to the database, some additional information is also added.
\item \textbf{EVENT} stores information on interesting events detected by the on-line algorithm
\end{itemize}

The other two tables are used for storing information from \texttt{piman} and \texttt{mount} modules :
\begin{itemize}
\item \textbf{PIMANCOMMAND} stores information about commands  executed by the \texttt{piman} program
\item \textbf{MOUNTSPEED} stores information about mount tracking speeds
\end{itemize}
Second part of \piname database is star catalog and will be described in
details in Section \ref{sec_star_catalog}.

\subsubsection{DAQ configuration}
\label{sec_daq_cfg}

The \texttt{DAQ} can be configured by means of configuration files. Every parameter has a default
value which is hard coded in the program code. In most cases this value is proper for typical
system settings. However, the default value can be overwritten by values 
read from configuration files. 
The priority of loading configuration files is the following ( starting from
the highest ) :

\begin{itemize}
\item Some parameters may be overwritten by the options passed to the
program from the command line
\item Program looks for file \texttt{ccd.cfg} in the  directory where it was started ( current directory ).
In case it is found the settings from this configuration file are loaded.
\item In case local \texttt{ccd.cfg} file is not found, program looks for global
configuration file \texttt{\$NDIR/cfg/ccd.cfg} and if found loads settings from this file
\end{itemize}

In case non of the \texttt{ccd.cfg} files is found the default values defined in
program code are used. They may not be optimal for current system
configuration so specific \texttt{ccd.cfg} file should be provided and
required parameter values should be placed there.
It is possible to change values of parameters during the data collection 
by executing command \texttt{change\_param} from \texttt{pishell} program ( Tab. \ref{tab_daq_exp_functions} )

\subsubsection{Observation Strategy}
\label{sec_obsstrategy}
The experiment is mainly devoted to optical flashes related to GRB, thus 
observation planing depends on pointing direction of GRB detecting satellites.
The final version of the detector covering 2 steradians will cover FOV of 
the SWIFT BAT or GLAST LAT detector and pointing of cameras will depend 
on pointing direction of these satellites. In the case of the prototype, FOV is 
much smaller, but observation strategy is very similar. The system tries 
to point the cameras to the center of FOV of one of the satellites which are
capable of determining the burst position.
Before June 2006 the primary targets were HETE and INTEGRAL, after this date 
the highest priority target is SWIFT satellite. 
The observation plan is generated automatically in form of \texttt{pish} script (Appendix \ref{lis_pish_old}). 
Currently the telescope pointing is performed dynamically. Pointing information is obtained from GCN server through the software socket during the
system operation and it is used to point the telescope.
The \texttt{piman} program executes pointing command every half an hour. It calls procedure 
which finds the best target to be observed. The best target is chosen from
list of targets sorted in order of priority, every target is checked and 
in case it satisfies several constraints :

\begin{itemize}
\item h > 28$^\circ$
\item distance from the Moon > 30$^\circ$
\end{itemize}

it is chosen as the best target. The primary target is SWIFT, secondary target is INTEGRAL
 and next targets are objects from the list of interesting astrophysical objects (Tab. \ref{tab_pointing_objects}) like
blasars, AGNs etc. The list was compiled mainly from list of interesting objects
used by Global Telescope Network (GTN) \cite{gtn_web_page}.
In case none of the targets on the list can be observed as the reserve target the field on the East
(closest to (az,h)=(90$^\circ$,28$^\circ$) position) is chosen to ensure the longest observation time.
Always when new position of SWIFT is sent via software socket the \texttt{piman} program 
executes re-pointing procedure immediately after finishing the current exposure.
In order to optimize photometry, the telescope does not observe arbitrary positions. Instead
it finds closest field from the predefined list and telescope points to the center of this field.
In this way each star is usually measured almost in the same position of the CCD chip.
The above pointing procedure is performed during most of the night. Twice a night 
the whole sky scan is performed by the system to cover all sky, which takes about $2 \times 1$ hour.
The normal observation strategy described above can be interrupted on
receiving the GCN alert about GRB with known position. When GCN module 
receives this kind of alert it sends it to \texttt{piman} module. In case
the event can be observed ( it is above the horizon ) the normal program is
postponed and system points the cameras towards the burst position. After half an
hour systems returns to the normal observation plan.

\subsubsection{Remote system control}
The "Pi of the Sky" prototype is installed in LCO and full detector will also
be installed in a remote location. This imposes specific requirements for the
system. The most important one is failure-free hardware. Another obvious 
requirement is that the system must be controllable via the Internet.
Most of the system functionality can be controlled via the ssh protocol
by logging to remote host and executing programs. 
However, nightly operation control does not require logging to remote host, 
most important information about the system performance is
copied to local machines and is available by the WWW interface. This 
information is basically : 
\begin{itemize}
\item Database records from on-line tables containing information about
images and results from the on-line algorithm ( Sec. \ref{online_db_structure} )
\item Status files of all system components 
\item Log files of most crucial system components
\item Parts of images containing events found by the on-line algorithm
\item Some of collected sky images ( converted to jpg format )
\end{itemize}
Under normal conditions the system does not require human attention, all jobs
to be performed are started automatically from \texttt{crontab}.
At given time night observation script is generated, configuration files and
data folders are prepared, system is started and it runs until morning.
The person on duty does not have to watch the system for the whole night.
In order to warn about system problems special alert system was designed.
In case status file from any module is not updated for long time or contains
information about problems, e-mail or SMS is sent to person on duty.
In such case probably human reaction will be necessary and will require
logging to system controlling computer.
The action would depend on the type of the problem, sometimes it is
enough to execute \texttt{piman} command from \texttt{pishell} or restart one of the modules, while 
in case of mount server problems calibration procedure may be required.

\section{Full Pi of the Sky detector}
\label{sec_full_pi}

\subsection{General Idea}
The full version of the detector is currently under construction. 
The final system will consist of 2 sets of 16 cameras. Each camera will cover
$20^\circ \times 20^\circ$ FOV, resulting in 2 steradians coverage of each set. 
The FOV of 2 steradians corresponds to FOV of the BAT detector on board of the SWIFT satellite
(\cite{swift_bat_www}, \cite{swift_bat_paper}) , it also matches FOV of the LAT detector on the GLAST satellite (\cite{glast_www}, \cite{glast_lat_www}),
which will be lunched in the beginning of 2008. 
These two sets will be installed in different places separated by distance of
$\approx$ 100 km. This separation distance will allow to reject optical
flashes caused by near Earth objects ( mostly satellites), using the parallax effect. 
The stereo observations have already been tested in LCO (see Sec. \ref{sec_slt}).
The parallax will allow to reject near Earth flashing objects up to orbit of the Moon.

\subsection{Mounts}
\begin{figure}[!htbp]
  \begin{center}
    \leavevmode
    \ifpdf
      \includegraphics[width=4in,height=4in]{white_mount_4x4_small_res.gif}
      \includegraphics[width=4in,height=4in]{new_mount2.gif}
    \else
      \includegraphics[width=4in,height=4in]{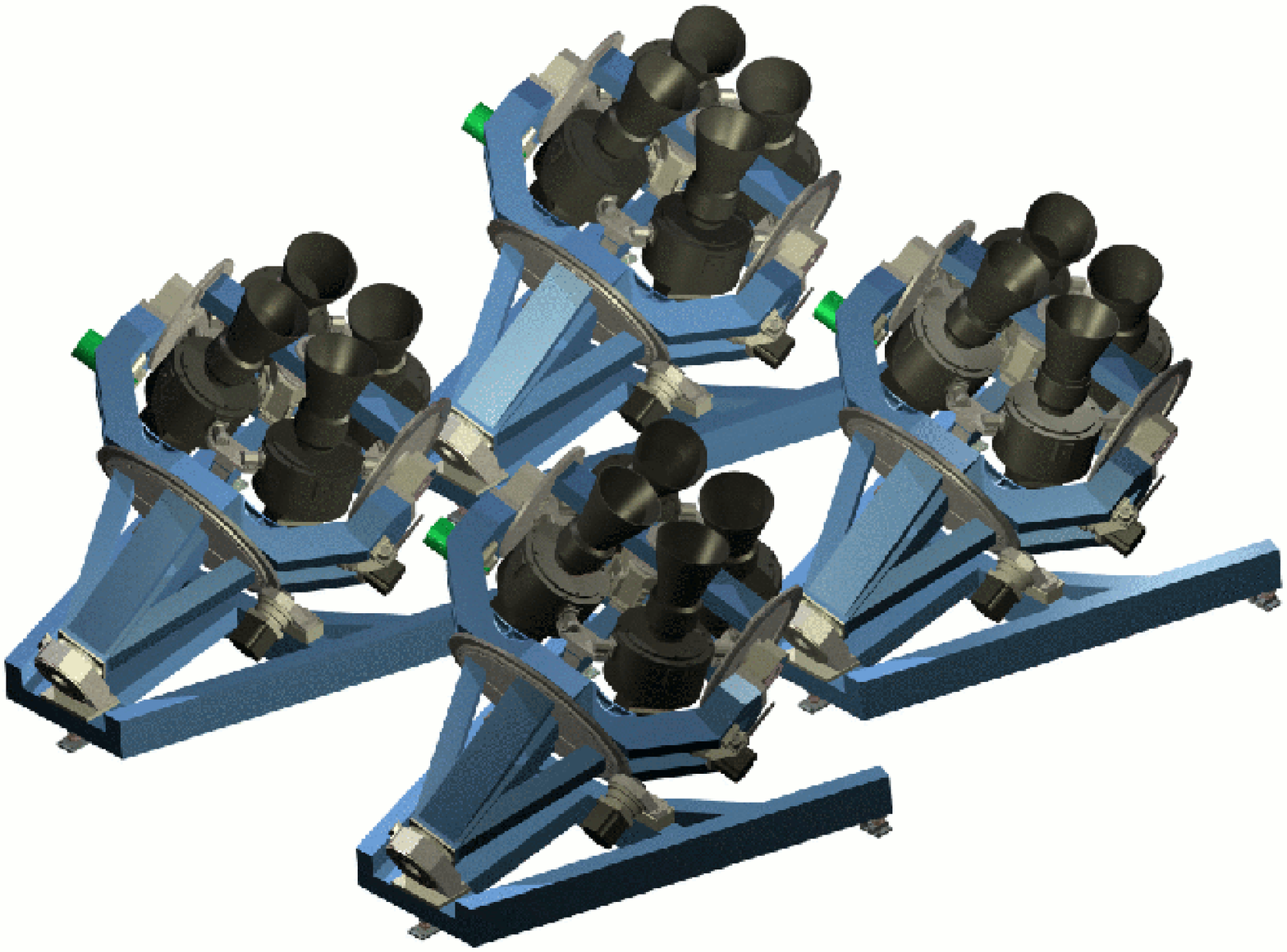}
		\includegraphics[width=4in,height=4in]{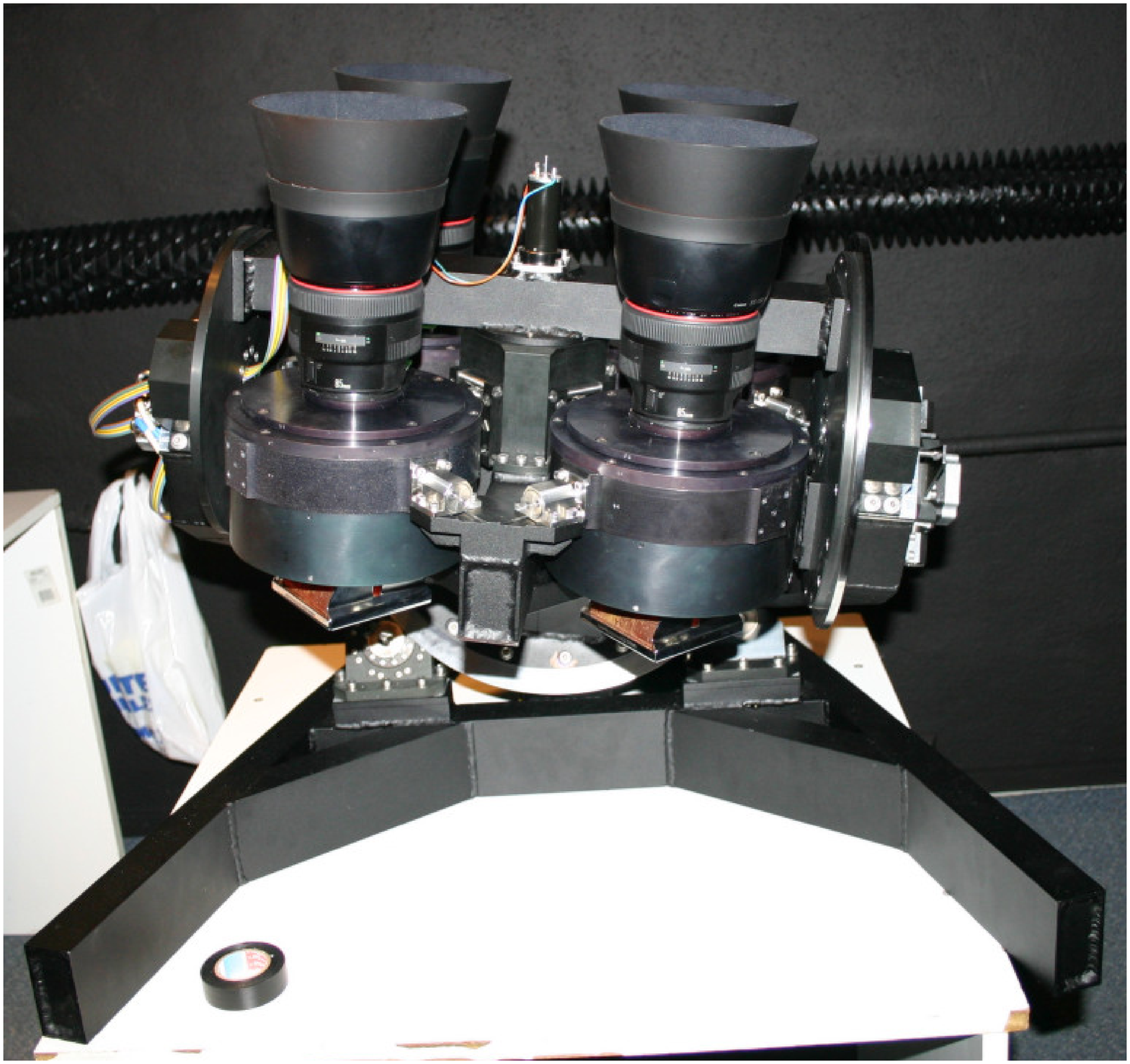}
    \fi
    \caption{Design of new mount for full "Pi of the Sky" system (upper
plot) and fully assembled mount in reality (lower plot)}                                                                                
    \label{fig_new_mount_design}
  \end{center}
\end{figure}

The mount design was based on the mount used in the prototype, but it was
redesigned and improved (Fig. \ref{fig_new_mount_design}). In the new design 4 cameras are installed
on a single mount. They can work in two modes so called \textit{wide} and \textit{deep}.
During normal operation cameras will work in the \textit{wide} mode, looking at
neighboring fields in the sky and covering FOV$\approx 40^\circ \times 40^\circ$ ( single mount ).
In order to obtain higher limiting magnitude and improve precision of the photometry
all cameras on the mount can be pointed to the same position in the so called 
\textit{deep} mode. This strategy can be used in case of GCN alert about GRB when it is important 
to increase the range of the system by averaging many images of same field
from different cameras.\\
Hardware improvements in the mount design include also usage of harmonic drives 
to improve pointing precision. Better control of the position will be 
provided by 13 bit encoders which will be used instead of potentiometers.
New design of the mounts includes changes in the control system. New mounts
will be controlled via the Controller Area Network (CAN) interface. In order to ensure flexibility, CAN
to Ethernet converters will be used and all commands from PC computer will be passed
through the Ethernet interface. This allows much more flexible system, where 
every mount obtains IP address and can be controlled by any PC computer in the cluster. 
The system schematic is shown in Figure \ref{fig_fullpi_schema}.

\subsection{Cameras}
\begin{figure}[!htbp]
  \begin{center}
    \leavevmode
    \ifpdf
      \includegraphics[width=5.5in]{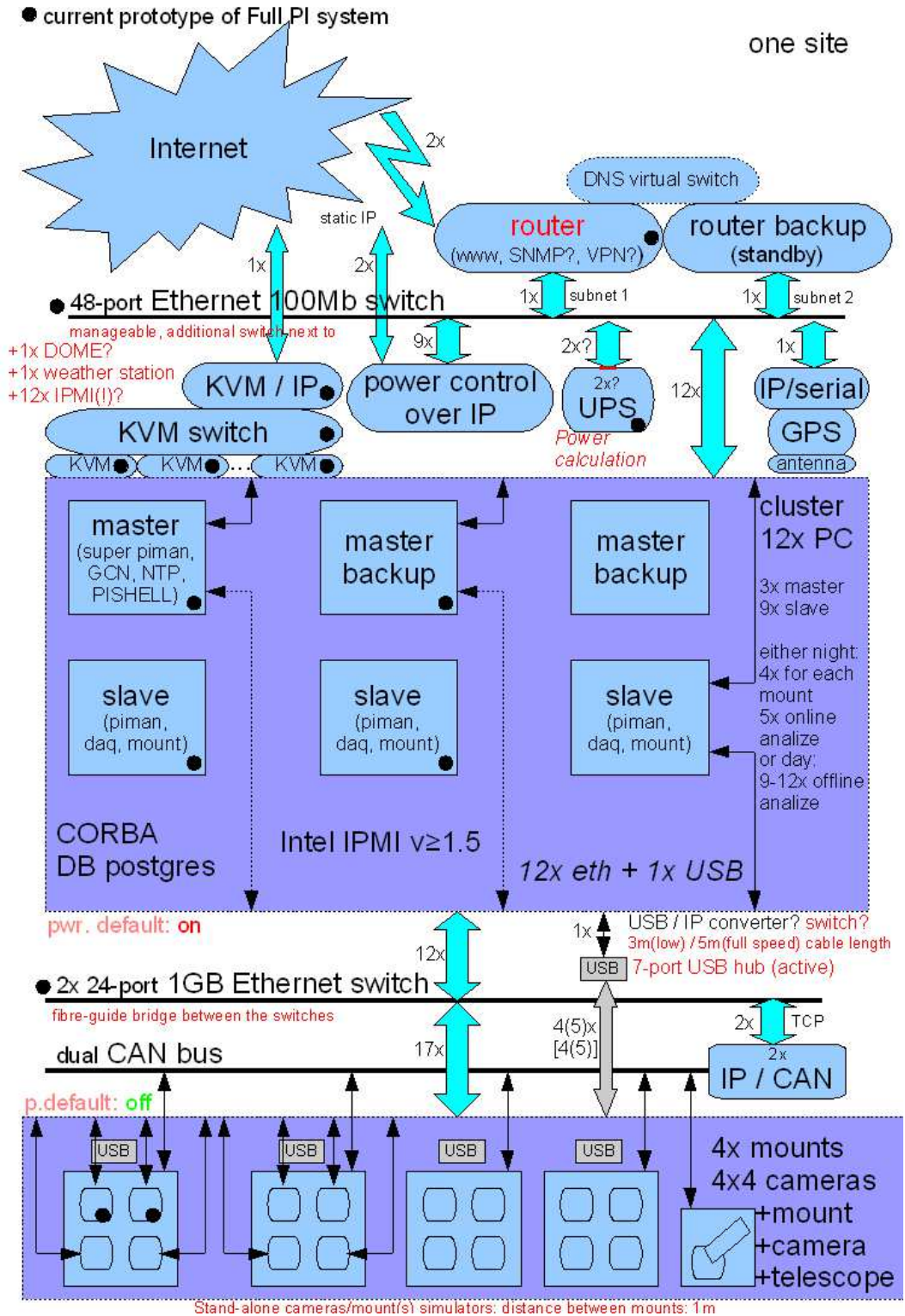}
    \else
      \includegraphics[width=5.5in]{full_pi/fullpi3.eps}
    \fi
    \caption{Design of the full "Pi of the Sky" system}                                                                                
    \label{fig_fullpi_schema}
  \end{center}
\end{figure}

Some improvements were also introduced in the final design of cameras. The major change is the Ethernet interface,
which was added to decouple cameras from the dedicated computer. 
The major disadvantage of the USB2.0 interface used in the prototype cameras was that 
they had to be connected directly to the controlling PC computer.
In case of the computer crash there was no way to use cameras until computer
was repaired or replaced.
This is not a problem in the case of the Ethernet interface, since the cameras can be
connected to Ethernet switch and be part of local area network.
In this configuration they can be controlled from any PC in the local network ( Fig. \ref{fig_fullpi_schema} ).
This ensures that crash of a single computer will not make any camera unusable.
The second major modification was the choice of the STA0820 CCD chip. 
There were also minor improvements. Stack of Peltier junctions was used to cool
cameras down to 40$^\circ$ C below ambient temperature. 
Lifetime of the shutter was also increased by special breaking algorithm
implemented in electronic which reduces the shock caused by the shutter opening and closing.

\subsection{Computer Cluster}
As it was described above the final system will consist of two sites with 4
mounts, each carrying 4 cameras. It was established during tests of the
prototype that data collection and analysis requires about 1 CPU per camera. This 
allows for data collection and on-line analysis during a night and off-line
data reduction and analysis during a day. 
This implies that each set of cameras requires 16 CPUs to handle the system and 
data analysis. Instead of 16 computers, 4 machines with four
core architecture will be used. In any case they will form a cluster of
computers, which must be efficiently managed. The system architecture is
presented in Figure \ref{fig_fullpi_schema}.
The main idea of this cluster is that none computer is unique and in case
 any of the machines crashes the system must remain fully functional and 
other PC will take over the tasks of the crashed computer.





\chapter{Data Analysis}
\label{sec_dataanalysis}
\ifpdf
    \graphicspath{{DataAnalysis/DataAnalysisFigs/PNG/}{DataAnalysis/DataAnalysisFigs/PDF/}{DataAnalysis/DataAnalysisFigs/}}
\else
    \graphicspath{{DataAnalysis/DataAnalysisFigs/EPS/}{DataAnalysis/DataAnalysisFigs/}}
\fi

\def\deg{$^\circ$}

Data analysis in the "Pi of the Sky" experiment consists of on-line and off-line parts.
The on-line data analysis is required to control performance of the detector, but 
it is also responsible for finding optical flashes in timescale 10 or 22 sec in real time.
Fast identification of optical flashes gives a possibility to distribute
alerts in the community for follow-up observations.
The off-line data analysis acts on the reduced data. The reduction
pipeline consists of three main stages : photometry, astrometry and
cataloging to the database. Final, reduced data consists of star brightness
measurements stored in the database, which provides easy and effective access.
Off-line analysis algorithms act on the database, there are several algorithms
developed for different purposes. In this thesis two off-line algorithms implemented
by author will be described in detail. These are flare recognition algorithm,
looking for brightness increase of existing stars and algorithm for finding new objects in the sky.
The database provides easy data access for broad spectrum of data analysis. 
Except algorithms presented in this thesis there are also other 
off-line algorithms implemented ( \cite{scan_nova}, \cite{mbiskup_var} ).

\section{On-line data reduction}
\label{sec_online_reduction}
On-line data analysis is required for detector control. The most important
task is astrometry, it is a transformation of chip coordinates (x,y) to
celestial coordinates ($\lambda,\delta$) : \\

\begin{equation}
	T: (x,y) \rightarrow (\lambda,\delta)
\label{eq_astro_transform}
\end{equation}

In order to obtain this transformation the following steps are performed :
\begin{itemize}
\item Fast image reduction - subtraction of dark image ( this is step is also
required by on-line analysis algorithm )
\item Photometry - identification of stars in an image and determination of instrumental magnitudes
and chip coordinates (x,y)
\item Astrometry - determination of non-linear transformation T:(x,y) $\rightarrow$ ($\lambda,\delta$)
It is iterational minimization procedure comparing stars identified in the
image by photometry with reference stars from external star catalog.
This procedure will be described in more detail in Section \ref{sec_image_reduction}.
\end{itemize}
After finding astrometry transform it is possible to calculate celestial
coordinates of any object on an image from its (x,y) coordinates. 
The ($\lambda,\delta$) coordinates of image center are calculated and
compared against the expected position and can be used to correct mount
tracking speeds ( so called \texttt{autoguiding} procedure ).
Before analyzing an image with on-line algorithm fast image processing is
performed. The first step is dark image subtraction. 
In the next step, image is transformed by special transformation called \texttt{laplace}\footnote{because it resembles a discrete version of Laplace operator}.
Value of each pixel is calculated as simple function of several surrounding 
pixels. Values in pixels just around transformed pixels are summed and values in 
other pixels far from it are subtracted. The idea of this transformation is
to calculate simple aperture brightness for every pixel.

\begin{figure}[!htbp]
  \begin{center}
    \leavevmode
    \ifpdf
      \includegraphics[width=6in]{laplace_type.gif}
    \else
      \includegraphics[width=6in]{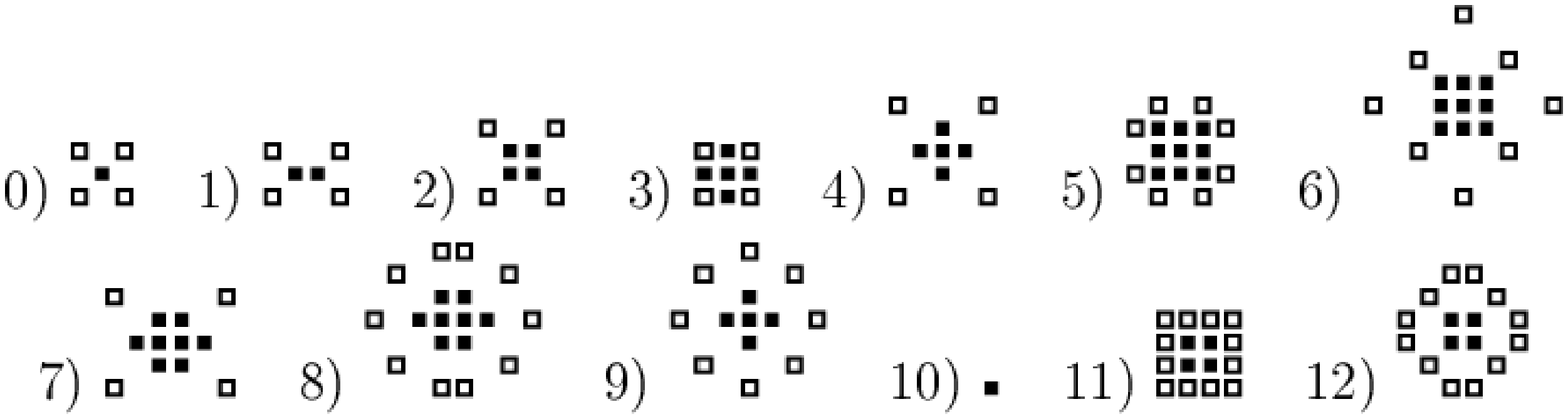}
    \fi
    \caption{Different \texttt{laplace} types tested in on-line flash recognition algorithms}
    \label{fig_laplacjans}
  \end{center}
\end{figure}

\begin{figure}[!htbp]
  \begin{center}
    \leavevmode
    \ifpdf
      \includegraphics[width=6in]{before_and_after_laplace.gif}
    \else
      \includegraphics[width=6in]{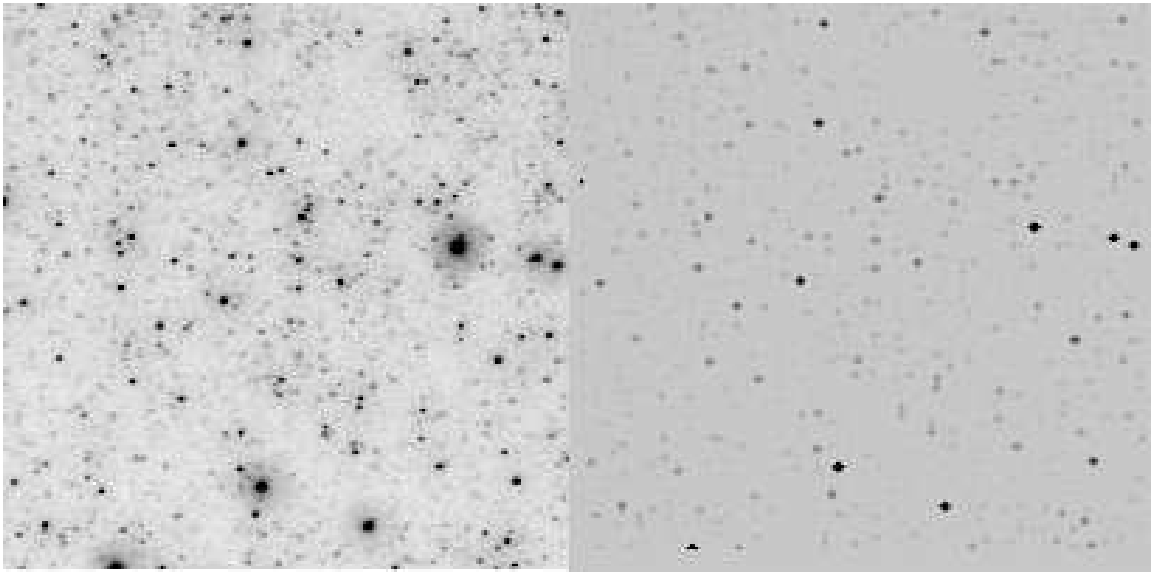}
    \fi
    \caption{Sky image before and after applying the \texttt{laplace} filter}
    \label{fig_before_and_after_laplace}
  \end{center}
\end{figure}

\begin{figure}[!htbp]
  \begin{center}
    \leavevmode
    \ifpdf
      \includegraphics[width=6in]{g12_distr.gif}
    \else
      \includegraphics[width=6in]{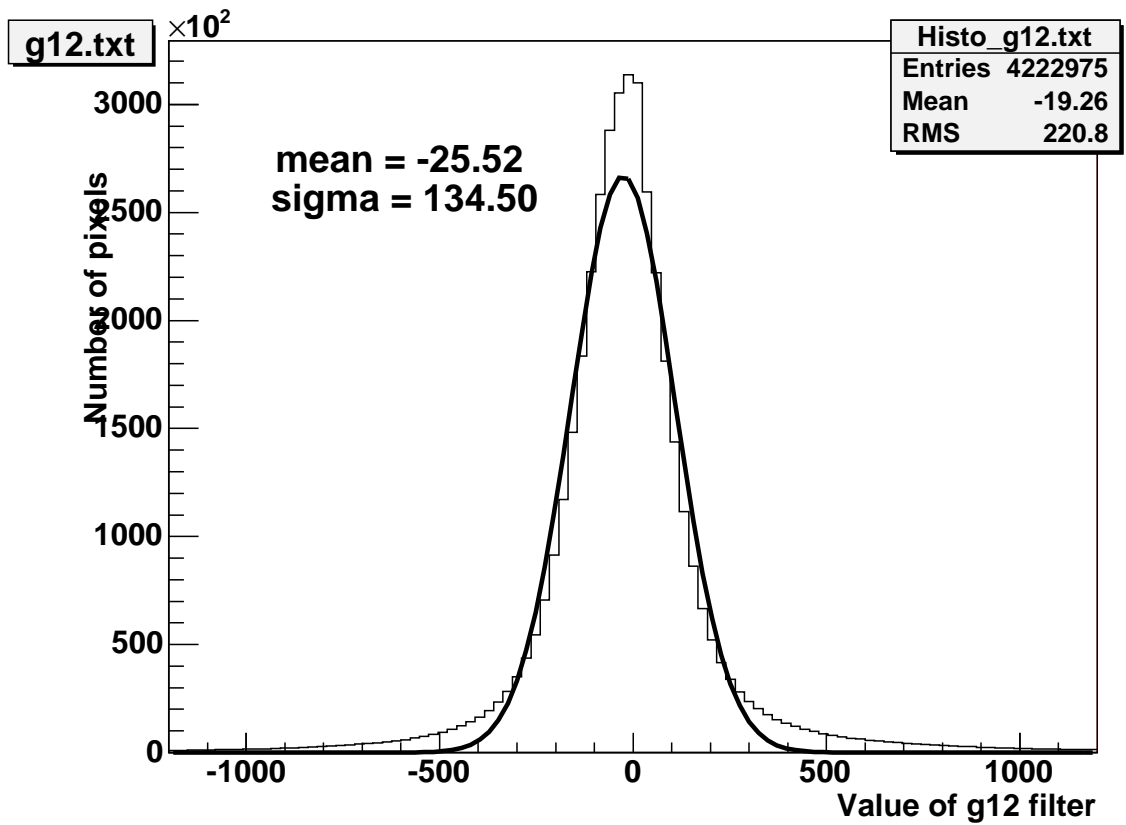}
    \fi
    \caption{Distribution of \texttt{laplace} 12 values on a single image}
    \label{fig_laplace_distr}
  \end{center}
\end{figure}

Several types of filters which were tested are shown in Figure \ref{fig_laplacjans}.
Images before and after applying the \texttt{g54 laplace} filter ( aperture 4 in Fig. \ref{fig_laplacjans} ) are shown in Figure \ref{fig_before_and_after_laplace}
it can be clearly seen that stars are sharper on the filtered image.
Finally, one filter was chosen according to lowest ratio of $\sigma_{g} / g_{avg}$ 
value calculated for faint stars, where $g_{avg}$ denotes average of maximum \texttt{laplace}
value for given star and $\sigma_{g}$ stands for its dispersion.
For old prototype version with Carl Zeiss f=50mm lenses \texttt{laplace} 4 (\texttt{g54}) was used and for the Canon f=85mm \texttt{laplace} 12 is used.
On-line algorithm is based on transformed images, distribution of pixel values after
such transformation is centered around zero ( Fig. \ref{fig_laplace_distr} ).
For every collected image a Gaussian curve is fitted and signal threshold $T_n$ is calculated 
as multiplicity of sigma value (  typically $T_n$ = 5$\sigma_B$ or 6$\sigma_B$ ).

\section{On-line flash recognition algorithms}
\label{sec_online_flashes_algo}

The aim of this type of algorithms is to find optical flashes occurring in
single image time scale ( 10 sec ). The signature of such events is the
following. Interesting event is an object which appears in a new image of
the sky and was not present in the previous image of the same field which was
taken a moment before. 
However, this simple idea is not so simple in practical realization. Most of
the events found by such simple comparison of two images are due to
background. The crucial task of the efficient algorithm is to 
reject most of the background without too much loss of the signal.
It is realized by a multilevel triggering system based on ideas
used in particle physics. 
Every image consists of $4 \cdot 10^6$ pixels suspected of being potential
interesting event. Image should be analysed while next image is being
collected which takes $\approx 12$s in the current configuration. This means
that the algorithm must be fast.
First trigger levels are simple and fast, they reject big amount of event
candidates with simple criteria. Higher trigger levels have more time and 
can use more sophisticated criteria to reject background events.

\subsection{First Level Trigger}
This level of algorithm must handle the highest data rate, so it must be
very fast and simple. It should preserve most of the signal and reject big fraction of non-interesting pixels.
At this stage flash-like events in single camera are identified and saved to
log files and optionally to the database. The following list of cuts are applied to every
pixel in the image : \\
\begin{itemize}

\item[$\bullet$] \boldmath{$T_n$} - this cut selects stars on new image by requiring
signal in the analysed pixel. The condition for signal presence is : 

\begin{equation}
N(x,y) > T_n
\label{eq_tn_cut_cond}
\end{equation}

where N(x,y) is ADU value in pixel (x,y) of the new image and the threshold $T_n$ is specified by DAQ
configuration parameters in multiplicities of $\sigma_B$ ( see Tab. \ref{tab_fltparams} and Fig. \ref{fig_laplace_distr} ).
The goal of this cut is selection of all stars in new image.

\item[$\bullet$] \boldmath{$T_v$} - this cut rejects constant stars. It
requires that there is no signal on the previous frame. "Previous frame" in this case 
means not just one single image, but average of several previous images.
The condition imposed on value of pixel in the "previous image" is the following :

\begin{equation}
	P(x,y) < T_v
\label{eq_tv_cut_cond}
\end{equation}

where P(x,y) is the value in pixel (x,y) on the average of $N_{aver}$ previous images.
Pixels remaining after this cut should be new objects which appeared on new
image and were not present on previous images. 

\item[$\bullet$] \textbf{MinLaplace} - rejects pixels which have value on previous
image lower then minimal accepted value ( $T_{MinLap}$ ). This cut allows to reject edges of
bright stars where values of pixels after \texttt{laplace} filter often become negative, but 
can also vary to values exceeding $T_n$.

\item[$\bullet$] \textbf{IfMoreAfterTv} - rejects the whole image if number of pixels
accepted after $T_v$ cut exceeds certain limit $N_{MaxTv}$.
This cut allows to reject images with big number of events which are due to system
error, Moon light or clouds. The image is bad, all events are certainly
rubbish so they are rejected and no further analysis of this image is performed.

\item[$\bullet$] \textbf{LocalMax} - requires that pixel value is a local maximum. This
cut allows to choose only one pixel of star like object for further analysis.

\item[$\bullet$] \textbf{SkipOverlaps} - checks number of accepted pixels in
certain radius $R_{overlap}$ from current pixel and leaves only 
one event and skips the others. This cut narrows the number of pixels 
to be analysed which are related to the same object to a single one.

%
\item[$\bullet$] \textbf{Shape} - object shape indicator S is calculated. It is defined as :

\begin{equation}
	S = \frac{S_{cluster}}{S_{circle}}
\label{eq_shape_def}
\end{equation}

where $S_{cluster}$ is area of cluster and $S_{circle}$ is the area of the
smallest circle circumscribed on this cluster. 
Cluster is defined as group of pixels around current pixel
with values satisfying $N(x,y) \geq T_{cluster}$.
Shape is required not to be elongated by imposing :

\begin{equation}
		S > T_{shape}
\label{eq_shape_cond}
\end{equation}

The idea of shape calculation is shown in Figure \ref{fig_stars_shape_distr}. The distribution of shape
value for stars in single image is shown in Figure \ref{fig_stars_shape_distr}.

\begin{figure}[!htbp]
  \begin{center}
    \leavevmode
    \ifpdf
      \includegraphics[width=2.7in,height=2.7in]{shape.gif}
		\includegraphics[width=2.7in,height=2.7in]{shape_allevents.gif}
    \else
      \includegraphics[width=2.7in,height=2.7in]{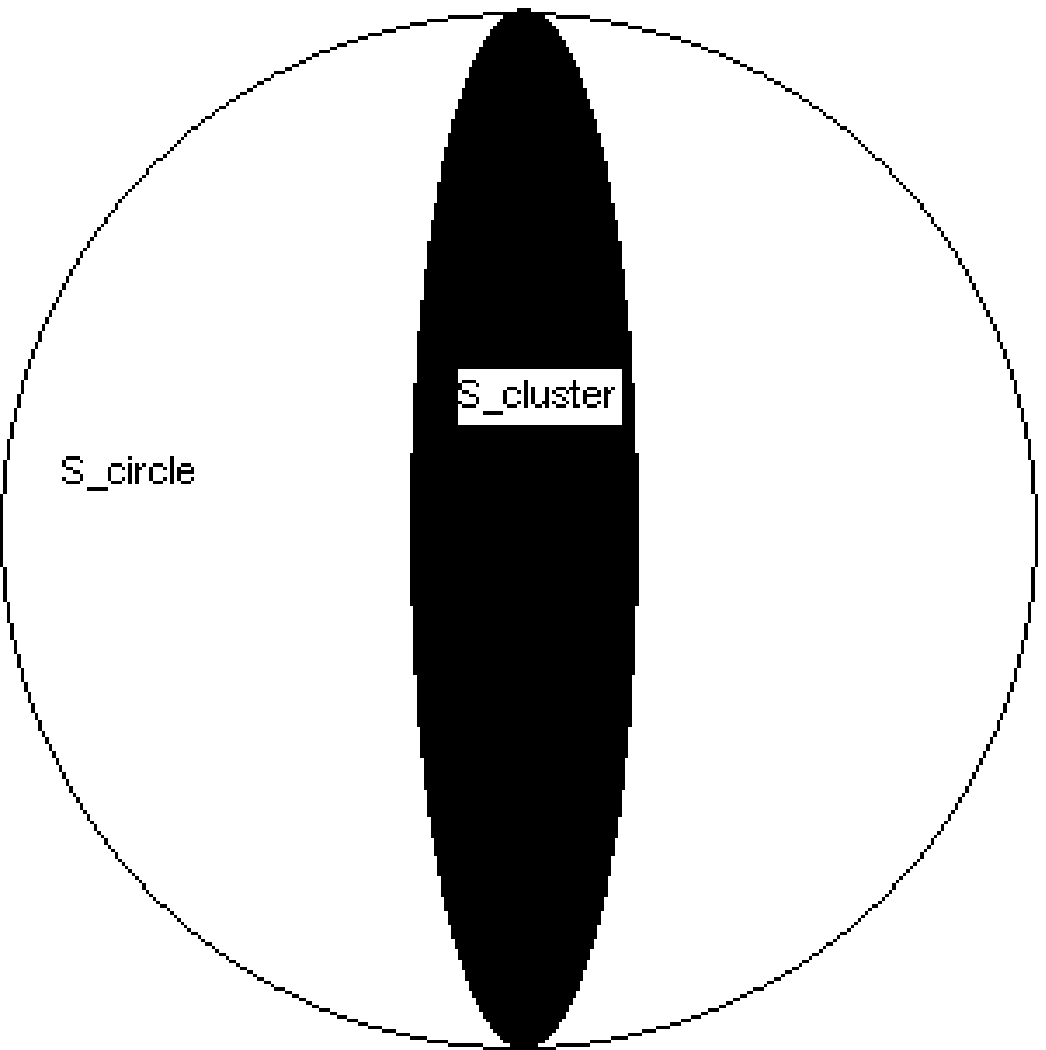}
		\includegraphics[width=2.7in,height=2.7in]{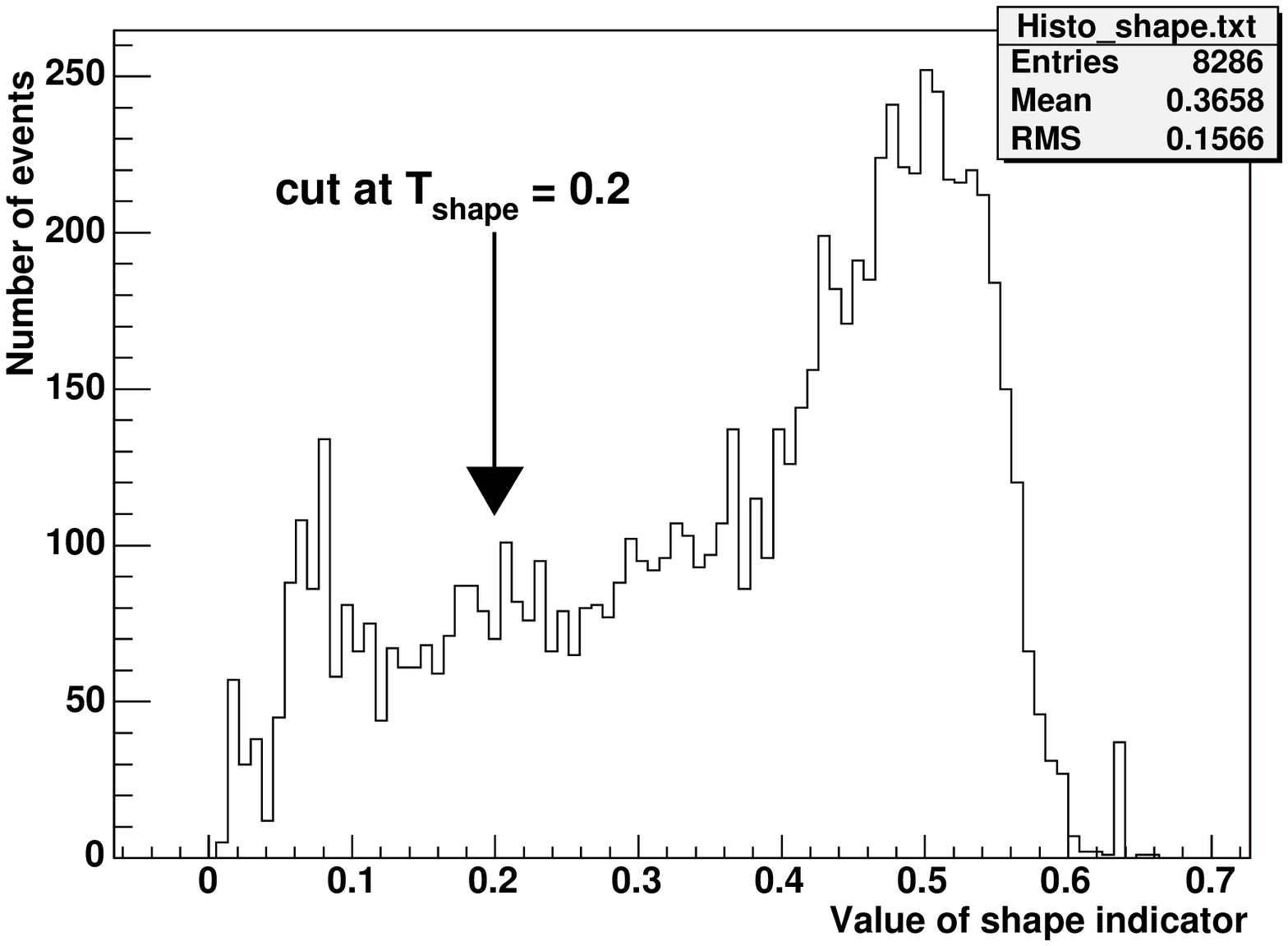}
    \fi
    \caption{Idea of shape indicator calculation (left) and its distribution for events from a single night and camera (right)}
    \label{fig_stars_shape_distr}
  \end{center}
\end{figure}

\item[$\bullet$] \textbf{BlackPixels} - this cut rejects pixels which have much  
smaller signal then neighboring pixels causing \texttt{laplace} filter to be high
due to underestimation of the background level. The following requirement is imposed
on every pixel :

\begin{equation}
	\frac{Min(P_{-})}{\sum P_{-}} \geq T_{black}
\label{eq_black_cond}
\end{equation}

where $P_{-}$ is value of pixel entering the \texttt{laplace} function with minus sign.
Black pixels on reduced image can be due to CCD defects or temperature
fluctuations, but are rather very seldom ( Tab. \ref{tab_fltcuts} ).

\item[$\bullet$] \textbf{HotPixels} - due to CCD chip defects some pixels can give 
much higher signal then normal good pixels. Such effects should generally be
subtracted by the dark image subtraction. However, sometimes it is not
enough, because new hot pixels can appear temporary during a night and 
become quiet again later. The main reason for this are excitations
from cosmic ray hits. Two ways of rejecting such events have been implemented. 
The first one is calculation of average value in pixel on previous images,
the imposed criteria is :

\begin{equation}
	\frac{\sum_{i=-1}^{N_{aver}}{P_i(x,y)}}{N_{aver}} < T_{hot}
\label{eq_hotpixel_cond}
\end{equation}

where $P_i(x,y)$ is ADU value in pixel in image \textit{i} before current one.
In the case of tracking mount this cut is neutral due to $T_v$ cut which is stronger.
Second anti-hotpixel cut is rejection of pixels by the list of known hot-pixels. 
This list is updated regularly when new defects are found. This is done
"manually" by running report which shows events described as hot pixel. In case
certain pixel is regularly giving false alerts it can be added to list of known
hot pixels which is stored in the database ( Fig. \ref{fig_starcat} ).

\item[$\bullet$] \textbf{IfMore} - checks nearby events in distance of $R_{ifmore}$ pixels, in
case number of events exceeds limit of $N_{ifmore}$ all nearby events are rejected.
This cut allows to reject events caused by planes or satellites.
\end{itemize}

Parameters for algorithms are passed as described in the Section \ref{sec_daq_cfg}.
The parameters used in First Level Trigger (FLT) are listed in Table \ref{tab_fltparams}.  

\begin{table}[htbp]
\begin{center}
\begin{tiny}
\begin{tabular}{|c|c|c|c|c|}
\hline
\textbf{Cut} & \textbf{Parameter Name} & \textbf{\begin{minipage}{1.8cm}{Value for confirmation on next}\end{minipage}} & \textbf{\begin{minipage}{1.8cm}Value for \hspace{3mm} coincidence\end{minipage}} & \textbf{Notes} \\
& & & & \\
\hline 
LaplaceType & LaplaceType & 12 & 12 & For Carl Zeiss f=50 \texttt{laplace}=4 \\
\hline
$T_n$ & $T_n$ & 6 & 5 & For Carl Zeiss f=50 $T_n$=5  \\ 
\hline
$T_v$ & $T_v$ & 2 & 2 & For Carl Zeiss f=50 $T_v$=2.5 \\
      & $N_{aver}$ & 7 & 7 & Number of averaged previous images \\
\hline
MinLaplace & $T_{MinLap}$ & 0 & 0 & - \\
\hline
IfMoreAfterTv & $N_{MaxTv}$ & 3000 & 3000 & - \\
\hline
SkipOverlaps & SkipOverlaps & 1 & 1 & enable/disable \\
SkipOverlaps & $R_{overlap}$ & 4 & 4 & radius in which overlaps are skipped \\
\hline
Shape & $T_{shape}$ & 0.2 & 0.2 & - \\
\hline
BlackPixels & $T_{black}$ & 0.5 & 0.5 & - \\
\hline
HotPixels & $T_{hot}$ & 3.6 & 3.6 & Threshold in anti-hotpixel cut \\
\hline
IfMore & $N_{IfMore}$ & 20 & 20 & \\
       & $R_{IfMore}$ & 150 & 150 & \\
\hline
\end{tabular}
\end{tiny}
\caption{FLT parameters summary, real names of parameters used in
configuration files can be found in Table \ref{tab_fltparams_trans} in Appendix \ref{sec_appendix2}}
\label{tab_fltparams}
\end{center}
\end{table}

The output from FLT is basically a list of events from single camera. These
events  are saved to a log file allevents\_N.log ( N stands for index of camera ).  
Optionally they can be saved to database to provide easy access for further analysis. 
Table \ref{tab_fltcuts} shows background rejection efficiency of subsequent FLT cuts.

\begin{table}[htbp]
\begin{center}
\begin{tiny}
\begin{tabular}{|c|c|c|c|}
\hline
\textbf{Cut} & \textbf{\% of all events} & \textbf{\% of events from previous level} & \textbf{Number of events} \\
\hline 
All & - & - & $3.212 \cdot 10^9$ \\
\hline
Tn & $2.06 \cdot 10^{-2}$ & 2.06\% & 66434214 \\
\hline
Tv & $2 \cdot 10^{-5}$ & 0.1\% & 68345 \\
\hline
MinTv & $2 \cdot 10^{-5}$  & 96.99\% & 66293 \\
\hline
Overlap & $5 \cdot 10^{-6}$ & 25.79\% & 17100 \\
\hline
Black & $5 \cdot 10^{-6}$ & 100\% & 17100 \\
\hline
Hot & $5 \cdot 10^{-6}$ & 100\% & 17100 \\
\hline
IfMore & $5 \cdot 10^{-6}$ & 99.85\% & 17075 \\
\hline
Coinc & $2.5 \cdot 10^{-6}$ & 46.28\% & 8099 \\
\hline
Satellites Catalog & $2.49 \cdot 10^{-6}$ & 98.96\% & 8015 \\
\hline
Star Catalog & $2.32 \cdot 10^{-6}$ & 93.11\% & 7463 \\
\hline
Shape & $2.25 \cdot 10^{-6}$ & 97.10\% & 7247 \\
\hline
Tracks & $1.55 \cdot 10^{-9}$ & 0.06\% & 5 \\
\hline
Accepted & $1.55 \cdot 10^{-9}$ & 0.06\% & 5 \\ 
\hline
\end{tabular}
\end{tiny}
\caption{Number of events remaining after subsequent cuts of coincidence algorithm for night 2006-05-27/28}
\label{tab_fltcuts}
\end{center}
\end{table}

\subsection{Second Level Trigger} 
\label{sec_slt}

The action at this level depends on the type of the system setup. Generally three 
configurations are possible :\\
\begin{itemize}
\item[$\bullet$] two cameras on a single mount working in coincidence. 
In this configuration events found by the first camera are verified in the 
corresponding image from the second camera. Only events present in images from both cameras are 
accepted. This configuration is realized by the prototype system in LCO. 

\item[$\bullet$] single camera, coincidence is replaced by confirmation of
signal on next image. This version of algorithm was also realized in LCO
when one of the cameras was not working due to technical problems.
This can be used as cross-check algorithm for two cameras working in coincidence. 

\item[$\bullet$] two cameras in separate locations working in coincidence.
This will be realized in the full version of the system. Cameras will be
paired, both cameras in the pair will observe the same field in the sky.
Spatial and time coincidence of the flash in both cameras will be required.
\end{itemize}


\begin{figure}[!htbp]
  \begin{center}
    \leavevmode
    \ifpdf
      \includegraphics[width=6in]{cosmic01666.gif}
		\includegraphics[width=6in]{cosmic01118.gif}
    \else
      \includegraphics[width=6in]{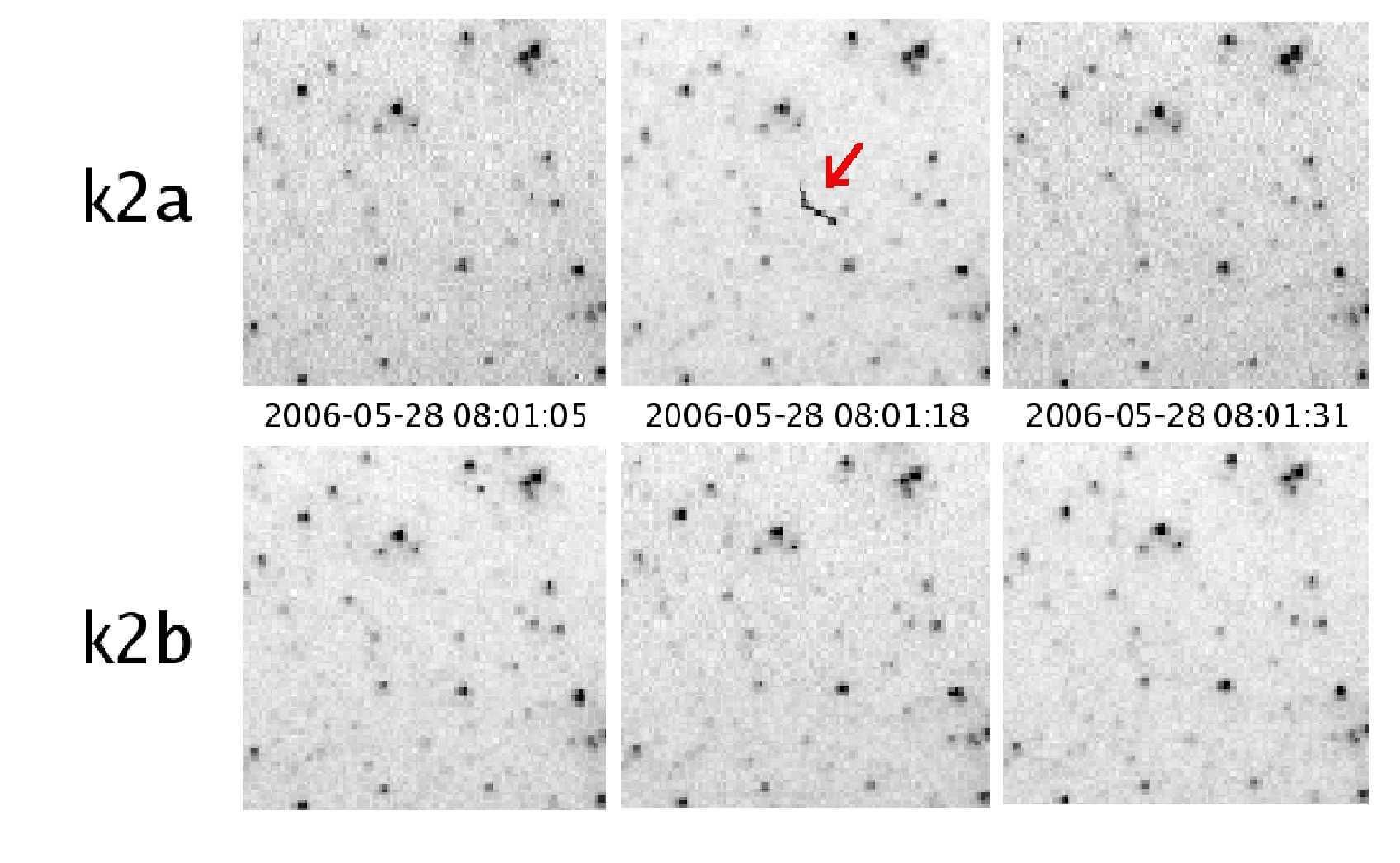}
		\includegraphics[width=6in]{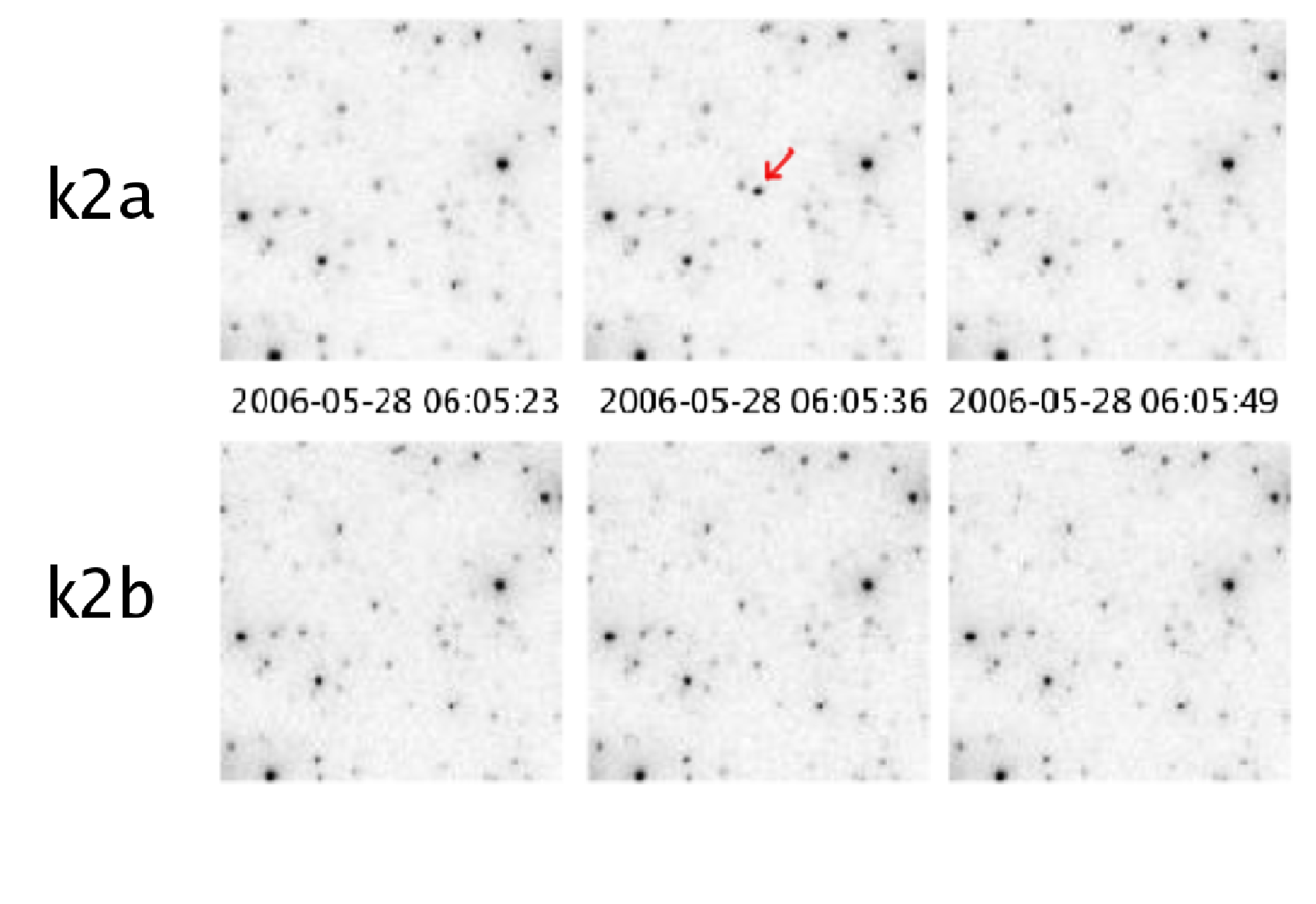}
    \fi
    \caption{Examples of events caused by cosmic ray hitting the CCD chip, with PSF easy to distinguish from stars (upper image) and very similar to PSF of stars (lower image)}
    \label{fig_cosmics}
  \end{center}
\end{figure}

\begin{figure}[!htbp]
  \begin{center}
    \leavevmode
    \ifpdf
      \includegraphics[width=6in]{coincidence_radius.gif}
    \else
      \includegraphics[width=6in]{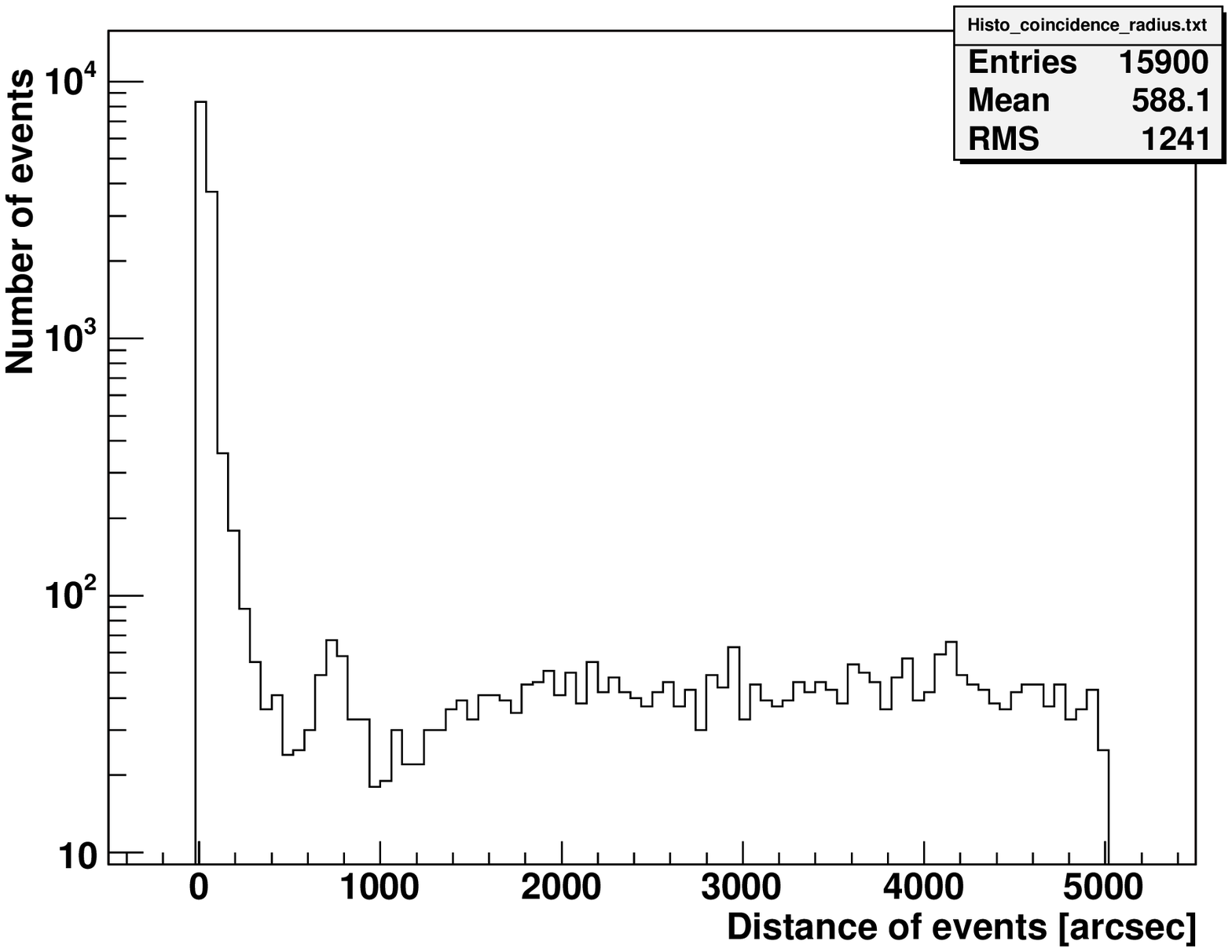}
    \fi
    \caption{Distribution of angular distances between events from corresponding images collected by cameras k2a and k2b during night
2006-05-28/29. Events with the distance < 250 arcsec are accepted.}
    \label{fig_coinc_radius}
  \end{center}
\end{figure}

In any case coincidence requirement is one of the most important cuts.
The main goal is rejection of cosmic rays hitting the CCD chip and imitating astrophysical
flashes (Fig. \ref{fig_cosmics}). In many cases cosmic rays have Point
Spread Function (PSF) completely different then PSF of stars and they could
be rejected by a shape recognition procedure. However, in some cases
they are very similar to PSF of the stars. Even if this is a 
very small fraction of all cosmic ray events this would cause all flashes 
found by the algorithm to be uncertain. A way of definite rejection of all 
cosmic rays events is required for credible flash recognition algorithm.
Probability that different cosmic ray particles will hit two chips in
the same time and in the same positions ( with respect to stars ) is negligible. 
Coincidence is also very effective way of rejecting background events due to sky
background fluctuations, edges of bright stars and clouds.
In the prototype version cameras take images parallely so the only parameter
is the angular distance of events in both cameras, currently used default value is 
$R_{coinc} = 150$ arcsec ( 250 arcsec was used for Carl Zeiss f=50mm lenses ). 
It was determined from distribution of angular distances
of corresponding stars in both cameras ( Fig. \ref{fig_coinc_radius} ).

\begin{figure}[!htbp]
  \begin{center}
    \leavevmode
    \ifpdf
      \includegraphics[width=6in]{parallax_new.gif}
    \else
      \includegraphics[width=6in]{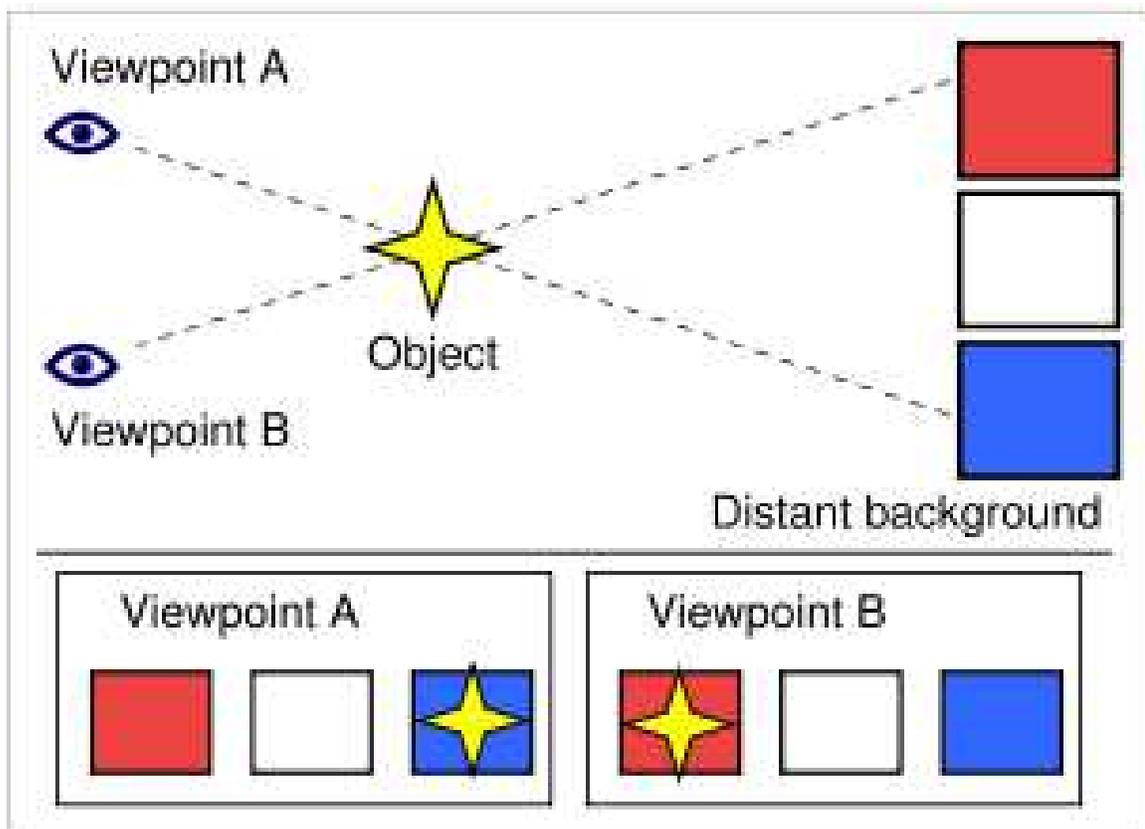}
    \fi
    \caption{A simplified example of parallax ( image taken from \cite{wiki_parallax} ) }
    \label{fig_parallax}
  \end{center}
\end{figure}

In the case of coincidence between cameras in separated sites, the
significance of coincidence is even bigger. In this configuration 
it is possible to reject close Earth flashing objects using the parallax ( Fig. \ref{fig_parallax} ).
Artificial satellites orbiting the Earth may rotate and sometimes reflex Sun light
causing flash-like events. The best method to distinct such kind of flashes is to use parallax.
In the prototype version of the experiment two cameras are installed on
single mount. However, this method was tested by coincidence with the RDOT
experiment \cite{rdot_web_page} located on La Silla at the distance of $\approx 30$ km. 

\begin{figure}[!htbp]
  \begin{center}
    \leavevmode
    \ifpdf
      \includegraphics[width=6in]{pi_vs_rdot_20050205.gif}
    \else
      \includegraphics[width=6in]{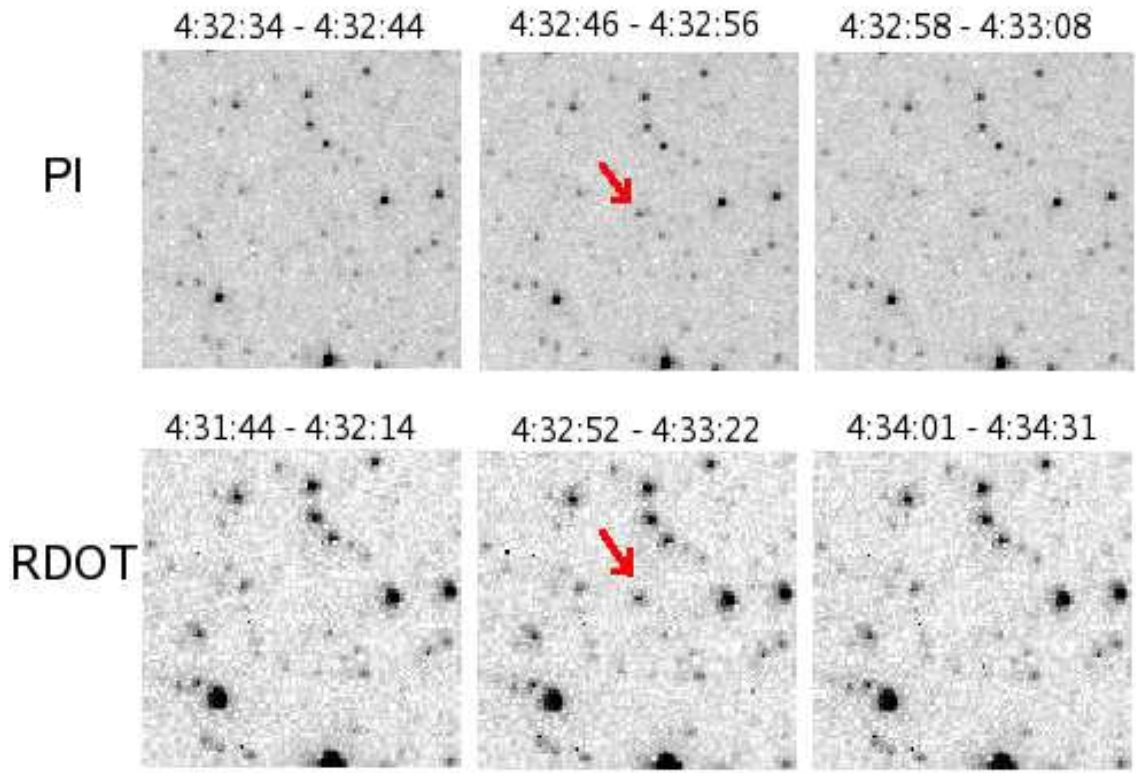}
    \fi
    \caption{Stereo observations of near Earth satellite by experiments PI and RDOT}
    \label{fig_rdot_parallax}
  \end{center}
\end{figure}

\begin{figure}[!htbp]
  \begin{center}
    \leavevmode
    \ifpdf
      \includegraphics[width=6in]{sat_radius_distr.gif}
		\includegraphics[width=6in]{20070526_vs_mc.gif}
    \else
      \includegraphics[width=6in]{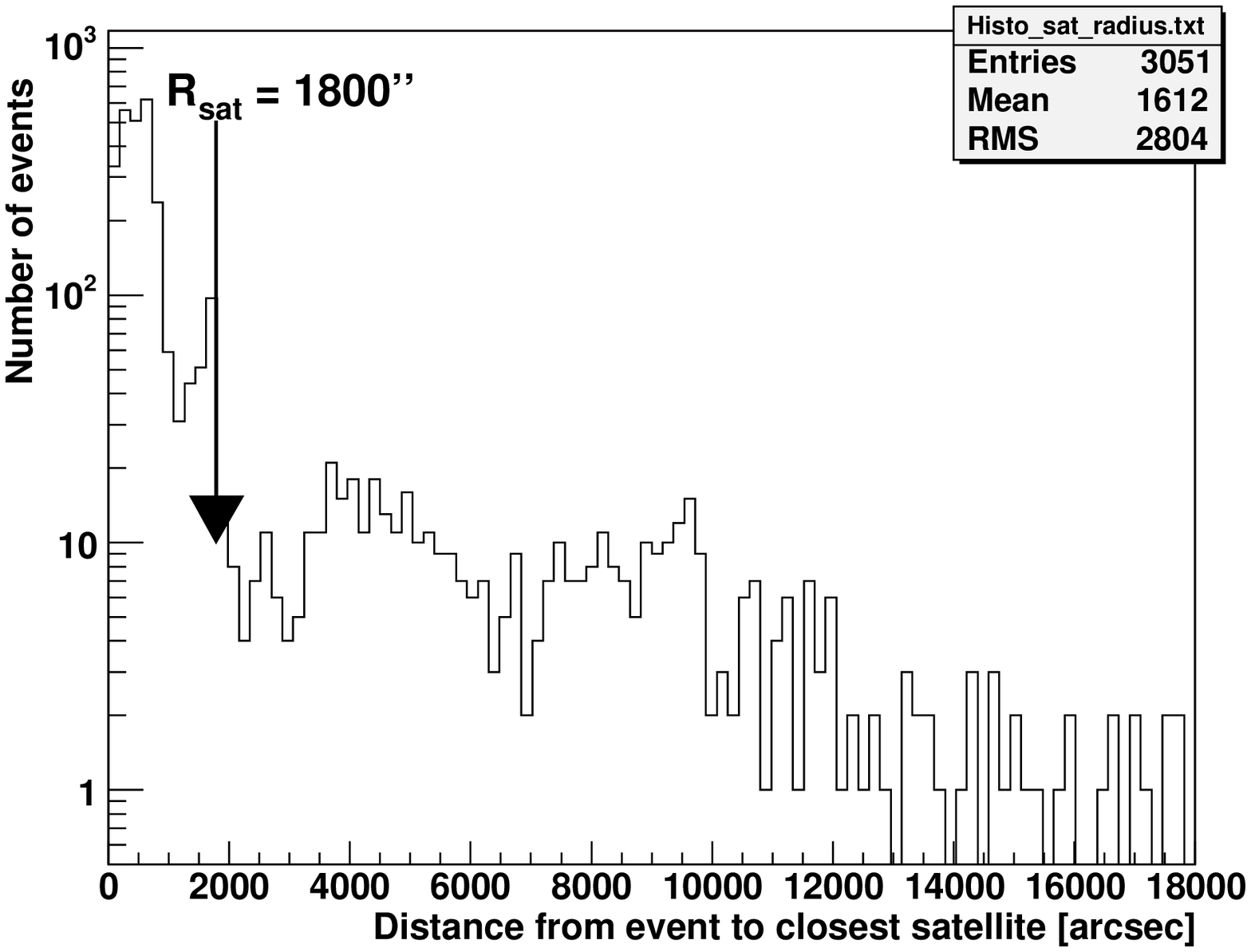}
		\includegraphics[width=6in]{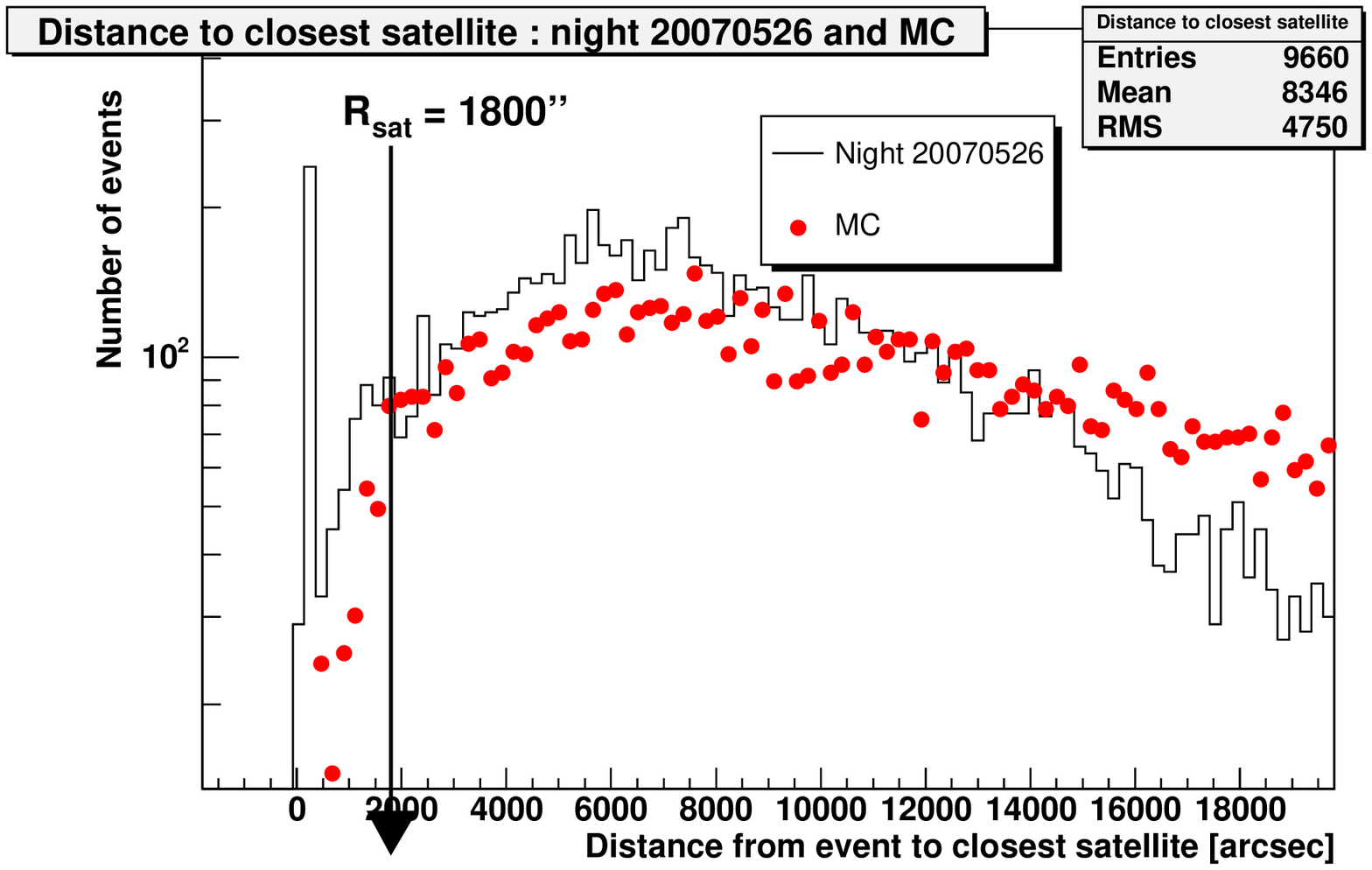}
    \fi
    \caption{Distribution of distance from flash event candidate to the closest satellite from the catalog. For events found by coincidence algorithm during night 2004.10.28/29 (upper plot)  and for single camera events from night 2007.05.26/27 (lower plot).  }
    \label{fig_satradius_distr}
  \end{center}
\end{figure}

Image collected by the prototype containing an optical flash was compared with 
image of same area of the sky taken by RDOT telescope at the same time. 
It can be clearly seen in Figure \ref{fig_rdot_parallax} that optical flash 
is visible in different positions with respect to the stars. Requirement 
of spatial coincidence would reject such event. It is probably caused by a
satellite in distance of $< 25000$ km from Earth.\\
This method has been tested for a few nights only. During normal operation it is  
not possible for the prototype to use parallax to reject flashing satellites. 
In order to reject most of such events a database of known satellites is used. 
It is retrieved every night from the Internet and contains $\approx 10000$ orbital elements. 
For every image, positions of all satellites in the database are calculated and every flash candidate is verified. 
In case it is closer then $R_{sat} = 0.5^\circ$ from any of the satellites it
is rejected. The rejection radius was determined from distribution of angular distances 
from flashes to closest satellite from the database which is clearly peaked around zero ( see Fig. \ref{fig_satradius_distr}). \\
The red dots on this plot represent distribution of distances to the closest
satellite from the catalog for randomly distributed flashes.
A clear peak at $R_{sat} < 0.5^\circ$ is visible, it corresponds to real satellites.

%

\begin{figure}[!htbp]
  \begin{center}
    \leavevmode
    \ifpdf
      \includegraphics[width=6in]{20050116.gif}
    \else
      \includegraphics[width=6in]{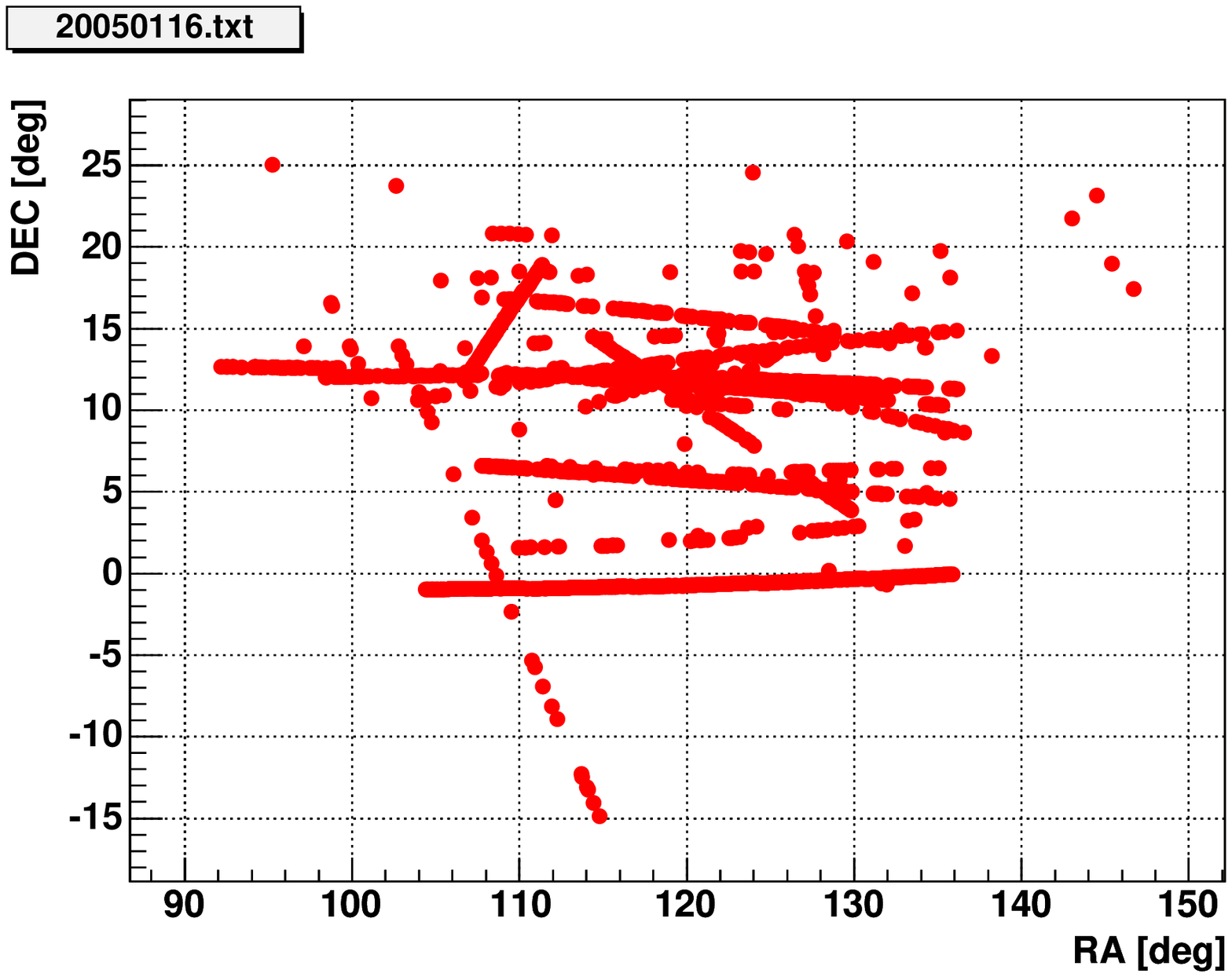}
    \fi
    \caption{Events rejected by track cut during single night 2005-01-16/17}
    \label{fig_night_tracks}
  \end{center}
\end{figure}

The orbital elements databases are not complete, there are many satellites
which are not included ( e.g. spy satellites ). In order to reject such kind of objects event candidates  
from many consecutive images are examined against track conditions. 
In case it is possible to fit track to set of events from
different images and velocity of object is constant all events on the
track are rejected. This rejects big fraction of flashing satellites and
planes ( see Fig. \ref{fig_night_tracks} ), however it is possible that rarely flashing ( rotating ) satellites
survive this cut. 

\begin{figure}[!htbp]
  \begin{center}
    \leavevmode
    \ifpdf
		\includegraphics[width=2.7in,height=2.7in]{20070308_confirm.gif}
		\includegraphics[width=2.7in,height=2.7in]{20060527_coinc_k2a.gif}
    \else
      \includegraphics[width=2.7in,height=2.7in]{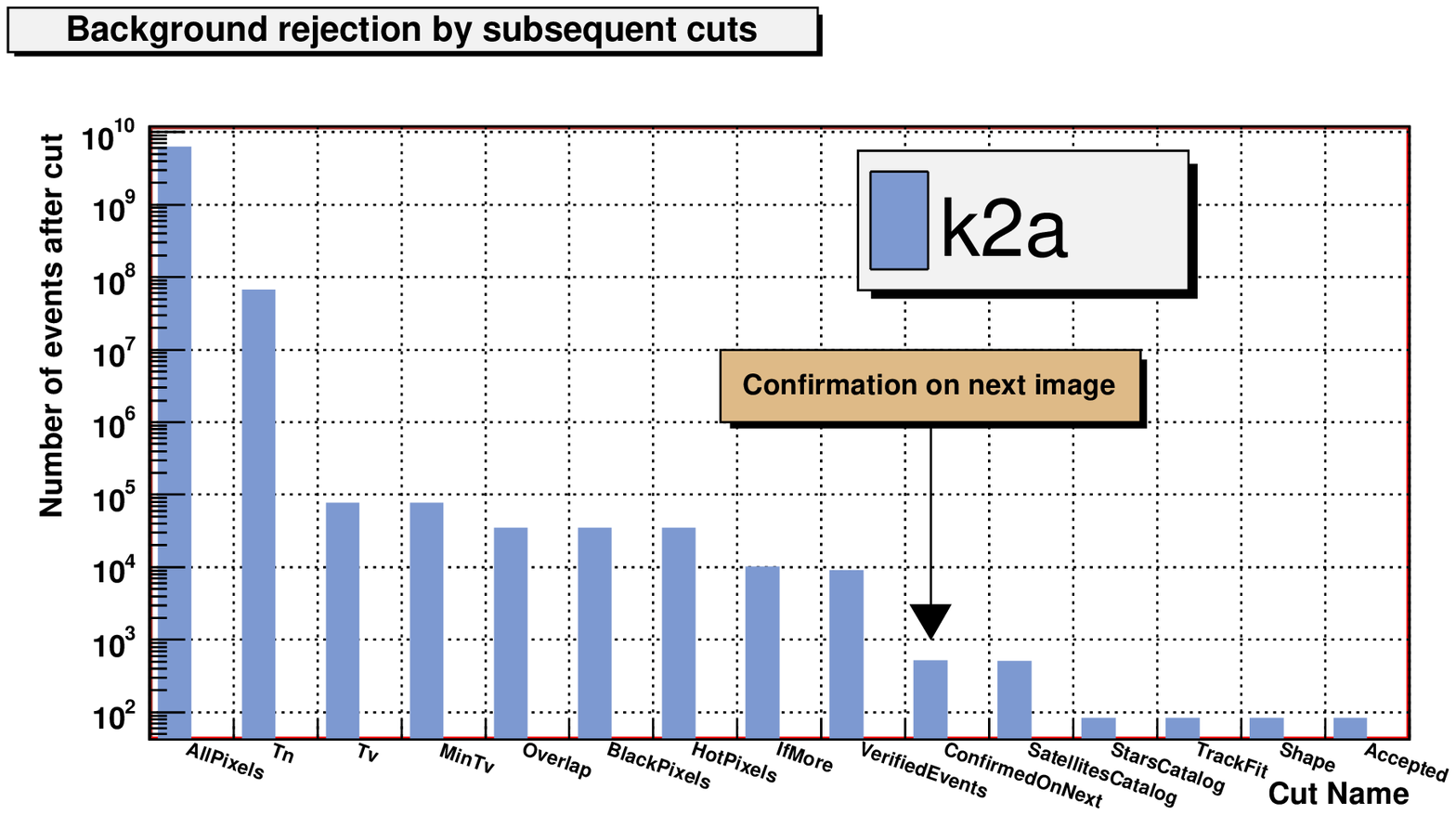}
		\includegraphics[width=2.7in,height=2.7in]{20060527_coinc_k2a.eps} 
    \fi
    \caption{Background rejection efficiency of algorithm requiring confirmation of flash on next image (left) and coincidence of two cameras ( right )}
    \label{fig_backgr_cuts_eff}
  \end{center}
\end{figure}

\begin{figure}[!htbp]
  \begin{center}
    \leavevmode
    \ifpdf
      \includegraphics[width=5in]{online_algo.gif}
    \else
      \includegraphics[width=5in,height=8in]{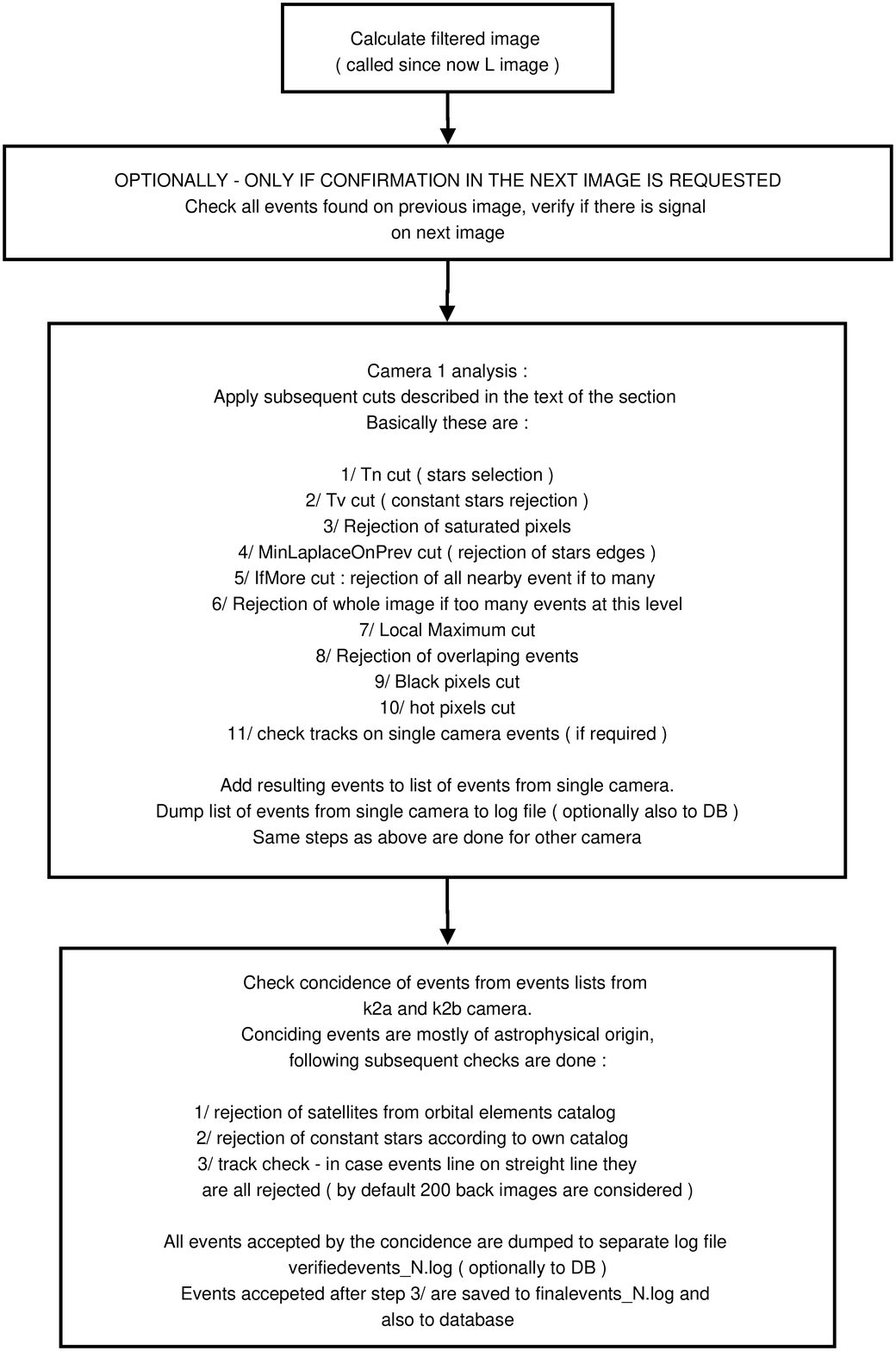}
    \fi
    \caption{Block diagram of on-line flash recognition algorithm}
    \label{fig_online_algo}
  \end{center}
\end{figure}

At this level of the trigger each event
candidate is checked against the catalog of constant stars. TYCHO-2 star catalog \cite{tycho} 
is used for this purpose. The event candidate is rejected in case there is a 
star brighter then $Mag_{max}$ in radius $R_{star}$ (see Tab. \ref{tab_sltparams}).
Stars can imitate flashes mainly due to clouds. When cloud moves and uncovers a star the FLT identifies
such an event as flash. 
All events accepted by coincidence are saved to files \texttt{verifiedevents\_N.log} and 
optionally to the database. 
Events accepted by the SLT are saved to \texttt{finalevents\_N.log} and to the database.
Parameters used in the SLT are listed in Table \ref{tab_sltparams}.
Block diagram of on-line flash recognition algorithm is shown in Figure \ref{fig_online_algo}.
Rejection efficiency of subsequent FLT and SLT cuts are show in Figure \ref{fig_backgr_cuts_eff}.

\begin{table}[htbp]
\begin{center}
\begin{tiny}
\begin{tabular}{|c|c|c|c|}
\hline
\textbf{Cut} & \textbf{Parameter Name} & \textbf{Current Value} & \textbf{Notes} \\
\hline 
Coincidence & $R_{coinc}$ & 150'' & It was 250'' for objectives f=50mm \\
\hline
Confirmation on next & $N_{confirm}$ & 1 & in case of 2 cameras working it is 0 \\
\hline
Satellite Catalog & $R_{sat}$ & 1800'' & angular distance from satellite to reject event \\
\hline
Star Catalog & $R_{star}$ & 120'' & angular distance from catalog star to reject event \\
             & $Mag_{max}$ & 13 & minimum brightness of used catalog stars \\
\hline
Track & $N_{track}$ & 200 & number of subsequent images used for track fit \\
      &             &     & \\
      & $\chi^2_{add}$ & 700 & \begin{minipage}{6cm}{Maximum allowed distance ( in pixels squared ) from existing track to match event to this track }\end{minipage} \\
		&             &     & \\
      & $\chi^2_{seed}$ & 100 & Maximum value of $\chi^2$ to initialize new track \\
	   &             &     & \\
\hline
\end{tabular}
\end{tiny}
\caption{SLT parameters summary, real names of parameters used in
configuration files can be found in Table \ref{tab_sltparams_trans} in Appendix \ref{sec_appendix2}}
\label{tab_sltparams}
\end{center}
\end{table}

\subsection{Third Level Trigger}
The first two levels of the trigger retain very small number of events, on average
it is not more then 10 per night. It depends strongly on weather conditions and in case of 
cloudy night this number can reach hundreds, but in this case fast inspection 
can reject most of them. However, in the full system the number of events will 
be 16 time larger reaching 100-200 per night and this will be much more
difficult to inspect. For this reason the third level of the
trigger has been implemented. It checks final events accepted by previous levels
which ensures that only small number of events will be examined.
Thus it is possible to implement much more sophisticated and time consuming algorithms to check
every event. Current implementation of the TLT consists of the following cuts :\\

\begin{itemize}
\item[$\bullet$] Comparison of event on both cameras and require signal to
be similar on both cameras, by imposing condition :  $L_{min} / L_{max} < L_{diff}$ 
\item[$\bullet$] Checks sphericity again with optionally more strict criteria $Shape < T_{shape}^{TLT}$
\item[$\bullet$] Simple Hough transform \footnote{Hough transform is a technique 
of image transform from (x,y) to cylindrical (r,$\phi$) coordinates in order to find particular shapes in an image} - uses small image part surrounding event. It
finds pixels with signal above certain level $T_{hough}$ and creates
distribution of $\phi$ coordinate ( $\phi = atan( (y-y0)/(x-y0) )$ ).
In case this distribution has significant peak this means it is probably due
to straight line from a plane or a satellite ( Fig. \ref{fig_hough_phi} )

\begin{figure}[!htbp]
  \begin{center}
    \leavevmode
    \ifpdf
		\includegraphics[width=2.8in,height=2.8in]{eventframe1340_camera0_eventno5_currframe1340.gif}
      \includegraphics[width=2.8in,height=2.8in]{hough.gif}
    \else
		\includegraphics[width=2.8in,height=2.8in]{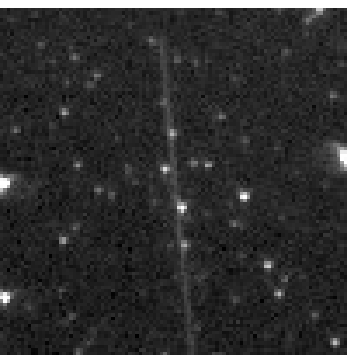}
      \includegraphics[width=2.8in,height=2.8in]{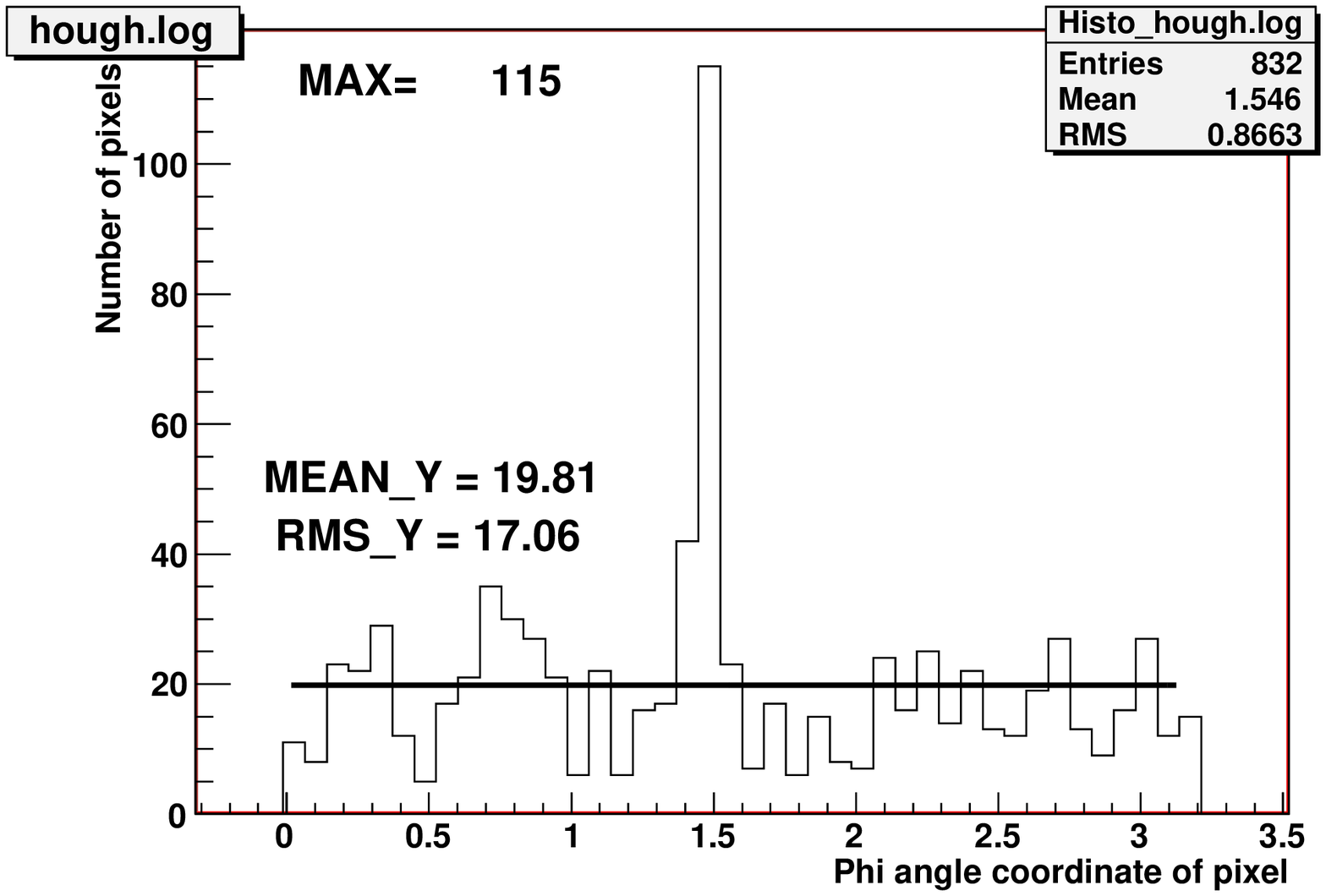}
    \fi
    \caption{Original event image (left image) and distribution of $\phi$ angle coordinate for background event (right image)}
    \label{fig_hough_phi}
  \end{center}
\end{figure}

Event will be considered as straight line if the maximum of distribution of angle $\phi$
satisfies the following condition :

\begin{equation}
\begin{split}
N_{max}(\phi) > MEAN_Y + T_{hough\_distr} * RMS_y \hspace{2mm} \\
AND \hspace{2mm} \\
N_{max}(\phi) > T_{hough\_height} * MEAN_Y
\end{split}
\label{eq_hough_cond}
\end{equation}

\item[$\bullet$] Track check : checks event against fitted tracks if the 
event was not correctly merged to existing track by the on-line algorithm
\item[$\bullet$] Algorithm on parts : algorithm is re-run on small parts
around the event, with less strict threshold $T_n^{TLT}= 4\cdot \sigma_B$. 
The event is rejected if tracks are found and it belongs to one of these tracks.
\item[$\bullet$] Cloud check : checks number of stars in the full image and
rejects event if $N_{stars} < 8000$.
\item[$\bullet$] Frame line : in case line can be fitted to events from a
single frame rejected by Hough transform, all events from the same frame and not 
yet rejected, are matched to this line. In case event matches this line it is rejected
\item[$\bullet$] All line check : checks if straight line can be fitted to
events from final list. It rejects final events matching this line.
\end{itemize}

\begin{table}[htbp]
\begin{center}
\begin{tiny}
\begin{tabular}{|c|c|c|c|}
\hline
\textbf{Cut} & \textbf{Parameter Name} & \textbf{Current Value} & \textbf{Notes} \\
\hline
Hough Transform & $T_{hough}$ & 1.5 & Threshold for choosing pixels \\
Hough Transform & $T_{hough\_distr}$ & 4.5 & 
\begin{minipage}{4cm}
Threshold for peak in $\phi$ distribution ( in $\sigma$ )
\end{minipage} \\
Hough Transform & $T_{hough\_height}$ & 2.0 & 
\begin{minipage}{4cm}
Threshold for minimum peak height ( in multiplicity of mean value )
\end{minipage} \\
 & & & \\
\hline
clouds cut & $N_{tstars}$ & 40000 & typical number of stars on whole image \\
clouds cut & $R_{clouds}$ & 0.2 & 
\begin{minipage}{4cm}
reject if when number of stars in the image < $R_{clouds} \cdot N_{tstars}$ 
\end{minipage} \\
 & & & \\
\hline
Event Comparison & $L_{diff}$ & 0.25 &  requires similar signal on both cameras \\
 & & & \\
\hline
 & & & \\
Algorithm on parts & $T_n^{TLT}$ & 4.00 & more loose then normal $T_n$ \\
\hline
\end{tabular}
\end{tiny}
\caption{TLT parameters summary, real names of parameters used in
configuration files can be found in Table \ref{tab_tltparams_trans} in Appendix \ref{sec_appendix2}}
\label{tab_tltparams}
\end{center}
\end{table}

Events rejected by this level of algorithm are not excluded from final list
of alerts claimed by the system. They are only flagged, this flag can be
useful indication for a person inspecting all night events.
Results of TLT are saved to log file and database. Parameters used in TLT are listed in Table \ref{tab_tltparams}. 

\subsection{Optimization of algorithm parameters}
\label{sec_opt_of_algo}

Optimization of algorithm parameters was a very important step of algorithm
development. Algorithm parameters can be changed by settings in the \texttt{ccd.cfg}
file as described in section \ref{sec_daq_cfg}.
Tables \ref{tab_fltparams} and \ref{tab_sltparams} list most important parameters of algorithms on
first and second levels of the trigger.

\begin{figure}[!htbp]
  \begin{center}
    \leavevmode
    \ifpdf
		\includegraphics[width=2.8in,height=2.8in]{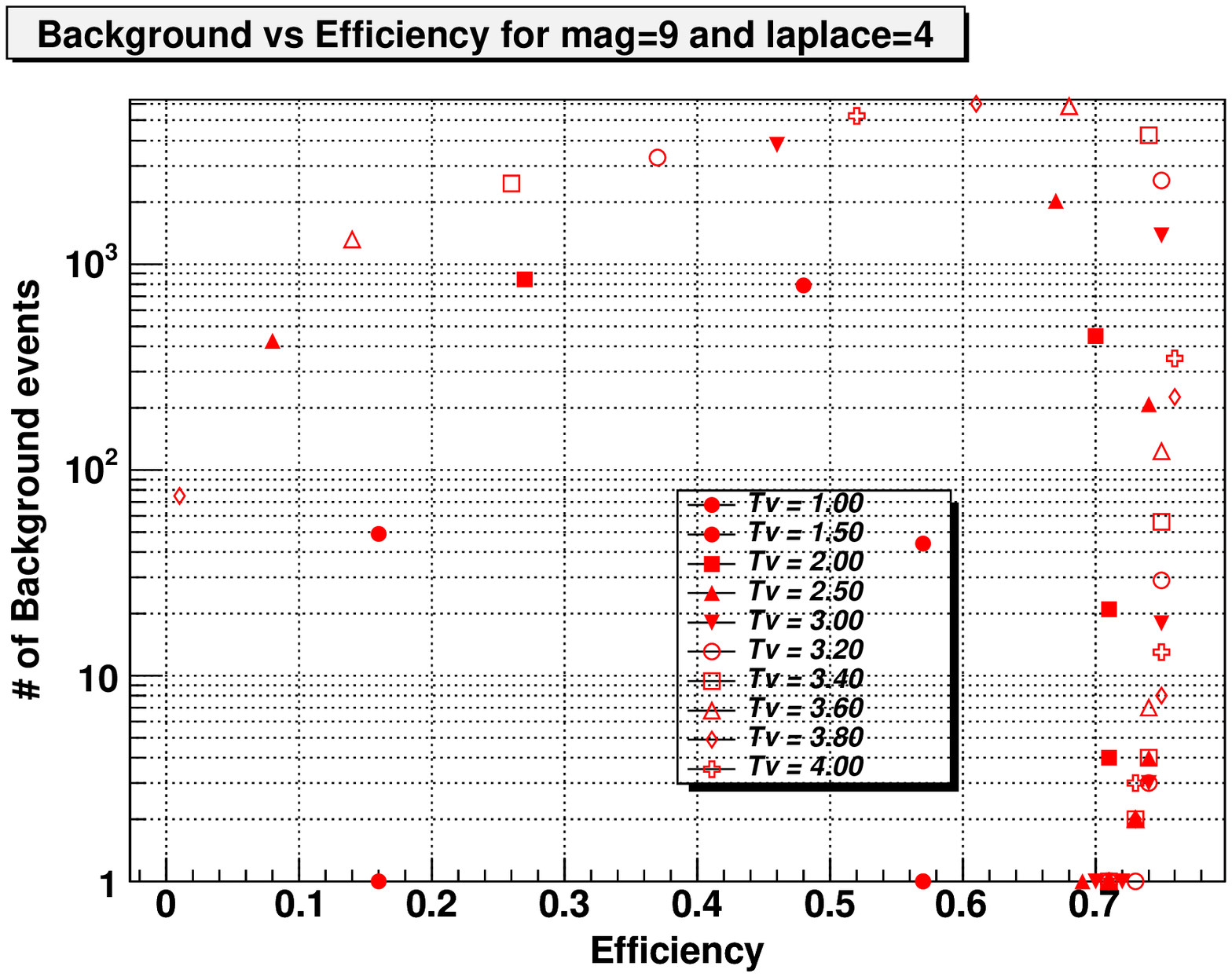}
		\includegraphics[width=2.8in,height=2.8in]{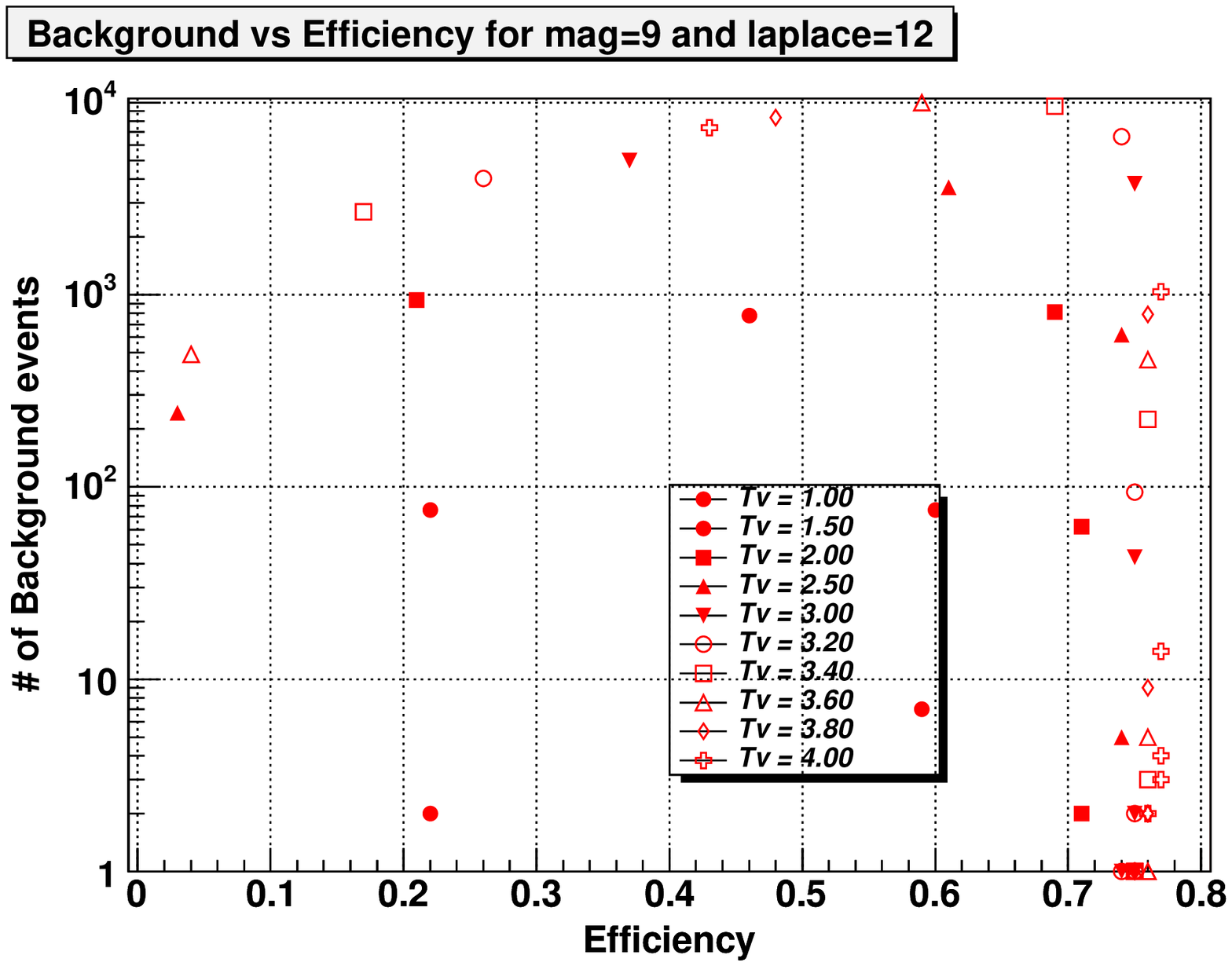}
    \else
		\includegraphics[width=2.8in,height=2.8in]{EffBkg/ccdsingle/mag9/laplace4/eff_bkg_9.00_lap4.eps}
		\includegraphics[width=2.8in,height=2.8in]{EffBkg/ccdsingle/mag9/laplace12/eff_bkg_9.00_lap12.eps}
    \fi
    \caption{Efficiency of $9^m$ flash recognition and background rejection 
				 by an algorithm requiring confirmation of event on next image. 
				 Test was performed on data from night 2007-04-25/26 for laplace=4 (left plot) and laplace=12 (right plot)}
    \label{fig_eff_vs_bkg_9mag}
  \end{center}
\end{figure}

\begin{figure}[!htbp]
  \begin{center}
    \leavevmode
    \ifpdf
		\includegraphics[width=2.8in,height=2.8in]{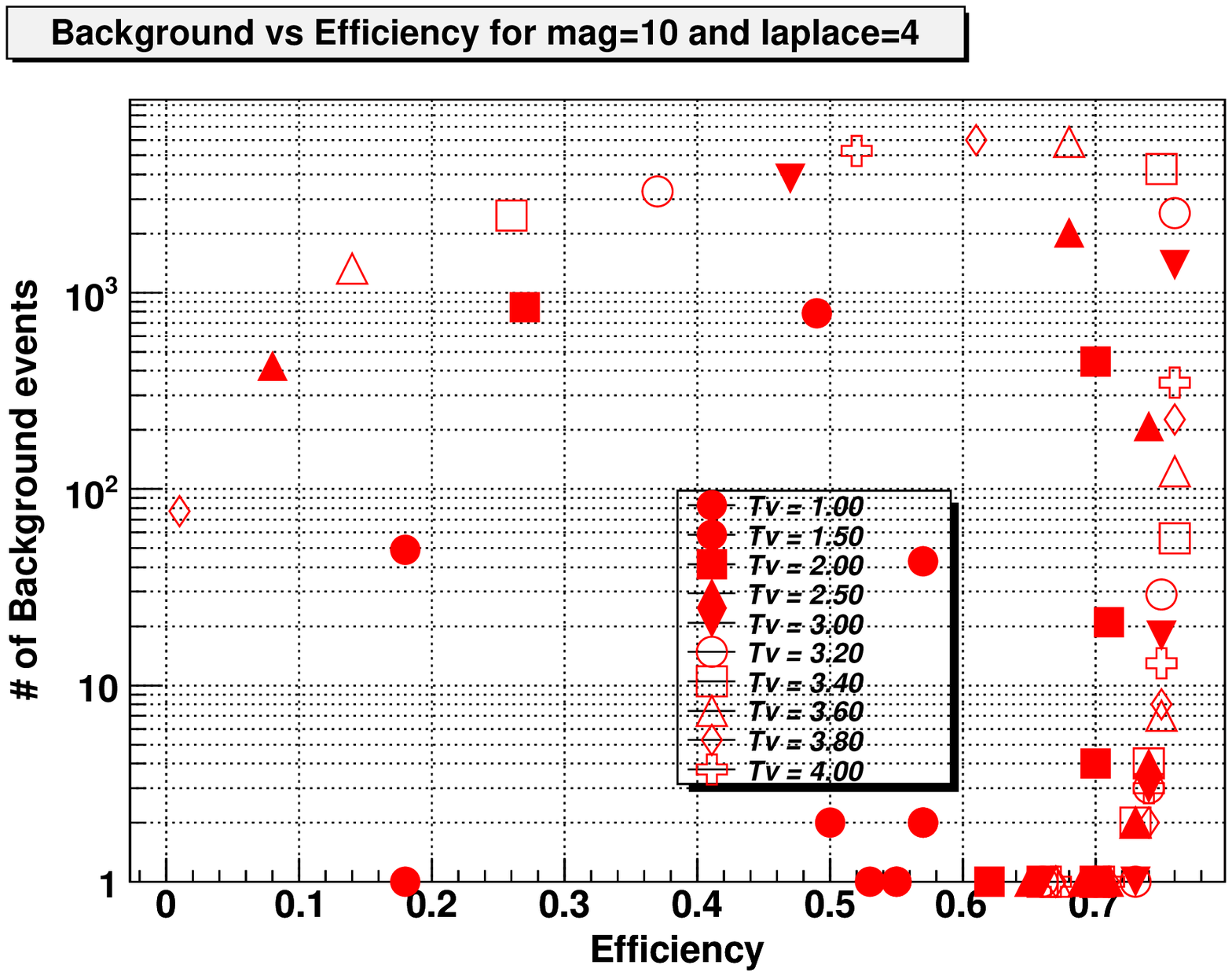}
		\includegraphics[width=2.8in,height=2.8in]{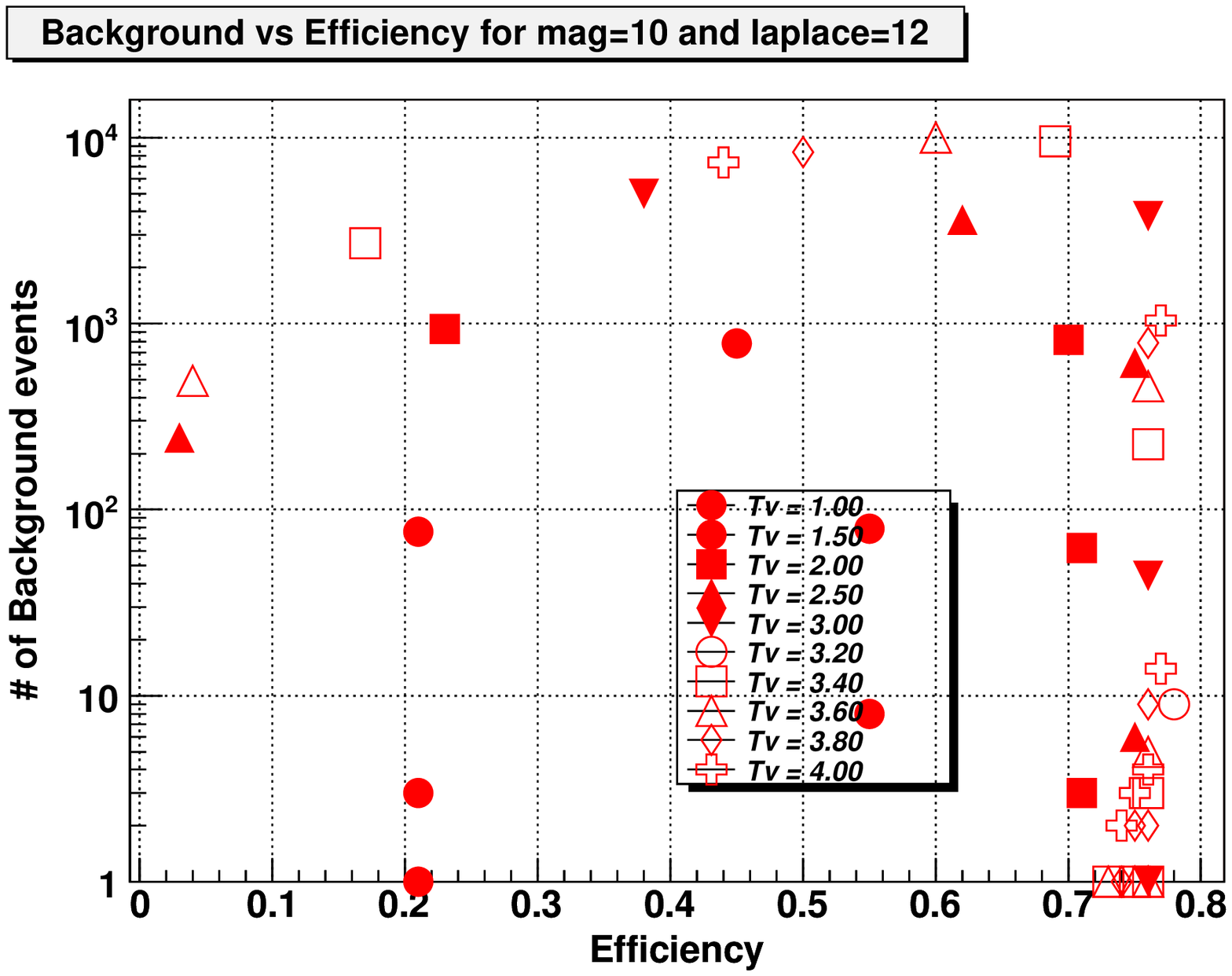}
    \else
		\includegraphics[width=2.8in,height=2.8in]{EffBkg/ccdsingle/mag10/eff_bkg_10.00_lap4.eps}
		\includegraphics[width=2.8in,height=2.8in]{EffBkg/ccdsingle/mag10/eff_bkg_10.00_lap12.eps}
    \fi
    \caption{Efficiency of $10^m$ flash recognition and background rejection 
				 by an algorithm requiring confirmation of event on next image.
				 Test was performed on data from night 2007-04-25/26 for laplace=4 (left plot) and laplace=12 (right plot)}
    \label{fig_eff_vs_bkg_10mag}
  \end{center}
\end{figure}

Some parameters were optimized by specific studies and the others were
optimized by running algorithms on sky data with simulated optical flashes. 
Testing of algorithms was performed on regular sky images. 
Exactly the same software was used, but instead of reading images from the camera like during regular night observations, 
the images were read from the \texttt{fits} files stored on a disk. 
Images were analysed and found events were considered as background events. 
Optical flashes were simulated in such a way
that samples of stars of given brightness were extracted from images and pasted in an image in
random positions. Single test of given parameters set was performed 
on all images from a single night, program was analyzing subsequent images
and one generated flash was added in every image.
After all images were analysed the number of background events $N_{bkg}$
was calculated, as those which were found by the algorithm, but were  not generated.
Generated events were checked on the list of identified events and 
number $N_{genident}$ of those generated and identified was determined.
Efficiency of identifying optical flashes of given magnitude was determined
as  : \\
\begin{equation}
	\epsilon ( mag ) = \frac{N_{genident}(mag)}{N_{generated}(mag)} \\
\label{eq_eff_calc}
\end{equation}
Every set of tested parameters was plotted as a single point in the plot of $\epsilon$ vs $N_{bkg}$.
Different points on this plot show values determined for different settings
of the algorithm. For every star magnitudo separate plot was created. 
Figures \ref{fig_eff_vs_bkg_9mag} and \ref{fig_eff_vs_bkg_10mag} show results of
efficiency and background rejection tests for algorithm requiring confirmation of flash in the next image 
performed on $\approx$500 images from night 2007.04.25/26.
These plots were created for simulated flashes of $9^m$ and $10^m$ respectively. The best values of efficiency are also shown
in Table \ref{tab_best_eff_vs_bkg}. 
According to these results optimal parameter values were chosen ( see Tables \ref{tab_fltparams} and \ref{tab_sltparams} ).

\begin{table}[htbp]
\begin{center}
\begin{tabular}{|c|c|c|c|c|c|}
\hline
\textbf{Magnitude} & \textbf{Laplace} & \textbf{Tv} & \textbf{Tn} & \textbf{Eff} & \textbf{Bkg} \\
\hline
\hline
9 & 4 & 2.0 & 4.0 & 0.71 & 21 \\
9 & 4 & 2.0 & 5.0 & 0.71 & 4 \\
9 & 4 & 2.0 & 6.0 & 0.70 & 0 \\
9 & 4 & 2.5 & 4.0 & 0.74 & 209 \\
9 & 4 & 2.5 & 5.0 & 0.74 & 4 \\
9 & 4 & 2.5 & 6.0 & 0.73 & 2 \\
9 & 4 & 3.0 & 5.0 & 0.75 & 18 \\
9 & 4 & 3.0 & 6.0 & 0.74 & 3 \\
\hline
\hline
9 & 12 & 2.0 & 5.0 & 0.71 & 2 \\
9 & 12 & 2.0 & 6.0 & 0.71 & 0 \\
9 & 12 & 2.5 & 5.0 & 0.74 & 5 \\
9 & 12 & 2.5 & 5.0 & 0.74 & 0 \\
9 & 12 & 3.0 & 5.0 & 0.75 & 43 \\
9 & 12 & 3.0 & 6.0 & 0.75 & 2 \\
\hline
\end{tabular}
\caption{Best values of efficiency of confirmation on next image algorithm obtained for simulated flashes of brightness 9$^m$, tested on 500 images from night 2007-04-25/26.}
\label{tab_best_eff_vs_bkg}
\end{center}
\end{table}

\begin{figure}[!htbp]
  \begin{center}
    \leavevmode
    \ifpdf
		\includegraphics[width=2.8in,height=2.8in]{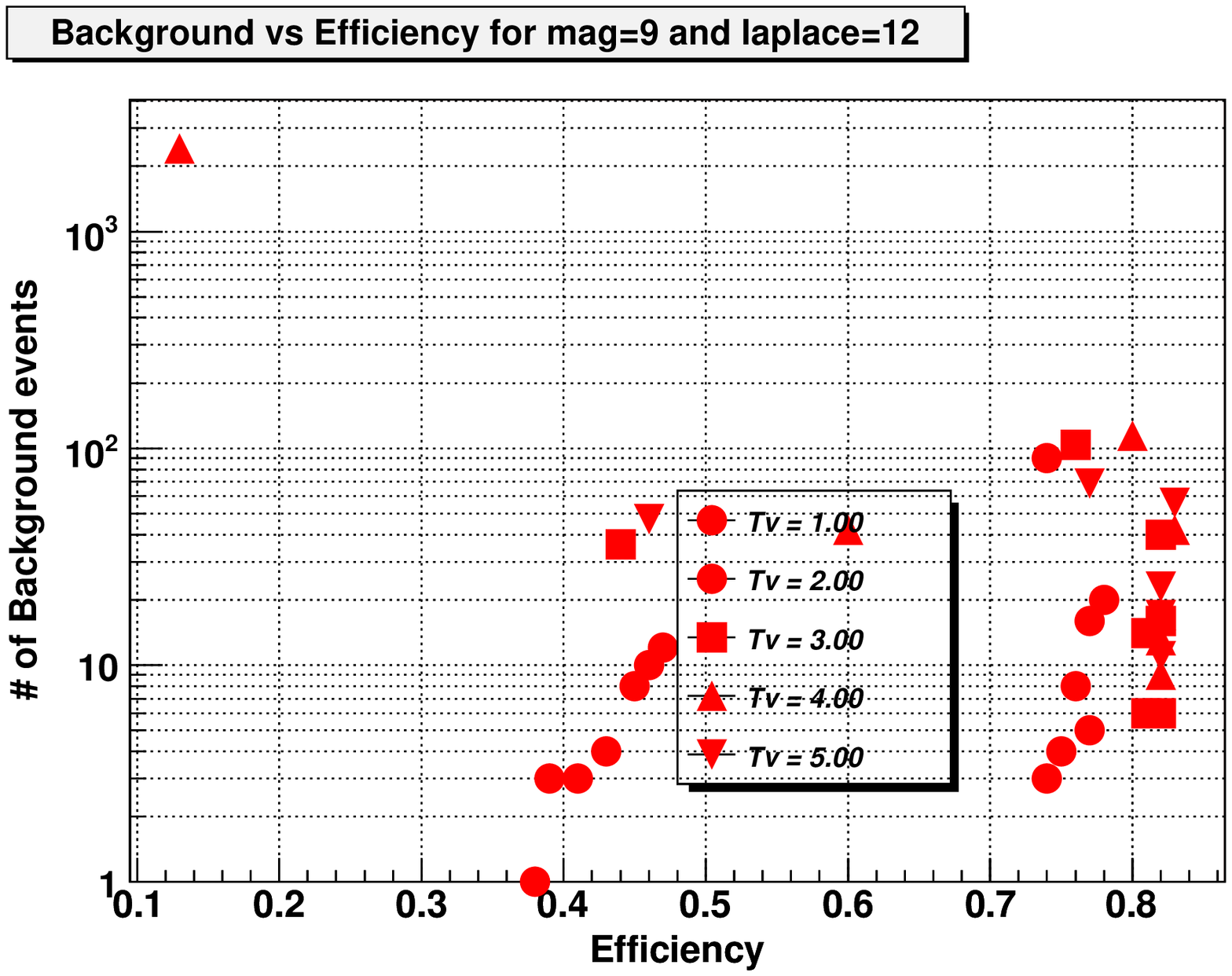}
		\includegraphics[width=2.8in,height=2.8in]{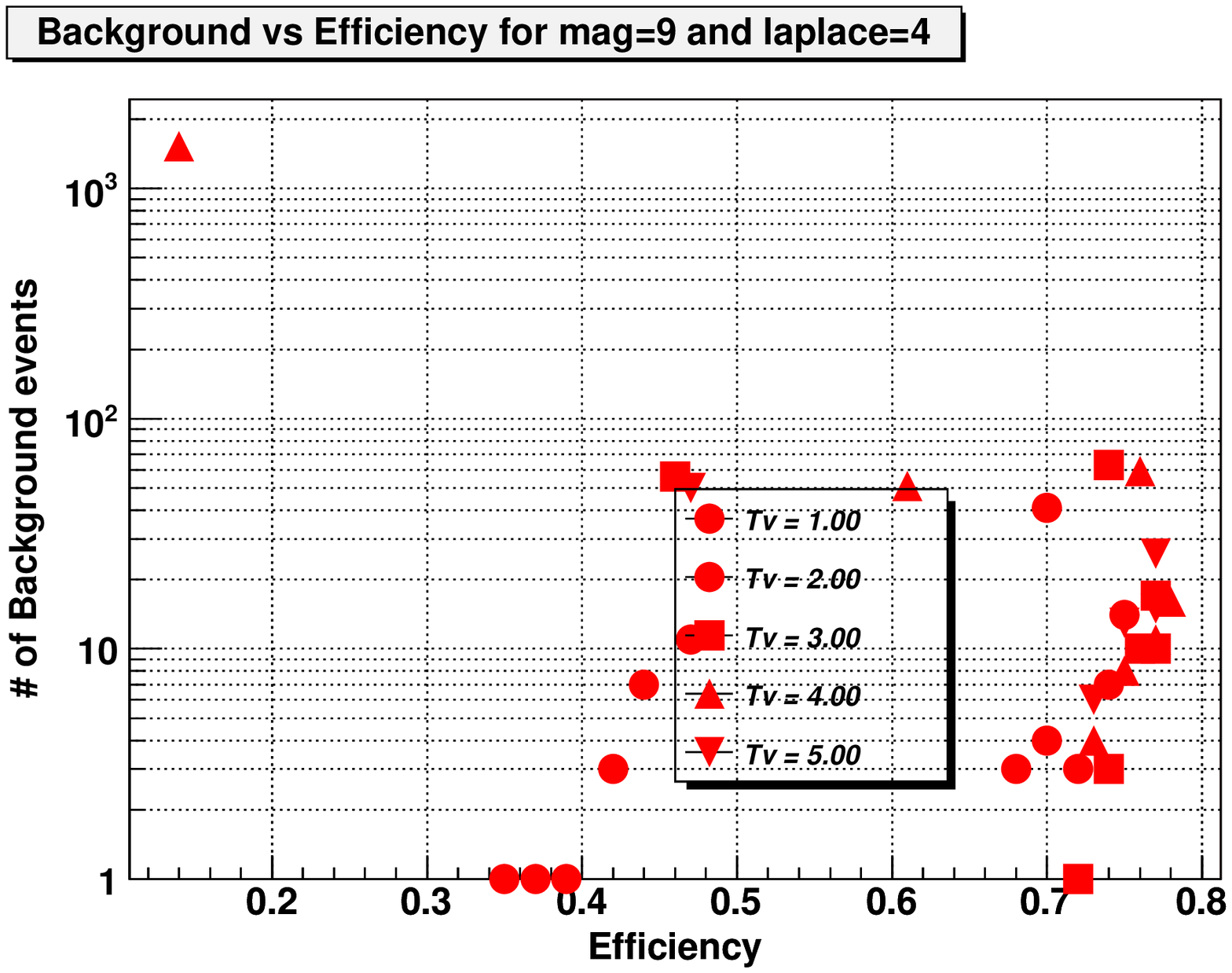}
    \else
		\includegraphics[width=2.8in,height=2.8in]{EffBkg/ccddouble/mag9/laplace12/eff_bkg_9.00_lap12.eps}
		\includegraphics[width=2.8in,height=2.8in]{EffBkg/ccddouble/mag9/laplace4/eff_bkg_9.00_lap4.eps}
    \fi
    \caption{Results of efficiency of $9^m$ flash recognition and background rejection of coincidence algorithm. Tests were performed
				 on data from night 2006-05-26/27 for laplace=4 (left plot) and laplace=12 (right plot)}
    \label{fig_fig_eff_vs_bkg_coinc_9mag}
  \end{center}
\end{figure}

The efficiency and background tests were also performed for the coincidence
algorithm. The Figures \ref{fig_fig_eff_vs_bkg_coinc_9mag} show best points of $T_n$ and
$T_v$ thresholds. 

\begin{table}[htbp]
\begin{center}
\begin{tabular}{|c|c|c|c|c|c|}
\hline
\textbf{Magnitude} & \textbf{Laplace} & \textbf{Tv} & \textbf{Tn} & \textbf{Eff} & \textbf{Bkg} \\
\hline
\hline
9 & 4 & 2.0 & 5.0 & 0.75 & 14 \\
9 & 4 & 2.0 & 6.0 & 0.74 & 7 \\
9 & 4 & 3.0 & 5.0 & 0.74 & 63 \\
9 & 4 & 3.0 & 6.0 & 0.77 & 17 \\
\hline
\hline
9 & 12 & 2.0 & 5.0 & 0.78 & 20 \\
9 & 12 & 2.0 & 6.0 & 0.77 & 5 \\
9 & 12 & 3.0 & 6.0 & 0.82 & 40 \\
9 & 12 & 3.0 & 7.0 & 0.82 & 16 \\
\hline
\end{tabular}
\caption{Best values of efficiency of coincidence algorithm obtained for simulated flashes of brightness 9$^m$, tested on 1952 images from night 2006-05-26/27.}
\label{tab_best_eff_vs_bkg}
\end{center}
\end{table}

\begin{figure}[!htbp]
  \begin{center}
    \leavevmode
    \ifpdf
		\includegraphics[width=2.8in,height=2.8in]{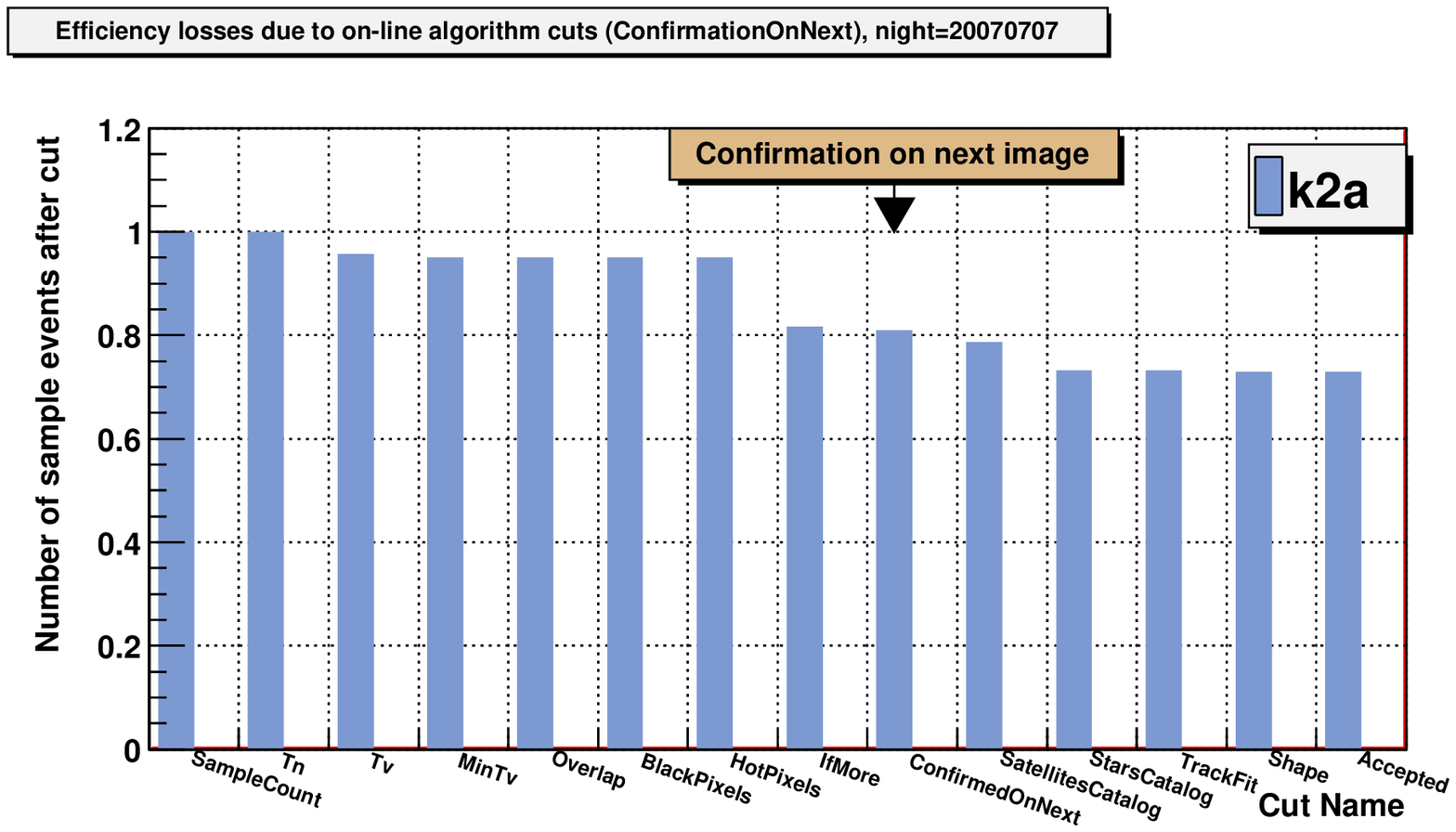}
		\includegraphics[width=2.8in,height=2.8in]{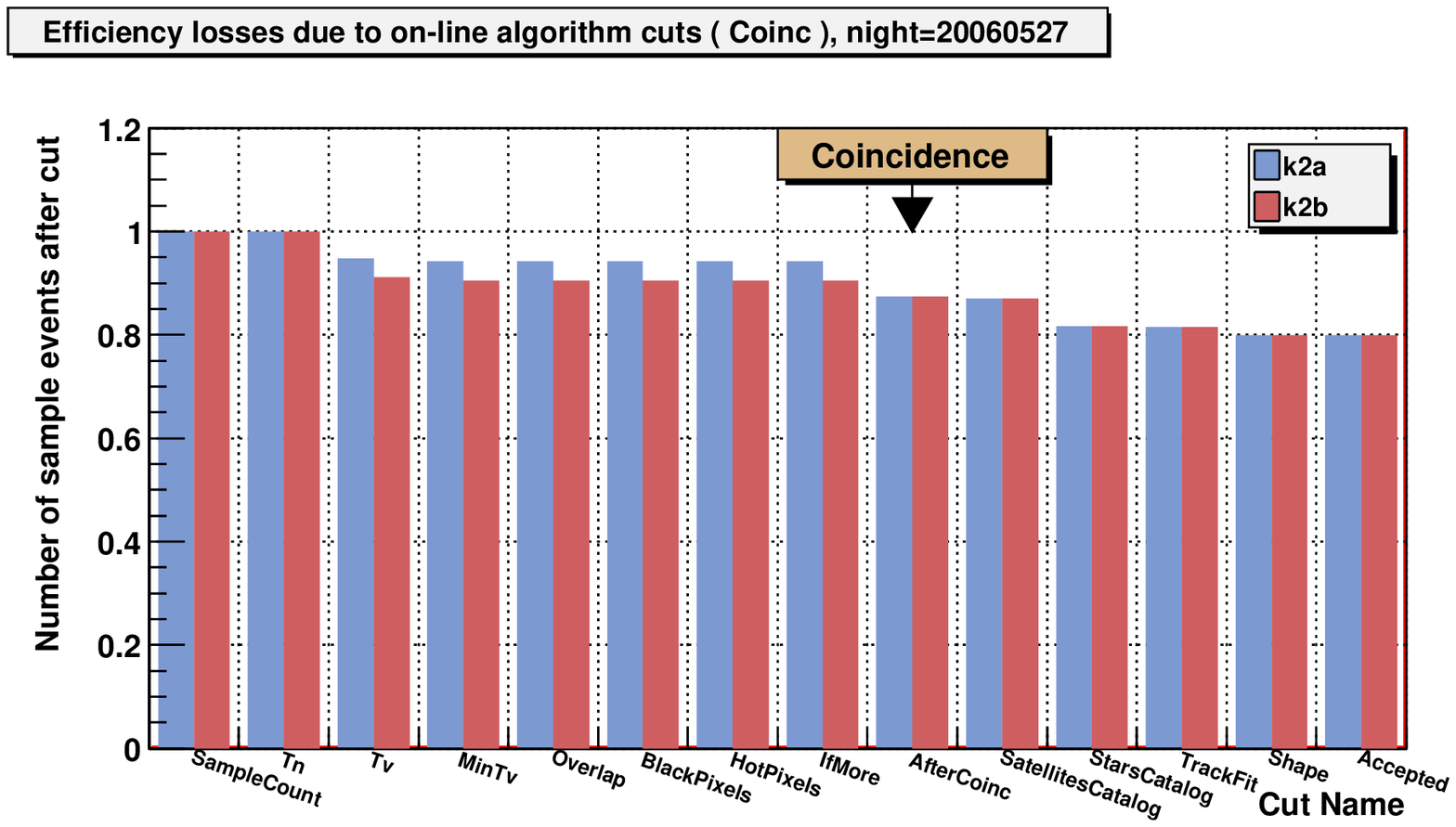}
    \else
		\includegraphics[width=2.8in,height=2.8in]{EffLoss/ccdsingle/20070707_confirm.eps}
		\includegraphics[width=2.8in,height=2.8in]{EffLoss/ccddouble/20060527_coinc_k2a.eps}
    \fi
    \caption{Efficiency losses due to subsequent cuts of on-line algorithm : confirmation on next image (left plot), coincidence (right plot)}
    \label{fig_efflosses}
  \end{center}
\end{figure}

The efficiency losses due to subsequent cuts of on-line algorithm was tested.
It was done by counting how many of generated samples were rejected by 
on-line algorithm cuts. Figure \ref{fig_efflosses} shows efficiency losses 
 due to subsequent cuts for data collected during single night.

\begin{figure}[!htbp]
  \begin{center}
    \leavevmode
    \ifpdf
		\includegraphics[width=2.8in,height=2.8in]{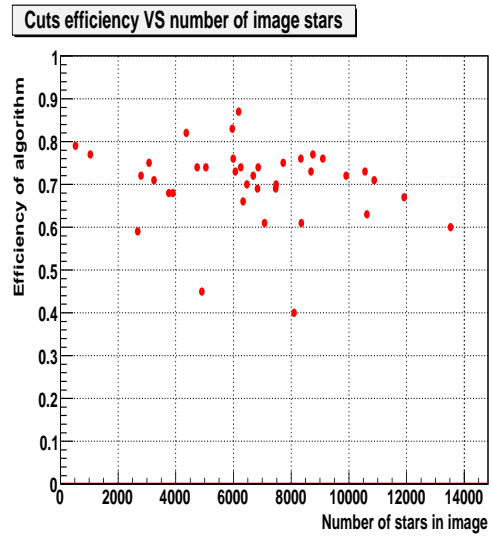}
    \else
		\includegraphics[width=2.8in,height=2.8in]{EffBkg/ccdsingle/EffVsNight/eff_vs_stars.eps}
    \fi
    \caption{Acceptance of on-line algorithm cuts in function of average number of stars on all images collected during a single night. Each point represents efficiency during a single night.}
    \label{fig_eff_vs_night}
  \end{center}
\end{figure}

The efficiency of on-line algorithm cuts was determined for several
different nights ( Fig. \ref{fig_eff_vs_night} ). The mean efficiency is $\approx 70$\%.
The testing procedure which pasts samples of stars into image does not take
clouds into account. In fact this procedure allows to estimate efficiency
of all cuts after Tn cut. This is because sample is pasted in an image on
top of clouds, so there is no possibility to have large loss of efficiency 
due to clouds in such kind of testing procedure.
The average efficiency of Tn cut was estimated as $\epsilon_{Tn}\approx$0.49 with usage of the TYCHO-2 star catalog
and cataloging procedure ( see Sec. \ref{sec_red_eff} ). This gives average
efficiency of flash identification algorithm  $\epsilon_{algo}\approx$0.35.

\subsection{Sources of background}
\begin{figure}[!htbp]
  \begin{center}
    \leavevmode
    \ifpdf
      \includegraphics[width=6in]{cosmic20040915_x2.gif}
    \else
      \includegraphics[width=6in]{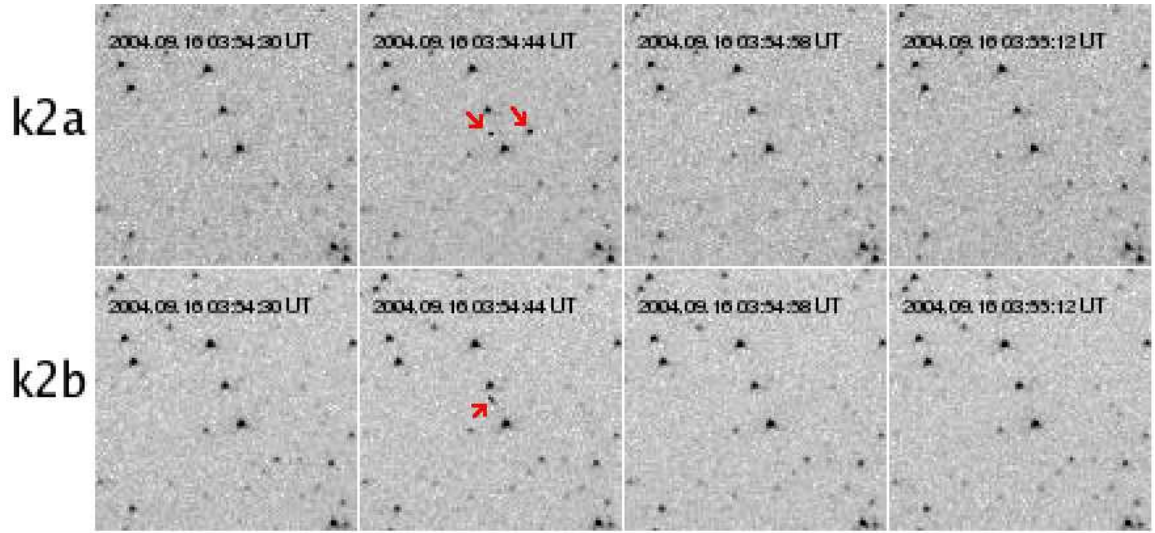}
    \fi
    \caption{Rare example of coincidence of two cosmic ray hits}
    \label{fig_bkg_cosmic_coinc}
  \end{center}
\end{figure}

\begin{figure}[!htbp]
  \begin{center}
    \leavevmode
    \ifpdf
      \includegraphics[width=6in]{plane20040912.gif}
    \else
      \includegraphics[width=6in]{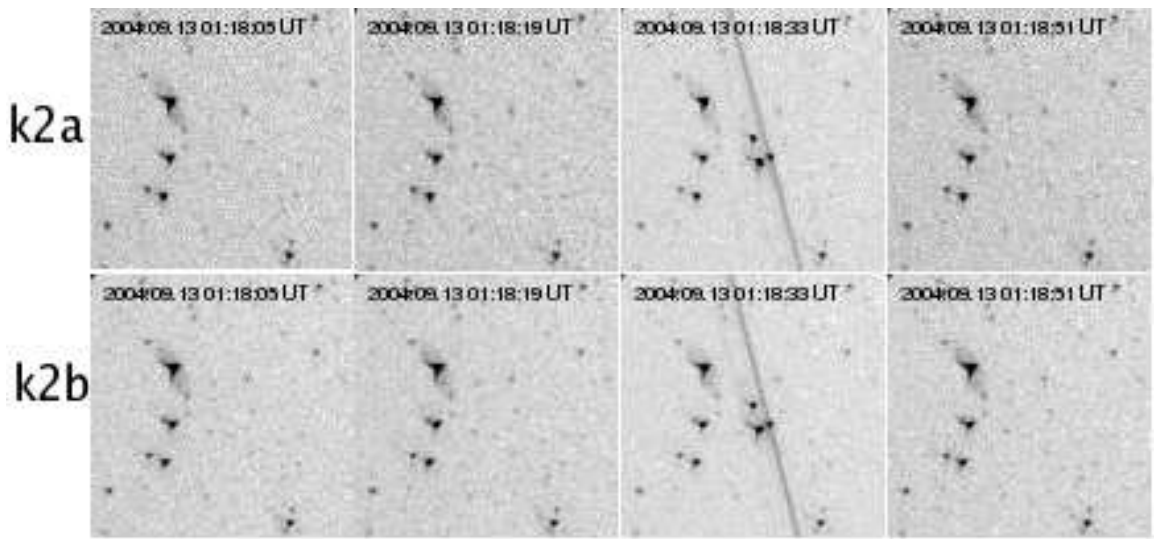}
    \fi
    \caption{Plane-like background event}
    \label{fig_bkg_plane}
  \end{center}
\end{figure}

\begin{figure}[!htbp]
  \begin{center}
    \leavevmode
    \ifpdf
      \includegraphics[width=6in]{meteor20050501.gif}
    \else
      \includegraphics[width=6in]{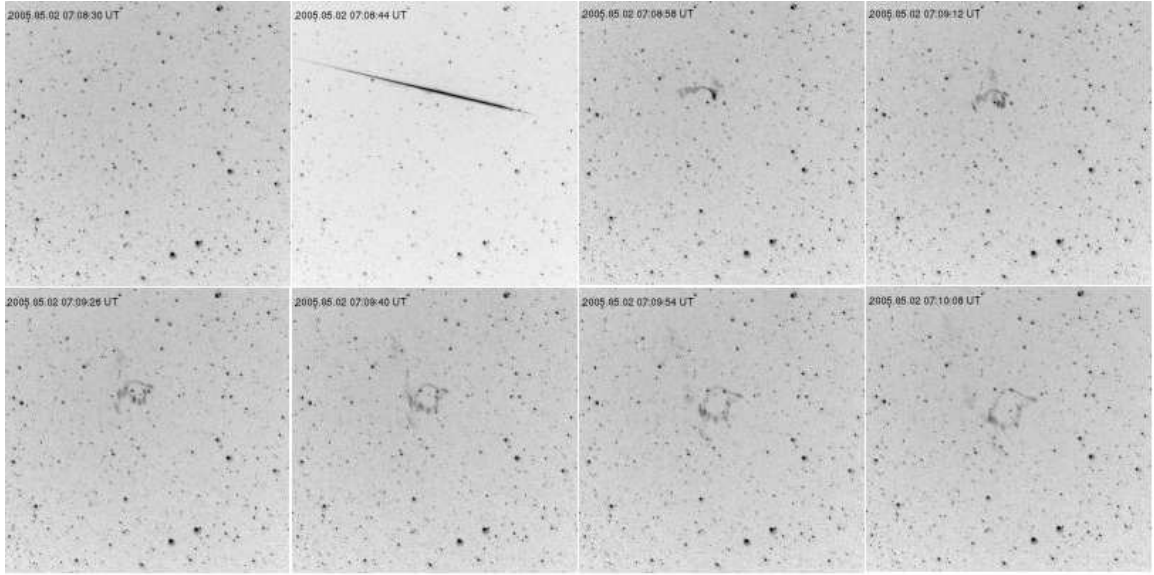}
    \fi
    \caption{Meteor trace blown by the wind}
    \label{fig_bkg_meteor}
  \end{center}
\end{figure}

\begin{figure}[!htbp]
  \begin{center}
    \leavevmode
    \ifpdf
      \includegraphics[width=6in]{flotilla20060915.gif}
    \else
      \includegraphics[width=6in]{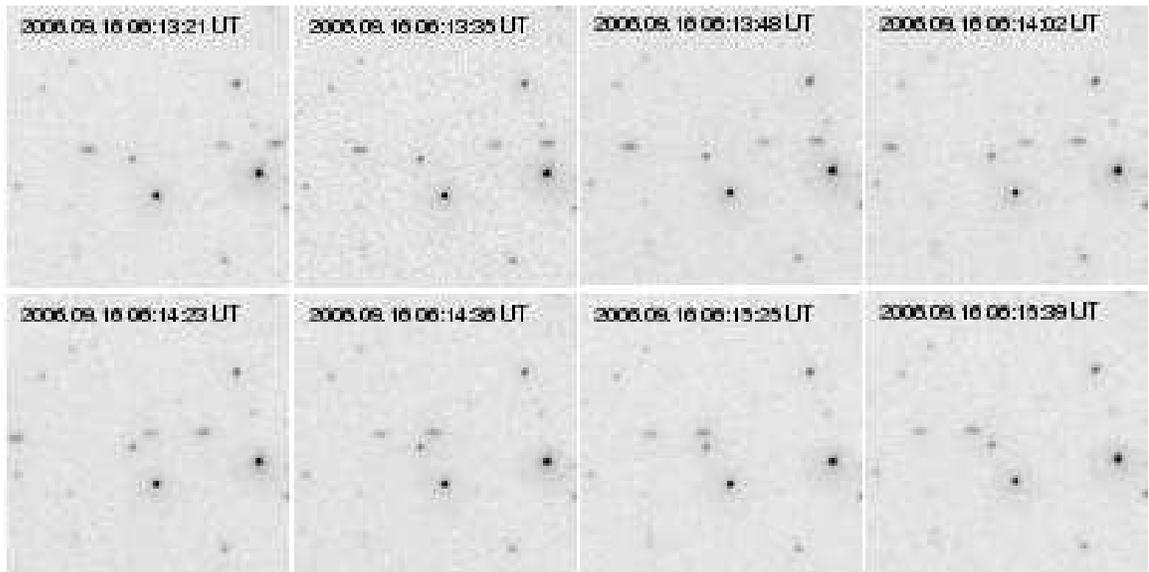}
    \fi
    \caption{Flotilla of artificial satellites}
    \label{fig_bkg_satellites}
  \end{center}
\end{figure}

Final list of events from one night of the prototype work did not exceed 100 events
, but typically remained on the level of 10 events per night. 
These events were mainly due to background events. The main sources of
the background were flashing satellites, planes and meteors. 
There was also background due to clouds, in this case usually number of 
events on cloudy images was big so it was easy to simply reject whole image.
Summary of background events statistics is given in Table
\ref{tab_online_background}. This table is divided into periods and 
types of algorithms in the cases where more then one algorithm was running.
Example of background event images are given in Figures
\ref{fig_bkg_cosmic_coinc}, \ref{fig_bkg_plane}, \ref{fig_bkg_meteor} and \ref{fig_bkg_satellites}.

\begin{landscape}
\begin{table}[htbp]
\begin{center}
\begin{tiny}
\begin{tabular}{|c|c|c|c|c|c|c|c|c|c|c|c|c|c|}
\hline
\textbf{Period} & \textbf{Algo} & \textbf{Flash}  & \textbf{Satellite} & \textbf{\begin{minipage}{1.2cm}{Flash or satellite}\end{minipage}} & \textbf{Plane} & \textbf{\begin{minipage}{1.cm}{Satellite or plane}\end{minipage}} & \textbf{Other}\footnotemark & \textbf{Meteor} & \textbf{Clouds} & \textbf{Hot Pixel} & \textbf{\begin{minipage}{1.2cm}{Saturated Star}\end{minipage}} & \textbf{\begin{minipage}{1.2cm}{Opened Shutter}\end{minipage}} & \textbf{\begin{minipage}{1.cm}{System Error}\end{minipage}} \\
\hline 
 & & & & & & & & & & & & & \\
\hline
2004.06.25 - 2005.01.20 & coinc. & 70      & 572 & 29 & 392 & 134 & 435 &  - &  - & - & - & - & - \\
\hline
2004.06.25 - 2005.08.09 & coinc. & 125     & 961 & 53 & 819 & 455 & 1810 & - &  - & - & - & - & - \\
\hline
2005.04.21 - 2005.08.09 & conf. next & 23 &  31 &  8 &  -  &   2 & 3676 & - &  - & - & - & - & - \\
\hline
2006.05.20 - 2006.08.08 & coinc. & 42 & 352 & 10 & 262 & 324 & 1968 & - &  - & - & - & - & - \\
\hline
2006.05.20 - 2006.12.31 & conf. next & 3 & 118 & 2 & 6 & 12 & 4817 & 0 & 113995 & 168 & 12 & 60 & 64 \\
\hline
2006.10.01 - 2006.12.31 & conf. next & 1 & 2 & 1 & - & 0 & 51 & 0 & 113995 & 168 & 12 & 60 & 64 \\
\hline
2007.01.01 - 2007.05.29 & conf. next & 2 & 363 & 7 & - & 7 & 46725 & 0 & 34729 & 2151 & 298 & 329 & 64445 \\
\hline
\end{tabular}
\end{tiny}
\caption{Statistics of classification of events from on-line algorithm in period 2006.06 - 2007.06}
\label{tab_online_background}
\end{center}
\end{table}
\footnotetext{Before 2006-10-01 Other was category for all later types in table}
\end{landscape}

\newpage

\subsection{Final verification of events}
\label{sec_flashes_verification}
The final list of accepted events is very small. However, the prototype does
not have a possibility to definitely reject all background events. 
The most difficult part of the background are flashing satellites. 
They are mostly rejected by catalog and track cuts described in the previous 
sections, but some objects are not cataloged and flash too rarely to 
be rejected by these criteria. 
Final events can be evaluated by several checks. In the case when object suddenly
appears and remains visible in the next several images in the same position 
it is very probable that it is not a satellite. Assuming the object is
Earth's satellite the following formula can be derived : \\

\begin{equation}
R > \left( \frac{\sqrt{G \cdot M} \cdot \delta t }{\alpha} \right)^{(2/3)} \approx 16700 \cdot \left( \frac{\delta t_{sec}}{\alpha_{arcmin}} \right)^{(2/3)} [km]
\label{many_pixels_contraint} 
\end{equation}

where $\alpha$ is the angular distance of the object in consecutive 2 images and 
 $\delta$t is the time separation of images. They can be substituted in 
arcmin and seconds respectively if formula on the right is used.
In the case of the prototype $\alpha \approx 0.6'$ which corresponds to a single
pixel and $\delta t \approx 12s$ which corresponds to time separation of 2
subsequent images ( $T_{exposure} + T_{dead}$= 10s $+$ 2s ). The minimal
distance of the object visible on 2 consecutive images in the same position
derived from these values is D $\approx$ 123 000 km.
For comparison, the geostationary orbit is $R_{geostat} \approx $42160 km
( from the Earth center ).

\begin{figure}[!htbp]
  \begin{center}
    \leavevmode
    \ifpdf
      \includegraphics[width=6in]{sat_distance_20041212.gif}
    \else
      \includegraphics[width=6in]{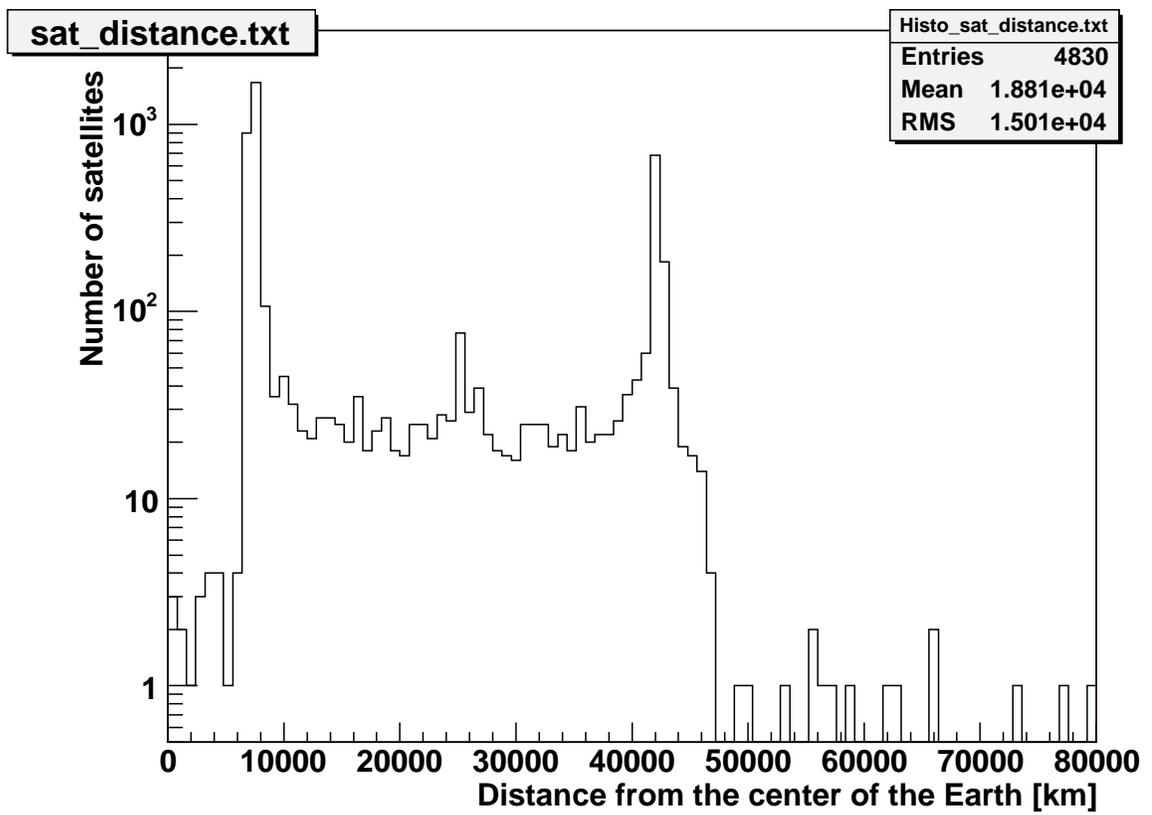}
    \fi
    \caption{Distribution of distances of artificial satellites from the center of the Earth. Peak at $\approx$ 42000 km is due to geostationary satellites.}
    \label{fig_sat_distance_distr}
  \end{center}
\end{figure}

\begin{figure}[!htbp]
  \begin{center}
    \leavevmode
    \ifpdf
		\includegraphics[width=6in]{satcone.gif}
      \includegraphics[width=6in]{earth_shadow.gif}
    \else
		\includegraphics[width=6in]{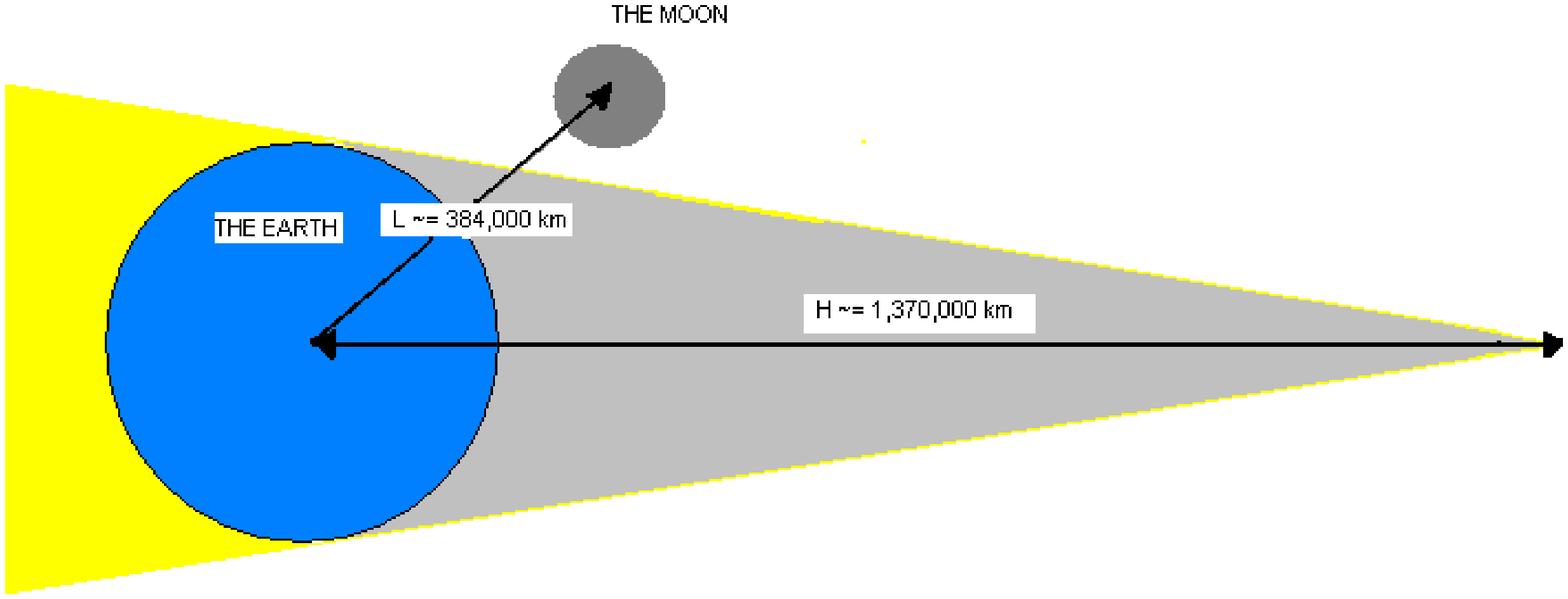}
      \includegraphics[width=6in]{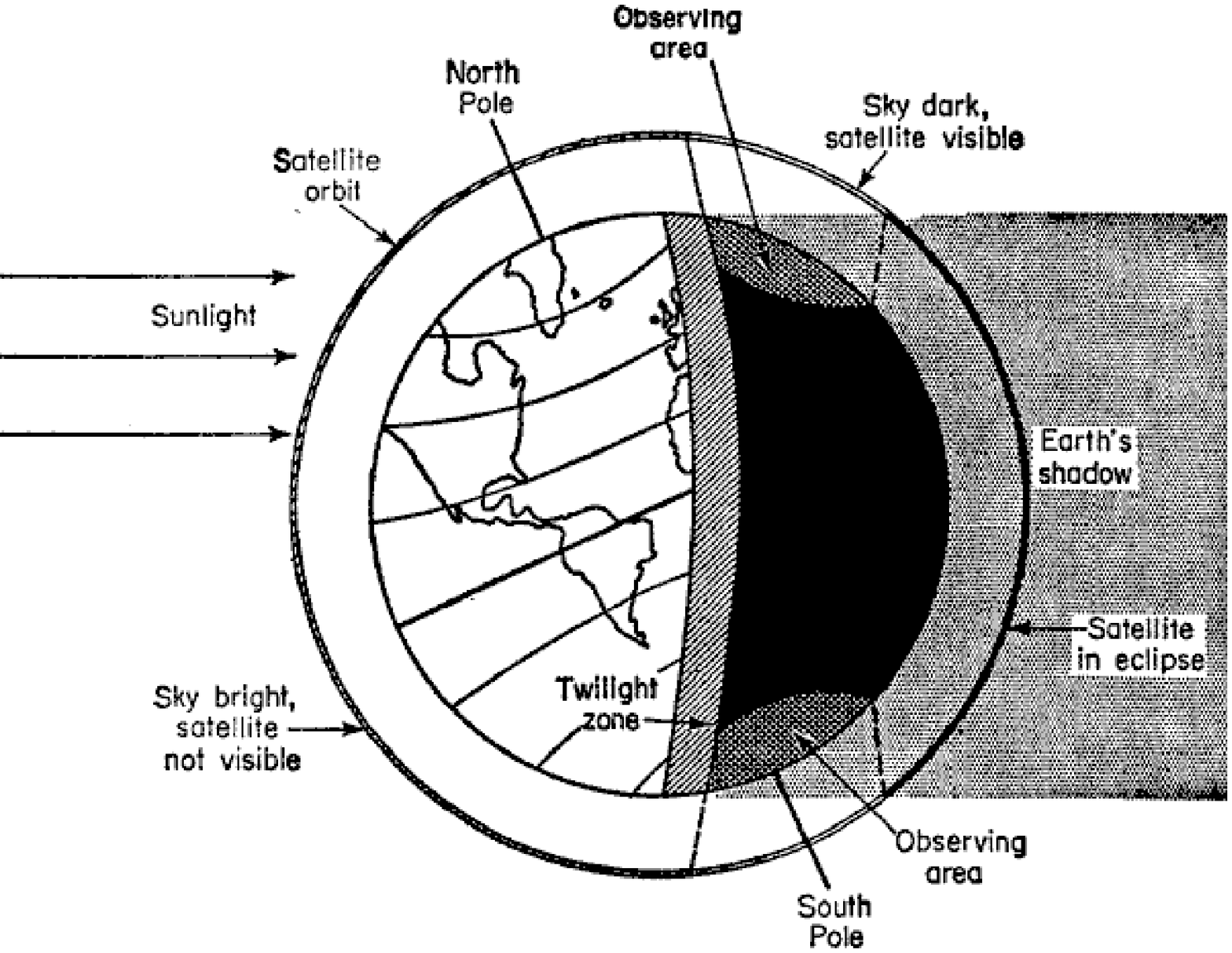}
    \fi
    \caption{Positions where satellites can reflect sun light}
    \label{fig_earth_shadow}
  \end{center}
\end{figure}

The distribution of the distance from the Earth to satellites in the catalog
is shown in Figure \ref{fig_sat_distance_distr}. Peak from geostationary 
satellites is clearly visible. There are not many satellites 
more distant then 50000 km, which supports "double image" events. 
However, it is possible that these flashes are caused by spacecrafts on the 
long missions which are very far objects and can also reflect sun light
towards the Earth. The probability of such events is very small.
The above check can not be applied to events visible only in a single image. 
For this class of events another check was implemented. Flashing satellites
can only reflect sun light when they are outside the Earth shadow cone and 
not on the illuminated side of the Earth (Fig. \ref{fig_earth_shadow}).

Knowing the position and time of the flash it is possible to calculate the
satellite's minimal distance from the Earth to be outside of the shadow
cone which is required for the satellite to cause a flash-like event.
The Earth shadow cone size is $H_{cone} \approx$ 1,37 mln km ( Fig. \ref{fig_earth_shadow} ), which is more then 
Moon's orbit size ( $R_{moon}$ $\approx$ 400000 km). In case any of the flash candidates would have $D_{cone} > R_{moon}$
 it would be most probably an event of astrophysical origin.

\section{Off-line data analysis}
Off-line data analysis acts on data reduced and cataloged to the database. 
The reduction chain consists of several steps which will be described in
detail in the next subsection. After this chain, star brightness measurements 
are stored in the database which is optimized for fast access.
The structure of the database will be described in subsection \ref{sec_star_catalog}.
In the last subsection algorithms for detecting brightness increases and 
new object appearance will be described. 

\subsection{Reduction pipeline}
The aim of the reduction procedure is to reduce raw data stored as images 
in \texttt{fits} files into essential data describing stars coordinates and brightness.
This allows to reduce the amount of data by factor of $\approx 10$ in the case of
single images and $\approx 100$ in the case of 20 averaged images. 

\subsubsection{Image reduction}
\label{sec_image_reduction}

Every image collected during a night is processed in the same way. 
Off-line data analysis described in this thesis was performed on data obtained by
averaging 20 subsequent images. There are also other reduction pipelines 
for reducing single image and scan images ( averaging 3 scan images ), but 
almost all of the steps are the same. The main difference is that in pipelines 
acting on averaged images there is an additional step calculating 
average of specified number of images.
Image reduction consists of the following steps :
\begin{itemize}
\item image averaging - it is present in reduction pipelines acting on
averaged images. In the case of \texttt{aver20} pipeline 20 subsequent images are averaged 
 and in the case of pipeline \texttt{scan3} 3 subsequent images are averaged.
Image coordinates are controlled and in case they change, average chain
is stopped not to allow for averaging images from different positions.
In the case of single image reduction the image averaging step is skipped.
\item Dark frame subtraction - the reason for dark frame subtraction was already
described in Section \ref{sec_online_reduction}.
As described in Section \ref{system_nightly_performance} in order to reduce
fluctuations the dark image is calculated as a median of several dark images.
This step allows to subtract signal offset produced by dark current and electronics.
It also reduces the effect of hot pixels.
\item Division by flat image - this step allows to correct for
non-uniformity of the optics and differences between pixels amplifications.
Standard way of finding this correction is taking images of
uniformly illuminated field. It is usually the sky just after dusk or just
before dawn, when the sky is bright and stars are not visible.
An alternative way is to use uniformly illuminated screen. In case of wide
field observations it is very difficult to obtain proper flat image.
It is due to difficulty of obtaining uniformly illuminated field of size 
of FOV $\ge 20^\circ$. The evening sky just after sunset is uniformly illuminated in
scale of arc minutes, but in scale of degrees non uniformities due to sky
gradient are significant. Due to this problem flat image is obtained by taking
images of evening sky with the mount tracking switched off. After taking many images and
calculating median image stars are eliminated, finally the image is normalized
to one. This procedure requires collection of many ($\geq$200) images so it is
performed rarely and for most of the time the same flat image is used in
analysis.
\end{itemize}

After the above operations the image is ready for photometry. The photometry 
is a procedure which finds stars in the image and determines their chip coordinates
(x,y) and brightness. 
In the "Pi of the Sky" data analysis two photometry procedures are
used depending on the type of reduction pipeline :
\begin{itemize}
\item ASAS photometry - aperture photometry adopted from ASAS experiment \cite{ASAS}.
It is rather slow so cannot be used for reduction of all single
images from a night. It is used for photometry of 20 averaged images ( $\approx 20 \cdot 12 = 240$ sec timescale ) 
and in reduction of scan images ( 3 images averaged ).
\item Fast photometry - it is fast, aperture photometry algorithm. Simple
aperture is used to calculate star brightness. This photometry is used 
on-line by \texttt{DAQ} to perform astrometry every 300 sec and in reduction of all
night images ( ASAS photometry is too slow for this purpose )

\begin{figure}[!htbp]
  \begin{center}
    \leavevmode
    \ifpdf
      \includegraphics[width=4in,height=4in]{apert9x9+9.gif}
    \else
      \includegraphics[width=4in,height=4in]{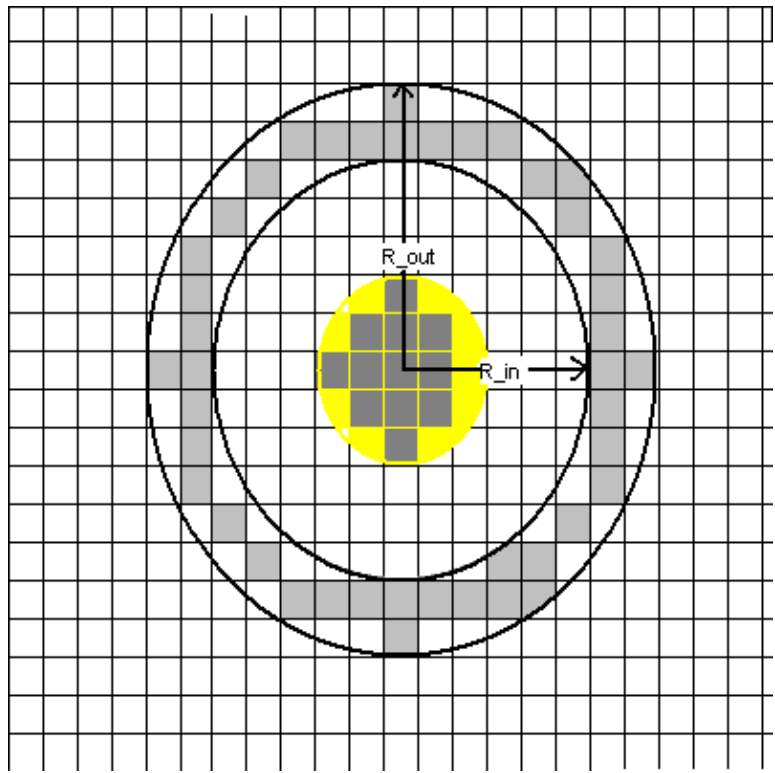}
    \fi
    \caption{Aperture used in fast photometry algorithm}
    \label{fig_fast_aperture}
  \end{center}
\end{figure}

Aperture used for brightness calculation is shown in Figure \ref{fig_fast_aperture}. 
Final brightness of star is determined as :\\
\begin{equation}
	I = \sum_{i}^{N_+} P_{+} - N_{+} \cdot B_{sky}
\label{eq_piphoto_mag}
\end{equation}
where $B_{sky}$ stands for average value of sky background near analysed
star. This value is obtained as median value of pixels in gray ring around
the star ( Fig. \ref{fig_fast_aperture} ). $P_{+}$ denotes signal pixels, shown as
dark gray pixels inside the star, they form 3x3 square around the maximum
brightness pixel with 3 most bright pixels contiguous to the square sides.
This procedure is not good for dense fields where sky background can be
calculated incorrectly due to stars entering the "background ring" and also 
$\sum_{i}^{N_+} P_{+}$ is affected by overlapping stars.
Instrumental magnitude is calculated from the formula : \\
\begin{equation}
	m_{\pi} = - 2,5 log(I) 
\label{eq_mag_def}
\end{equation}
Star coordinates (x,y) are determined as centroid of cluster of pixels according
to the formula : \\
\begin{equation}
	X_{star} = \frac{\sum_{cluster} x_i \cdot V_i}{\sum_{cluster} V_i} \hspace{2mm},\hspace{2mm}
	Y_{star} = \frac{\sum_{cluster} y_i \cdot V_i}{\sum_{cluster} V_i}
\label{eq_star_center}
\end{equation}
Cluster of pixels is determined as pixels around the star which satisfy
P(x,y) > $T_{cluster} = 3.5 \cdot \sigma$.
\end{itemize}

Both photometry procedures write resulting list of stars with (x,y)
coordinates and magnitudes to output \texttt{mag} files. The format of this 
files is similar to \texttt{fits} format. They consist of header which is taken from
\texttt{fits} header with some additional fields added, after the header section, 
list of stars in binary format is written. The \texttt{mag} files are input files for astrometry procedure. \\
This procedure finds transformation T:(x,y) $\rightarrow (\lambda,\delta)$, the 
transformation T is described by the following formula :\\

\begin{equation}
\lambda = \sum_{i,j \leq O} P_{ij} \cdot x^i \cdot y^i \hspace{2mm},\hspace{2mm}
\delta = \sum_{i,j \leq O} Q_{ij} \cdot x^i \cdot y^i
\label{eq_astro_transform} 
\end{equation}

Where O is the order of the transformation, in current configuration O=4 
and due to this fact coefficients $P_{14},P_{23},P_{24},P_{32},P_{33},P_{34},P_{41},P_{42},P_{43},P_{44}=0$
and also corresponding Q coefficients are zero.
It allows to calculate equatorial coordinates for any chip coordinate (x,y). 
Astrometry requires input information about image center position, 
pixscale \footnote{pixscale is an angular size of the CCD pixel} and rotation angle of the image with respect to the celestial
coordinates, these settings are read from header of \texttt{mag} file.
The astrometry procedure was adopted from the ASAS experiment, it is an
iteration procedure where stars in \texttt{mag} file are matched against catalog
stars in given position in the sky. Star catalog currently used in the
procedure is based on TYCHO catalog, however any star catalog can be used
instead. The procedure consists of the following main steps : \\

\begin{itemize}
\item loading of \texttt{mag} file
\item read stars from the reference star catalog
\item estimation of shift from the expected position $(\lambda,\delta)_{mount}$
and real position $(\lambda,\delta)$ using the correlation function between
image stars and stars in the catalog
\end{itemize}

After the above initialization steps the iterational procedure begins, every 
iteration consists of the following main steps :\\

\begin{itemize}
\item recalculation of reference and image stars coordinates to standard coordinates (
with respect to image center )
\item matching of image stars from \texttt{mag} file against reference stars
\item determination of transformation parameters by using the method of 
singular value decomposition (SVD).
\item check the convergence condition requiring astrometry error $\delta \alpha < \delta \alpha_{MAX}$
\item recalculation of image center coordinates $(\lambda,\delta)_{center}$
\end{itemize}

The iteration steps are repeated until the convergence conditions are satisfied.
In the case astrometry procedure converges, for all stars in the \texttt{mag}
file coordinates $(\lambda,\delta)$ are calculated from the formula \ref{eq_astro_transform}.
The results are saved to \texttt{ast} file which consists of same information as \texttt{mag} 
file with additional fields for $(\lambda,\delta)$.
All night images are processed in the same way and every sky image ( \texttt{fits}
file ) has a corresponding \texttt{ast} file.

\subsubsection{Star Catalog}
\label{sec_star_catalog}

\begin{figure}[!htbp]
  \begin{center}
    \leavevmode
    \ifpdf
      \includegraphics[width=6in,height=8.5in]{star_catalog_all.gif}
    \else
      \includegraphics[width=6in,height=8.5in]{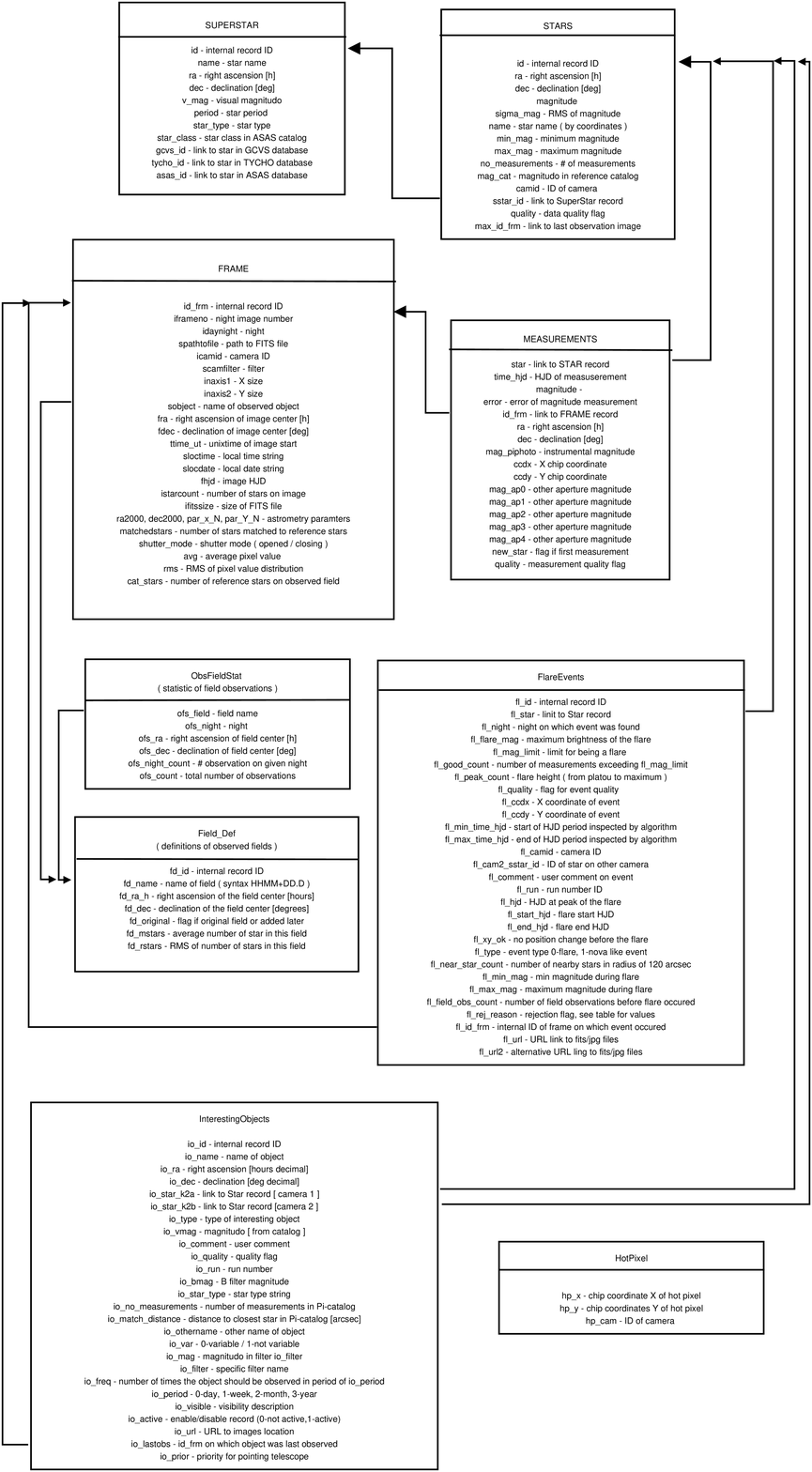}
    \fi
    \caption{Structure of star catalog database}
    \label{fig_starcat}
  \end{center}
\end{figure}

The star catalog is developed as relational PostgreSQL database \cite{POSTGRES}. 
The database structure is shown in Figure \ref{fig_starcat}.
It consists of tables described earlier in Section \ref{online_db_structure} with
additional tables :
\begin{itemize}
\item \textbf{Star} - this table contains all objects observed by given camera. 
Same physical star can be observed by many cameras so same physical star 
has N$\geq$1 records in the table \textbf{Star}, where N stands for number of cameras by
which this star was observed.
\item \textbf{Measurements} - this table stores information on every observation of
the star. It is linked to table \textbf{Star} by reference field \texttt{star}, it is also linked
by field  \texttt{id\_frm} to \textbf{Frame} on which the star was observed. 
\item \textbf{SuperStar} - it is a table containing real physical stars. In case 
star is observed by different cameras it has multiple records in table \textbf{Star},
but only one record in the table \textbf{SuperStar}. Every record in the table \textbf{Star} is linked
to corresponding \textbf{SuperStar} record by field \texttt{sstar\_id}. The relation between
\textbf{SuperStar} and \textbf{Star} table is one to many.
\item \textbf{ObsFieldStat} - statistical table containing information on number 
of images collected for a specific field
\item \textbf{Field\_Def} - definitions of fields observed by the system
\end{itemize} 

Star catalog database can be huge, after year of data collection it can reach 50-200 GB ( \texttt{aver20} database ), so in order to 
be efficiently used it must be optimized. An important element of the database structure are indexes. 
They allow searching indexed fields by fast binary search algorithm.
The most important database queries were optimized by creating indexes on fields used in conditional statements.
Another optimization performed on the database is placing \textbf{Measurements}
 records for a given star in the same physical location
on the disk. This is very important for fast reading of star light curves. 
This optimization is executed by PostgreSQL command CLUSTER which must be
called from time to time after large amount of new data is added to catalog.
Also table \textbf{Star} is optimized by CLUSTER command according to the celestial
coordinates in order to optimize reading of stars from database during cataloging.
The most important indexes are listed in Table \ref{tab_indexes_in_db}.
There are also database optimizations on PostgreSQL server configuration
level which set parameter value proper for huge database.

\begin{table}[htbp]
\begin{center}
\begin{tabular}{|c|c|c|c|}
\hline
\textbf{Index Name} & \textbf{Table} & \textbf{Indexed Fields} & \textbf{Clustered} \\
\hline
stars\_id\_index & STARS & ID & no \\
stars\_ra\_dec\_index & STARS & ra,dec & yes \\
stars\_dec\_index & STARS & dec & no \\
superstar\_gcvs\_id\_index & SUPERSTAR & gcvs\_id & no \\
measurements\_star\_index & MEASUREMENTS & star & yes \\
measurements\_id\_frm\_index & MEASUREMENTS & id\_frm & no \\
\hline 
\end{tabular}
\caption{Indexes most important for optimizing star catalog}
\label{tab_indexes_in_db}
\end{center}
\end{table}

\subsubsection{Cataloging procedure}
\label{sec_cataloging}

Reduced data is a set of \texttt{ast} files, these files have to be loaded to the
database structure described in the previous section. This task is called
cataloging and is performed by \texttt{piaddast2} program. 
Except of loading data to the database this program normalizes star
magnitudes according to V filter magnitudes in catalog of reference stars.
The block diagram of the \texttt{piaddast2} program is shown in Figure
\ref{fig_piaddast2}. Generally, cataloging procedure reads stars already
observed before from the database and matches new observations to these stars.
Subsequent \texttt{ast} files are organized in the memory until observation field 
changes which triggers dump of data from memory to the database and
selection of stars for new position. Every \texttt{ast} file is processed in the following way :

\begin{figure}[!htbp]
  \begin{center}
    \leavevmode
    \ifpdf
      \includegraphics[width=6in,height=8.5in]{piaddast2.gif}
    \else
      \includegraphics[width=6in,height=8.5in]{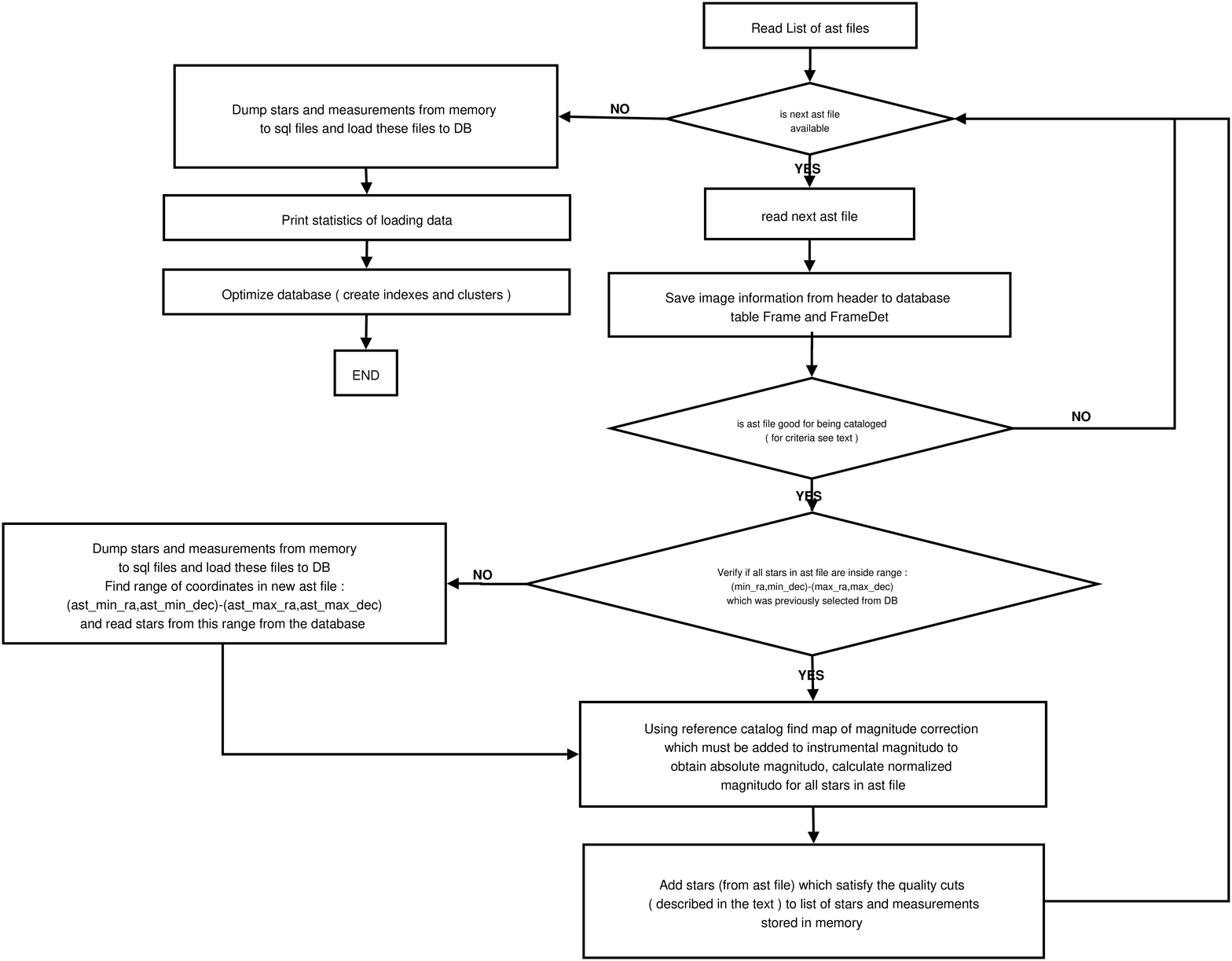}
    \fi
    \caption{Block diagram of the cataloging program}
    \label{fig_piaddast2}
  \end{center}
\end{figure}

\begin{itemize}
\item Read next \texttt{ast} file, determine range of celestial coordinates in \texttt{ast} file
	$(\lambda_{min}^{ast},\delta_{min}^{ast})-(\lambda_{max}^{ast},\delta_{max}^{ast})$
\item Save image header information to database tables \textbf{FRAME} and \textbf{FRAME\_DET}.
\item Check if all stars in \texttt{ast} file have coordinates in the range $(\lambda_{min}^{prev},\delta_{min}^{prev})-(\lambda_{max}^{prev},\delta_{max}^{prev})$. 
In case there are stars from outside this range program dumps stars and measurements stored in memory to \texttt{sql}
files and loads \texttt{sql} files to the database, then selects stars from database with coordinates matching 
range of coordinates in new \texttt{ast} file ( as determined in first step )
\item Check if the image represented by the \texttt{ast} file is good enough to be cataloged, the following criteria are
checked :\\
	\begin{itemize}
	\item number of images used for averaging ( in the case of cataloging averaged images )
	  for 20 averaged images the lower limit is 10 images and for \texttt{scan3} pipeline lower limit is 3. \\
	\item check if $N_{star}^{image} \geq N_{star}^{MIN} ( = 5000 ) $ , in
	  case number of stars is lower it means that it is probably cloudy image ( Fig. \ref{fig_image_stars} ) . 
	\item check if number of stars is not too high : $N_{star}^{image}  \leq N_{star}^{MAX} ( = 70000 ) $ ( Fig. \ref{fig_image_stars} ), which indicates bad image.
	\item check if average astrometry error is not too large : $A_{err} \leq A_{err}^{max} ( = 0.3 )$ ( Fig. \ref{fig_avg_ast_err} ) 
	\end{itemize}

\begin{figure}[!htbp]
  \begin{center}
    \leavevmode
    \ifpdf
      \includegraphics[width=4in,height=4in]{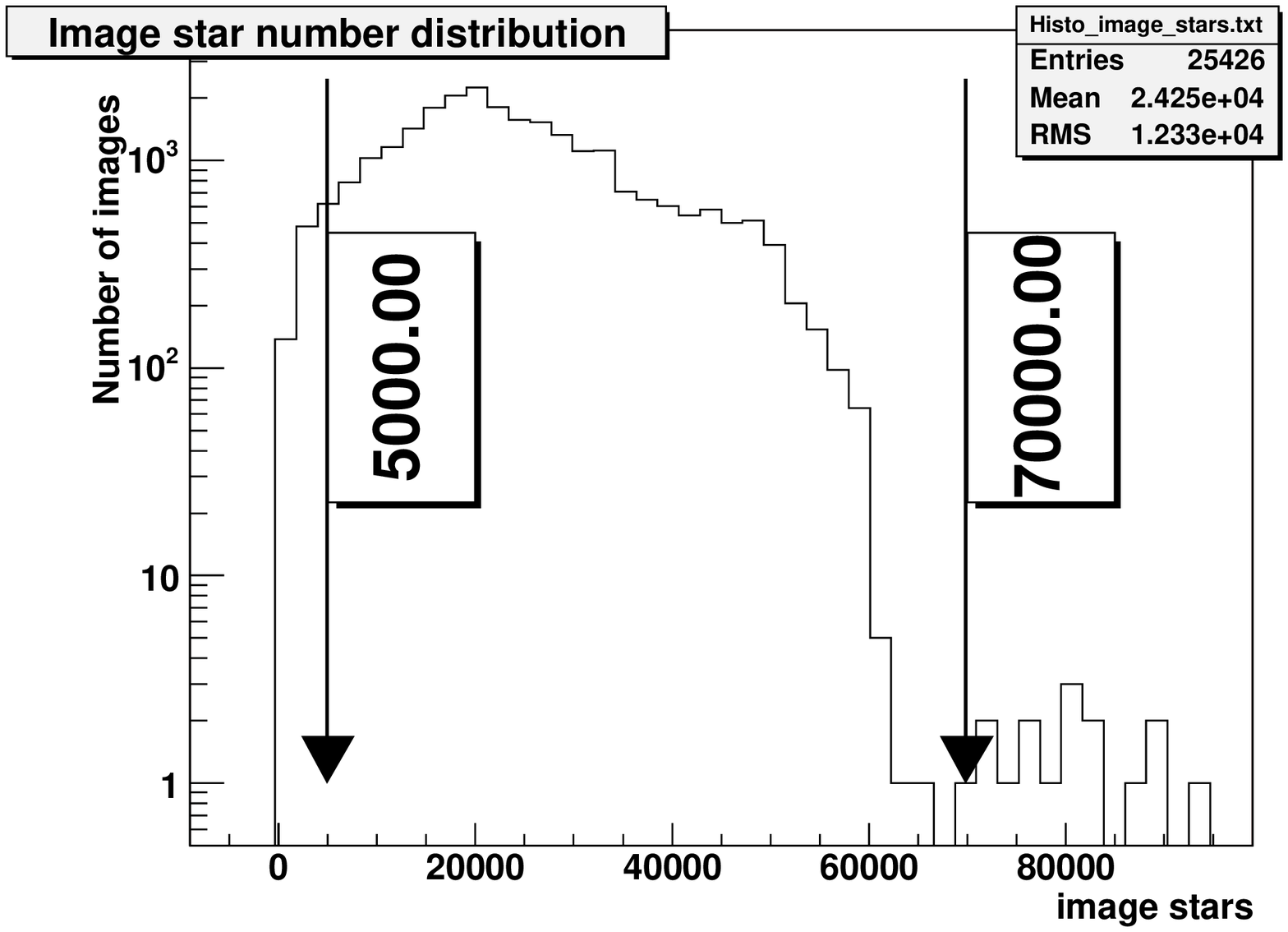}

		\includegraphics[width=4in,height=4in]{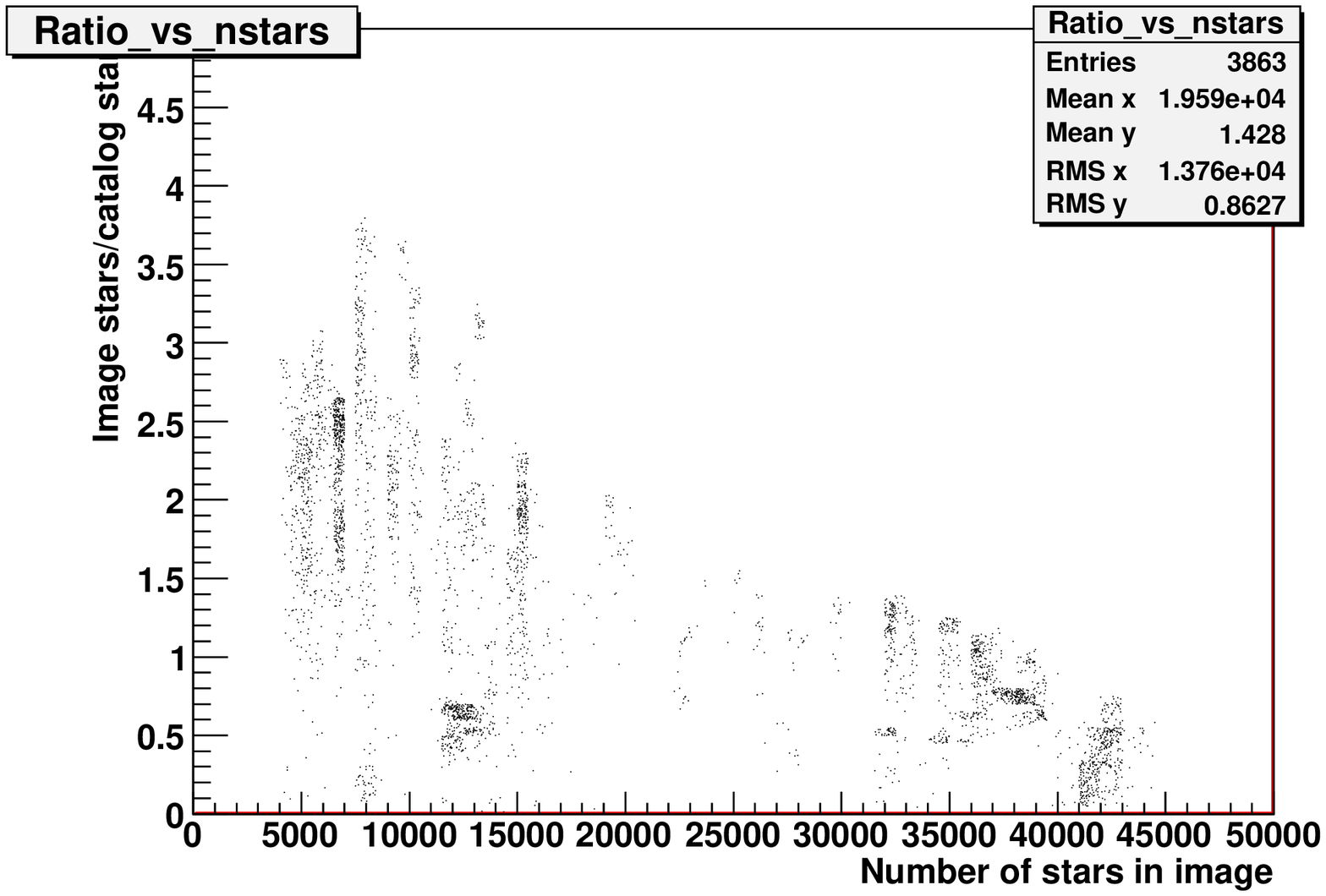}
    \else
      \includegraphics[width=4in,height=4in]{cat_cuts/image_stars.eps}

		\includegraphics[width=4in,height=4in]{cat_cuts/ratio_vs_nstars.eps}
    \fi
    \caption{Distribution of number of stars in an image (upper plot) and distribution of ratio of number of stars in an image to number of catalog stars in the observed field (lower plot)}
    \label{fig_image_stars}
  \end{center}
\end{figure}

\begin{figure}[!htbp]
  \begin{center}
    \leavevmode
    \ifpdf
      \includegraphics[width=4in,height=4in]{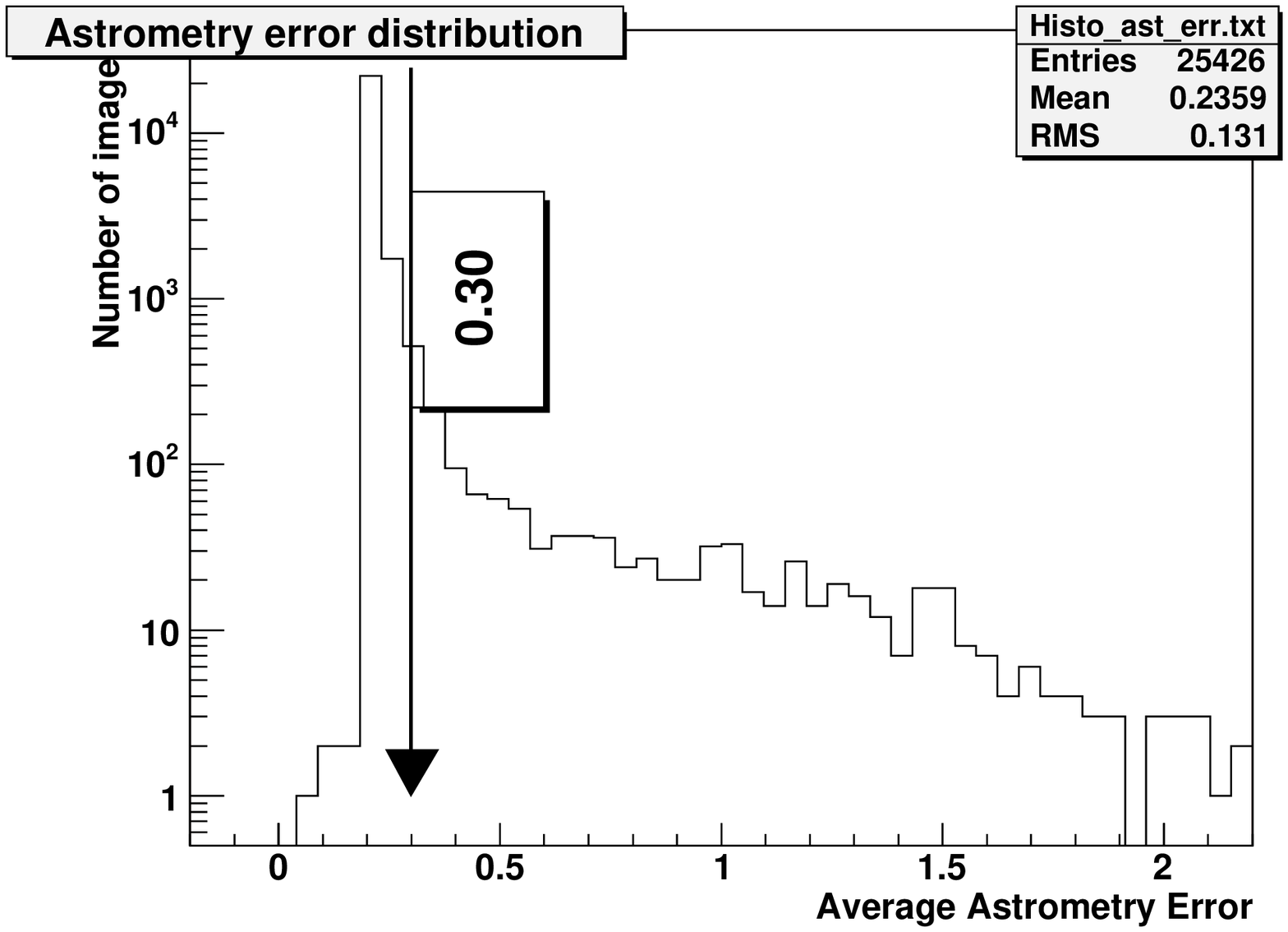}
    \else
      \includegraphics[width=4in,height=4in]{cat_cuts/ast_err_log.eps}
    \fi
    \caption{Distribution of average astrometry error in an image}
    \label{fig_avg_ast_err}
  \end{center}
\end{figure}

In case all criteria are satisfied \texttt{ast} file is accepted and stars are
cataloged, otherwise it is skipped. 
\item Each star in ast file is examined against certain quality cuts :\\
	\begin{enumerate}
	\item Star altitude is required to by h>$h_{min}$=15$^\circ$, in order to
reject measurements close to horizon which are of very poor quality
	\end{enumerate}
\item Star magnitudes are normalized by comparing with the catalog of reference
stars. Matching allows to create correction image ( Fig. \ref{fig_corrimage} )
which can be used to normalize magnitudes of all stars in the image.
The normalization is performed in the following way :\\

	\begin{itemize}
	\item catalog of reference stars is read to memory
	\item for each star in \texttt{ast} file program finds a corresponding star in 
	  the reference catalog. In case it is found the field $mag_{cat}$ is filled with
	  the magnitude of the star from the reference catalog. 
	  Correction for this star is calculated as :\\
\begin{equation}
			\delta_{mag} = mag_{cat} - mag_{\pi}
\label{eq_corr_value}
\end{equation}
	  this is value which must be added to instrumental magnitude to obtain 
	  normalized one. Typically about 50\% of \texttt{ast} stars have corresponding 
	  star in the currently used reference catalog (TYCHO).
	\item after matching procedure, corrections values are known only for pixels
	  where stars matched to catalog stars are present. 
  	  In order to calculate correction for each star in the image,
     correction values are calculated for all pixels of the image by 
     extrapolating values determined for reference stars.
     Correction value C(x,y) for each pixel in the image is calculated as average 
	  of corrections for nearby reference stars weighted by distance to pixel where 
	  this star was observed :
     
     \begin{equation}
		  C(x,y) = \sum_{ref. stars \hspace{1mm} : R<R_{max}} w( \sqrt{ (x-x_i)^2+(y-y_i)^2 } ) \cdot C_(x_i,y_i)
	  \label{eq_corr_image_formula}
	  \end{equation} 
		
	  An example of resulting correction image is shown in Figure \ref{fig_corrimage}.
	\item normalized magnitude is calculated for every star in the image according to formula :\\
		\begin{equation}
		mag_{norm} = mag_{cat} + C(x,y)
		\label{eq_mag_norm}
		\end{equation}
		where (x,y) are chip coordinates of the normalized star.
	\end{itemize}

\begin{figure}[!htbp]
  \begin{center}
    \leavevmode
    \ifpdf
      \includegraphics[width=6in]{20070302_corr_smooth_18.gif}
    \else
      \includegraphics[width=6in]{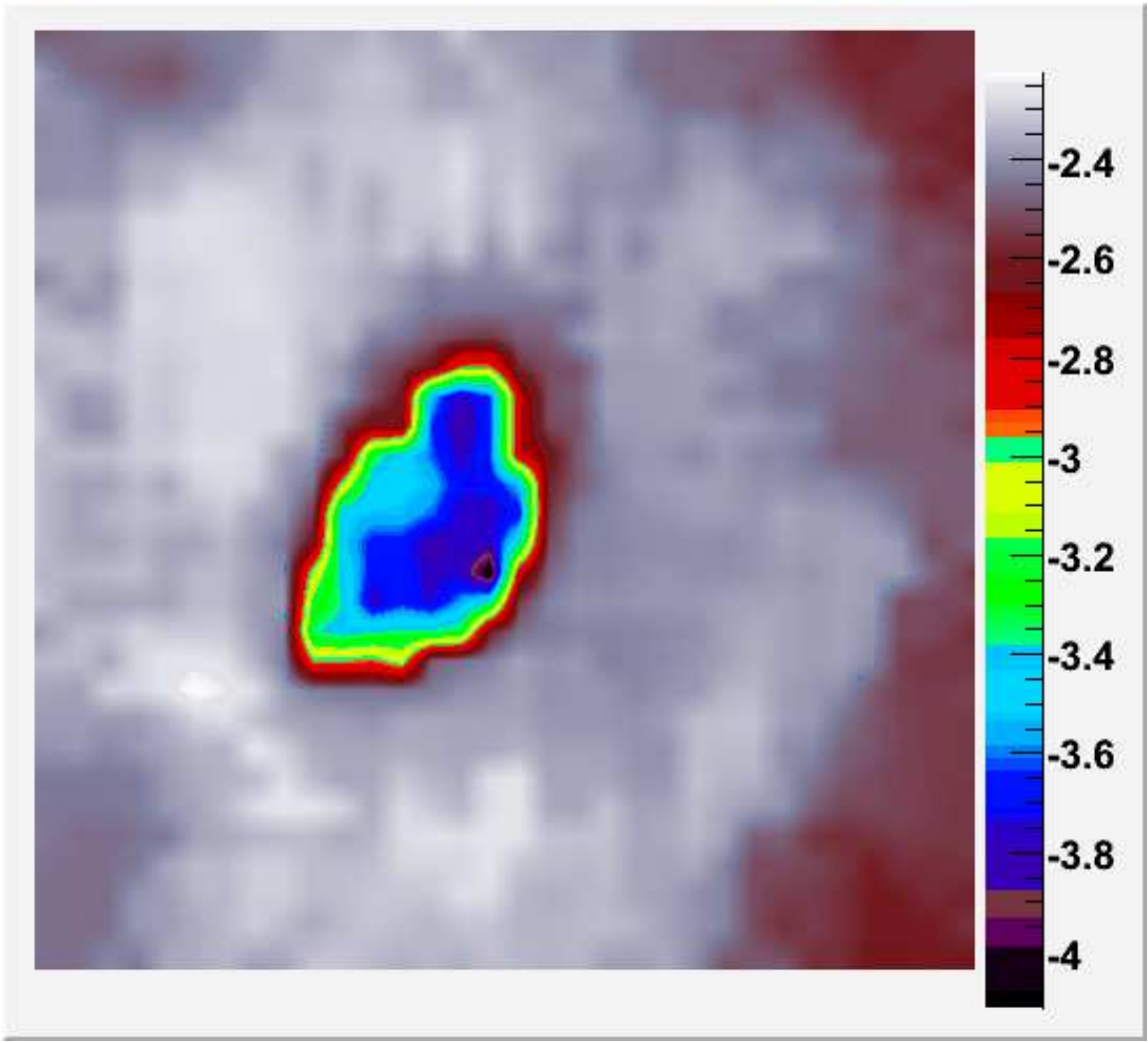}
    \fi
    \caption{Correction image obtained in cataloging to normalize instrumental magnitudes to magnitudes of stars in the reference catalog.}
    \label{fig_corrimage}
  \end{center}
\end{figure}

\item Stars from the \texttt{ast} file are matched to stars read from the database ( see step 3 ) and kept in the memory. 
In case given star is found on the list of stars kept in the memory, the star measurement from current \texttt{ast} file is linked to the list of
measurements of this star. In case this star was not yet observed, it is
added to the list of stars in the memory and flagged as a new object in the catalog.
After all stars in \texttt{ast} file are processed next \texttt{ast} file is read and process starts from
the beginning ( step 1 ). The procedure continues until all \texttt{ast} files are processed.
\end{itemize}

After the above steps all good quality data in \texttt{ast} files is saved to 
the database. However, there are several technical details which should be 
mentioned here. First of all, as can be seen in Figure \ref{fig_starcat} 
the \textbf{Star} table has statistical fields like magnitude, dispersion of magnitudo (\texttt{sigma\_mag}), \texttt{no\_measurements}
which are not updated during data loading, but they are very useful in further 
data analysis. These fields are recalculated by 
\texttt{pg/sql} procedure \texttt{ReCalcNight} after the data is loaded to the database. 
The second step is optimization of the database, before loading of
night \texttt{ast} files most of the indexes on tables \textbf{Star} and \textbf{Measurements} must be
dropped in order to load data efficiently. After loading is finished these
indexes are re-created. After loading new data its location on the disk must be
re-organized in order to provide fast access. It is realized by PostgreSQL 
command \texttt{CLUSTER}.
After all optimizations and recalculations are finished the database is unlocked and 
off-line algorithms can be executed on new night data.

\subsubsection{Efficiency, purity and precision of observations}
\label{sec_red_eff}

Efficiency and purity of reduction and cataloging was tested by the
following procedure :\\
\begin{enumerate}
\item Initialization of database - star catalog database was initialized
with stars from TYCHO-2 star catalog
\item Initial stars were flagged as TYCHO-2 stars in the database
\item Test data was loaded to catalog initialized with  \mbox{TYCHO-2} stars
\end{enumerate}

After this steps database was filled with data and there was an easy way to
distinguish stars which are present in TYCHO-2 star catalog from those which
are not and were observed only in "Pi of the Sky" images.
The test was performed on two samples of images one from night 2007-04-25/26
taken with shutter in normal mode and second sample were images from night
2007-05-12/13 taken with shutter permanently opened.
Total star identification efficiency which is defined as : \\

\begin{equation}
\epsilon_{\pi-red} = \frac{N_{TYCHO2/PI}}{N_{TYCHO2}}
\label{eq_red_eff}
\end{equation}

where $N_{TYCHO2/PI}$ stands for number of TYCHO-2 stars identified by "Pi
of the Sky" photometry and $N_{TYCHO2}$ is number of TYCHO-2 stars in the
observed field.
The Table \ref{tab_red_eff_total} shows total efficiencies for single image
and set of images. 

\begin{table}[htbp]
\begin{center}
\begin{tabular}{|c|c|c|c|}
\hline
\textbf{Night} & \textbf{Number of images} & \textbf{Magnitude range} & \textbf{Efficiency} \\
\hline
2007.04.25/26 & 1 & 0 - 12 & 0.74 \\
2007.04.25/26 & 41 & 0 - 12 &  0.84 \\
2007.04.25/26 & 41 & 0 - 15 & 0.77 \\
\hline
2007.05.12/13 & 1 & 0 - 12 & 0.78 \\
2007.05.12/13 & 41 & 0 - 12 & 0.87 \\
2007.05.12/13 & 41 & 0 - 15 & 0.82 \\
\hline
\end{tabular}
\caption{Total efficiencies for night 2007.04.25/26 ( data collected with
shutter in normal mode ) and 2007.05.12/13 ( data collected with permanently
opened shutter ) }
\label{tab_red_eff_total}
\end{center}
\end{table}

\begin{figure}[!htbp]
  \begin{center}
    \leavevmode
    \ifpdf
      \includegraphics[width=2.7in,height=2.7in]{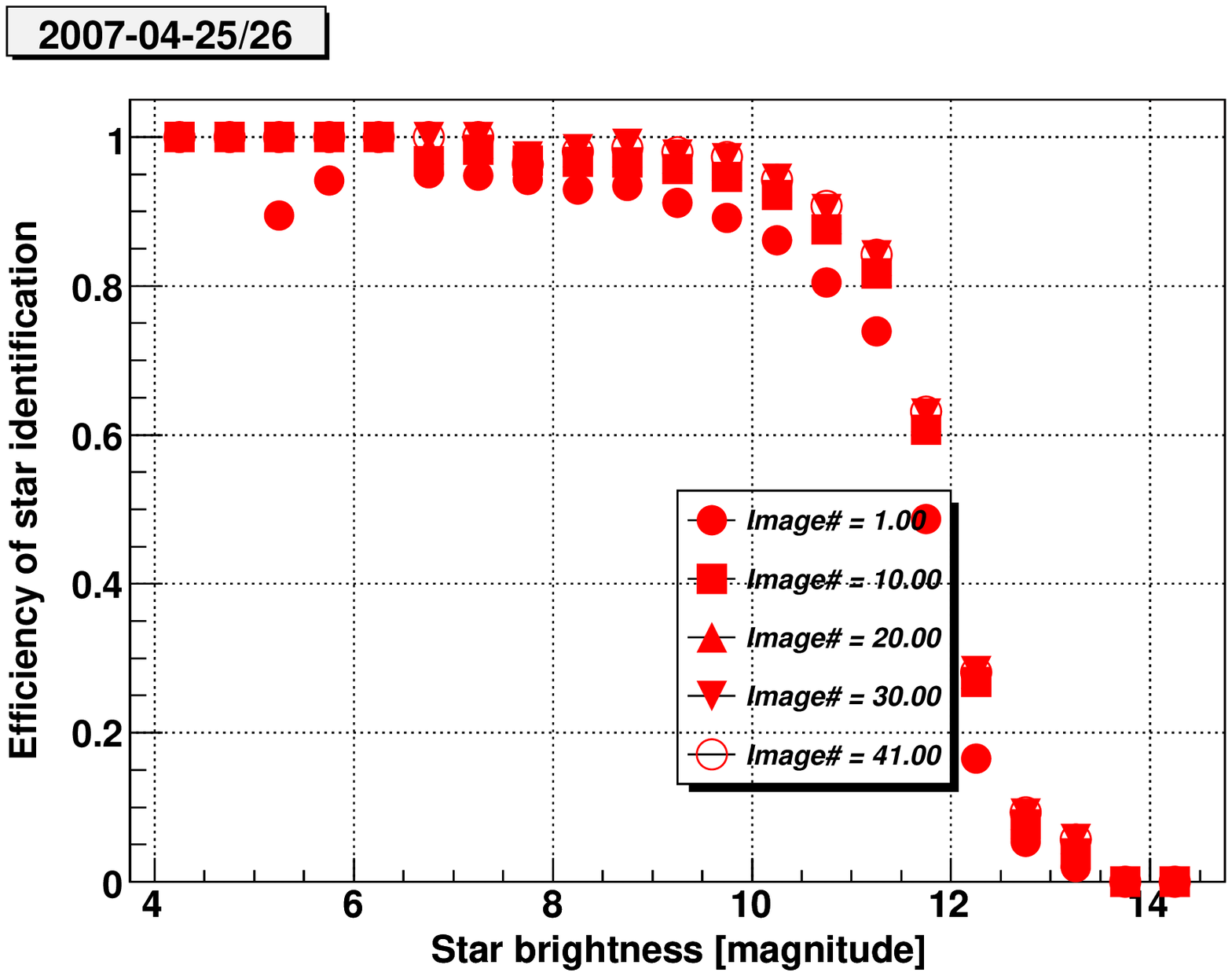}
		\includegraphics[width=2.7in,height=2.7in]{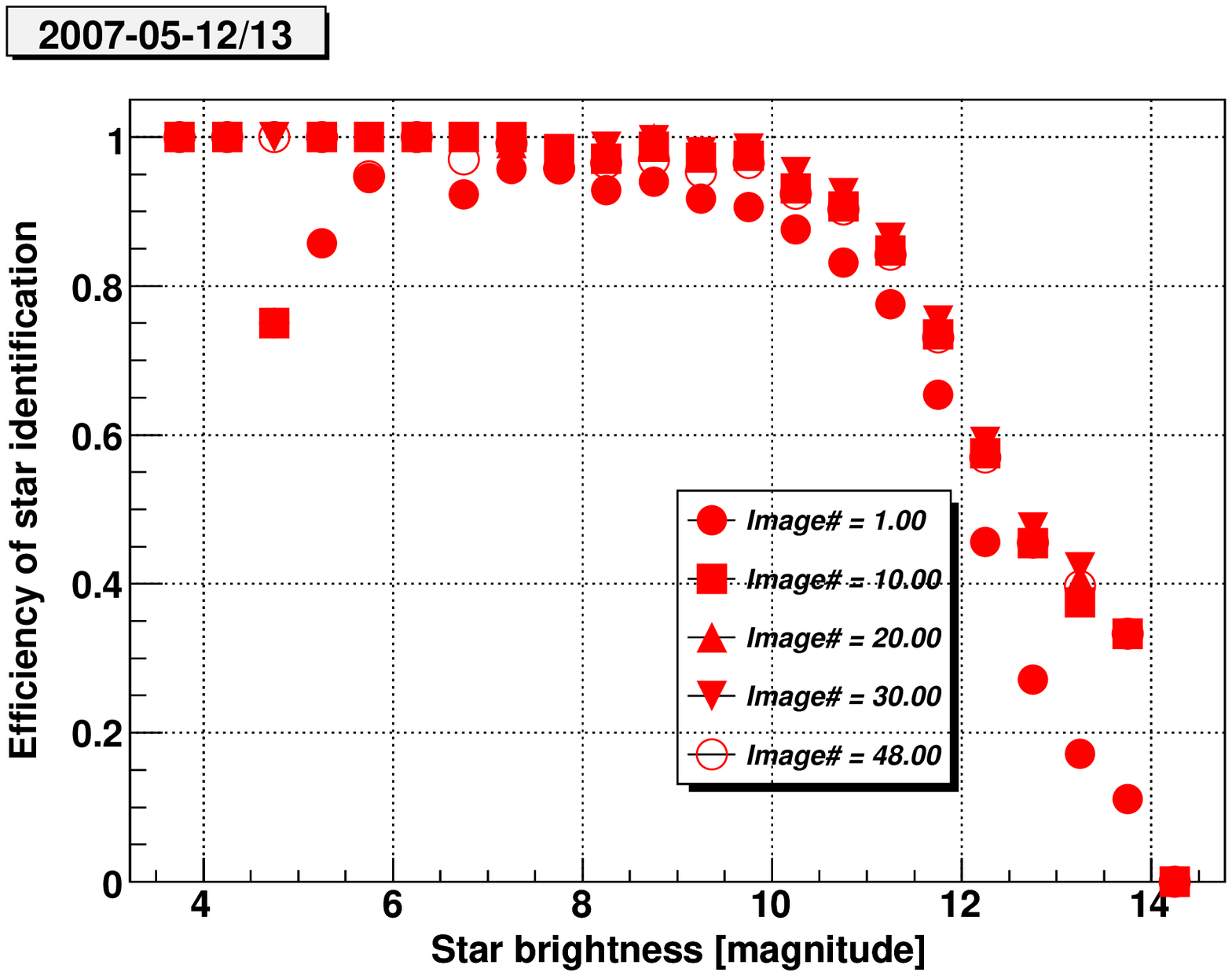}		
    \else
      \includegraphics[width=2.7in,height=2.7in]{PhotoEff/eff_vs_mag/20070425_shutter_normal.eps}
		\includegraphics[width=2.7in,height=2.7in]{PhotoEff/eff_vs_mag/20070512_shutter_open.eps}		
    \fi
    \caption{Efficiency of TYCHO-2 stars identification in function of magnitude for data collected during night 2007.04.25/26 with shutter in normal (open/close) mode (left plot ) and data collected with shutter permanently opened during night 2007.05.12/13 ( right plot ) }
    \label{fig_eff_vs_mag}
  \end{center}
\end{figure}

The Figure \ref{fig_eff_vs_mag} shows the efficiency of
star identification in function of star brightness for data collected with
permanently opened shutter ( 2007.05.12 ) and shutter in normal open/close
mode ( 2007.04.25 ).

\begin{figure}[!htbp]
  \begin{center}
    \leavevmode
    \ifpdf
		\includegraphics[width=2.7in,height=2.7in]{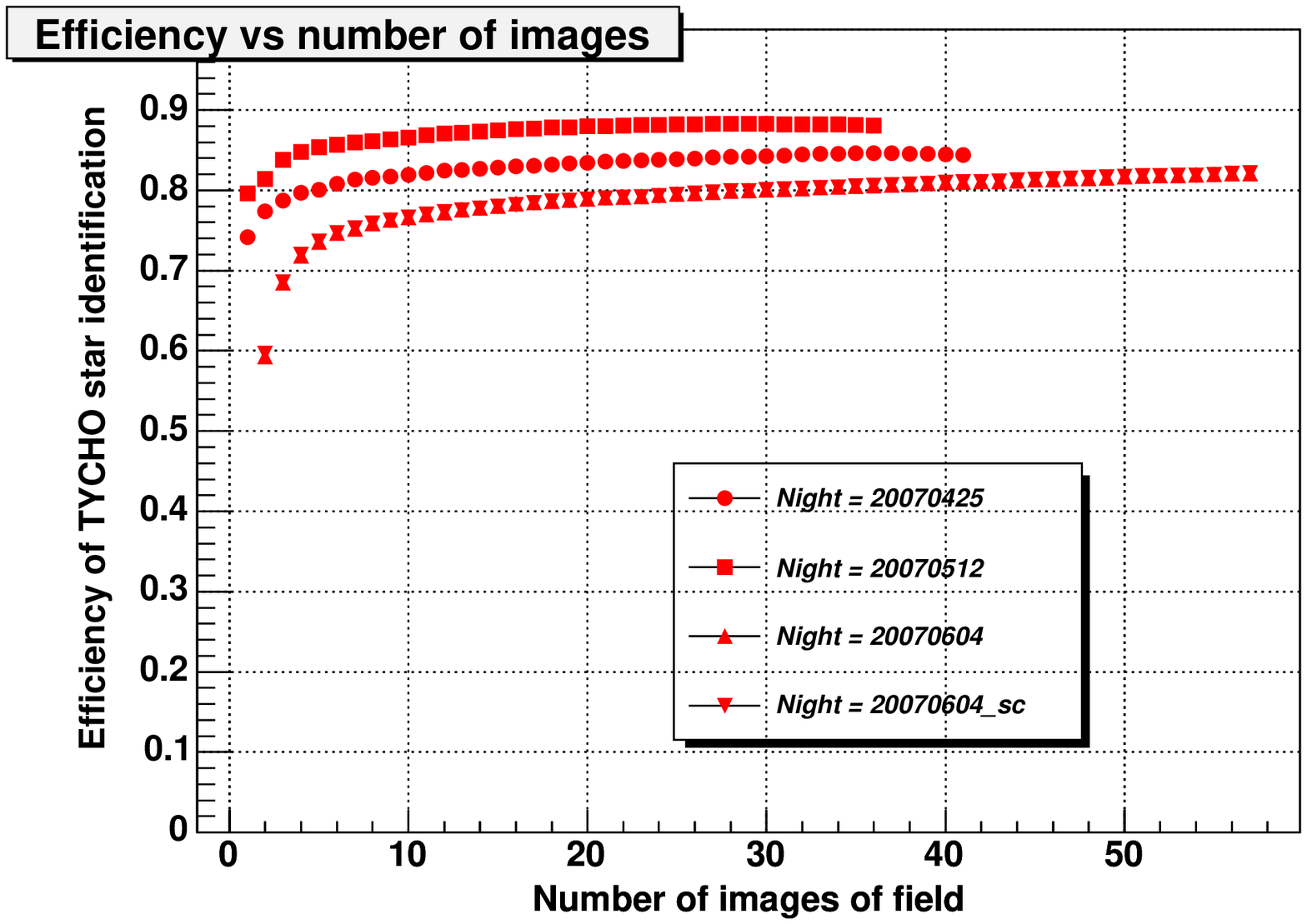}
    \else
		\includegraphics[width=2.7in,height=2.7in]{PhotoEff/eff_vs_image/eff_vs_image_many_nights.eps}
    \fi
    \caption{Average efficiency ( stars 0-12$^m$ ) of TYCHO-2 stars identification in function of
number of field observations for 3 different nights and after applying
correction of shutter opened effect.}
    \label{fig_eff_vs_image_many_nights}
  \end{center}
\end{figure}

The Figure \ref{fig_eff_vs_image_many_nights} 
shows efficiency in function of number of field observations. It can be seen
that at least 10 images of field must be catalogued before running nova
identification algorithm and the safe limit is 20 exposures. This ensures
that nova candidates will not be mainly due to normal faint stars added to catalog
for the first time. The best way would be to initialize star catalog with
 all objects up to 12-13$^m$, which is planned to be done.

\begin{figure}[!htbp]
  \begin{center}
    \leavevmode
    \ifpdf
		\includegraphics[width=2.7in,height=2.7in]{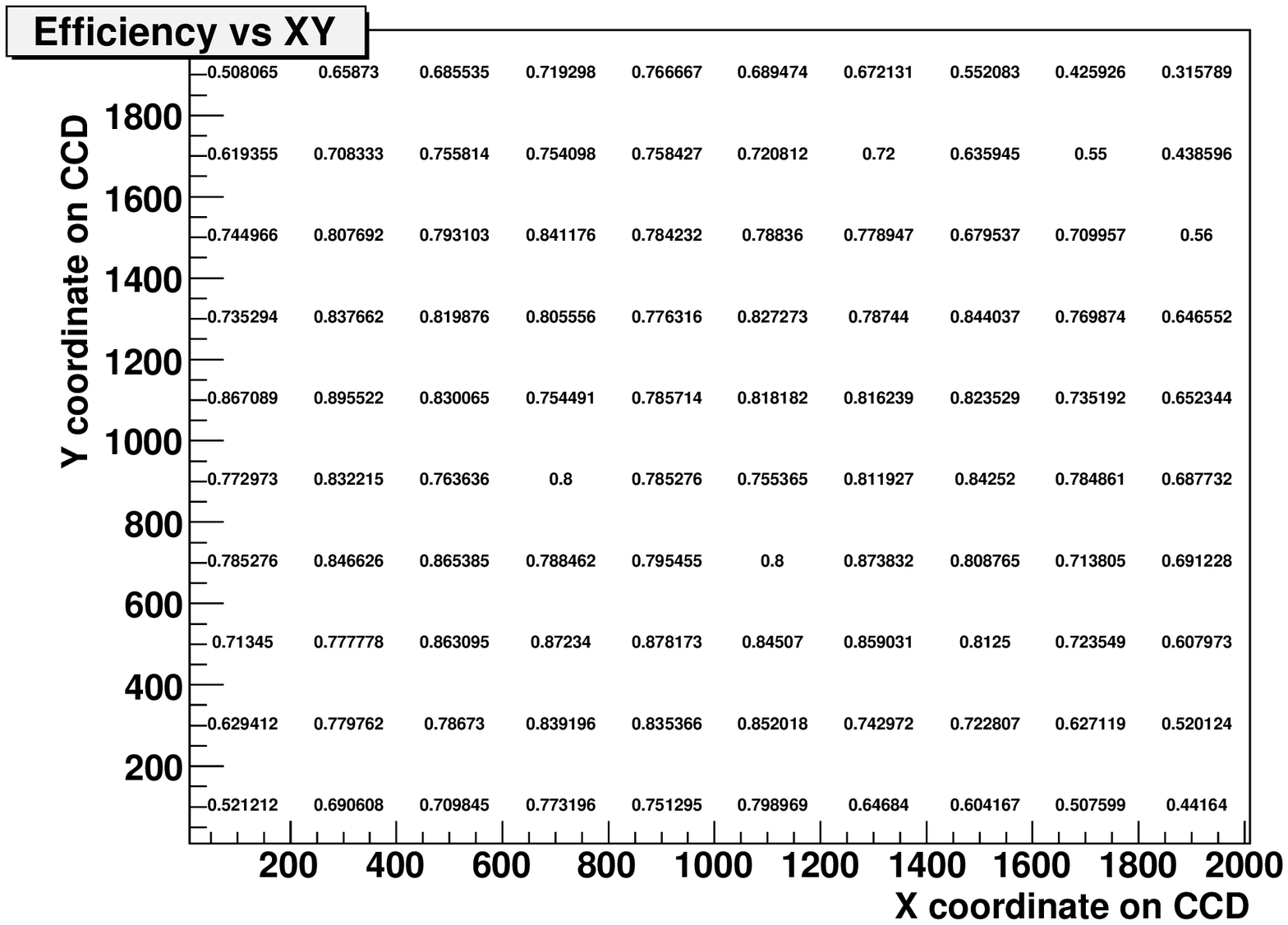}
		\includegraphics[width=2.7in,height=2.7in]{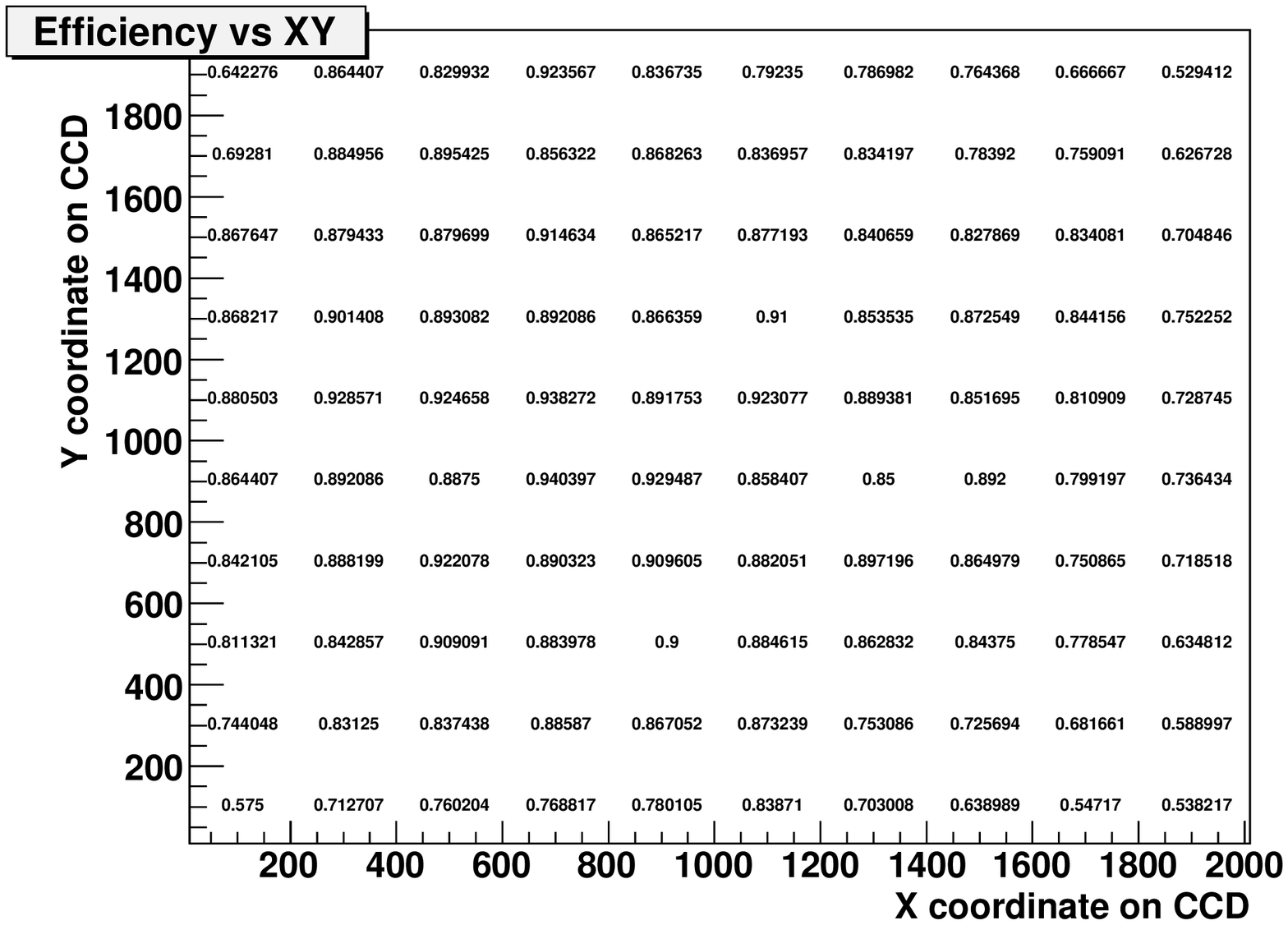}		
    \else
		\includegraphics[width=2.7in,height=2.7in]{PhotoEff/eff_vs_xy/20070425_shutter_normal_eff.eps}
		\includegraphics[width=2.7in,height=2.7in]{PhotoEff/eff_vs_xy/20070512_shutter_open_eff.eps}		
    \fi
    \caption{Efficiency of TYCHO-2 stars identification in function of star position (x,y) on the CCD chip.
				 Left plot shows efficiency of single average of 20 images from
				 night 2007.04.25/26 collected with shutter in normal mode and right plot
				 shows efficiency on single image from night 2007.05.12/13 collected with permanently opened shutter}
    \label{fig_eff_vs_xy}
  \end{center}
\end{figure}

\begin{figure}[!htbp]
  \begin{center}
    \leavevmode
    \ifpdf
		\includegraphics[width=2.7in,height=2.7in]{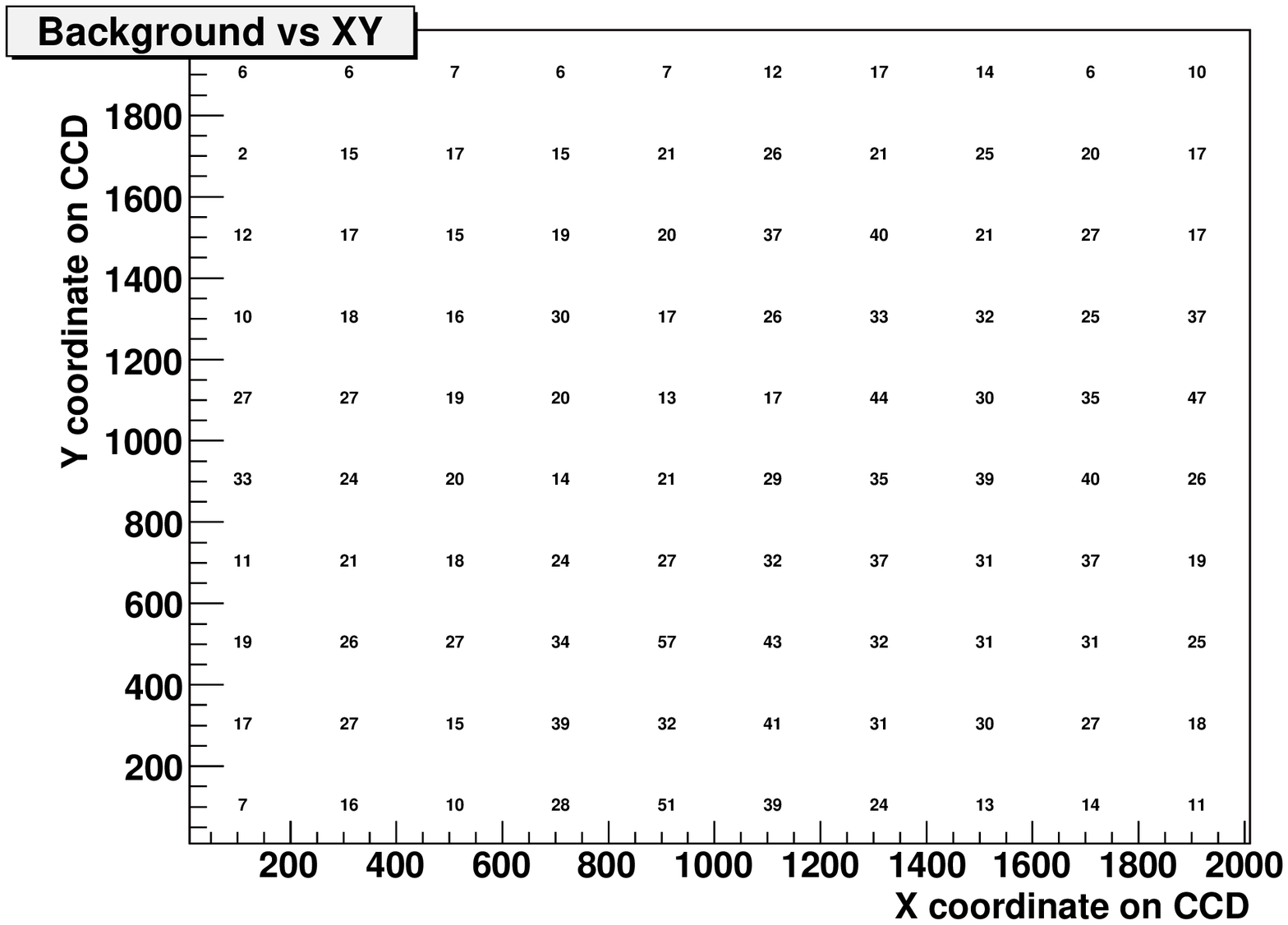}
		\includegraphics[width=2.7in,height=2.7in]{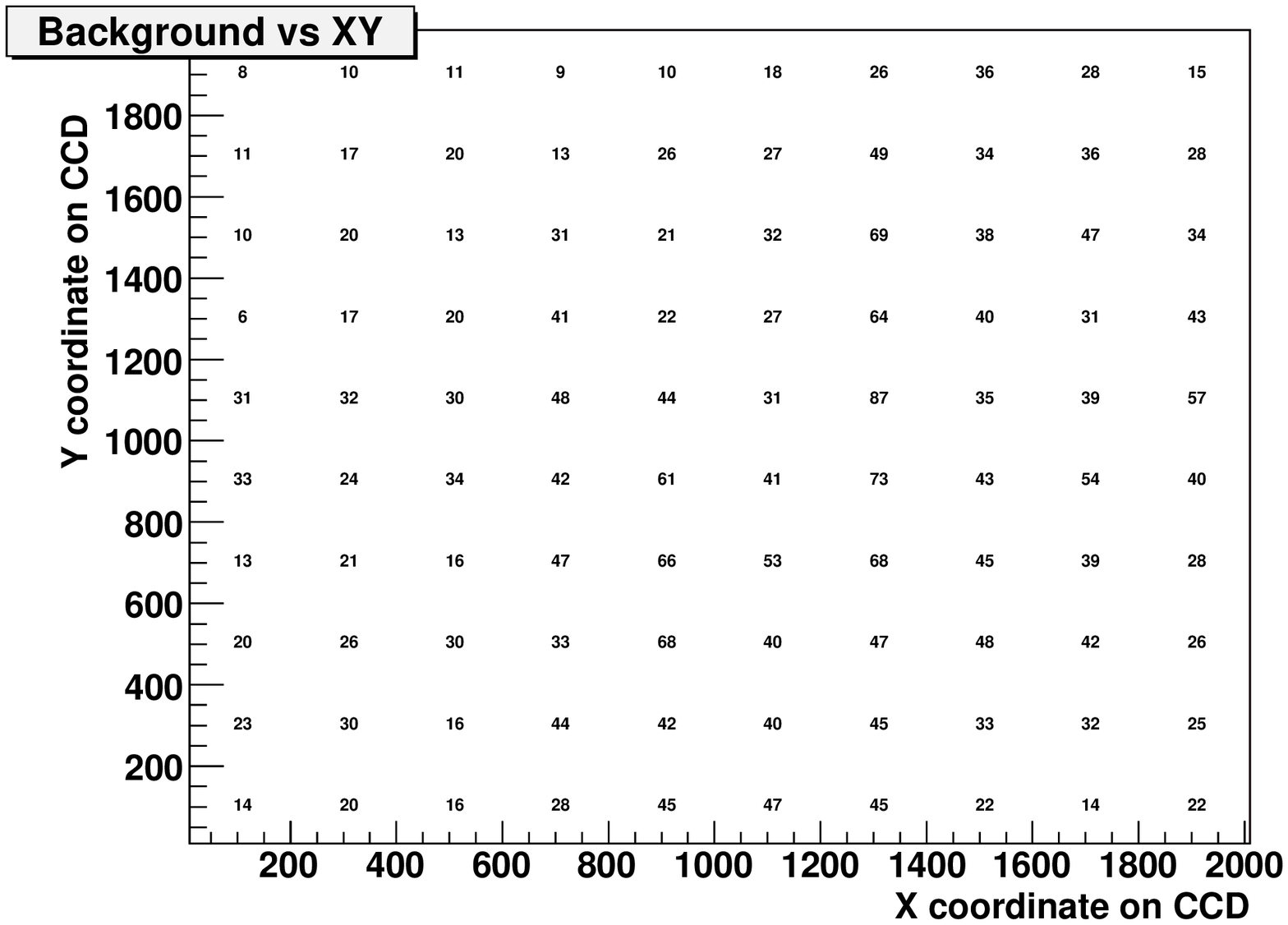}		
    \else
		\includegraphics[width=2.7in,height=2.7in]{PhotoEff/eff_vs_xy/20070425_shutter_normal_bkg.eps}
		\includegraphics[width=2.7in,height=2.7in]{PhotoEff/eff_vs_xy/20070512_shutter_open_bkg.eps}		
    \fi
    \caption{Number of observed objects not existing in TYCHO-2 catalog in function of star position (x,y) on the CCD chip.
				 Left plot shows objects found on single average of 20 images from
				 night 2007.04.25/26 collected with shutter in normal mode and right plot
				 shows objects found on single image from night 2007.05.12/13 collected with permanently opened shutter}
    \label{fig_bkg_vs_xy}
  \end{center}
\end{figure}

The Figures \ref{fig_eff_vs_xy} and \ref{fig_bkg_vs_xy} show efficiency and background in function of chip coordinates.
It is clear that efficiency and background drop in the corners of the CCD chip and reach the
highest values in the center of the chip which is due to the fact that 
less light reaches corners of CCD because of properties of the optical system.

\begin{figure}[!htbp]
  \begin{center}
    \leavevmode
    \ifpdf
		\includegraphics[width=2.7in,height=2.7in]{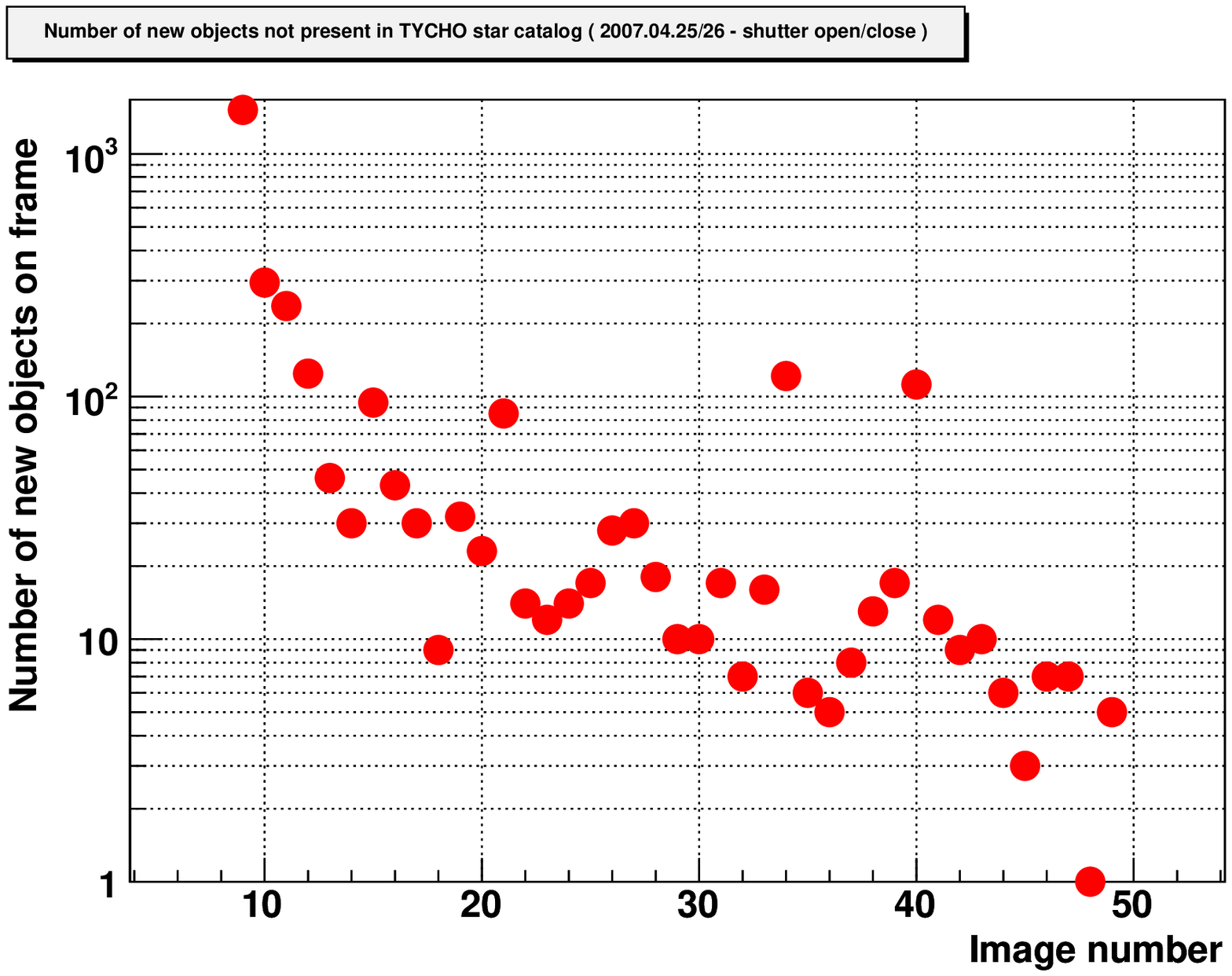}
		\includegraphics[width=2.7in,height=2.7in]{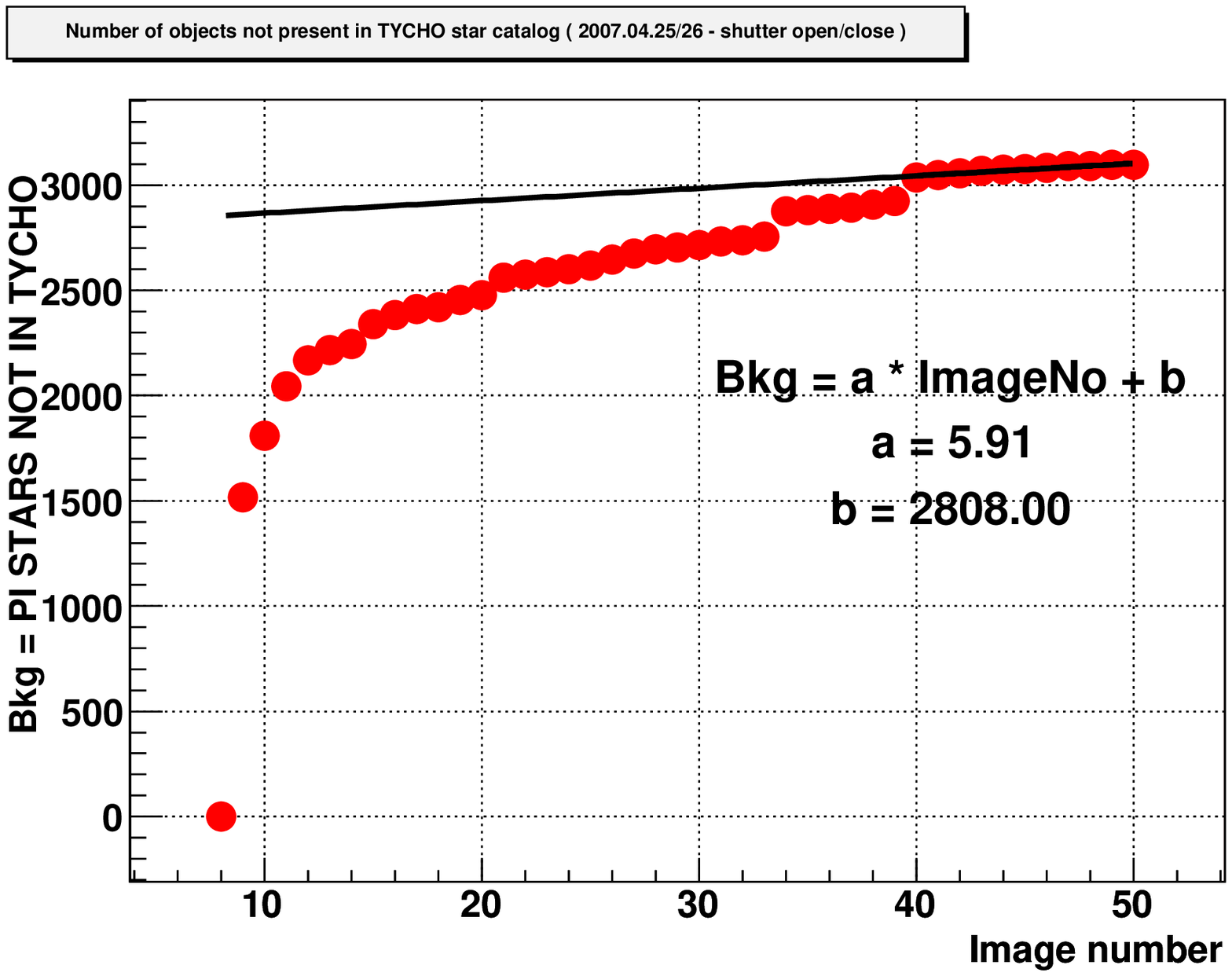}
    \else
		\includegraphics[width=2.7in,height=2.7in]{PhotoEff/bkg_vs_image/20070425/new_vs_i.eps}
		\includegraphics[width=2.7in,height=2.7in]{PhotoEff/bkg_vs_image/20070425/bkg_vs_i.eps}
    \fi
    \caption{Number of new objects added to catalog on subsequent images
(left plot) and total number of new objects added to catalog in function of
image number right plot. The data was collected on field 0800$+$20 ( average number of stars ~ 18000 ) 
during night 2007.04.25/26 with shutter in normal (open/close) mode }
    \label{fig_new_vs_imageno_20070425}
  \end{center}
\end{figure}

\begin{figure}[!htbp]
  \begin{center}
    \leavevmode
    \ifpdf
		\includegraphics[width=2.7in,height=5.4in]{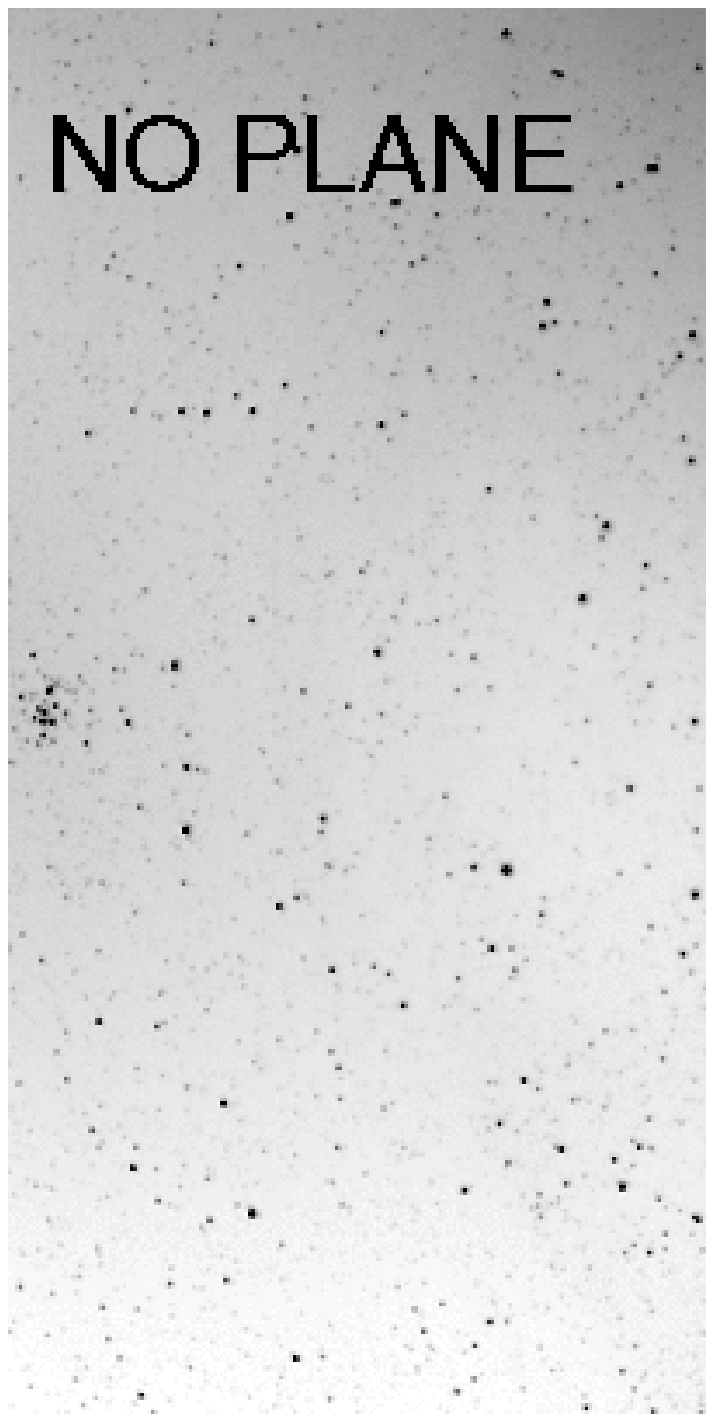}
		\includegraphics[width=2.7in,height=5.4in]{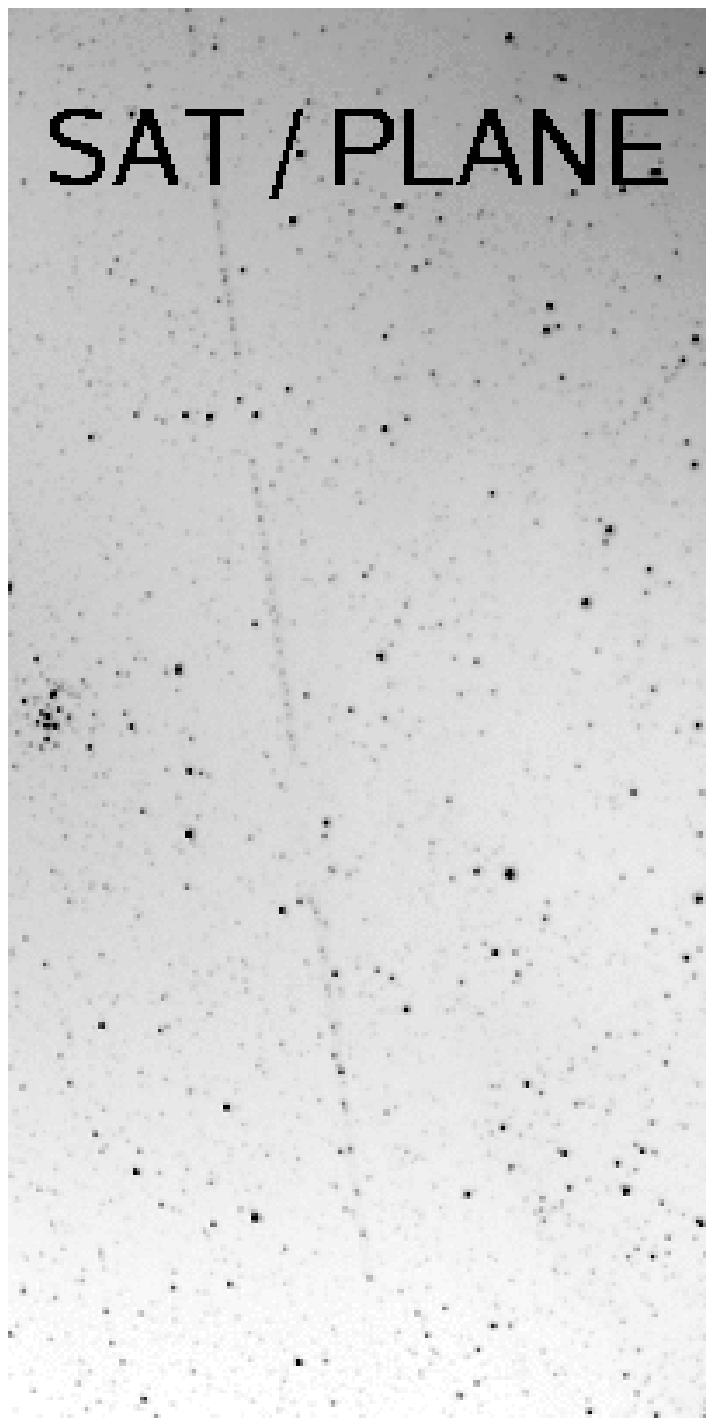}
    \else
		\includegraphics[width=2.7in,height=5.4in]{PhotoEff/bkg_vs_image/20070425/sat/no_plane.eps}
		\includegraphics[width=2.7in,height=5.4in]{PhotoEff/bkg_vs_image/20070425/sat/plane.eps}
    \fi
    \caption{The reason for step visible on previous image at frame id=34. Part of image without satellite (or plane) trace is visible on left
image and with the trace on right image. This trace causes addition of new objects to star catalog}
    \label{fig_red_bkg_plane}
  \end{center}
\end{figure}

\begin{figure}[!htbp]
  \begin{center}
    \leavevmode
    \ifpdf
		\includegraphics[width=2.7in,height=2.7in]{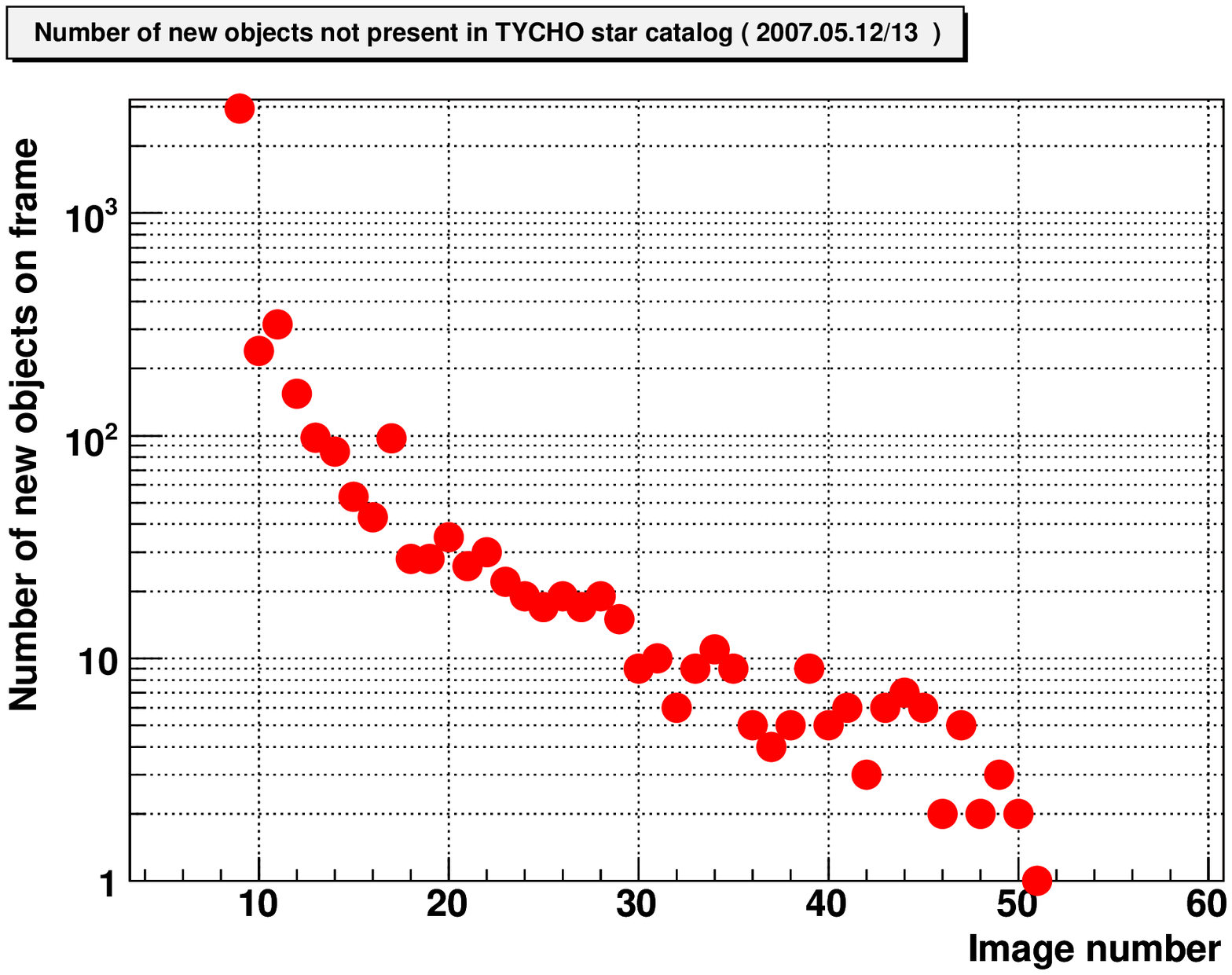}
		\includegraphics[width=2.7in,height=2.7in]{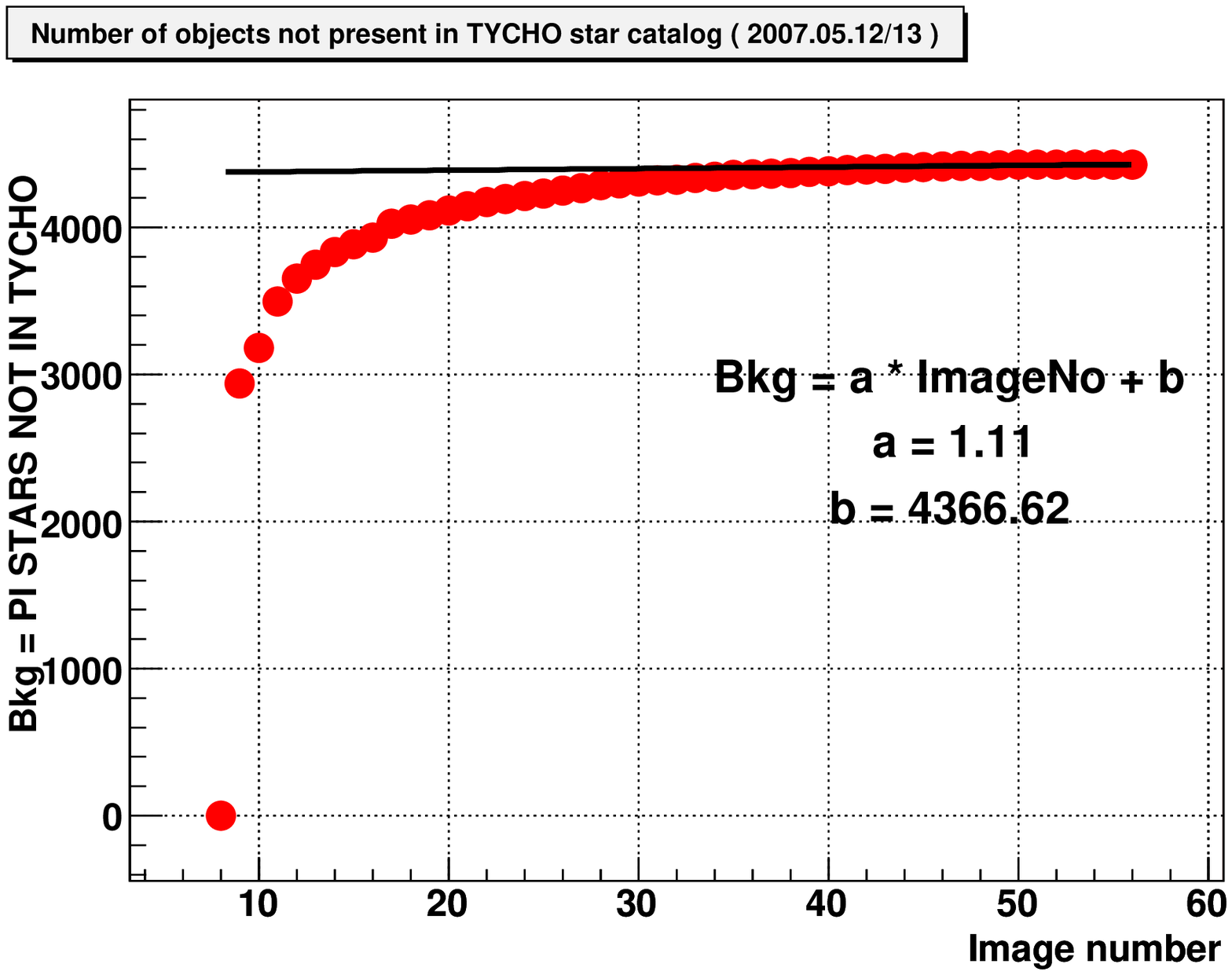}
    \else
		\includegraphics[width=2.7in,height=2.7in]{PhotoEff/bkg_vs_image/20070512/new_vs_i.eps}
		\includegraphics[width=2.7in,height=2.7in]{PhotoEff/bkg_vs_image/20070512/bkg_vs_i.eps}
    \fi
    \caption{Number of new objects added to catalog in function of number of images
(left plot) and total number of new objects added to catalog in function of
image number right plot. The data was collected on field 0800$+$20 ( average number of stars ~ 18000 )
during night 2007.05.12/13 with shutter permanently opened. }
    \label{fig_new_vs_imageno_20070512}
  \end{center}
\end{figure}

The situation is slightly more difficult in the case of purity
determination. The objects in the TYCHO-2 star catalog are only stars.
Objects observed by "Pi of the Sky" which are not present in the TYCHO-2
catalog are not only background events. They can be non-star objects like
galaxies etc. However, after many observations of given field all objects in
the range of the telescope should already be observed and new objects
appearing in the catalog after many field observations can be considered as
background ( assuming signal events are very seldom ). 
The Figure \ref{fig_new_vs_imageno_20070425} shows number of new objects
added to star catalog on subsequent images of the same field and also total
number of new objects in function of number of field observations is shown.
The data presented on this plot was collected during night 2007.04.25/26 with
shutter in normal open/close mode. The steps at frame id=40 and id=34 are
due to trace of satellite ( or plane ) which is shown in Figure \ref{fig_red_bkg_plane}.
It is clear that after many ( >40 ) field observations number of new objects
added to star catalog on every image is very small and equals few events per
image, unless background event ( like satellite, plane etc ) is observed.
For comparison the same plots for data collected on the same field with shutter permanently
opened ( night 2007.05.12/13 ) is shown in Figure \ref{fig_new_vs_imageno_20070512}.
The conditions during both night were similar, the number of objects added to
star catalog was much higher during night when data was collected with permanently
opened shutter.

\begin{figure}[!htbp]
  \begin{center}
    \leavevmode
    \ifpdf
		\includegraphics[width=2.7in,height=2.7in]{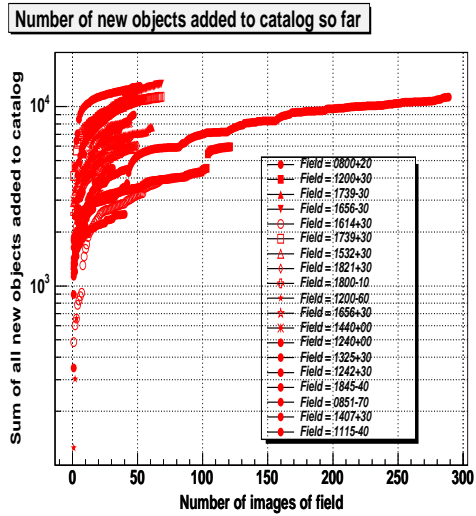}
    \else
		\includegraphics[width=2.7in,height=2.7in]{PhotoEff/bkg_vs_image/bkg_vs_i_all.eps}
    \fi
    \caption{Number of new objects added to catalog from the beginning to
given frame number. Data for different fields collected during many nights is
shown.}
    \label{fig_bkg_vs_image_many_nights}
  \end{center}
\end{figure}

Figure \ref{fig_bkg_vs_image_many_nights} shows cumulative number of events added to
catalog in function of image number for couple of fields observed during 
many nights. These plots indicate that the minimum number of field observations 
to consider new objects as potential nova candidate is 20-30.

\begin{figure}[!htbp]
  \begin{center}
    \leavevmode
    \ifpdf
		\includegraphics[width=2.7in,height=2.7in]{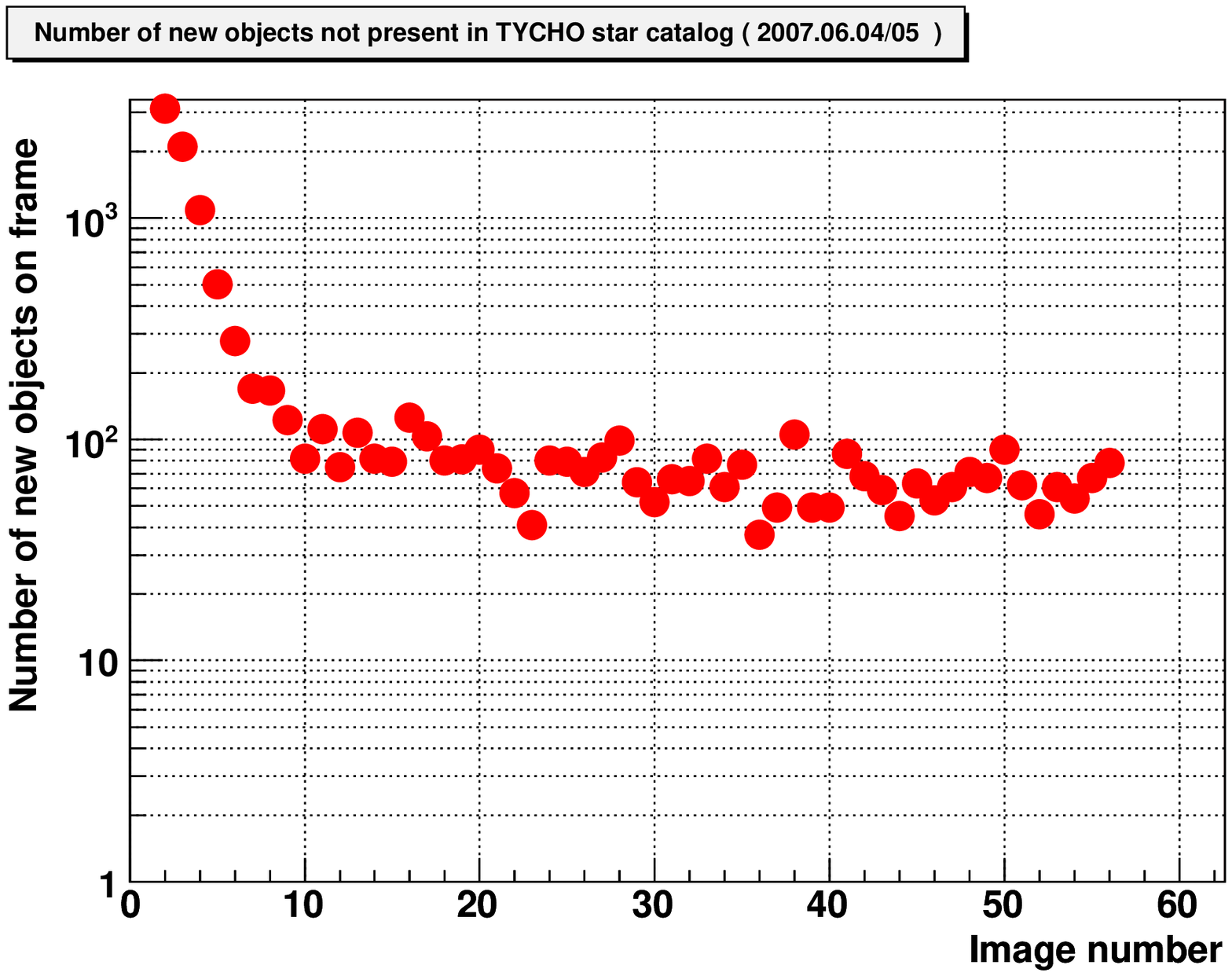}
		\includegraphics[width=2.7in,height=2.7in]{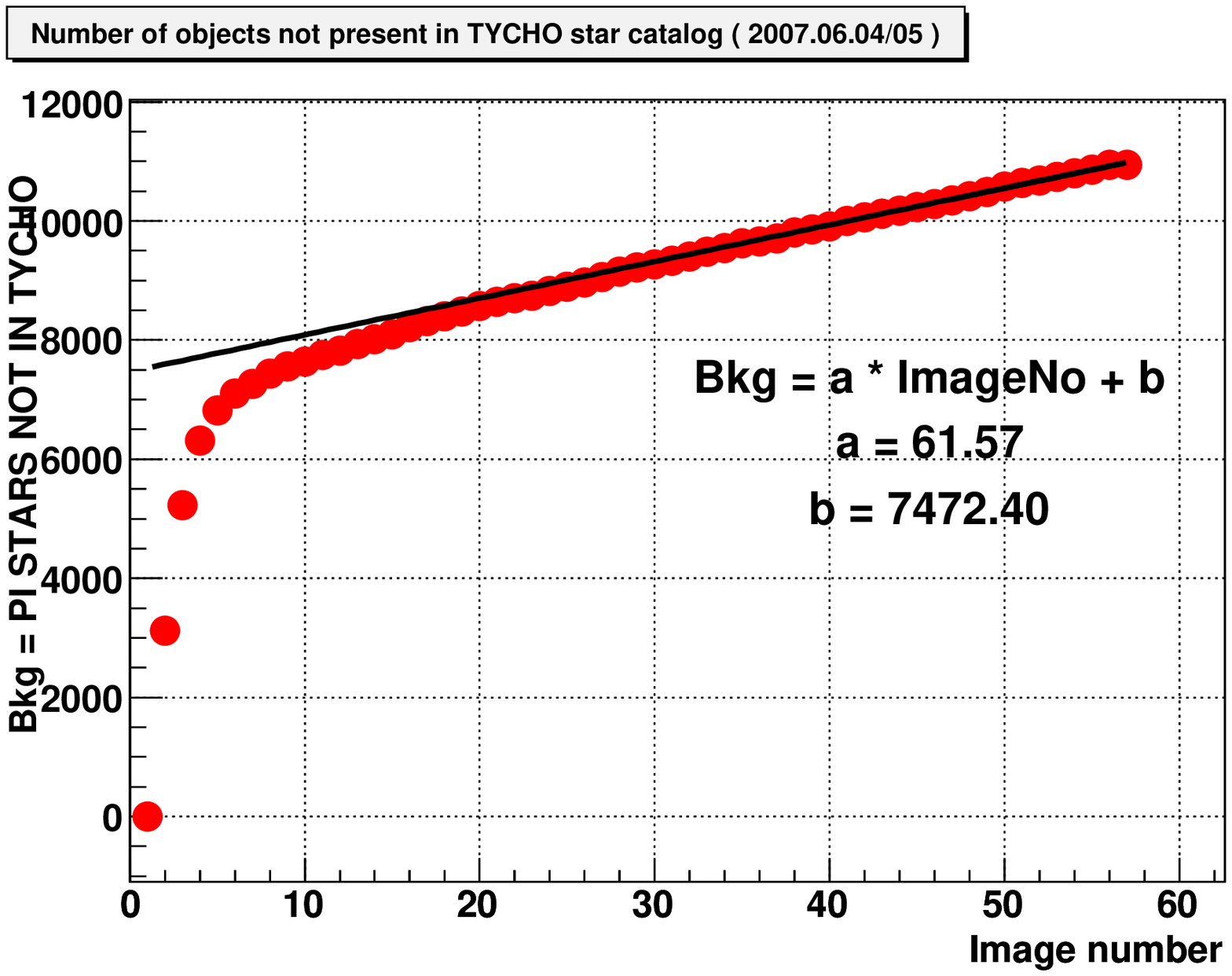}
    \else
		\includegraphics[width=2.7in,height=2.7in]{PhotoEff/bkg_vs_image/20070604/new_vs_i.eps}
		\includegraphics[width=2.7in,height=2.7in]{PhotoEff/bkg_vs_image/20070604/bkg_vs_i.eps}
    \fi
    \caption{Number of new objects added to catalog on subsequent images
(left plot) and total number of new objects added to catalog in function of
image number right plot. Data was collected during night 2007.06.04/05 with
permanently opened shutter, on field 0851-70 ( average number of stars ~ 33000 ) }
    \label{fig_new_vs_imageno_20070604}
  \end{center}
\end{figure}

\begin{figure}[!htbp]
  \begin{center}
    \leavevmode
    \ifpdf
		\includegraphics[width=2.7in,height=3.24in]{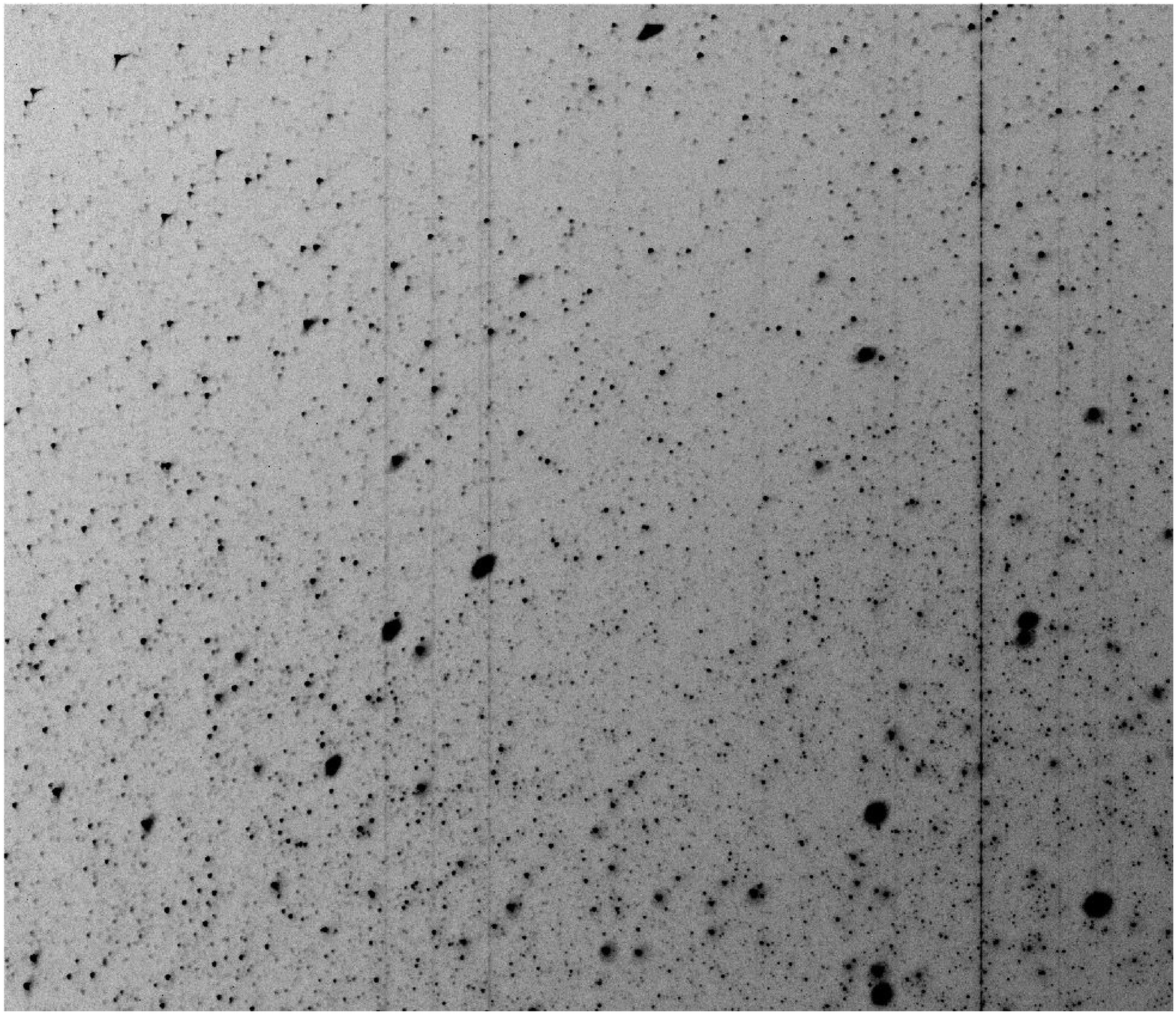}
		\includegraphics[width=2.7in,height=3.24in]{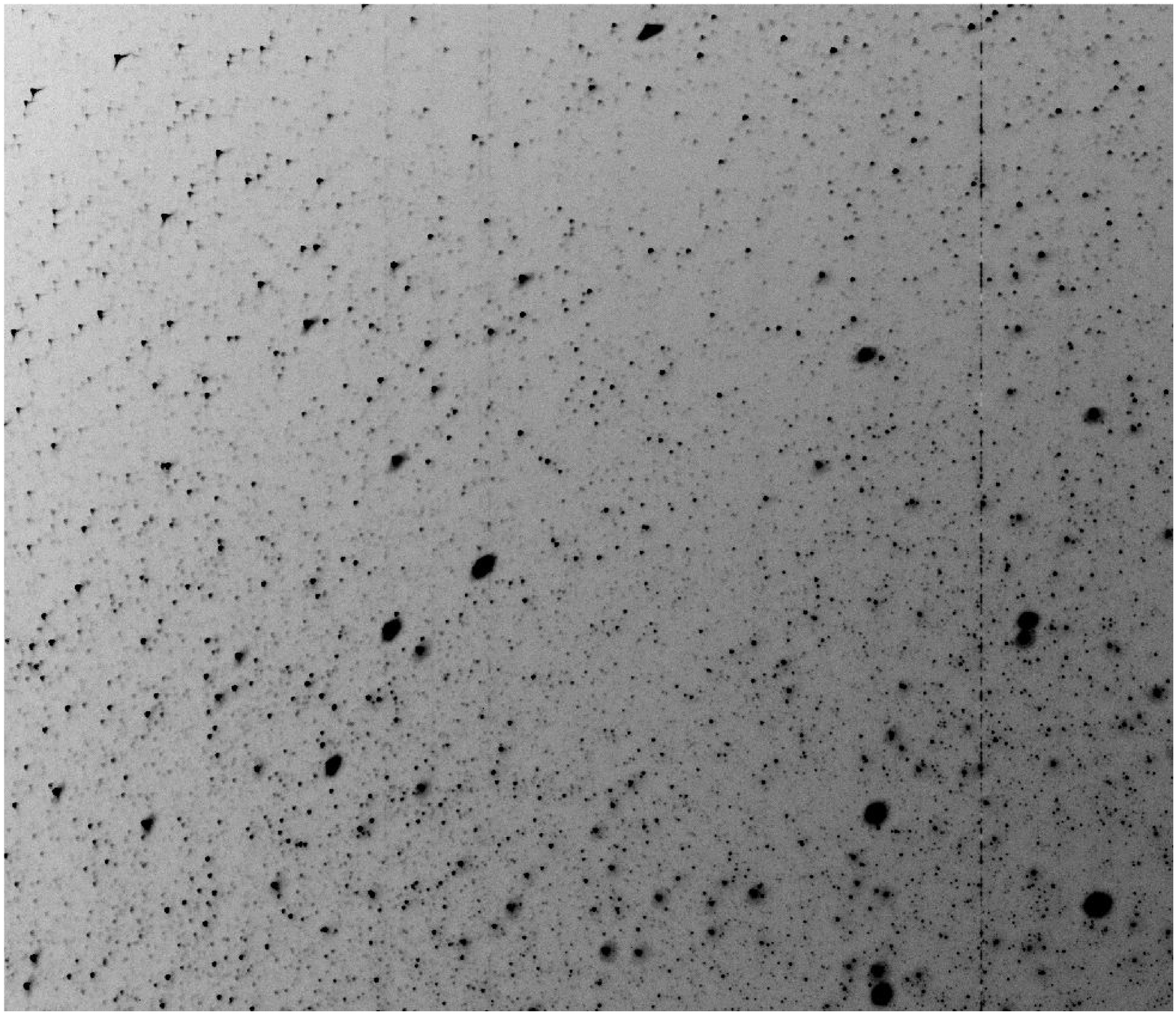}
    \else
		\includegraphics[width=2.7in,height=3.24in]{PhotoEff/OpenedShutterCorr/image100_1000_1200.eps}
		\includegraphics[width=2.7in,height=3.24in]{PhotoEff/OpenedShutterCorr/image100_1000_1200_sc.eps}
    \fi
    \caption{Image taken with permanently opened shutter before correction (left image)
and after correction (right image) }
    \label{fig_opened_shutter_effect}
  \end{center}
\end{figure}

In Figure \ref{fig_new_vs_imageno_20070604} results for field ( 0851-70 )
are shown, the number of background events added
to catalog on every image is much larger. Most of these objects are
artifacts coming from photometry of strip of charge which appears when
images are collected with permanently opened shutter ( see Fig. \ref{fig_opened_shutter_effect} ).
On field 0851-70 number of stars ( ~33000 ) is much higher then in field 0800$+$20 ( ~18000 )
which causes much higher number of stars causing significant "open shutter" strips.

\begin{figure}[!htbp]
  \begin{center}
    \leavevmode
    \ifpdf
		\includegraphics[width=2.7in,height=2.7in]{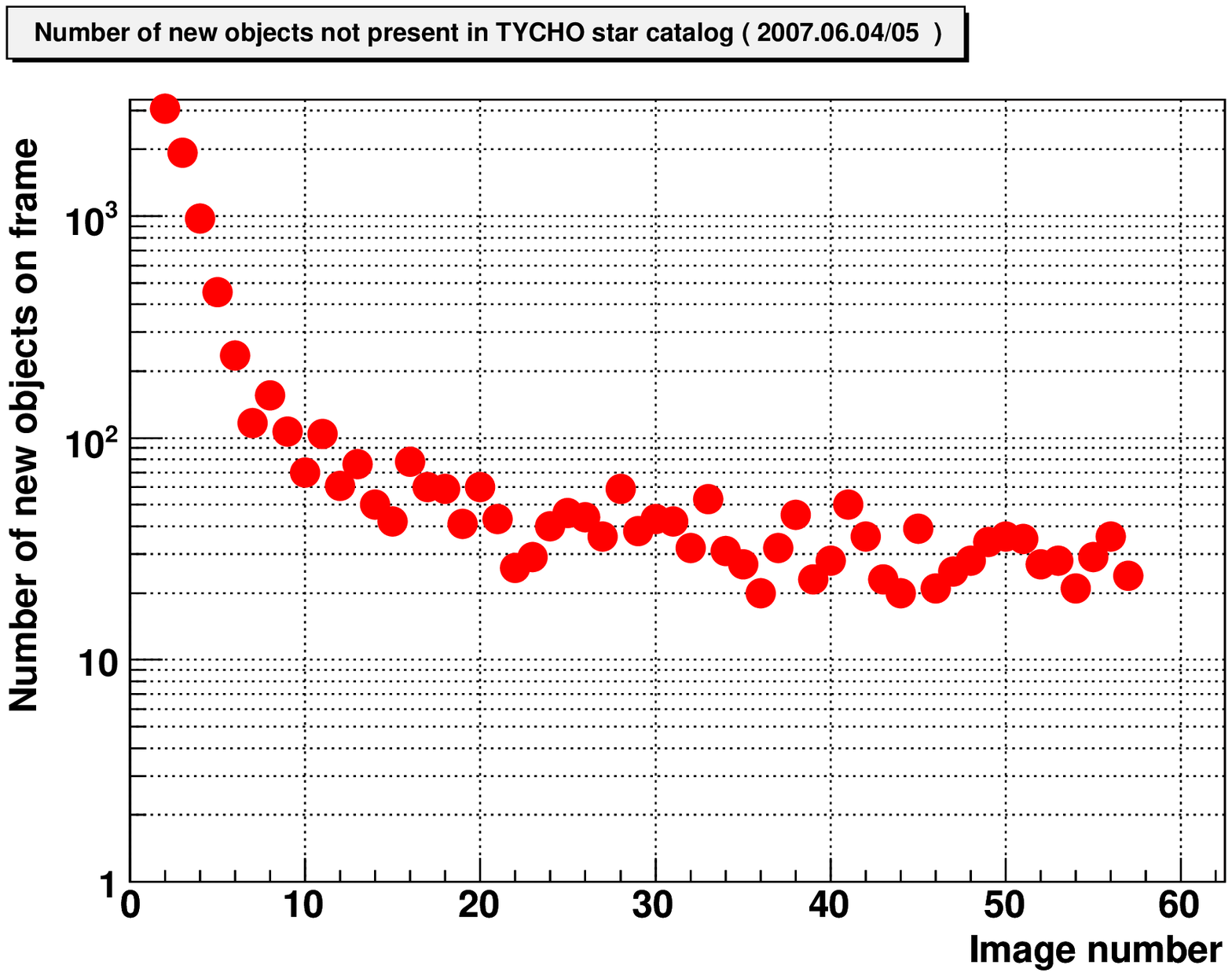}
		\includegraphics[width=2.7in,height=2.7in]{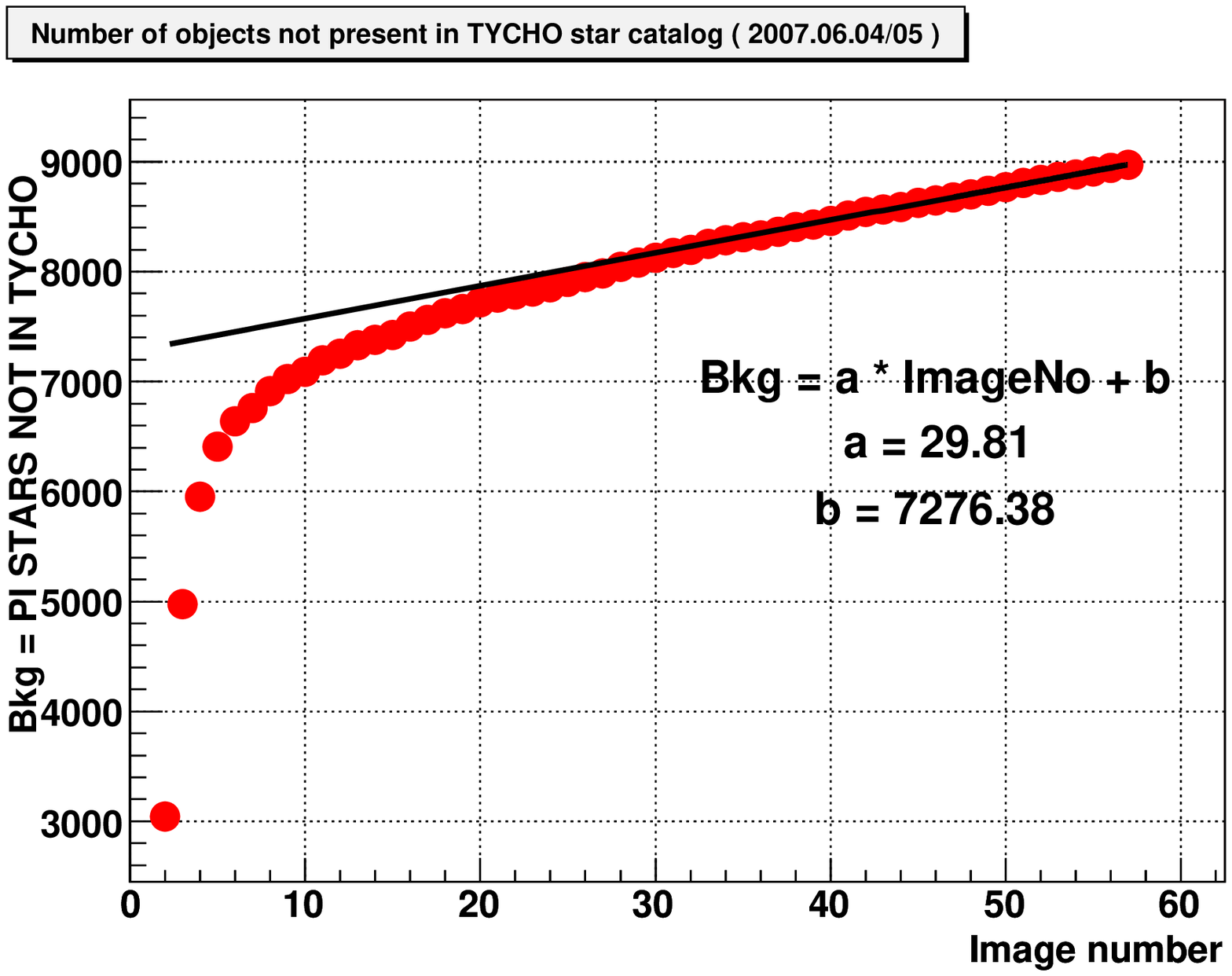}
    \else
		\includegraphics[width=2.7in,height=2.7in]{PhotoEff/bkg_vs_image/20070604_sc/new_vs_i.eps}
		\includegraphics[width=2.7in,height=2.7in]{PhotoEff/bkg_vs_image/20070604_sc/bkg_vs_i.eps}
    \fi
    \caption{Number of new objects added to catalog on subsequent images
(left plot) and total number of new objects added to catalog in function of
image number right plot. Data was collected during night 2007.06.04/05 with
shutter with opened shutter, on field 0851-70 and correction of opened shutter effect was applied }
    \label{fig_new_vs_imageno_20070604_sc}
  \end{center}
\end{figure}

Open shutter causes that charge is collected also in pixels above the
readout pixel during chip readout ( Fig. \ref{fig_opened_shutter_effect} ). It is possible to reduce this effect by
subtracting from every pixel fraction of values in pixels below. The image before and after
the correction is shown in Figure \ref{fig_opened_shutter_effect}. The
correction is not perfect, but significantly reduces this effect.
The data from night 2007.06.04/05 was corrected and cataloged, number of new objects
in function of image number is shown in Figure \ref{fig_new_vs_imageno_20070604_sc}.
The number of new objects in every image is reduced approximately by a
factor of 2. 

\begin{figure}[!htbp]
  \begin{center}
    \leavevmode
    \ifpdf
		\includegraphics[width=6in,height=3.0in]{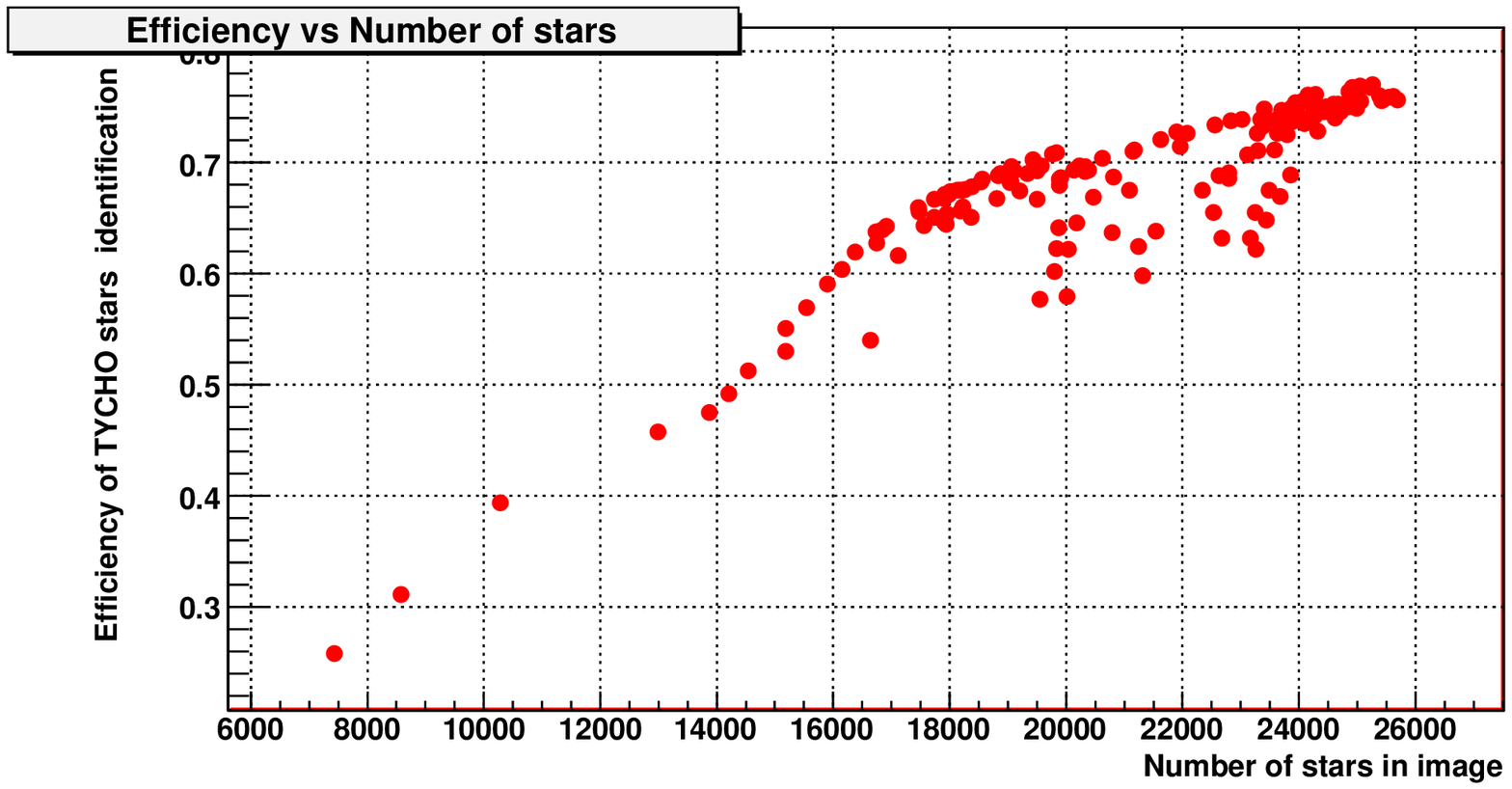}
    \else
		\includegraphics[width=6in,height=3.0in]{PhotoEff/eff_vs_stars/eff_vs_stars_aver20.eps}
    \fi
    \caption{Efficiency of star identification in function of number of
stars in image for sky field 0800$+$20. Star catalog obtained from images averaged over 20}
    \label{fig_eff_vs_stars}
  \end{center}
\end{figure}

\begin{figure}[!htbp]
  \begin{center}
    \leavevmode
    \ifpdf
		\includegraphics[width=6in,height=3.0in]{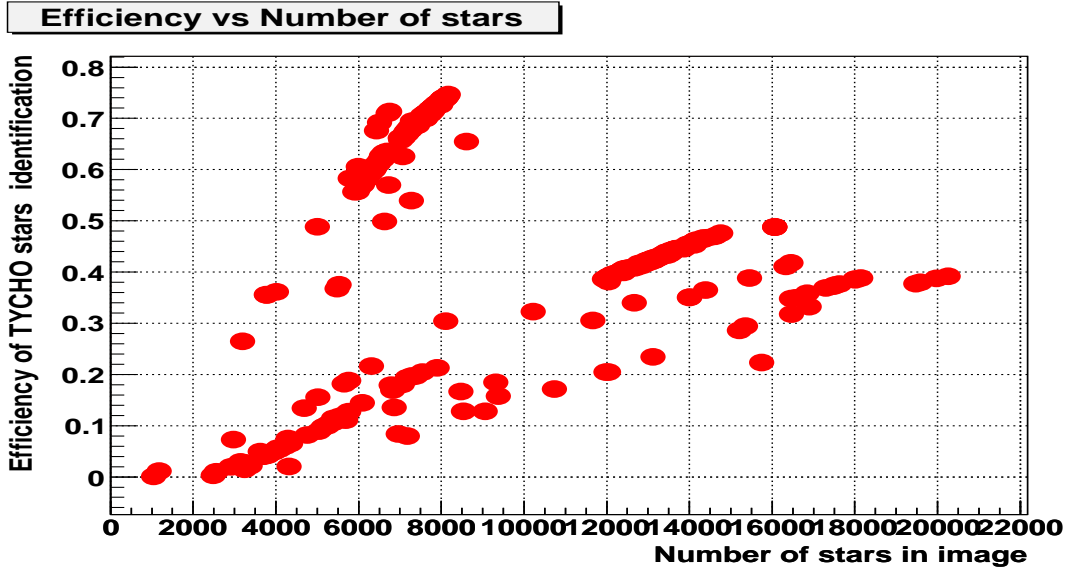}
    \else
		\includegraphics[width=6in,height=3.0in]{PhotoEff/eff_vs_ratio/eff_vs_stars.eps}
    \fi
    \caption{Efficiency of star identification in function of number of stars on single 10s image }
    \label{fig_eff_vs_stars_many_fields}
  \end{center}
\end{figure}

The efficiency of star identification can be parameterized in function of
number of stars in image. Figures \ref{fig_eff_vs_stars} and \ref{fig_eff_vs_stars_many_fields} show star identification efficiency in function of 
number of stars in image for star catalog created from images averaged over
20 and star catalog created from single images ( respectively ) . These plots were obtained for 
single sky field 0800$+$20. 

\begin{figure}[!htbp]
  \begin{center}
    \leavevmode
    \ifpdf
		\includegraphics[width=6in,height=3.0in]{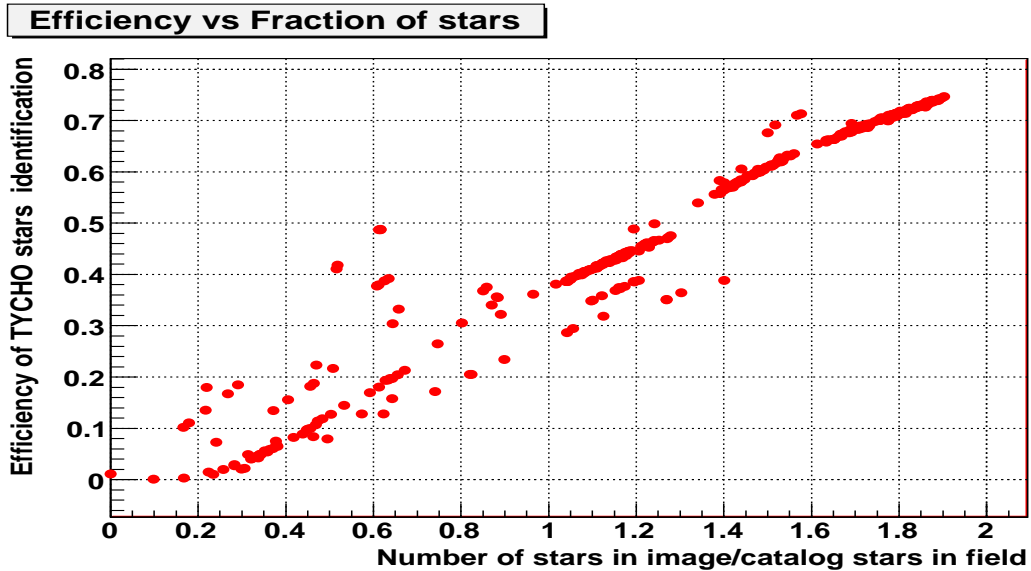}
    \else
		\includegraphics[width=6in,height=3.0in]{PhotoEff/eff_vs_ratio/eff_vs_fract.eps}
    \fi
    \caption{Efficiency of star identification in function of ratio $R_{cat}$}
    \label{fig_eff_vs_ratio}
  \end{center}
\end{figure}

However, as can be observed from Figures  \ref{fig_eff_vs_stars} and
\ref{fig_eff_vs_stars_many_fields} this is not the best parametrization.
Much better parametrization is efficiency in function of ratio $R_{cat}$,
defined as :

\begin{equation}
R_{cat} = \frac{N_{image\_stars}}{N_{cat\_stars}}
\label{eq_ratio_cat}
\end{equation}

where $N_{image\_stars}$ is number of stars in image and $N_{cat\_stars}$ is
number of stars found in reference catalog ( used in cataloging ) in the
observed field.
Star identification efficiency in function of ratio $R_{cat}$ is show in
Figure \ref{fig_eff_vs_ratio}, this efficiency was calculated for different
sky fields and different nights. As expected this dependency is linear.
The average efficiency of star identification on single 10s exposure was
determined by averaging efficiency from many single images collected during 
different nights ( every 25 image from single night was cataloged ). 
The resulting average efficiency for stars of brightness 0-12$^m$ is :

\begin{equation}
\epsilon_{star} \approx 0.49
\label{eq_aver_single_image_star_eff}
\end{equation}

This number can be used to calculate total efficiency of the on-line flash
recognition algorithm ( Sec. \ref{sec_opt_of_algo} ). Multiplication of
efficiency of algorithm cuts determined as $\epsilon_{cuts} \approx $70\% by efficiency of 
finding stars in image $\epsilon_{star} \approx $50\% gives the overall efficiency of the 
flash recognition algorithm $\epsilon_{on-line} \approx$35\% ( for stars 0-12$^m$ ).
It is larger for brighter stars and can reach nearly 70\% for stars 0-10$^m$.

\begin{figure}[!htbp]
  \begin{center}
    \leavevmode
    \ifpdf
		\includegraphics[width=6in]{sigmamag_vs_mag_20070606_aver20_2006_at_pi3.gif}
    \else
		\includegraphics[width=6in]{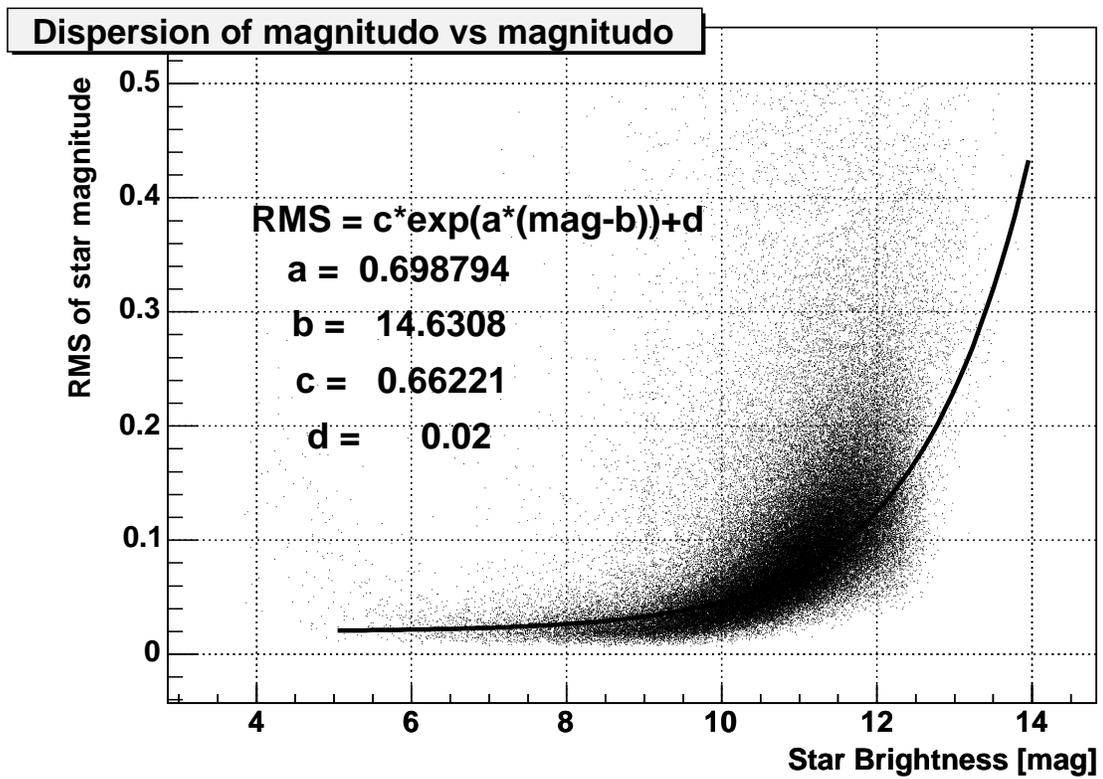}
    \fi
    \caption{Precision of star brightness measurements in the photometry on images obtained by averaging 20 single (10s) images.}
    \label{fig_sigmamag_vs_mag}
  \end{center}
\end{figure}

The precision of star brightness measurements can be determined by finding
dispersion of magnitude measurements for individual stars and plotting
dispersion in function of star brightness. The Figure \ref{fig_sigmamag_vs_mag} shows 
distribution of dispersion in function of star brightness for stars observed during
night 2007.06.06/07.

\subsection{Off-line algorithms}  
\label{sec_offline_algo}

Off-line algorithms act on data stored in the database. The data is
cataloged in the way described in previous section and it is stored in tables
\textbf{frame}, \textbf{star} and \textbf{measurements}. It is optimized for certain types of queries
which are executed by analyzing programs. The algorithms described here have been
implemented and tested on a star catalog created from images averaged over twenty - so
called \texttt{aver20} database. However, they can also be used to analyse
the catalog of single images data. Two algorithms developed by author will be
described. 
The first one looks for new objects appearing in the star catalog which
 correspond to new objects appearing in the sky.
Such kind of processes may be due to nova stars explosions or other
kinds of processes when object below detection limit suddenly increases its
brightness and appears as a new object.
The second algorithm looks for sudden increases of
stars brightness, such events can occur in flare stars, but this can
also happen in other objects like blasars or AGNs.
In both cases rejection of false event was the most important task. 
Both algorithms were implemented in \texttt{perl} scripting language.

\subsubsection{Nova identification algorithm}
\label{sec_nova_ident_algo}
This algorithm was developed to find new objects which were not present in
the star catalog before. The algorithm is invoked for data collected during 
specified night. It performs the analysis of all new objects added to the catalog during analysed
night. The objects which are expected to be found by this kind of
algorithm are those which are normally below detection threshold of "Pi of
the Sky" detector, but due to intrinsic reasons increase their brightness
and become possible to be detected. The astrophysical processes can 
be nova star explosion, supernova or GRB, but it can also be a flare star or
even variable star of large amplitude of brightness variations.
There is also large amount of background processes which must be efficiently
rejected. 
The algorithm relies on flag \texttt{Measurements.new\_star} which is filled during
cataloging. This field is true for the record which is the first measurement of 
the object. The algorithm is realized by \texttt{perl} script
\texttt{do\_flareevents.pl}. The main steps of the algorithm are the following :\\

\begin{enumerate}
\item connection to the database star catalog
\item selection of fields which were observed during the given night and have
total number of observations $N_{obsfield}^{total} > N_{obsfield}^{min}$ .
Only images from selected fields will be considered in next steps.
\item every field selected in the previous step is analysed now. The first step is
selection of all images of specific field from given night to list \texttt{id\_frm\_table}
\item images are verified against clouds and moonlight, average pixel value $P_{avg}$
in the image is compared with the limit $T_{avg}$, calculated as :
\begin{equation}
T_{avg}=<P_{avg}> + N_{avgcut} \cdot \sigma_{avg}
\label{eq_avg_cut}
\end{equation} 
values $<P_{avg}>$ and $\sigma_{avg}$ are obtained from distribution of
 average pixel values from all images collected so far.
In case the image does not satisfy condition 
$P_{avg} < T_{avg}$ it is rejected and skipped from further analysis.
Cloudy images are also rejected in cataloging phase by condition imposed on
the number of stars in image ( see Sec. \ref{sec_cataloging} ).

Every image from the analysed field kept in the array \texttt{id\_frm\_table} is now
analysed, and the following subsequent criteria are applied to every image :\\

\item all measurements from an image, flagged as new object ( with \texttt{Measurements.new\_star} = true )
are selected with additional conditions for brightness of the new
object \mbox{mag  $\leq mag_{min}$}  and for number of measurements \mbox{$N_{obs} \geq N_{obs}^{min}$}.
Every object accepted by the above criteria is analysed as potential
nova-like event and subsequent criteria are applied. 
\item verification of object in the second camera ( if available )
\item determination of the number of object observations during the given night ( in case 
the second camera is not working this is the best indication of the event quality )
\item The event is checked if there is no nearby ( R<$R_{bigstar}^{min}$ ) very bright star ( $mag_{bigstar} < mag_{bigstar}^{max}$)  
  in "Pi of the Sky" or TYCHO-2 \cite{tycho} catalog.
\item Despite the fact that the field was already observed at least
$N_{obsfield}^{min}$ times, it is possible ( most probably due to bad weather conditions
or faintness of the object ) that it is just a normal star which remained
undetected before. Thus every event is checked against the 
 TYCHO catalog and in case a star of brightness $mag < mag_{tycho}^{max}$ is found in distance R <
$R_{star}^{tycho}$ the event candidate is rejected.
\item In case the image was collected with permanently opened shutter, 
an additional check for bright stars below and above ( in CCD y coordinate ) 
is performed. In case bright star ( $mag < mag_{above}$ ) 
is found in the same image in the column close to the event candidate column ( $|X_{event} - X_{bigstar}| < \Delta X_{bigstar}^{max}$ ), 
then the event candidate is rejected.
\item Event is verified against the list of known hot pixels, which are stored
in database table \textbf{HotPixels} ( Fig. \ref{fig_starcat} )
\item In case coordinates of the event are close to one of the Solar System
planets it is rejected ( $D_{planet} < R_{planet}$ )
\item Dispersions of mount tracking speeds are checked for each image. In case they 
exceed certain limit, the image is flagged as bad quality image ( shifted image ). Events
found on such images are automatically rejected.
\end{enumerate}

All events which survived criteria 6) are saved to database table \textbf{FlareEvents}
( Fig. \ref{fig_starcat} ). Later cuts do not reject events definitely, only
rejection flag field \texttt{fl\_rej\_tycho} is set in the database. Its value
depends on the cut which rejected the event. 
Table \ref{tab_rejection_codes} lists possible values of rejection flag.

\begin{table}[htbp]
\begin{center}
\begin{tabular}{|c|c|}
\hline
\textbf{Rejection Flag Value} & \textbf{Description} \\
\hline
\hline
0 & accepted event \\
\hline
1 & star found in TYCHO \\
\hline
2 & nearby bright star found in TYCHO \\
\hline
3 & saturated star found above event ( $Y_{star} > Y_{event}$ ) \\
\hline
4 & saturated star found below event ( $Y_{star} < Y_{event}$ ) \\
\hline
5 & bright star found in \piname star catalog \\
\hline
6 & hot pixel \\
\hline
7 & sky background level > $T_{avg}$ \\
\hline
8 & anti planets cut \\
\hline
9 & image quality cut \\
\hline 
\end{tabular}
\caption{Rejection flags values}
\label{tab_rejection_codes}
\end{center}
\end{table}

Events presentation is implemented in form of \texttt{php} script selecting events
from the database. The additional difficulty is the fact that algorithm is
executed on the remote server and results are originally saved to the remote
database. The events data is copied to the local server ( Sec.
\ref{sec_data_synchronization} ) for convenience and also in order not to
overuse the Internet link to LCO by direct accessing of the remote database
multiple times.
The synchronization is executed once and all later analysis and inspections of events are
using events data from local server not from LCO which is much faster.

\begin{table}[htbp]
\begin{center}
\begin{tabular}{|c|c|c|}
\hline
\textbf{Parameter} & \textbf{Default Value} & \textbf{Script option}  \\
\hline
\hline
$N_{obsfield}^{min}$   & 50 & -min\_obs\_field \\
\hline
$mag_{min}$            & 11 & -mag \\
\hline
$mag_{bigstar}^{max}$  &  8 & -near\_bigstar\_max\_mag \\
\hline
$R_{bigstar}^{min}$    & 15*36 [arcsec] & - \\
\hline
$R_{star}^{tycho}$     & 150 [arcsec] & -near\_star\_radius\_arcsec \\
\hline
$mag_{tycho}^{max}$    & 13 & -max\_tycho\_star \\
\hline
$mag_{bigstar}^{above}$ & 5 & -max\_bigstar\_above \\
\hline
$\Delta X_{bigstar}^{max}$ & 10 & -max\_bigstar\_x\_dist \\
\hline
$N_{avgcut}$ & 3 & - \\
\hline 
$R_{planet}$ & 7200 [arcsec] & -reject\_planets \\
\hline 
\end{tabular}
\caption{Most important parameters of the nova identification algorithm}
\label{tab_nova_algo_params}
\end{center}
\end{table}

\begin{figure}[!htbp]
  \begin{center}
    \leavevmode
    \ifpdf
      \includegraphics[width=6in]{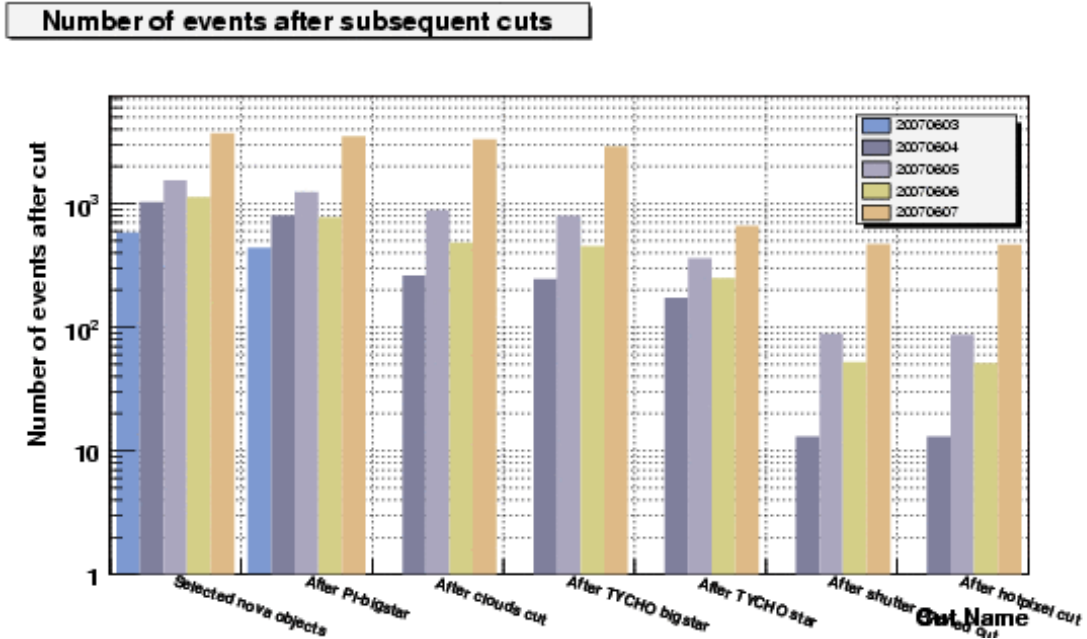}
    \else
      \includegraphics[width=6in]{cuts/nova/nova_algo_cuts.eps}
    \fi
    \caption{Number of events after subsequent cuts of nova identification algorithm. Results from five nights are shown.}
    \label{fig_nova_algo_cuts}
  \end{center}
\end{figure}

The cataloging program stores measurements of $10^4$ - $10^6$ stars per night and flags new objects with special flag. 
The nova identification algorithm analysis stars from a given night and
flagged as new objects in the catalog. Large part of the job is performed by the
cataloging program, the algorithm itself analysis $N_{new} \approx 10^4-10^5$ objects, which takes 1-3 hours.
Figure \ref{fig_nova_algo_cuts} shows number of events after subsequent cuts for several nights.

Efficiency of nova identification algorithm was tested. The following
procedure was used, given number of stars ( typically 100 / image ) of given
brightness were added to star catalog database. 
Algorithm was executed on data from night for which artificial stars with measurements were
added. The number $N_{gen}^{ident}$ of generated and accepted events was
determined and using this number efficiency was determined as : \\

\begin{equation}
	\epsilon_{nova} = \frac{N_{gen}^{ident}}{N_{gen}}
\label{eq_eff_nova}
\end{equation}

\begin{figure}[!htbp]
  \begin{center}
    \leavevmode
    \ifpdf
      \includegraphics[width=6in]{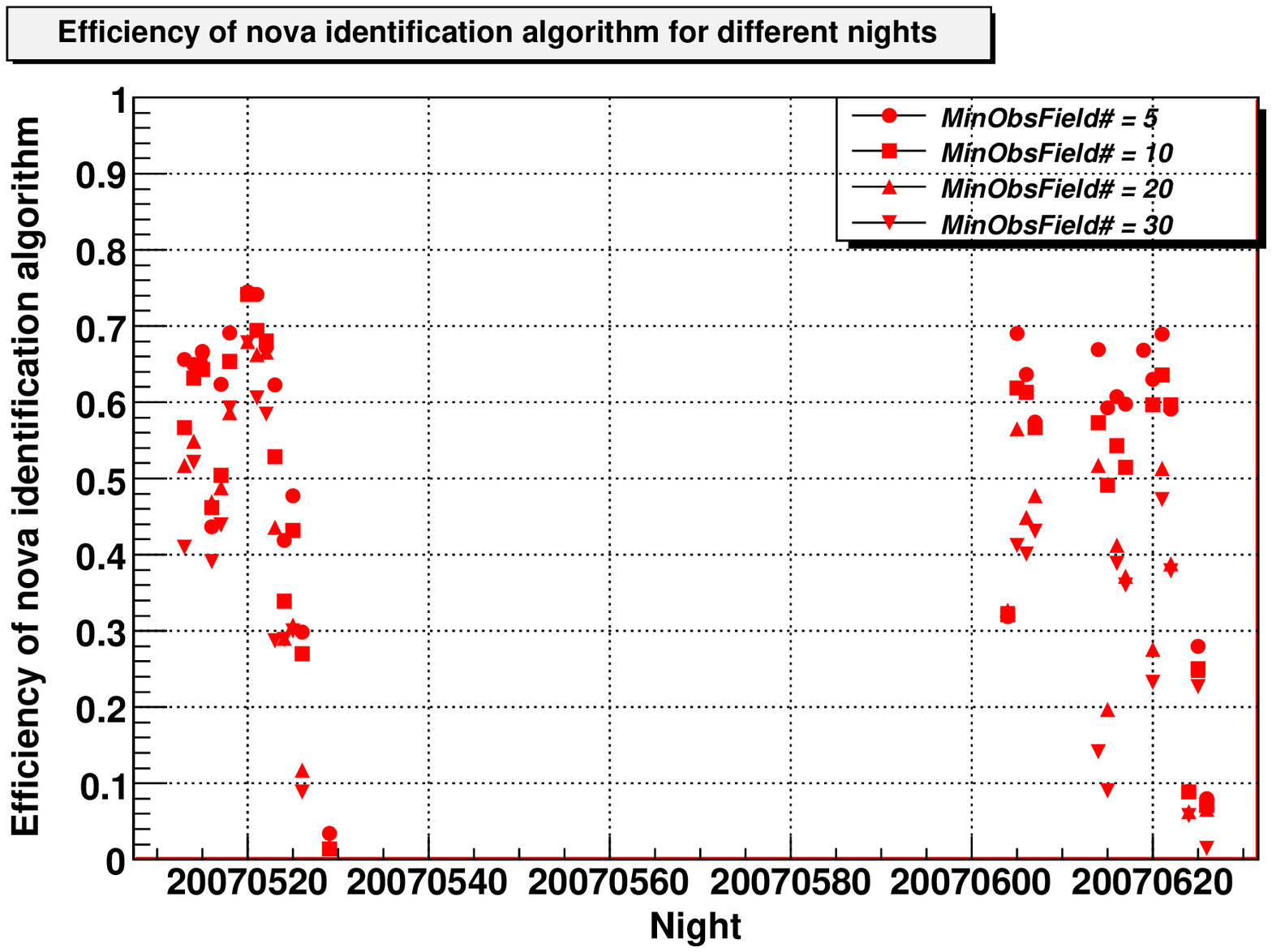}
    \else
      \includegraphics[width=6in]{nova_eff/novaeff_vs_night.eps}
    \fi
    \caption{Efficiency of nova identification algorithm for several nights.
Different values of minimum number of field observations requirement were
tested}
    \label{fig_novaeff_vs_night}
  \end{center}
\end{figure}

\begin{figure}[!htbp]
  \begin{center}
    \leavevmode
    \ifpdf
      \includegraphics[width=6in]{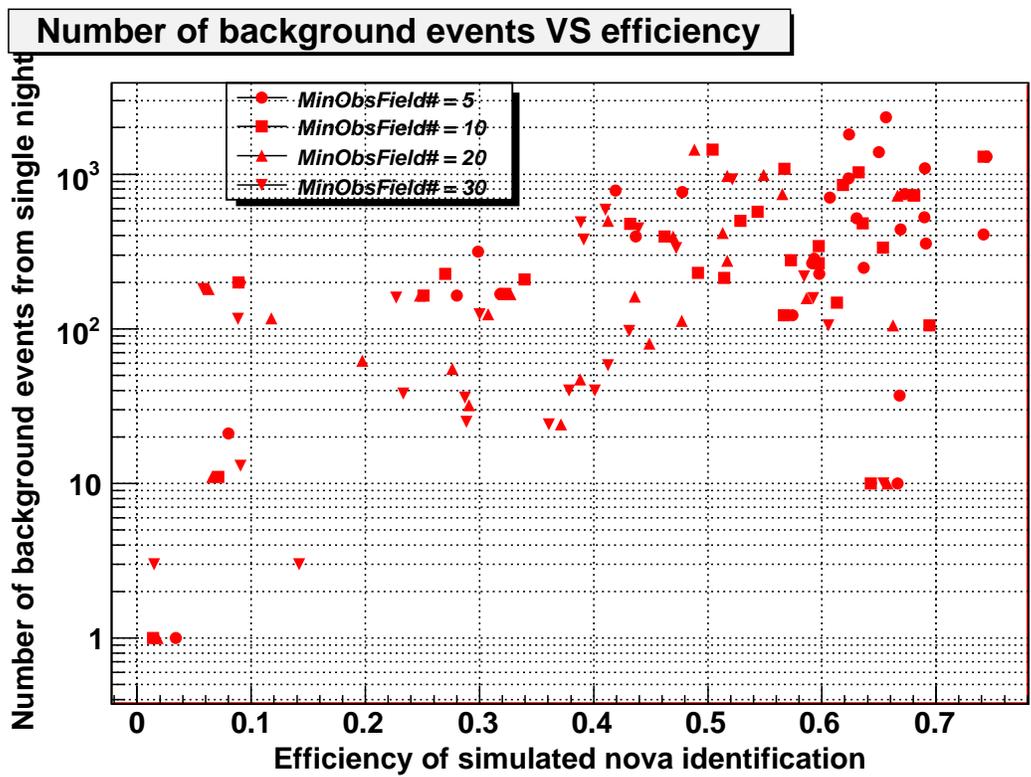}
    \else
      \includegraphics[width=6in]{nova_eff/novaeff_vs_bkg.eps}
    \fi
    \caption{Efficiency of nova identification algorithm vs number of
background events from single night}
    \label{fig_novaeff_vs_bkg}
  \end{center}
\end{figure}

\begin{figure}[!htbp]
  \begin{center}
    \leavevmode
    \ifpdf
		\includegraphics[width=7.2in,height=4in]{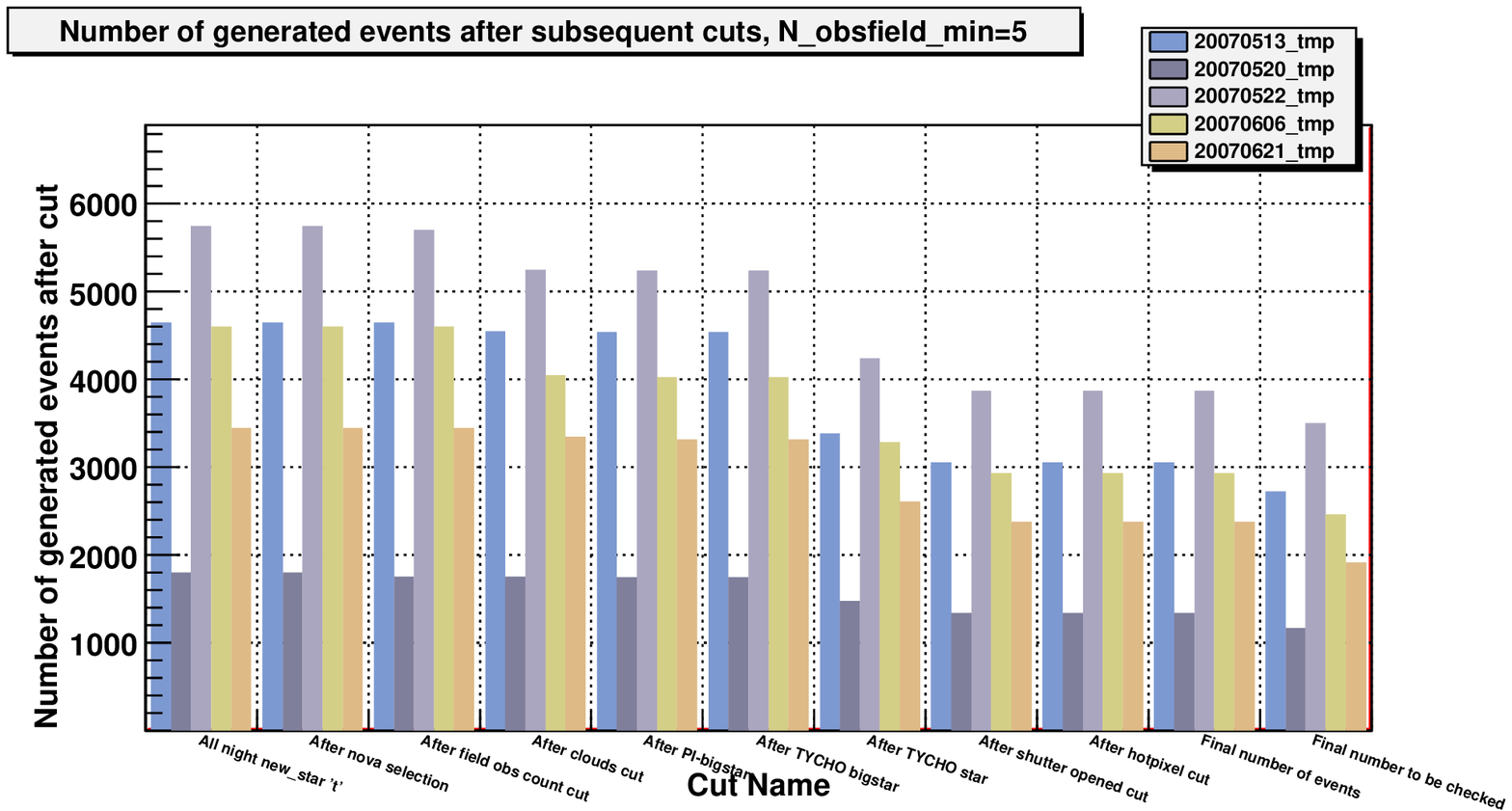}
		\includegraphics[width=7.2in,height=4in]{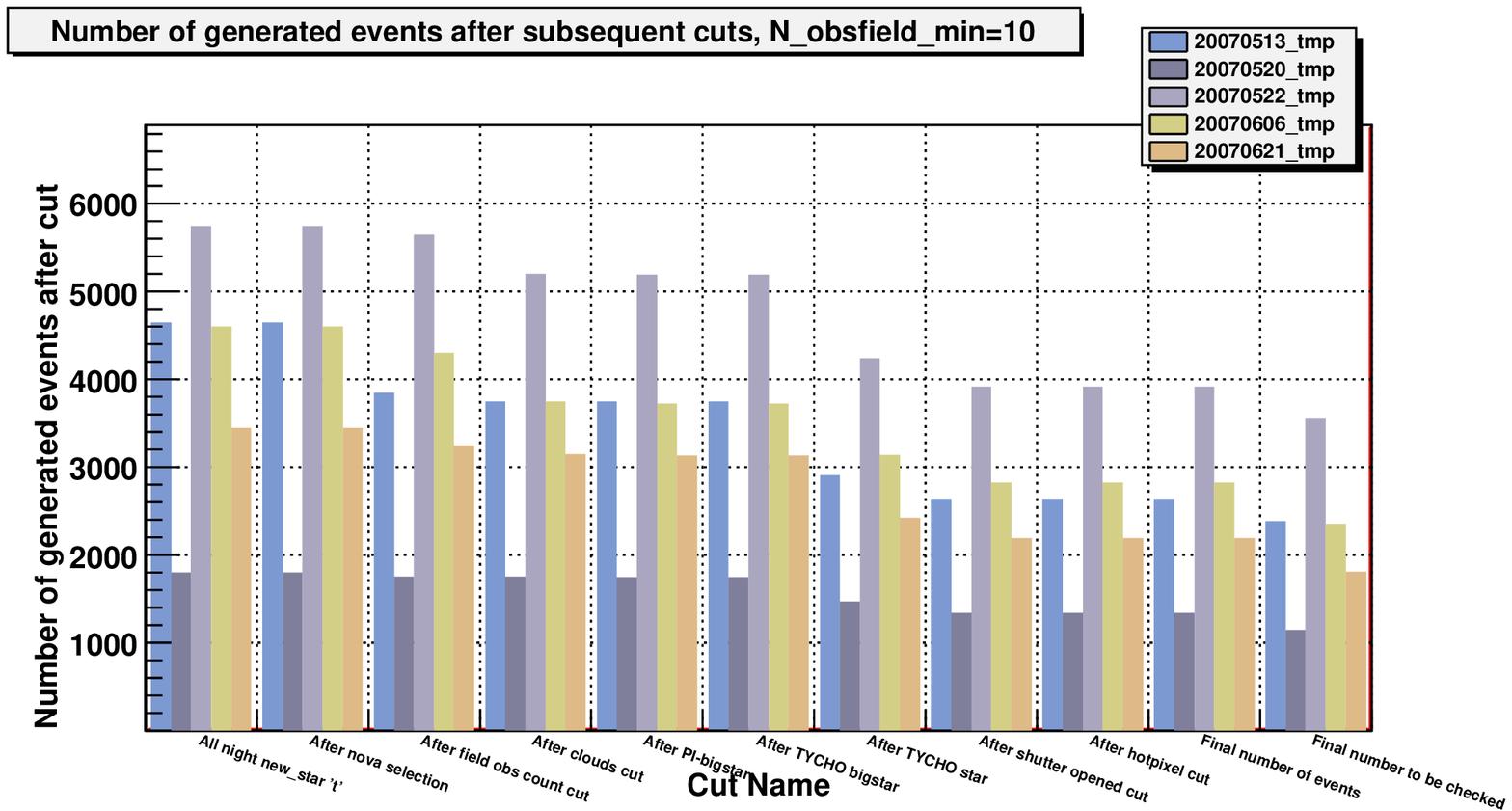}
    \else
		\includegraphics[width=7.2in,height=4in]{nova_eff/min_obs_field_compare/min_obs_field5.eps}
		\includegraphics[width=7.2in,height=4in]{nova_eff/min_obs_field_compare/min_obs_field10.eps}
    \fi
    \caption{Efficiency losses on subsequent cuts of algorithm for different
values of parameter $N_{obsfield}^{min}$=5,10}
    \label{fig_novaeff_loses1}
  \end{center}
\end{figure}

\newpage
\begin{figure}[!htbp]
  \begin{center}
    \leavevmode
    \ifpdf
		\includegraphics[width=7.2in,height=4in]{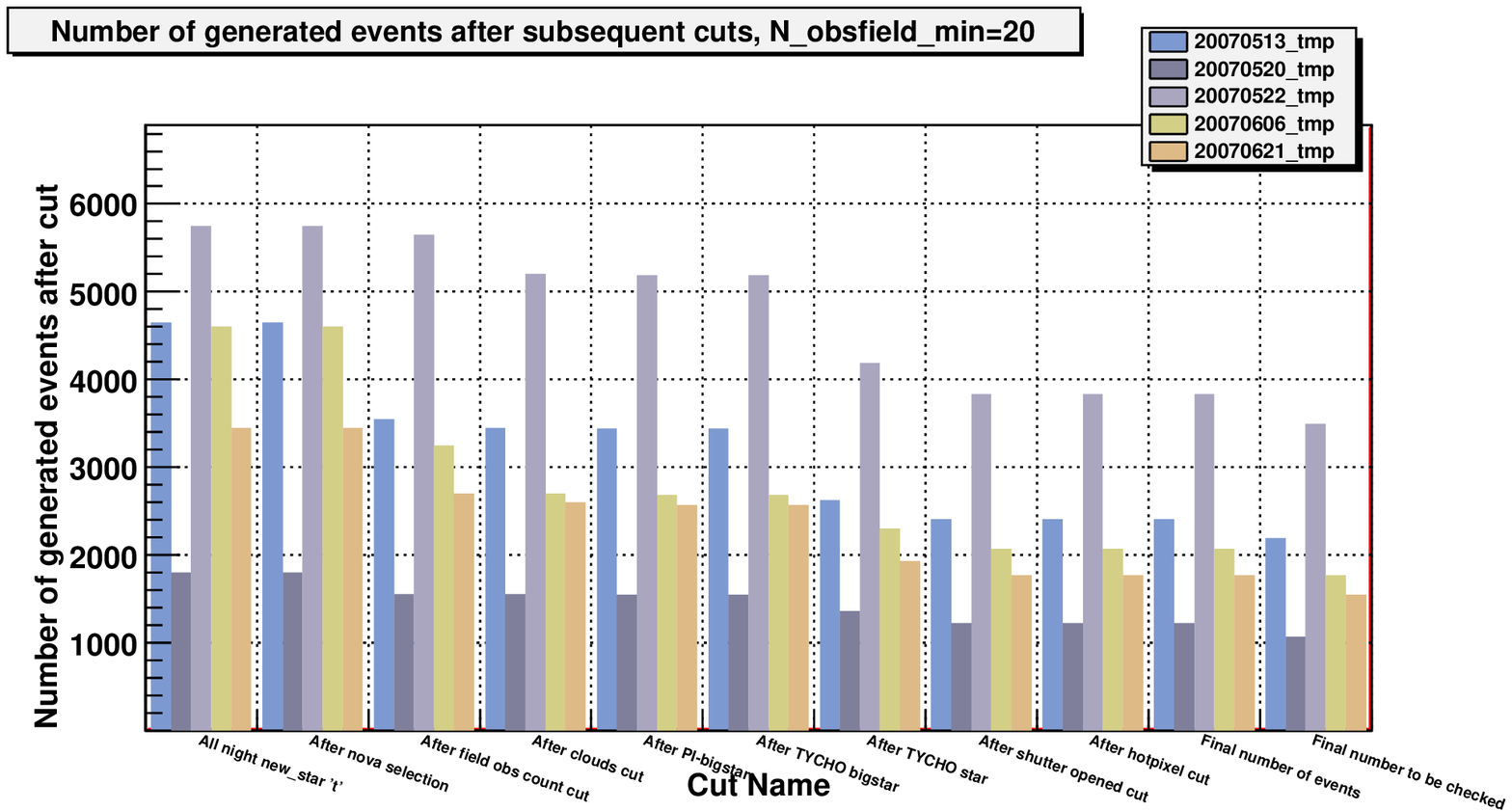}
		\includegraphics[width=7.2in,height=4in]{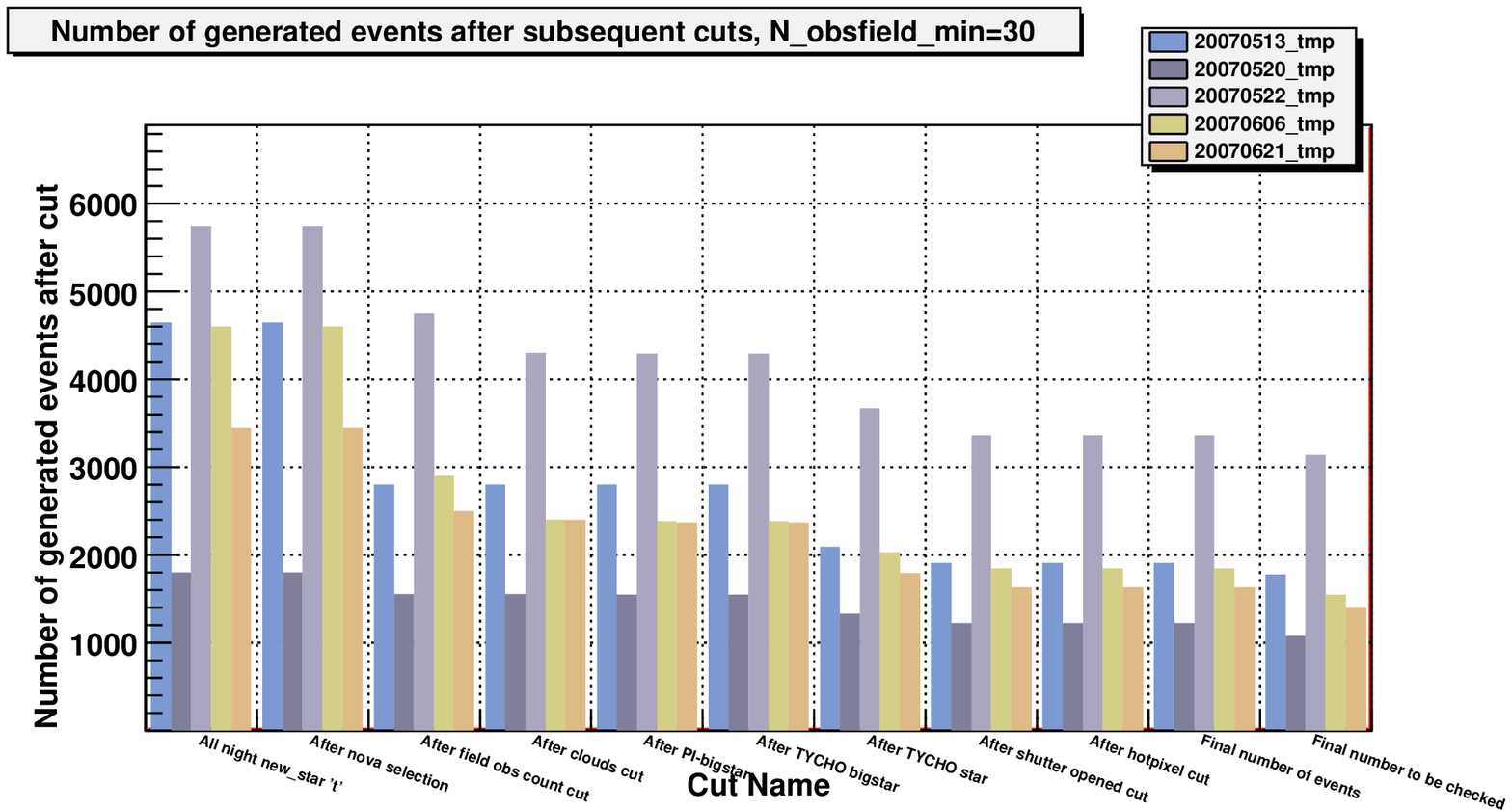}
    \else
		\includegraphics[width=7.2in,height=4in]{nova_eff/min_obs_field_compare/min_obs_field20.eps}
		\includegraphics[width=7.2in,height=4in]{nova_eff/min_obs_field_compare/min_obs_field30.eps}
    \fi
    \caption{Efficiency losses on subsequent cuts of algorithm for different
values of parameter $N_{obsfield}^{min}$=20,30}
    \label{fig_novaeff_loses2}
  \end{center}
\end{figure}

The Figure \ref{fig_novaeff_vs_night} shows efficiency for several
subsequent nights. The best efficiencies were obtained when only 5 earlier
observations of sky field were required. However, this is very small number
of field observations and many weak objects still remain undetected. This
implies large amount of background events, which can be seen in Figure
\ref{fig_novaeff_vs_bkg}.
The Figures \ref{fig_novaeff_loses1} and \ref{fig_novaeff_loses2} show efficiency losses on subsequent
nights for different values of $N_{obsfield}^{min}$ parameter. 

%
\begin{figure}[!htbp]
  \begin{center}
    \leavevmode
    \ifpdf
		\includegraphics[width=4in,height=4in]{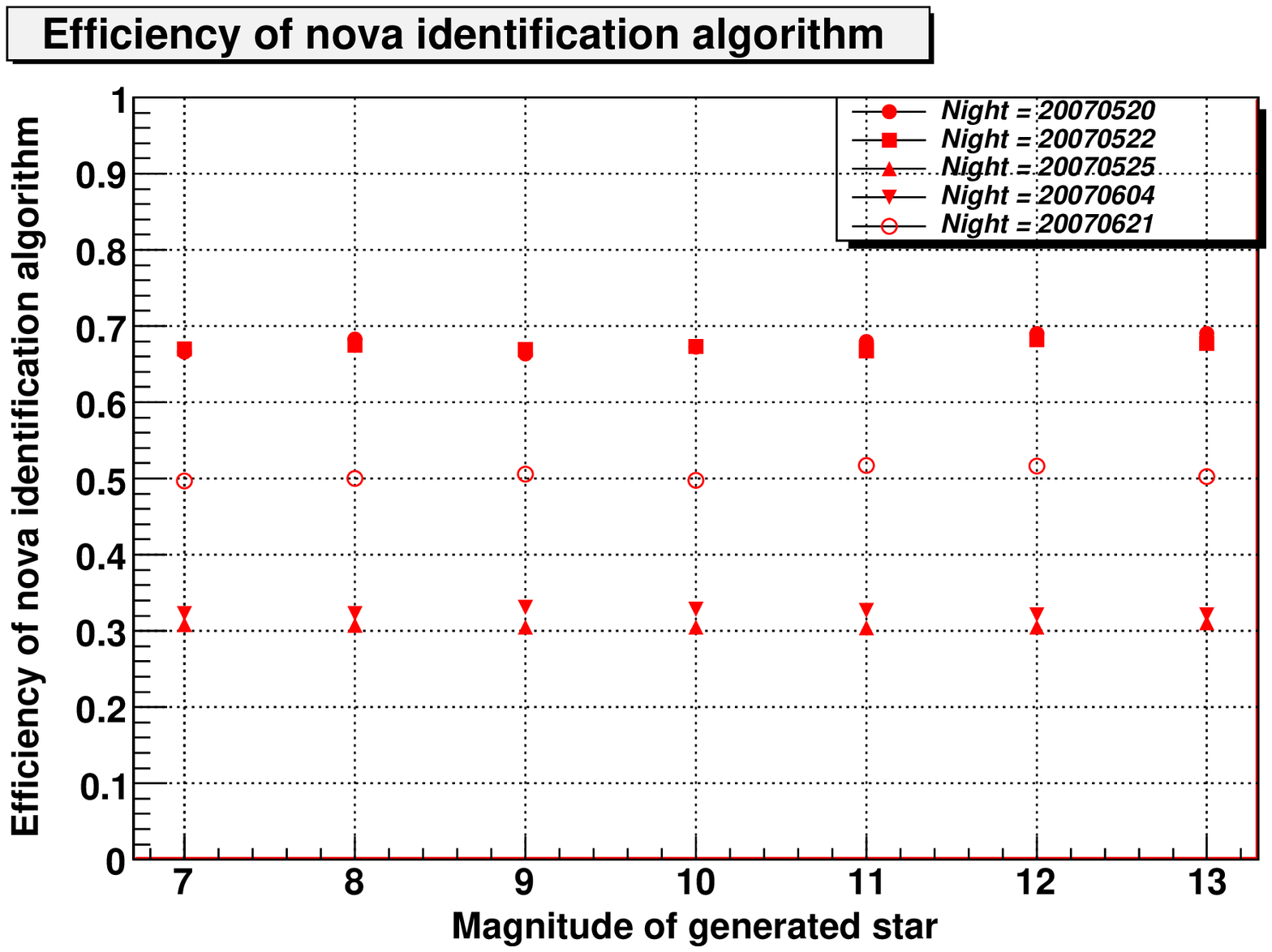}
		\includegraphics[width=4in,height=4in]{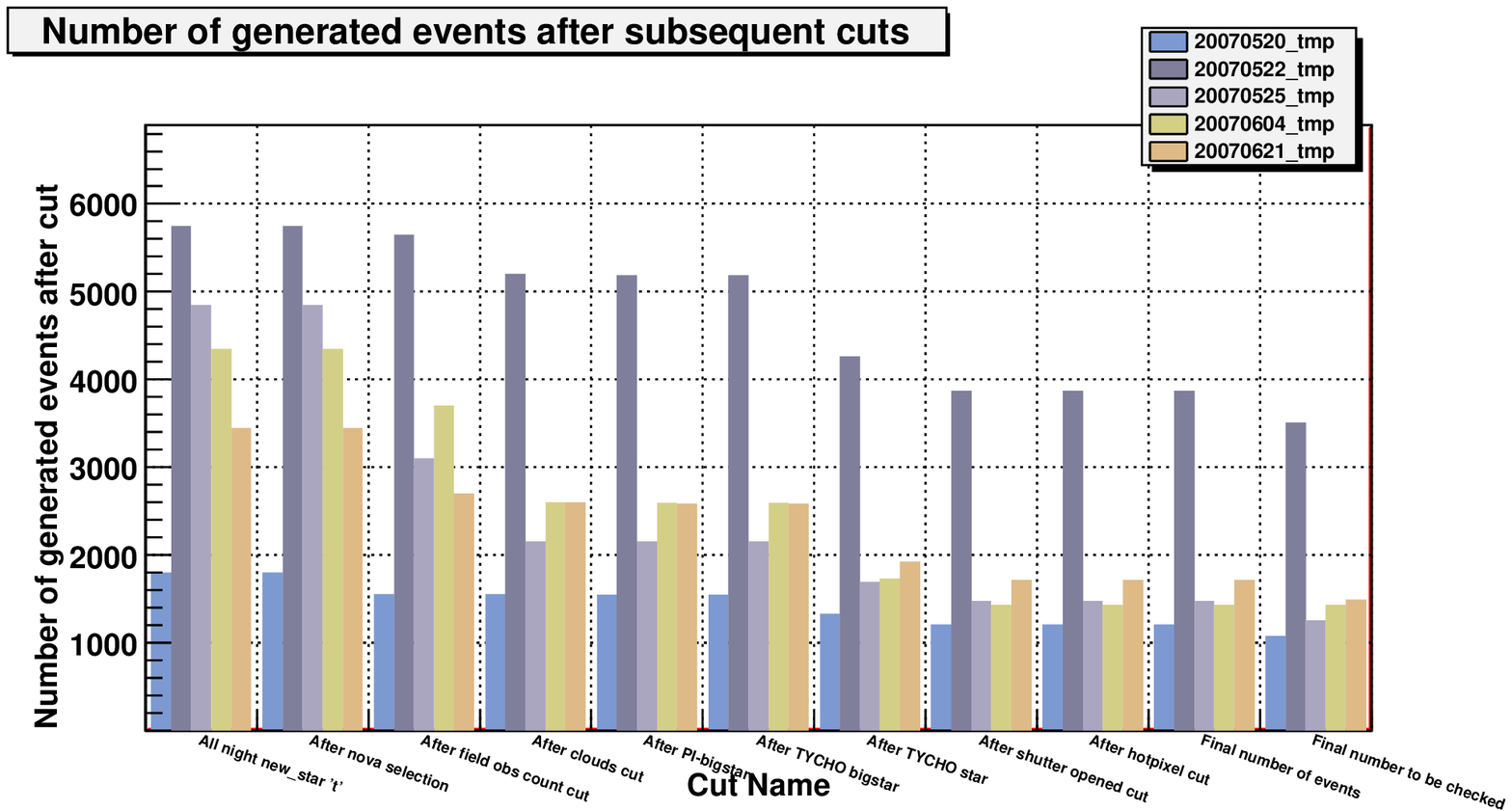}
    \else
		\includegraphics[width=4in,height=4in]{nova_eff/many_nights.eps}
		\includegraphics[width=4in,height=4in]{nova_eff/many_night_effloss_reason.eps}
    \fi
    \caption{Efficiency of nova identification algorithm for different brightness of object added to star
catalog, tested for different nights (upper plot). The lower plot shows number of
generated events after subsequent cuts.}
    \label{fig_novaeff_vs_mag}
  \end{center}
\end{figure}

\begin{figure}[!htbp]
  \begin{center}
    \leavevmode
    \ifpdf
		\includegraphics[width=7.2in,height=4in]{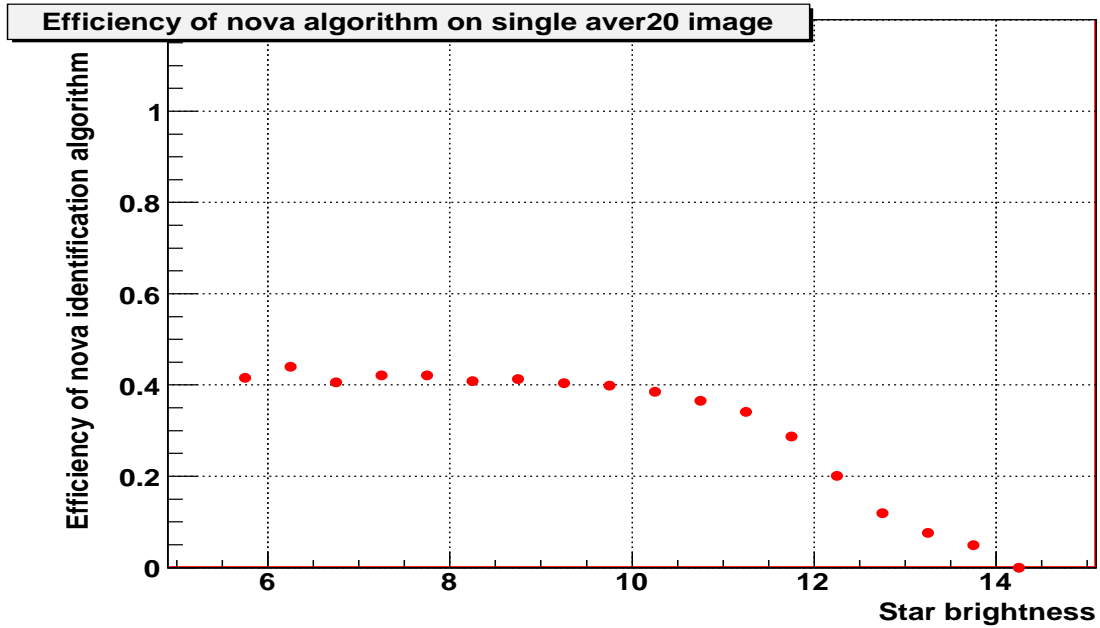}
    \else
		\includegraphics[width=6.0in,height=3.3in]{nova_eff/eff_on_single_aver20_image_total.eps}
    \fi
    \caption{Total efficiency of nova ( of brightness 0-12$^m$ ) identification algorithm on single aver20 image ( resulting from averaging 20 single images ).}
    \label{fig_overall_nova_eff}
  \end{center}
\end{figure}

\begin{figure}[!htbp]
  \begin{center}
    \leavevmode
    \ifpdf
		\includegraphics[width=7.2in,height=4in]{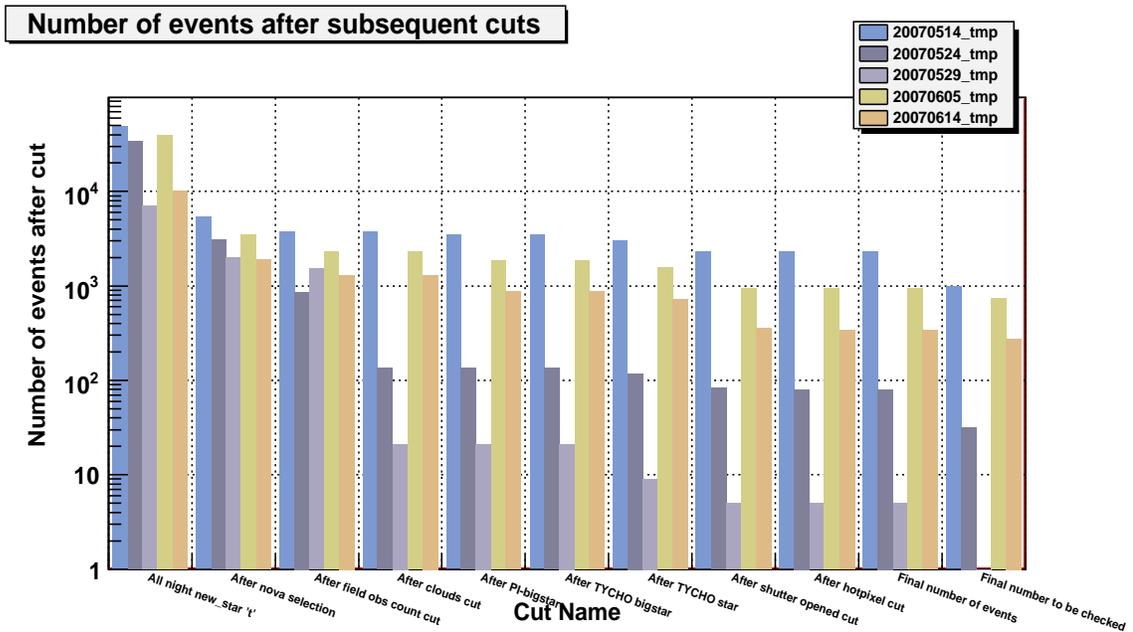}
    \else
		\includegraphics[width=6.0in,height=3.3in]{nova_eff/night_compare.eps}
    \fi
    \caption{Comparison of efficiency losses on subsequent cuts for 5 nights. During night 2007-05-29/30
the Moon was almost full and cut requiring low level of sky background 
rejected most of the generated events}
    \label{fig_novaeff_night_compare}
  \end{center}
\end{figure}

The Figure \ref{fig_novaeff_vs_mag} shows efficiency in
function of brightness of generated stars. It may be astonishing that
it is independent of star magnitude, however, it is clear that it should not.
The procedure described above tested only efficiency of subsequent cuts of
the algorithm which does not depend on star brightness. 
The efficiency is different for different nights mainly due to differences in Moon phase 
 and observed fields. Due to cut on number of field observations $N_{obsfield}>N_{obsfield}^{min}$ 
causing rejection of events generated in images of rarely observed fields.
The best way to eliminate dependency of the efficiency on number of field
observations would be to initialize star catalog with the complete catalog
of objects matching the range of "Pi of the Sky" telescope ( ~ 12-13$^m$ ).
Such a catalog should contain all types of objects ( star, galaxies etc ) in
the range of the telescope.
The overall efficiency of nova determination depends on the star brightness. This
dependency is "hidden" in the photometry procedure. The efficiency of star identification procedure was determined and
 described in Section \ref{sec_red_eff}. The result of  multiplication of photometry efficiency 
on single image (  result of average over 20 ) in
function of star brightness by average algorithm efficiency $\epsilon_{nova-algo} \approx$0.44 is
shown in Figure \ref{fig_overall_nova_eff}.
Comparison of efficiency losses on subsequent cuts for five nights is shown in Figure
\ref{fig_novaeff_night_compare}.

%

Important result of the tests is determination of the amount of background
events which must be verified daily. The Figure \ref{fig_novaeff_vs_bkg} shows
number of background events vs efficiency of the algorithm for several
different values of $N_{obsfield}^{min}$ parameter.
The final values of the parameters allow to efficiently suppress the
background so that number of events to checked does not exceed twenty events ( during clean sky night ).

\subsubsection{Flare identification algorithm}
\begin{figure}[!htbp]
  \begin{center}
    \leavevmode
    \ifpdf
      \includegraphics[width=2.8in,height=2.8in]{3413787_flare_20061127.gif}
		\includegraphics[width=2.8in,height=2.8in]{3413787_measurements_histo.gif}
    \else
      \includegraphics[width=2.8in,height=2.8in]{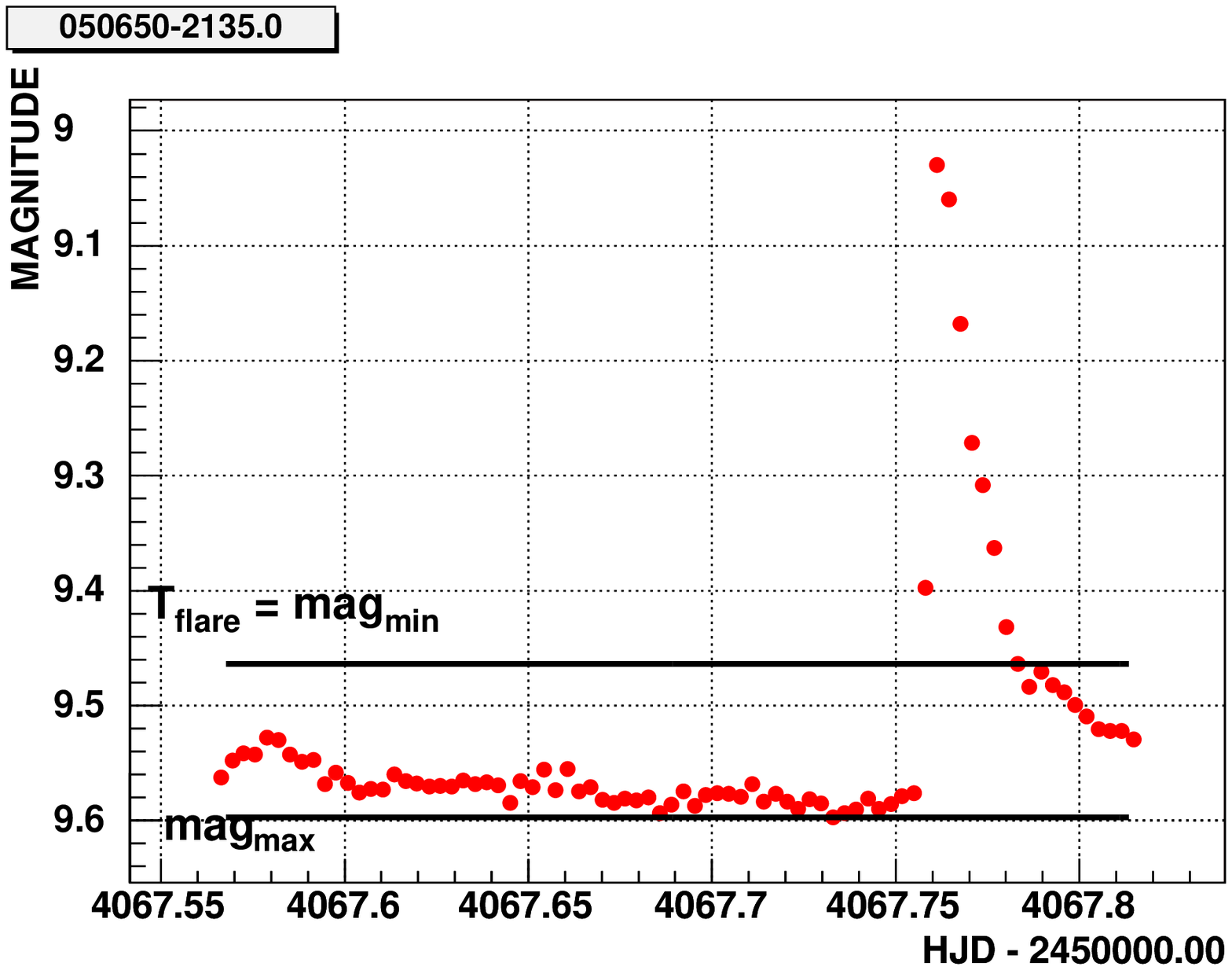}
		\includegraphics[width=2.8in,height=2.8in]{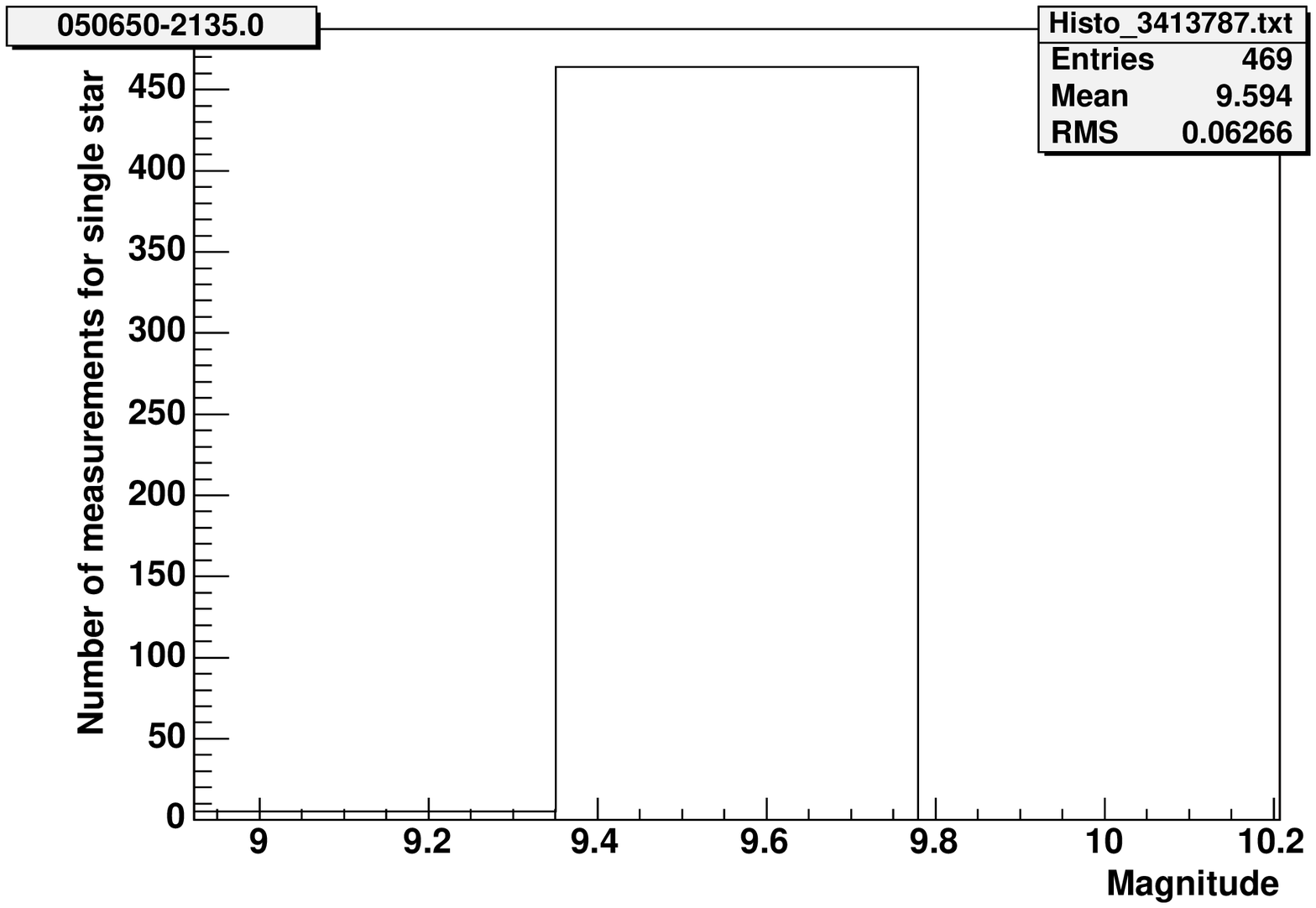}
    \fi
    \caption{Example of determined range of 85\% of measurements points ( area between horizontal lines in the left plot ) and distribution of all measurements for the same star (right plot)}
    \label{fig_measurements_strip}
  \end{center}
\end{figure}

\begin{figure}[!htbp]
  \begin{center}
    \leavevmode
    \ifpdf
		\includegraphics[width=4in]{208722300000.txt_histo.gif}
    \else
		\includegraphics[width=4in]{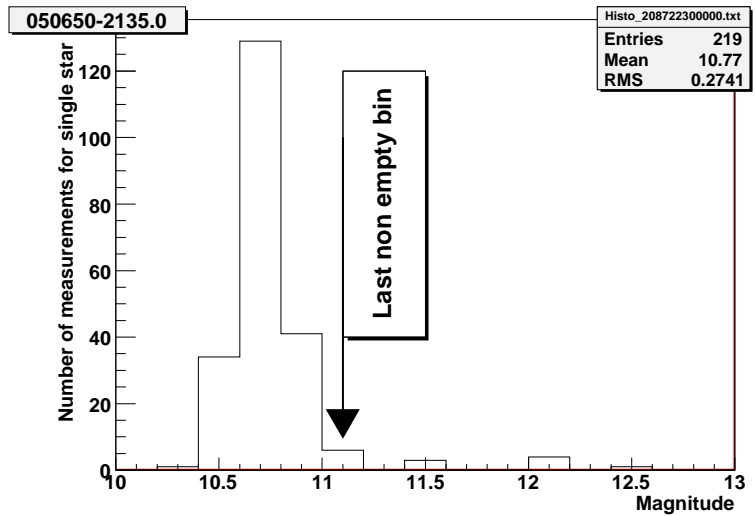}
    \fi
    \caption{The distribution of magnitude measurements for a single star an argument why not maximum magnitude is used as upper limit $mag_{max}$ in the algorithm}
    \label{fig_last_non_empty_bin}
  \end{center}
\end{figure}

This algorithm was developed to find outbursts of stars which are already in
the database star catalog, but manifest sudden increase of brightness. 
The algorithm is executed on data collected during specified night, it
is started after cataloging of new night is finished. Currently cataloging 
is performed on-line during data collection so analyzing script
( \texttt{find\_flares\_all\_nights.pl} ) can be launched in the morning.
The main steps of the algorithm are the following :\\
\begin{enumerate}
\item Selection of stars brighter then $mag_{min}$, observed during analysed night and having
total number of measurements $N_{obs} \geq N_{obs}^{min}$
\item Variability cut - star brightness range must satisfy condition : $mag_{MAX}-mag_{MIN} \geq T_{magdiff}$. 
The number of stars selected at this stage is of the order of $10^4-10^5$. It depends on
 quality of the night data and on $T_{magdiff}$ parameter.
This dependency is shown in Fig. \ref{fig_magdiff_histo}.

%
\begin{figure}[!htbp]
  \begin{center}
    \leavevmode
    \ifpdf
      \includegraphics[width=2.7in,height=2.7in]{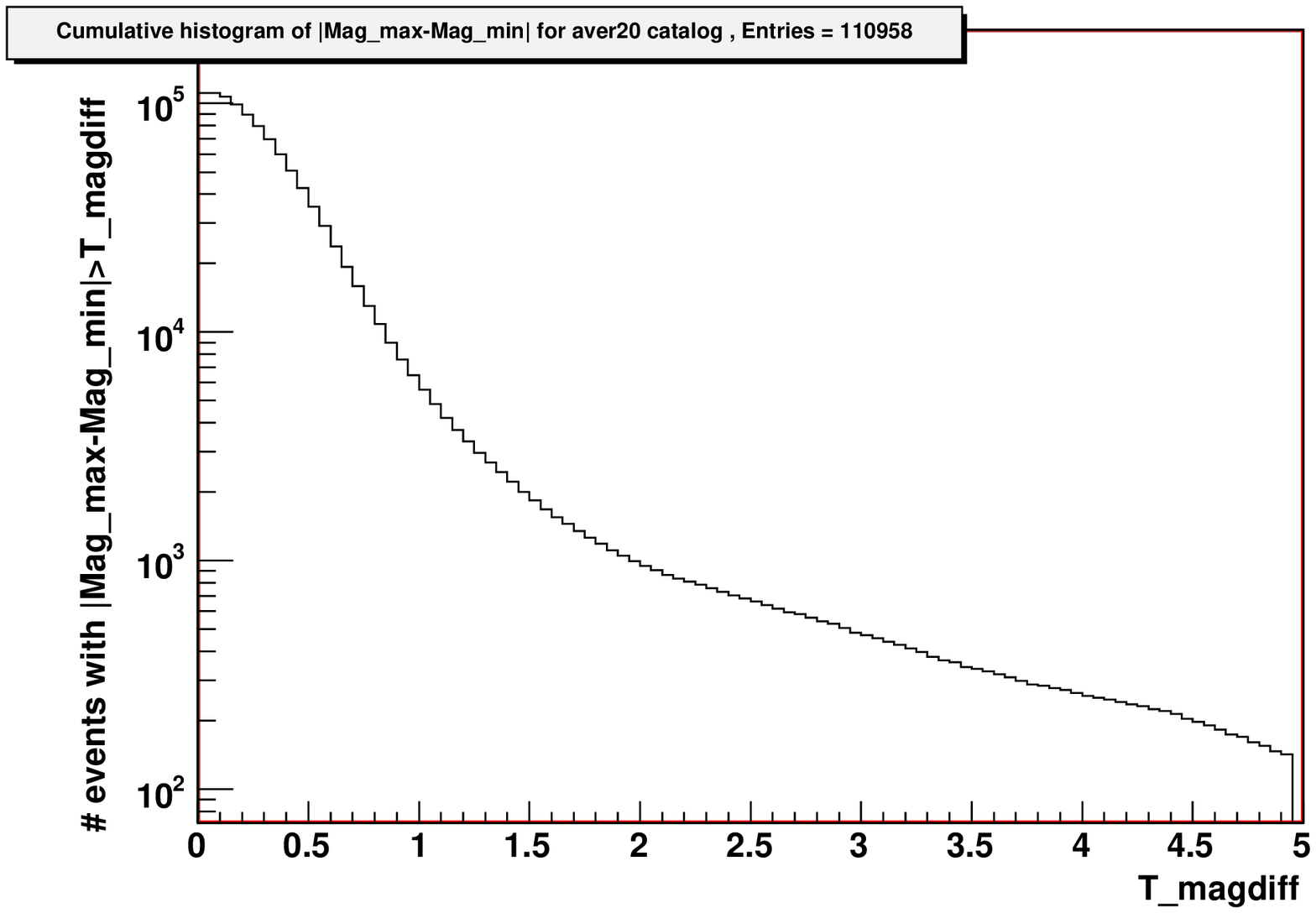}
		\includegraphics[width=2.7in,height=2.7in]{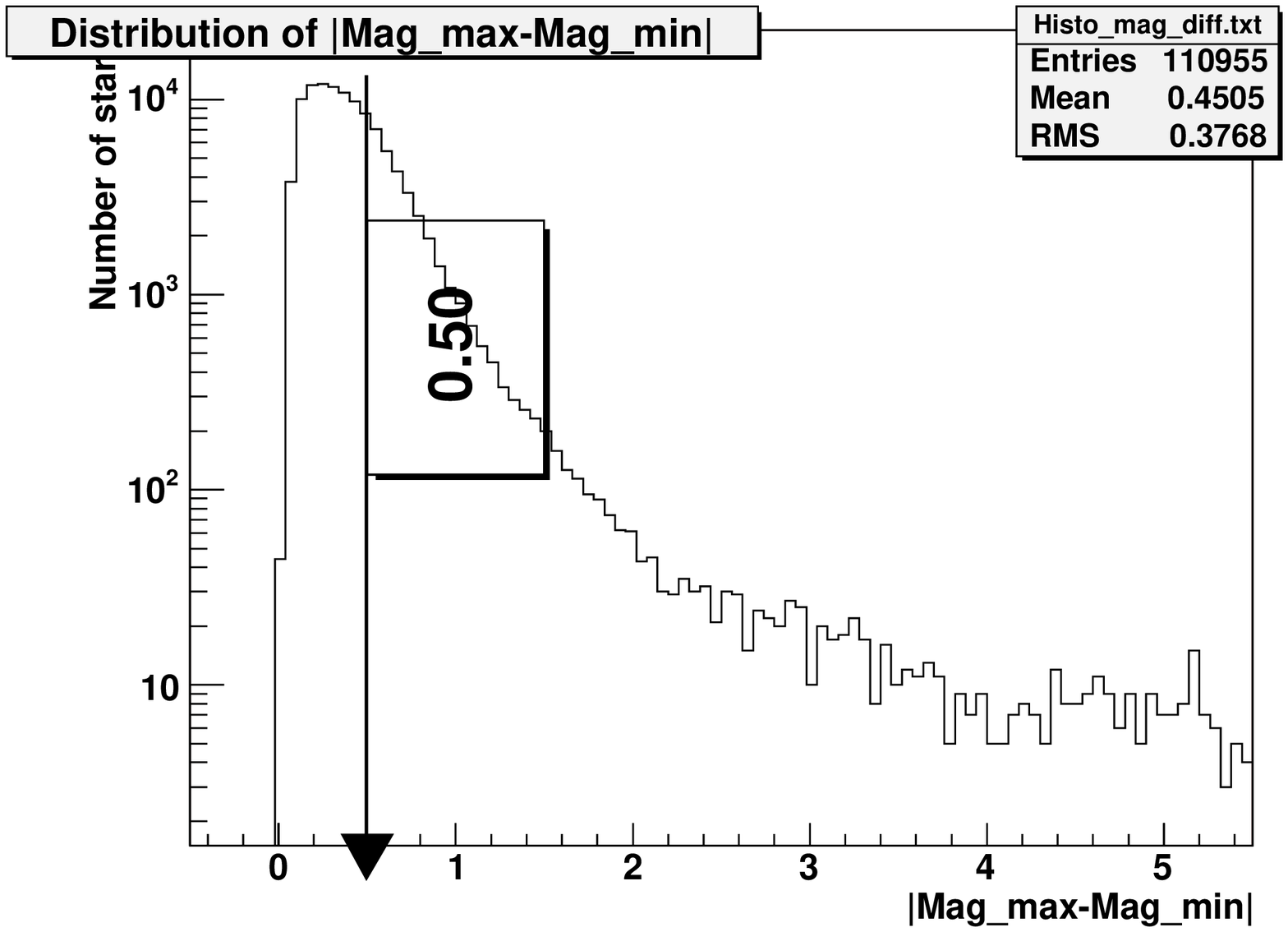}
    \else
		\includegraphics[width=2.7in,height=2.7in]{offline/flare/20070526_magdiff.eps}
		\includegraphics[width=2.7in,height=2.7in]{offline/flare/mag_diff_normal.eps}
    \fi
    \caption{Number of stars with $mag_{MAX}-mag_{MIN} \leq T_{magdiff}$ (left plot) and distribution of $mag_{MAX}-mag_{MIN}$ (right plot). Only stars satisfying condition on number of measurements are shown ( data from night 2007.05.26/27 ). }
    \label{fig_magdiff_histo}
  \end{center}
\end{figure}

For every selected star the following steps and criteria are performed :

\item Select measurements for given star
\item Check if number of measurements for the given night $N_{obs}^{night} \geq N_{minobs}^{night}$
\item Find upper limit of the magnitude range $mag_{max}$, which is defined
as maximum value of magnitudo in the "last" non empty bin \footnote{in the direction of increasing magnitude} of magnitudo
measurements distribution ( see Fig. \ref{fig_measurements_strip} ).
In most cases the value $mag_{max}$ is the same as maximum magnitude 
measurement for given star. However, in some cases (e.g. due to clouds) 
real "last" non empty bin can contain few outliers and can be separated
from most of the measurements ( see Fig. \ref{fig_last_non_empty_bin} ). In order to exclude those bad measurements, 
"last" non empty bin [M,M$+\Delta$M] is chosen as the one before an empty bin and
satisfying condition :

\begin{equation}
N( mag \leq M+\Delta M ) \geq 50\% \cdot N_{obs}
\label{eq_flare_range_cond}
\end{equation}

The maximum magnitude is found in the "last" non empty bin and used as upper limit $mag_{max}$. 
\item Find lower limit of magnitude $mag_{min}$ so that at least 85\% of all magnitude 
measurements for given star belong to range ($mag_{min}$,$mag_{max}$) ( see Fig. \ref{fig_measurements_strip} ).
Value of flare threshold is determined as $T_{flare}=mag_{min}$, all points brighter then this value 
are considered as belonging to an outburst event.
\item Find longest series of measurements with mag < $T_{flare}$ 
and require this set to have at least $N_{min}^{flare}$ points
\item Check maximum brightness measurement $M_{max}^{flare}$ in the flare series and 
determine its time $t^{flare}$
\item Verify if measurements within the series and before and after 
have the same chip coordinates (x,y), because due to calibration imperfectness star brightness
measurements can vary with the star position on the chip, causing background events with flare 
like signature ( visible as step on the light curve )
\item If the data was collected with permanently opened shutter verify if there is no star brighter then
$mag_{above}$ in the strip of $\Delta X_{bigstar}^{max}$ pixels from the analysed star.
Opened shutter causes that column obtains additional signal from stars with
$Y \leq Y_{0}$ and can generate background events. The rejection condition is the
same as in described earlier nova identification algorithm.
\item Check if there is no bright star nearby which could affect
measurements of flare suspected star. In case a star brighter then
$mag_{bigstar}^{near}$ is closer then $R_{bigstar}$ then event is rejected.
This condition is checked in the "Pi of the Sky" star catalog and also in TYCHO-2 \cite{tycho} star catalog
\item Verify if brightness increase is not due to hotpixel. Chip coordinates 
(x,y) are verified. The event is rejected if its (x,y) coordinates belong to a list of known CCD defects stored
in the database table \textbf{HotPixel} ( Fig. \ref{fig_starcat} ).
\item Peak height $\Delta M_{max}^{flare}$ over the average brightness level is determined and flare
event obtains quality flag according to this value, events with
$\Delta M_{max}^{flare}>0.4^m$ obtain quality=1 and those with $\Delta M_{max}^{flare}>1.0^m$ obtain quality=2
\item Finally event is accepted and saved to the database table \textbf{FlareEvent} with 
all information describing this event and calculated by the algorithm
\item Sky background level in the image is checked in the same way as in 
nova identification algorithm (Sec. \ref{sec_nova_ident_algo}).
\end{enumerate}

Algorithm parameters are listed in Table \ref{tab_flare_algo_params}.

%

\begin{table}[htbp]
\begin{center}
\begin{tabular}{|c|c|c|}
\hline
\textbf{Parameter} & \textbf{Default Value} & \textbf{Script option}  \\
\hline
$N_{obs}^{min}$   & 30 & -min\_points \\
\hline
$mag_{min}$       & 12 [mag] & -min\_mag\_star \\
\hline
$T_{magdiff}$     & 0.5 [mag] & -min\_max\_mag\_limit \\
\hline
$N_{minobs}^{night}$ & 20 & -min\_night\_points \\  
\hline
$N_{min}^{flare}$ &  3 & -min\_points\_above \\
\hline
$mag_{bigstar}^{strip}$ & 5 [mag] & -max\_bigstar\_above \\
\hline
$W_{strip}$       & 10 [pixels] & -max\_bigstar\_x\_dist \\
\hline
$mag_{bigstar}^{near}$ & 8 [mag]  & -near\_bigstar\_max\_mag \\
\hline
$R_{bigstar}$ & 15 [pixels] & -near\_bigstar\_distance \\
\hline 
\end{tabular}
\caption{Most important parameters of the flare identification algorithm}
\label{tab_flare_algo_params}
\end{center}
\end{table}

\begin{figure}[!htbp]
  \begin{center}
    \leavevmode
    \ifpdf
		\includegraphics[width=7.2in,height=4in]{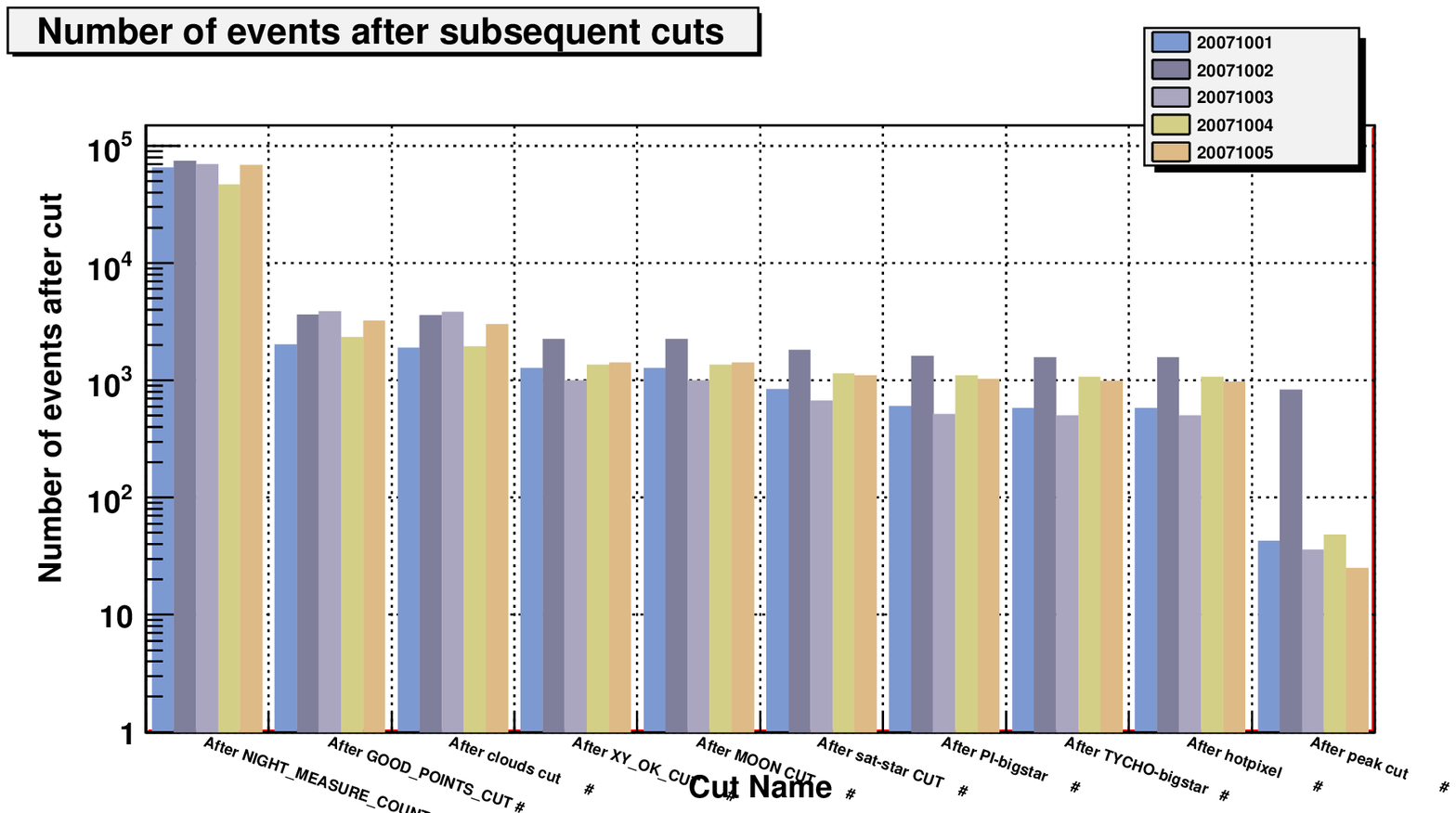}
    \else
		\includegraphics[width=6.0in,height=3.3in]{flare_algo/flare_algo_cuts.eps}
    \fi
    \caption{Number of events after subsequent cuts of the flare identification algorithm. The data from five nights is shown.}
    \label{fig_flare_algo_cuts}
  \end{center}
\end{figure}

Depending on the quality of the night, algorithm analysis $N_{star}<10^5$ stars which 
takes about 1-2 hours. Event numbers after subsequent cuts are shown in Figure \ref{fig_flare_algo_cuts}.

In order to determine efficiency of this algorithm flare events were generated. 
Events were generated for specified night and analysed by the same script
for flare identification used for analysis of the real data.
The simulation consists of the following main steps :

\begin{itemize}
\item Certain number $N_{flare}^{gen}$ of stars in the star catalog was selected ( specified by simulation parameter )
\item Time of flare start was randomly chosen from the series of measurements for a given night
\item Magnitude measurements after the starting time were replaced by values
obtained from flare parametrization
\item Star with generated points added was analysed and accepted or rejected by the algorithm
\end{itemize}

In order to simulate flare-like outburst simple linear/exponential parametrization was
fitted to real flare detected by the algorithm ( see Section \ref{offline_flashes} ) :

\begin{equation}
mag = \left\{
   \begin{array}{rl}
      m_0 - F_{max} \cdot \frac{t}{\Delta t} & \mbox{for }t\leq \Delta t  \\
      m_0 - F_{max} \cdot exp\left[ -\frac{t-\Delta t}{ \tau } \right] & \mbox{for }t\ge \Delta t
   \end{array}\right.
\label{eq_flare_param}
\end{equation}

\begin{figure}[!htbp]
  \begin{center}
    \leavevmode
    \ifpdf
		\includegraphics[width=6in,height=2in]{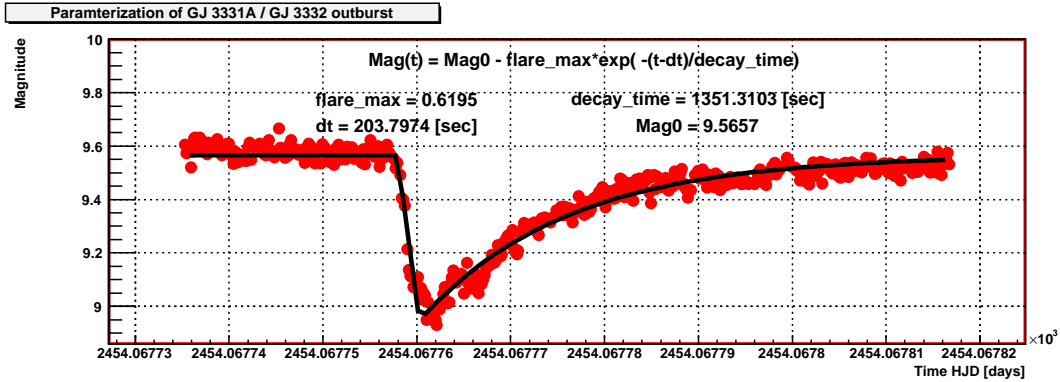}
    \else
		\includegraphics[width=6in,height=2in]{flare_eff/gj_lc_fit.eps}
    \fi
    \caption{Linear/Exponential parametrization of the flare-like outburst. Fit was
performed to real flare star GJ 3331A / GJ 3332 outburst which occurred on 2006.11.28 06:03 UT }
    \label{fig_flare_paramterization}
  \end{center}
\end{figure}

These parametrization is presented in Figure \ref{fig_flare_paramterization}.

\begin{figure}[!htbp]
  \begin{center}
    \leavevmode
    \ifpdf
		\includegraphics[width=2.7in,height=2.7in]{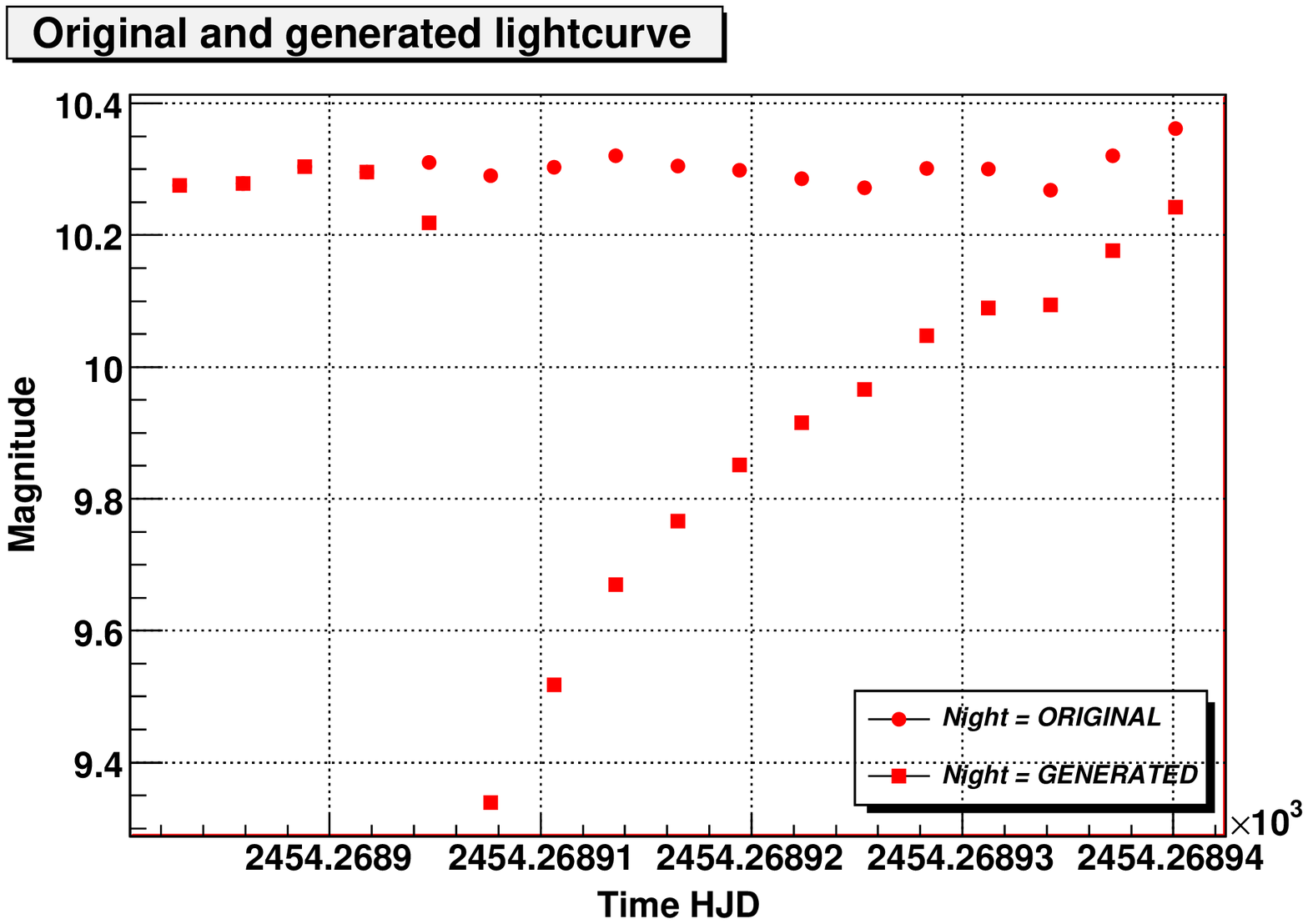}
		\includegraphics[width=2.7in,height=2.7in]{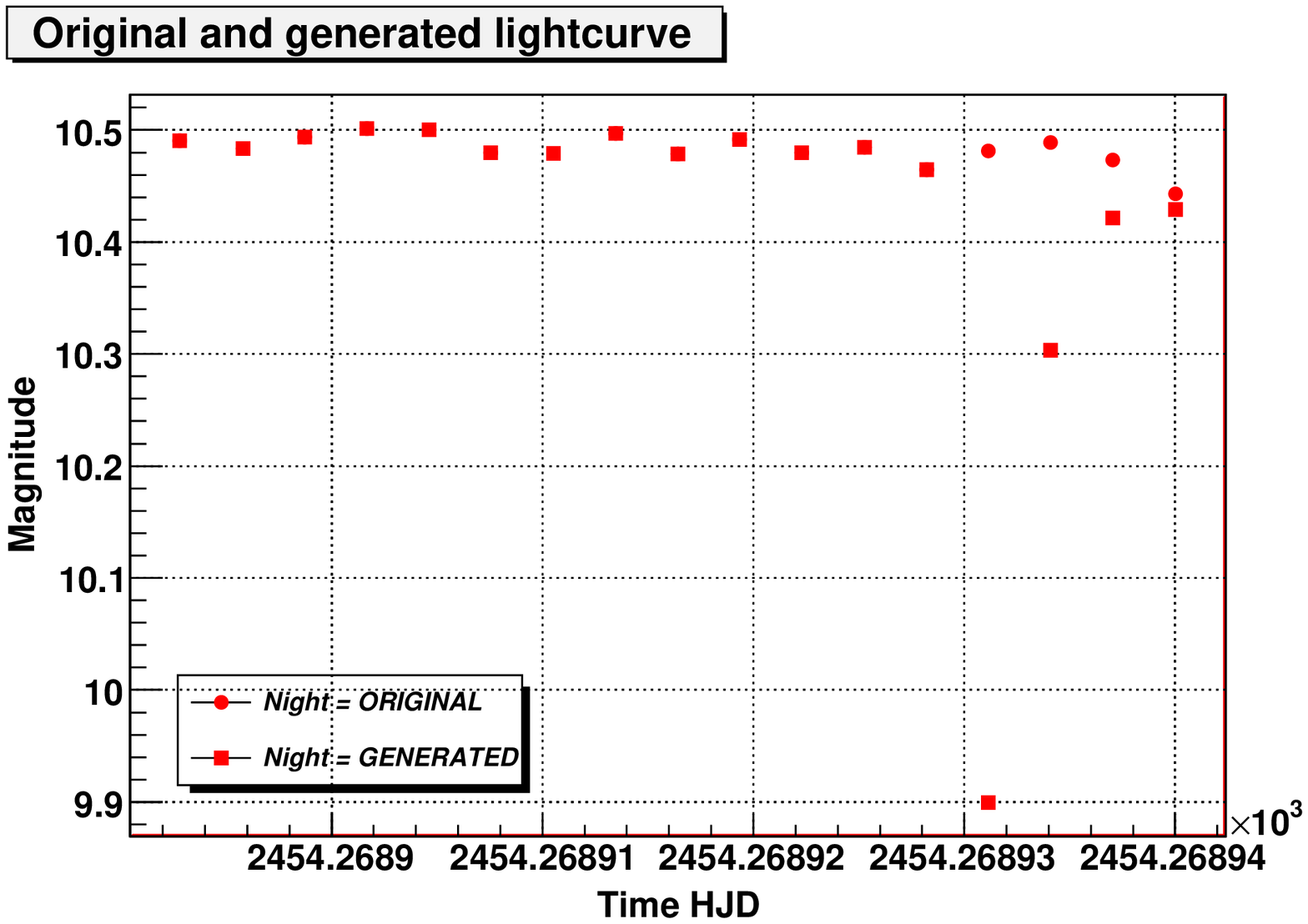}
    \else
		\includegraphics[width=2.7in,height=2.7in]{flare_eff/gen/1149554_both.eps}
		\includegraphics[width=2.7in,height=2.7in]{flare_eff/gen/1261271_both.eps}
    \fi
    \caption{Example of light curves with generated flare with outburst peak of
1$^m$ above average brightness level (left plot) and decay time of 200 sec (right plot), other parameters
of the light curves are the same as those fitted to GJ3331A/GJ3332 outburst}
    \label{fig_gen_flares}
  \end{center}
\end{figure}

Examples of generated flare light curves are shown in Figure \ref{fig_gen_flares}.
Flare identification efficiency was tested in function of 4 parameters used
to describe the flare GJ3331A/GJ3332 in fitted parametrization.

\begin{figure}[!htbp]
  \begin{center}
    \leavevmode
    \ifpdf
		\includegraphics[width=2.7in,height=2.7in]{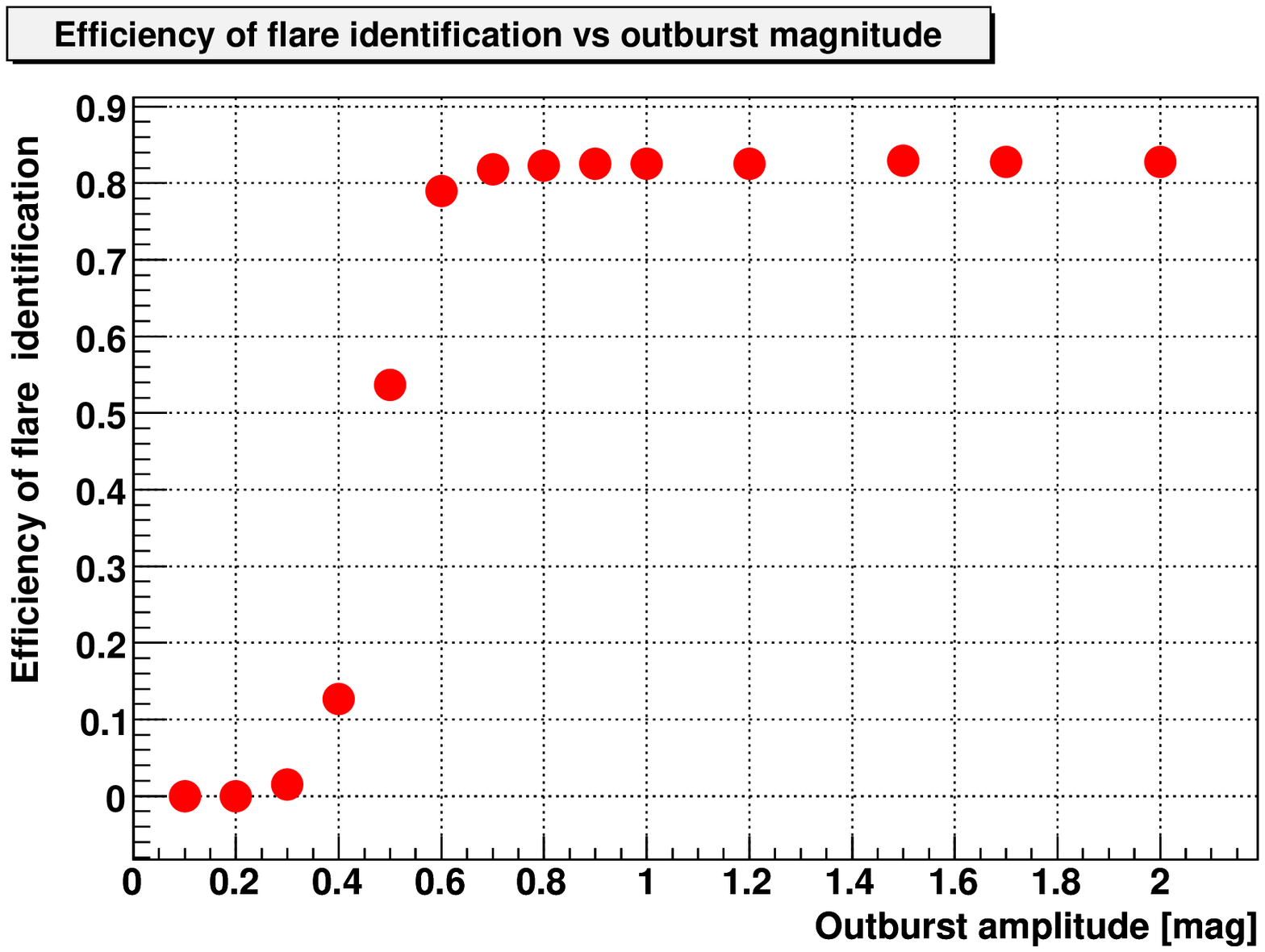}
		\includegraphics[width=2.7in,height=2.7in]{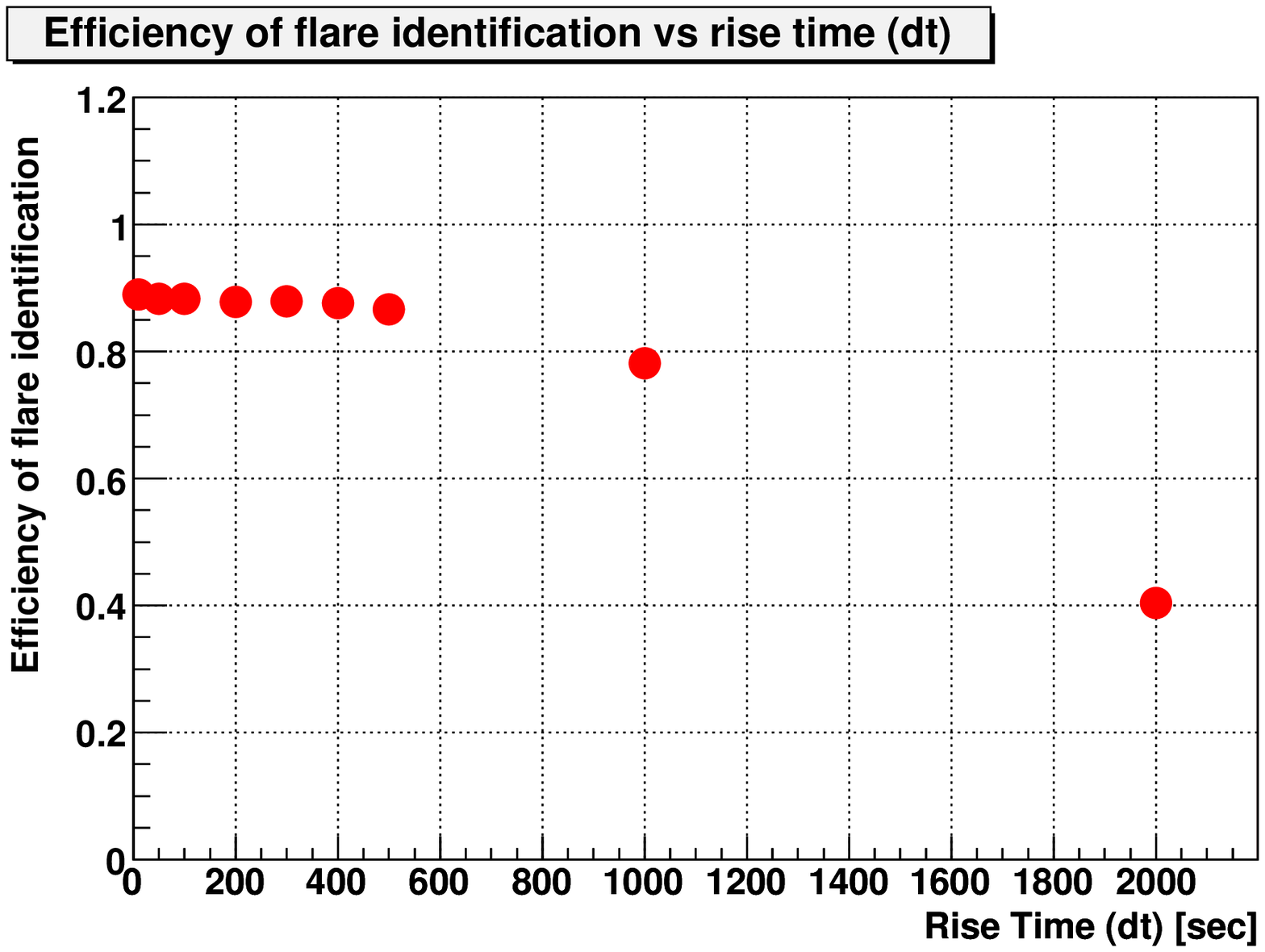}
		\includegraphics[width=2.7in,height=2.7in]{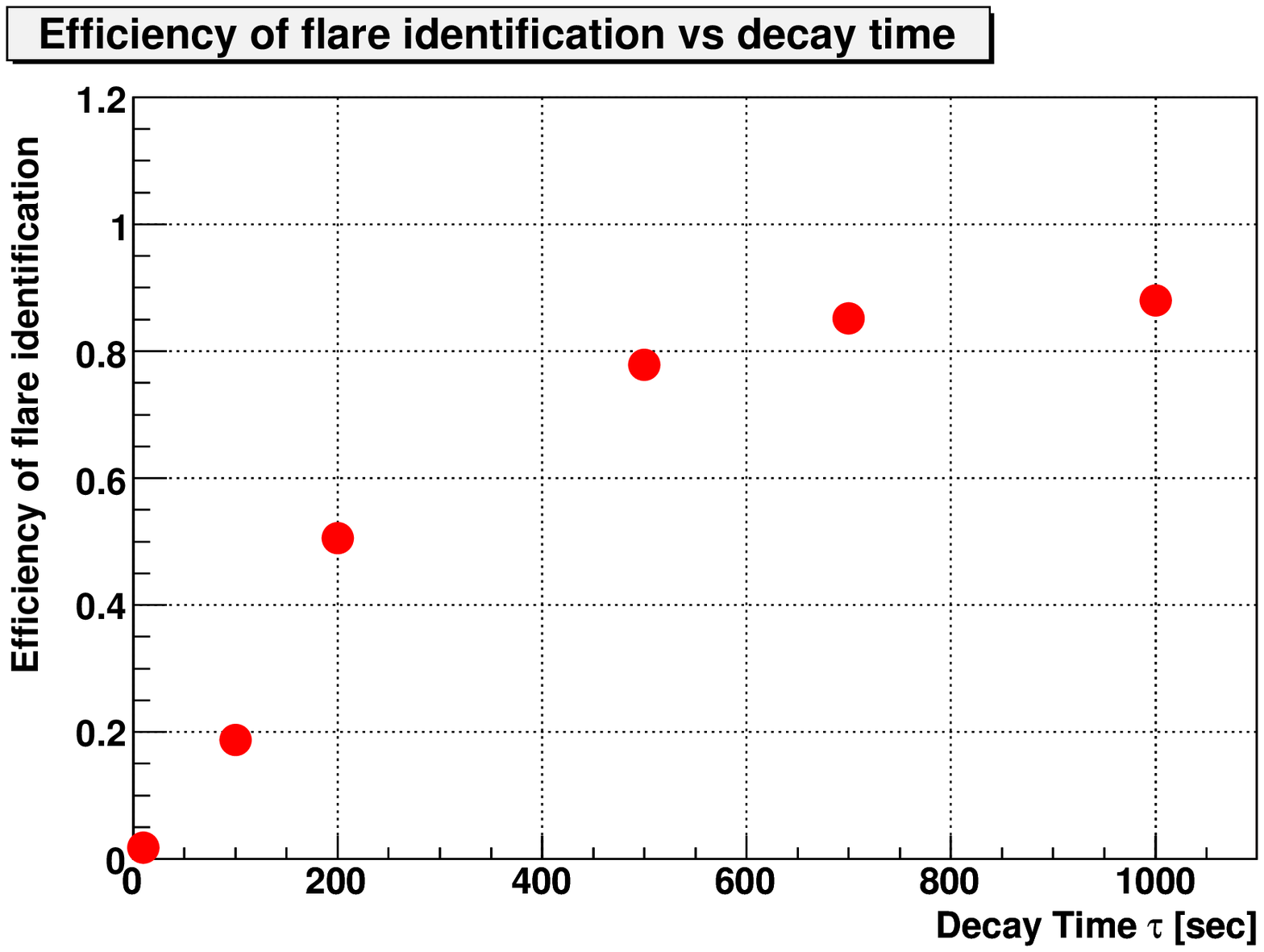}
		\includegraphics[width=2.7in,height=2.7in]{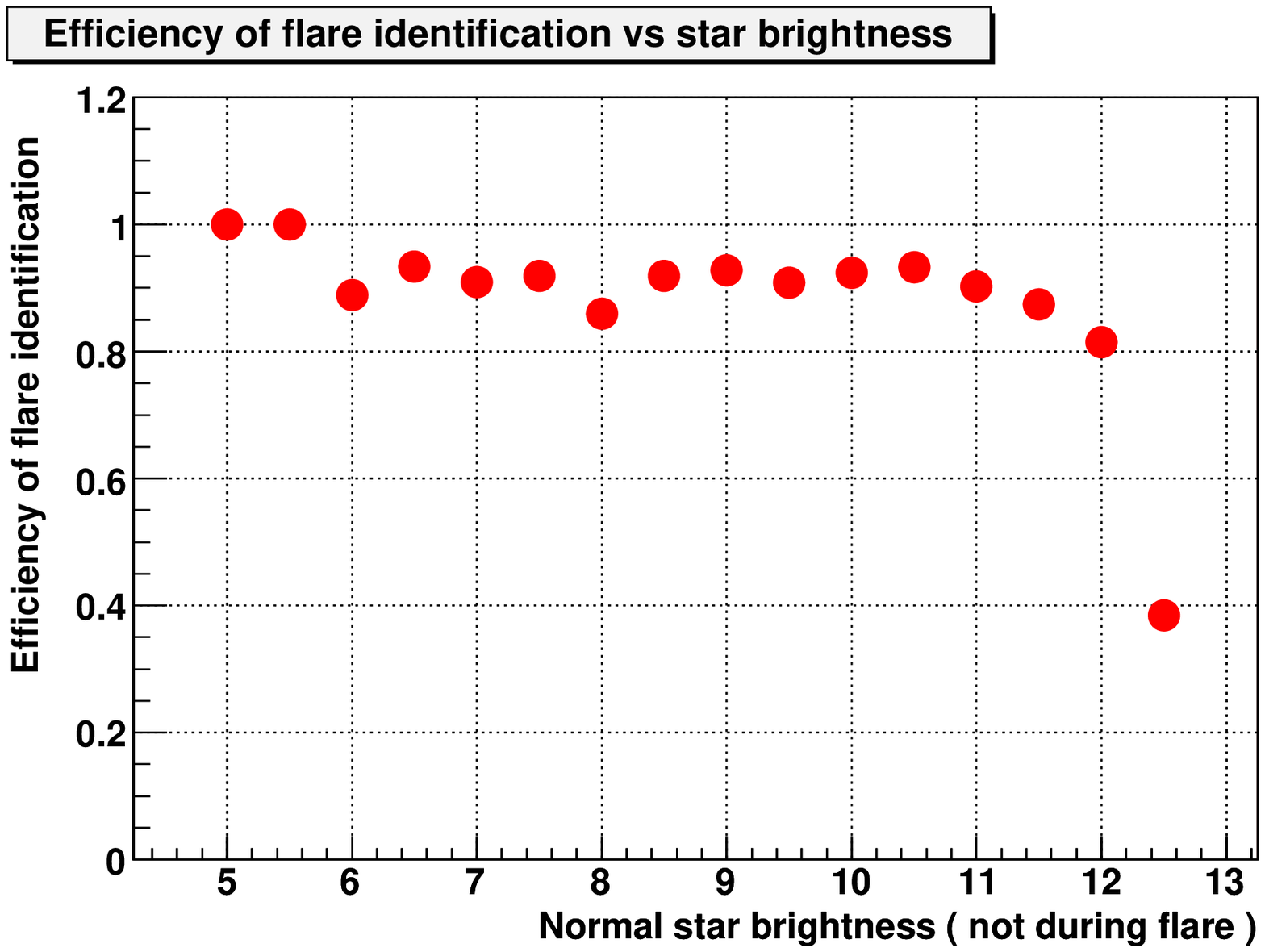}
    \else
		\includegraphics[width=2.7in,height=2.7in]{flare_eff/flare_eff_vs_flare_mag.eps}
		\includegraphics[width=2.7in,height=2.7in]{flare_eff/flare_eff_vs_rise_time.eps}
		\includegraphics[width=2.7in,height=2.7in]{flare_eff/flare_eff_vs_decay_time.eps}
		\includegraphics[width=2.7in,height=2.7in]{flare_eff/flare_eff_vs_mag0.eps}
    \fi
    \caption{Efficiency of flare identification in function of outburst
amplitude (left upper plot), rise time (right upper), decay time (left
bottom), star average magnitude (right bottom)}
    \label{fig_flare_eff}
  \end{center}
\end{figure}

The results of these tests are shown in Figure \ref{fig_flare_eff}.

\subsubsection{Data synchronization and presentation}
\label{sec_data_synchronization}

The cataloging and data analysis are performed on the remote server. In the
prototype version it is performed on pi1, pi2 and pi3 computers at LCO.
In order not to overuse the bandwidth only the final results are once copied to local server in
Warsaw. It is impossible to synchronize the entire database because it is too
large. However, results of algorithms can be synchronized with light curves of
corresponding stars. There is also possibility to specify interesting
objects which light curves should also be synchronized. 
After synchronization data is stored in the local database which is
accessible from the command line and script level, but also through the WWW
\texttt{php} interface.
Results of on-line algorithms are synchronized on-line during a night.
Besides database information also subrenders of the sky images containing events
are copied to local server allowing for quick verification of event candidates.
They are also available through the WWW interface.

\section{Periods when algorithms were working}
\label{sec_algo_usage_periods}

Before showing the results of the algorithms it is important to mention that 
not all algorithms were used during the full time of the prototype work.
Table \ref{tab_proto_periods} shows periods and configurations in which the prototype was working.
Because of technical problems shown in this table coincidence algorithm
could not work for all the time. Also not all the algorithms were ready to be
used since the very beginning. Table \ref{tab_algo_periods} shows which
algorithms were used in what periods. Some algorithms could also be run
off-line, but large amount of events could not be verified. 
Currently all off-line algorithms are working and events are systematically checked
after each night. \\

\begin{table}[htbp]
\begin{center}
\begin{tabular}{|c|c|c|}
\hline
\textbf{Algorithm Type} & \textbf{Working Periods} & \textbf{Notes} \\
\hline 
& & \\
On-line coincidence & \begin{minipage}{4.5cm}{\mbox{2004.06.25 - 2005.03.01 ,} 2006.05.20 - 2006.08.08}\end{minipage} & two cameras  \\ 
& & \\
\hline
& & \\
On-line confirmation on next image & \begin{minipage}{4.5cm}{2005.03.12 - 2005.08.09 , 2006.05.20 -}\end{minipage} & single camera \\ 
& & \\
\hline
Off-line nova on aver20 & 2006.07.21 -  & \\
\hline
Off-line flare on aver20 & 2006.07.21 -  & \\
\hline
\end{tabular}
\caption{Periods in which different algorithms were used}
\label{tab_algo_periods}
\end{center}
\end{table}



\chapter{Results}
\ifpdf
    \graphicspath{{Results/ResultsFigs/PNG/}{Results/ResultsFigs/PDF/}{Results/ResultsFigs/}}
\else
    \graphicspath{{Results/ResultsFigs/EPS/}{Results/ResultsFigs/}}
\fi
\def\deg{$^\circ$}



\section{Data from the prototype system}
\label{sec_proto_data}
%
%
%
Data from the prototype system have been collected in the time period from June 2004 until now 
( 22 October 2007 ) with 9 months break due to maintenance reasons. 
During this period the system collected data from 738 nights which is $\approx$80\% of 
nights in the period when system was operational. The remaining 20\% was lost
mainly due to clouds or some minor system failures which could be remotely 
solved after few hours or days. Table \ref{tab_proto_periods} lists data 
collection periods with description of major problems in the system.


\begin{table}[htbp]
\begin{center}
\begin{tabular}{|c|c|c| }
\hline
\textbf{Start Date} & \textbf{End Date} & \textbf{Period Description} \\
\hline
\hline
20040625 & 20050120 & both cameras k2a and k2b working correctly \\
\hline
20050120 &    -     & problems with k2b power supply and cooling \\
& & \\
\hline 
20050223 &    -     & k2b mostly not working \\
&  & \\
\hline
& & \\
20050419 &    -     & 
\begin{minipage}{9cm} 
k2b shutter mechanism broken, data collected with opened shutter 
\end{minipage} \\
\hline
20050702 & 20050809 & k2a cooling definitely broken, implies large noise \\
\hline
20050810 & 20060520 & system down due to maintenance reasons \\
\hline
20060520 & 20060808 & 
\begin{minipage}{9cm}
system working after cameras repair and upgrade, both cameras working
\end{minipage} \\
\hline
& & \\
20060809 &    -     & 
\begin{minipage}{9cm}
k2b camera broken due to electricity problems in LCO,
since then system works only with k2a camera 
\end{minipage} \\
& & \\
\hline
\end{tabular}
\caption{Data collection periods of the prototype, with crucial moments of
major failures listed}
\label{tab_proto_periods}
\end{center}
\end{table}

The observation time was calculated using the image information saved in the
database. The Table \ref{tab_obs_time} shows observation times for two 
versions of the prototype. This times were normalized to $4\pi$ coverage.
In the first period all sky images from nights with at least 100 images were 
taken into account, in the second period additional cut on number of stars 
on image $N_{stars} \geq 1000$ was imposed in order to ignore cloudy data. This
cut could not be used for the first period because this information was
added to the database during the system upgrade.

\begin{table}[htbp]
\begin{center}
\begin{tabular}{|c|c|c| }
\hline
\textbf{Time Period} & 
\begin{minipage}{5cm} 
\textbf{Normalized observation time [days]} 
\end{minipage} & \textbf{Lenses} \\
\hline
\hline
2004.06.23 - 2005.08.09 & 2.04 & Carl Zeiss f=50mm \\
\hline
2006.05.20 - 2007.08.28 & 0.78 & Canon f=85mm \\
\hline
\end{tabular}
\caption{Data collection periods of the prototype, with crucial moments of
major failures listed}
\label{tab_obs_time}
\end{center}
\end{table}

\section{Optical flashes in 10s and 22s timescales}
\label{sec_online_flashes}

Every night when the system was running, the flash identification algorithm was looking for optical flashes on-line. 
The final results of the algorithm were verified by a person on duty and 
events which could not be rejected by any criteria were flagged as "flashes".
The statistics of all optical flashes detected by the prototype is shown in
Table \ref{flash_stat}. Most of the flashes were visible in a single 10s
exposure in both cameras. All these events are listed in Tables
\ref{all_singles2004}, \ref{all_singles2005} and \ref{all_singles2006}. 
They contain also the $D_{cone}$ value ( Sec. \ref{sec_flashes_verification} ).
However, none of the events has this value sufficiently large to confirm 
astrophysical origin of the flash.
There are a few events which were observed on two consecutive images, but only
by a single camera ( when the second was not operational ). They are listed in Table
\ref{tab_double_flashes}. In this case also limit of distance $D_{move}$ derived from 
constant position of the flash is shown ( Sec. \ref{sec_flashes_verification} ).

\begin{table}[htbp]
\begin{center}
\begin{tabular}{|c|c|c|c| }
\hline 
\textbf{Period} & \begin{minipage}{4cm}\textbf{Algorithm}\end{minipage} & 
\begin{minipage}{2cm}
\textbf{Number of flashes} 
\end{minipage}
& 
\begin{minipage}{2cm}
\textbf{Exposure Time [days]}
\end{minipage} \\
\hline
\hline
2006.05.20 - 2007.08.28 & \begin{minipage}{4cm}Confirmation on next image\end{minipage} & 1 & 0.78 \\
\hline
2006.05.20 - 2006.08.22 & Coincidence & 35 & 0.14 \\
\hline
2004.06.25 - 2005.08.09 & Coincidence & 90 & 1.68 \\ 
\hline
2004.06.25 - 2005.08.09 & \begin{minipage}{4cm}Confirmation on next image\end{minipage} & 5 & 0.8 \\
\hline
2006.07.21 - 2007.09.06 & \begin{minipage}{4cm}Flash identification in 240 sec timescale\end{minipage} & 0 & 0.69 \\
\hline
\end{tabular}
\caption{Statistics of optical flashes detected by the prototype}
\label{flash_stat}
\end{center}
\end{table}

\begin{table}[htbp]
\begin{center}
\begin{tiny}
\begin{tabular}{|c|c|c|c|c|r@{.}l|c| }
\hline
\textbf{Date} & \textbf{UT Time} & \textbf{RA} & \textbf{Dec} & \boldmath{$D_{cone}$} & \multicolumn{2}{c|}{\boldmath{$M_{max}$}} & \textbf{coinciding alerts / sources} \\
\hline
\hline
2006.07.25 & 07:21:27 & 18h39m03.14s & -10\deg37'43'' & 26878 & 10&49 & \\ 
\hline
2006.07.20 & 07:24:02 & 18h41m08.83s & -10\deg43'39'' & 31843 & 9&78 & \\ 
\hline
2006.07.17 & 05:17:15 & 18h22m42.45s & -12\deg00'06'' & 25099 & \multicolumn{2}{c|}{-} & \\ 
\hline
2006.07.17 & 04:48:47 & 23h56m13.09s & -23\deg52'06'' & 11975 & 10&1 & \\ 
\hline
2006.07.17 & 03:46:37 & 19h18m36.07s & -10\deg50'05'' & 34274 & 8&8 & \\ 
\hline
2006.07.17 & 00:55:35 & 12h12m16.84s & -65\deg13'37'' & 7597 & 10&7 & \\ 
\hline
2006.07.15 & 05:57:24 & 19h42m01.54s & -09\deg26'54'' & 37745 & 10&5 & \\ 
\hline
2006.07.15 & 03:01:12 & 18h24m02.21s & -11\deg56'53'' & 18370 & 10&1 & \\ 
\hline
2006.07.14 & 07:29:28 & 18h46m13.81s & -10\deg48'56'' & 39594 & 10&2 & \\ 
\hline
2006.07.10 & 04:51:00 & 20h05m44.80s & -10\deg08'05'' & 29064 & 9&04 & \\ 
\hline
2006.07.10 & 00:18:02 & 12h42m52.55s & 22\deg36'27'' & 6981 & 10&81 & 
\begin{minipage}{4cm}
<1 arcmin from NGP9 F378-0444065 - Galaxy, 
1 arcmin from FIRST J124239.4+223536 - Radio-source
\end{minipage} \\
\hline
2006.07.03 & 08:42:04 & 01h02m50.01s & 04\deg29'27'' & 7380 & 9&76 & \\ 
\hline
2006.07.02 & 09:48:58 & 00h40m30.20s & 04\deg00'26'' & 6832 & 9&6 & \\ 
\hline
2006.07.02 & 08:15:12 & 01h42m27.20s & 09\deg34'42'' & 7413 & 10&45 & \\ 
\hline
2006.06.29 & 22:58:34 & 11h40m20.20s & 13\deg34'58'' & 6531 & 11&37 & \\ 
\hline
2006.06.29 & 22:49:16 & 11h38m18.76s & 11\deg50'08'' & 6493 & 8&05 & \\ 
\hline
2006.06.27 & 03:19:54 & 16h10m39.49s & -11\deg14'39'' & 14128 & 11&17 & \\ 
\hline
2006.06.27 & 02:24:34 & 18h04m16.33s & -08\deg04'15'' & 24909 & \multicolumn{2}{c|}{-} & \\ 
\hline
2006.06.27 & 02:19:55 & 17h59m55.77s & -07\deg56'07'' & 23450 & \multicolumn{2}{c|}{-} & \\ 
\hline
2006.06.26 & 03:51:28 & 16h39m03.75s & -11\deg40'55'' & 17909 & 11&11 & \\ 
\hline
2006.06.24 & 02:47:18 & 16h15m34.48s & -37\deg06'22'' & 13788 & 10&81 & \\ 
\hline
2006.06.23 & 09:13:04 & 23h41m41.95s & 00\deg01'38'' & 7174 & \multicolumn{2}{c|}{-} & 
\begin{minipage}{4cm}
7 arcsec from J234135.2-000145 Seyfert 1 Galaxy
\end{minipage} \\
\hline
2006.06.18 & 04:19:38 & 19h20m32.60s & -31\deg07'50'' & 22918 & 10&1 & \\ 
\hline
2006.06.14 & 09:48:41 & 22h17m47.85s & -06\deg26'54'' & 6873 & 9&7 & \\ 
\hline
2006.06.13 & 07:38:58 & 22h24m32.22s & -07\deg17'11'' & 8471 & 9&26 & \\ 
\hline
2006.06.12 & 07:53:49 & 22h01m06.08s & -10\deg47'12'' & 8405 & 10&5 & \\ 
\hline
2006.06.11 & 00:03:36 & 09h59m22.93s & 00\deg29'16'' & 6937 & 8&7 & 
\begin{minipage}{4cm} 
1 arcmin from [LPZ94] 279 -- Radio-source 
\end{minipage} \\
\hline
2006.06.10 & 04:57:53 & 18h32m04.61s & -10\deg04'08'' & 21981 & 8&93 & \\ 
\hline
2006.06.08 & 07:44:44 & 17h31m54.42s & -11\deg02'14'' & 22002 & 8&5 &
\begin{minipage}{4cm}
2 arcmin from IRAS 17291-1057 - Infra-Red source
\end{minipage} \\
\hline
2006.06.08 & 06:08:36 & 17h44m25.43s & -17\deg24'02'' & 30565 & \multicolumn{2}{c|}{-} & \\ 
\hline
2006.06.06 & 07:20:08 & 17h41m41.34s & -62\deg07'57'' & 11407 & \multicolumn{2}{c|}{-} & \\ 
\hline
2006.06.05 & 00:57:09 & 12h15m14.43s & -20\deg19'19'' & 7959 & \multicolumn{2}{c|}{-} & \\ 
\hline
2006.06.01 & 03:51:14 & 12h49m26.68s & 13\deg17'54'' & 10775 & \multicolumn{2}{c|}{-} & 
\begin{minipage}{4cm}
1 arcmin from FIRST J124926.0+131612 - Radio-source 
\end{minipage} \\
\hline
2006.05.22 & 05:21:25 & 16h39m21.25s & -09\deg35'11'' & 29140 & \multicolumn{2}{c|}{-} & \\ 
\hline
2006.05.21 & 05:06:28 & 15h38m06.76s & 12\deg49'19'' & 18605 & \multicolumn{2}{c|}{-} & \\ 
\hline
\end{tabular}
\end{tiny}
\hfill
\caption{Optical flashes identified by the on-line algorithm since June 2006 requiring
coincidence of flash on 2 cameras. These flashes were observed on single 10s
exposure}
\label{all_singles2006}
\end{center}
\end{table}

\begin{table}[htbp]
\begin{center}
\begin{tiny}
\begin{tabular}{|c|c|c|c|c|c|c| }
\hline
\textbf{Date} & \textbf{UT Time} & \textbf{RA} & \textbf{Dec} & \boldmath{$D_{cone}$} & \boldmath{$M_{max}$} & \textbf{coinciding alerts / sources} \\
\hline
\hline
2005.08.07 & 23:59:16 & 21h44m40s & -41\deg49' & 8444 & 8.3(?) & k2a not working \\ 
\hline
2005.08.07 & 06:52:52 & 21h59m20s & -11\deg52' & 21841 & 7.1(?) & k2a not working \\ 
\hline
2005.08.07 & 00:21:42 & 19h27m22s & -23\deg52' & 7833 & 8.8(?) & k2a not working \\ 
\hline
2005.08.01 & 01:56:25 & 17h40m45s & +14\deg22' & 9352 & 8.05 & \\ 
\hline
2005.06.30 & 09:39:38 & 19h02m58s & -18\deg40' & 11784 & 9.8 & \\ 
\hline
2005.06.02 & 03:35:18 & 11h42m51s & +01\deg57' & 9669 & 10.3 & \\ 
\hline
2005.06.01 & 23:34:14 & 12h41m14s & +10\deg44' & 6889 & 9.04 & 1arcmin from GRDG +11 54 Galaxy \\ 
\hline
2005.06.01 & 22:54:14 & 12h02m24s & +12\deg49' & 6553 & 10.03 & \\ 
\hline
2005.06.01 & 03:47:30 & 11h50m11s & +01\deg28' & 9983 & 11.2 & \\ 
\hline
2005.06.01 & 00:19:37 & 12h56m19s & +14\deg21' & 7486 & 9.0 & \\ 
\hline
2005.05.31 & 22:56:15 & 12h23m32s & +11\deg49' & 6572 & 10.0 & J122331.1+114752 radio \\ 
\hline
2005.05.31 & 04:12:52 & 11h58m51s & -02\deg02' & 10571 & 11.0 & Konus 4:27:26 \\ 
\hline
2005.05.31 & 00:01:46 & 11h54m52s & +15\deg07' & 7111 & 9.3 & \\ 
\hline
2005.05.27 & 23:02:13 & 10h15m13s & +14\deg20' & 6571 & 10.9 & \\ 
\hline
2005.05.27 & 22:55:12 & 10h52m40s & +28\deg37' & 6537 & 10.0 & \\ 
\hline
2005.05.27 & 22:53:34 & 10h51m3s & +24\deg42' & 6531 & 7.2 & \\ 
\hline
2005.05.27 & 05:36:11 & 16h56m42s & -10\deg27' & 29060 & 10.0 & \\ 
\hline
2005.05.25 & 03:57:10 & 16h01m30s & -08\deg01' & 36601 & 10.0 & \\ 
\hline
2005.05.20 & 08:33:30 & 15h33m51s & -32\deg21' & 18299 & 12.0(?) & \\ 
\hline
2005.05.15 & 23:04:50 & 13h39m42s & -39\deg34' & 6660 & 10.1 & \\ 
\hline
2005.05.14 & 03:28:22 & 13h28m41s & -15\deg57' & 15960 & 9.3 & \\ 
\hline
2005.05.14 & 03:19:27 & 13h18m45s & -21\deg30' & 14734 & 10.3 & \\ 
\hline
2005.05.14 & 02:35:02 & 13h33m57s & -27\deg26' & 13338 & 10.5 & \\ 
\hline
2005.05.12 & 04:59:08 & 12h47m39s & +10\deg24' & 14674 & 8.9 & \\ 
\hline
2005.05.11 & 03:05:50 & 11h13m34s & +12\deg15' & 10060 & 8.5 & \\ 
\hline
2005.05.06 & 05:53:02 & 14h02m13s & +00\deg35' & 28414 & 8.5(?) & \\ 
\hline
2005.05.01 & 10:09:25 & 18h49m16s & +09\deg17' & 6572 & - & \\ 
\hline
2005.04.24 & 06:24:24 & 16h48m39s & -10\deg18' & 11779 & 9.1 & \\ 
\hline
2005.04.01 & 06:12:52 & 11h16m47s & +06\deg59' & 26486 & 10.2 & GSC 00269-00778 \\ 
\hline
2005.03.30 & 04:45:08 & 11h00m03s & +14\deg38' & 19964 & 8.9 & \\ 
\hline
2005.02.16 & 00:32:49 & 10h00m19s & -13\deg16' & 6746 & 8.3 & \\ 
\hline
2005.02.15 & 08:01:54 & 10h08m17s & +13\deg10' & 23523 & 8.4 & \\ 
\hline
2005.02.15 & 04:54:47 & 11h00m10s & +06\deg16' & 17082 & 9.0 & \\ 
\hline
2005.02.12 & 03:24:46 & 09h45m31s & -14\deg43' & 9816 & 9.3 & \\ 
\hline
2005.02.04 & 07:07:33 & 09h51m46s & +12\deg19' & 12323 & 9.6 & \\ 
\hline
2005.01.31 & 00:35:47 & 07h00m11s & +11\deg37' & 6567 & 9.2 & \\ 
\hline
2005.01.28 & 07:09:05 & 06h48m39s & +13\deg23' & 17546 & 9.6 & \\ 
\hline
2005.01.17 & 04:01:01 & 08h49m00s & +17\deg12' & 21937 & 9.0 & \\ 
\hline
2005.01.10 & 00:54:15 & 05h10m31s & +13\deg40' & 6625 & 9.7 & \\ 
\hline
2005.01.09 & 02:23:36 & 06h58m53s & +17\deg11' & 10701 & 9.25 & \\ 
\hline
2005.01.07 & 02:52:50 & 06h02m36s & +09\deg06' & 8640 & 11.0 & \\ 
\hline
2005.01.03 & 06:34:48 & 07h59m55s & 18\deg00' & 10150 & 6.3 & \\ 
\hline
2005.01.02 & 06:41:28 & 05h59m41s & +20\deg53' & 30461 & 8.9 & Konus+Integral 6:27:27 \\ 
\hline
\end{tabular}
\end{tiny}
\hfill
\caption{Optical flashes identified by the on-line algorithm in 2005 requiring
coincidence of flash on 2 cameras. These flashes were observed on single 10s
exposure}
\label{all_singles2005}
\end{center}
\end{table}

\begin{table}[htbp]
\begin{center}
\begin{tiny}
\begin{tabular}{|c|c|c|c|c|r@{.}l|c| }
\hline
\textbf{Date} & \textbf{UT Time} & \textbf{RA} & \textbf{Dec} & \boldmath{$D_{cone}$} & \multicolumn{2}{c|}{\boldmath{$M_{max}$}} & \textbf{coinciding alerts / sources} \\
\hline
\hline
2004.12.27 & 07:22:21 & 05h48m30s & 00\deg00' & 8074 & 9&2 & \\ 
\hline
2004.12.27 & 03:51:03 & 04h49m40s & +19\deg58' & 11532 & 9&8 & \\ 
\hline
2004.12.22 & 02:22:12 & 07h07m03s & +17\deg41' & 20514 & 9&3 & \\ 
\hline
2004.12.20 & 02:24:46 & 05h56m24s & +03\deg10' & 8344 & 9&3 & \\ 
\hline
2004.12.19 & 02:46:43 & 05h03m58s & +07\deg34' & 8624 & 9&9 & \\ 
\hline
2004.12.18 & 05:53:37 & 05h48m38s & +12\deg26' & 11325 & 12&00 & \\ 
\hline
2004.12.14 & 02:29:21 & 05h32m09s & +14\deg29' & 11375 & 9&1 & \\ 
\hline
2004.12.11 & 00:47:21 & 03h13m14s & +06\deg09' & 6641 & 9&8 & \\ 
\hline
2004.12.04 & 07:34:46 & 05h19m21s & +14\deg34' & 7945 & 10&3 & \\ 
\hline
2004.12.04 & 06:28:54 & 05h05m28s & +21\deg15' & 18084 & 11&2 & \\ 
\hline
2004.12.04 & 02:05:22 & 03h01m51s & +05\deg57' & 7689 & 9&7 & galaxy [ZHG90] 0259+0545 \\ 
\hline
2004.12.02 & 05:13:41 & 04h07m24s & +20\deg27' & 35734 & 9&6 & \\ 
\hline
2004.12.01 & 01:28:47 & 01h34m19s & +14\deg51' & 7123 & 9&8 & \\ 
\hline
2004.11.30 & 04:09:42 & 03h21m04s & +19\deg01' & 15993 & 11&5(?) & \\ 
\hline
2004.11.28 & 04:27:16 & 01h27m28s & +10\deg48' & 10041 & 9&6 & \\ 
\hline
2004.11.28 & 03:20:15 & 03h04m44s & +05\deg42' & 9502 & 7&5 & \\ 
\hline
2004.11.26 & 05:58:15 & 05h10m12s & +14\deg40' & 11028 & 9&7 & \\ 
\hline
2004.11.24 & 07:39:33 & 06h54m30s & +18\deg49' & 7145 & 9&7 & \\ 
\hline
2004.11.23 & 02:39:10 & 05h14m29s & -07\deg06' & 9111 & 10&0(?) & \\ 
\hline
2004.11.19 & 00:53:08 & 03h14m50s & +03\deg20' & 7166 & 9&2 & \\ 
\hline
2004.11.18 & 07:53:55 & 04h22m13s & +12\deg31' & 7505 & 10&5 & \\ 
\hline
2004.11.18 & 01:36:09 & 03h10m22s & +03\deg12' & 8120 & 9&0 & \\ 
\hline
2004.11.18 & 00:25:36 & 01h30m17s & +11\deg44' & 6655 & 8&9(?) & \\ 
\hline
2004.11.17 & 03:31:17 & 02h48m36s & +11\deg13' & 13252 & 11&00 & \\ 
\hline
2004.11.12 & 02:51:31 & 02h27m12s & +22\deg06' & 23387 & 10&8 & \\ 
\hline
2004.11.11 & 04:46:43 & 02h08m07s & +10\deg13' & 16192 & 10&7 & \\ 
\hline
2004.11.11 & 01:57:49 & 04h03m47s & +18\deg33' & 39317 & 10&0 & \\ 
\hline
2004.11.06 & 06:02:58 & 03h47m24s & +10\deg22' & 12005 & 9&0 & \\ 
\hline
2004.11.05 & 07:49:20 & 02h34m32s & +05\deg47' & 9813 & 8&8 & \\ 
\hline
2004.11.03 & 01:22:40 & 00h12m32s & +00\deg49' & 7483 & 9&7 & Konus 1:26:43, Integral 1:27:11 \\ 
\hline
2004.11.02 & 07:14:15 & 01h23m53s & +01\deg13' & 13855 & 8&6 & Konus 7:27:56 \\ 
\hline
2004.11.01 & 05:49:34 & 23h47m44s & -04\deg23' & 10894 & 9&6 & \\ 
\hline
2004.11.01 & 05:22:28 & 23h20m39s & -05\deg43' & 10147 & 10&4 & \\ 
\hline
2004.10.31 & 06:15:37 & 23h24m40s & -01\deg12' & 11215 & 10&2 & \\ 
\hline
2004.10.30 & 05:22:06 & 00h42m54s & -29\deg30' & 9094 & 9&0(?/9.4) & \\ 
\hline
2004.10.28 & 05:56:42 & 03h34m40s & -08\deg15' & 11525 & 9&8 & \\ 
\hline
2004.10.24 & 03:40:00 & 00h36m31s & -05\deg10' & 11388 & 10&3 & Radio-source Cul 0034-054 \\ 
\hline
2004.10.21 & 02:36:00 & 01h02m16s & -04\deg17' & 11101 & 10&0 & \\ 
\hline
2004.10.11 & 07:37:09 & 03h12m08s & +10\deg39' & 8485 & 8&6 & \\ 
\hline
2004.10.11 & 06:15:13 & 02h54m57s & +18\deg16' & 13332 & 9&8 & \\ 
\hline
2004.10.05 & 04:18:40 & 01h24m04s & +14\deg17' & 41061 & 9&9 & \\ 
\hline
2004.10.02 & 02:20:50 & 23h05m26s & +12\deg36' & 13431 & 9&9 & \\ 
\hline
2004.10.01 & 04:09:12 & 21h46m26s & -08\deg18' & 11642 & 10&3 & \\ 
\hline
2004.09.30 & 00:49:38 & 23h19m50s & -02\deg36' & 8308 & 8&5 & \\ 
\hline
2004.09.27 & 04:20:55 & 03h29m33s & +14\deg08' & 12965 & 10&0 & \\ 
\hline
2004.09.17 & 03:50:25 & 23h42m55s & +31\deg44' & 20126 & 8&4 & \\ 
\hline
2004.09.13 & 03:33:52 & 00h14m45s & +02\deg50' & 40744 & - & \\ 
\hline
\end{tabular}
\end{tiny}
\hfill
\caption{Optical flashes identified by the on-line algorithm in 2004 requiring
coincidence of flash on 2 cameras. These flashes were observed on single 10s
exposure}
\label{all_singles2004}
\end{center}
\end{table}

\begin{table}[htbp]
\begin{center}
\begin{tiny}
\begin{tabular}{|c|c|c|c|c|c|c|r@{.}l|c| }
\hline
\textbf{ID} & \textbf{Date} & \textbf{UT Time} & \textbf{RA} & \textbf{Dec} & \boldmath{$D_{cone}$[km]} & \boldmath{$D_{move}$[km]} & \multicolumn{2}{c|}{\boldmath{$M_{max}$}} &
\begin{minipage}{3cm}
\textbf{External Alerts or Sources}
\end{minipage} \\
\hline
\hline
7 & 2006.10.10 & 02:44:43 & 00h8m57.80s & 34\deg51' & 18582 & 108964 & 11&7  &  \\ 
\hline
6 & 2005.06.03 & 06:29:57 & 19h37m14s & -50\deg47' & 11616 & 47268 & 10&5  & \begin{minipage}{3cm}Konus 6:29:01, excl. by Swift coords \end{minipage} \\ 
\hline
5 & 2005.05.31 & 04:33:17 & 17h43m48s & -19\deg55' & 28323 & 60048 & 6&1 & \begin{minipage}{3cm}Konus 4:27:26, excl. by IPN \end{minipage} \\ 
\hline
4 & 2005.04.16 & 00:49:59 & 11h03m11s & 11\deg06' & 8119 & 108964 & 10&5 & SN2003L NGC3506 \\ 
\hline
3 & 2005.04.04 & 05:37:59 & 11h49m37s & -05\deg31' & 30853 & 108964 & 10&3 & 
\begin{minipage}{3cm}
GSC 04937-00794or LCRS B114710.1-051705or 1RXS J114951.0-053041
\end{minipage} \\ 
\hline
2 & 2005.04.02 & 01:13:42 & 10h56m29s & +07\deg02' & 9082  & - & 9&0 & \begin{minipage}{3cm}V* CN Leo - flare star \end{minipage} \\ 
\hline
1 & 2005.03.31 & 01:36:46 & 11h52m53s & -05\deg59' & 12797 & 108964 & 10&0 & \begin{minipage}{3cm}several galaxies or GSC 04938-00378 \end{minipage} \\ 
\hline
\end{tabular}
\end{tiny}
\hfill
\caption{Optical flashes identified by the on-line algorithm requiring
signal at least on 2 consecutive images. Events 5 and 6 are probably due to
satellites. Value of $D_{move}$ in most cases was calculated for $\alpha_{arcsec}$=36'' corresponding angular size of 1 pixel.}
\label{tab_double_flashes}
\end{center}
\end{table}

\begin{figure}[!htbp]
  \begin{center}
    \leavevmode
    \ifpdf
      \includegraphics[width=2.7in,height=2.7in]{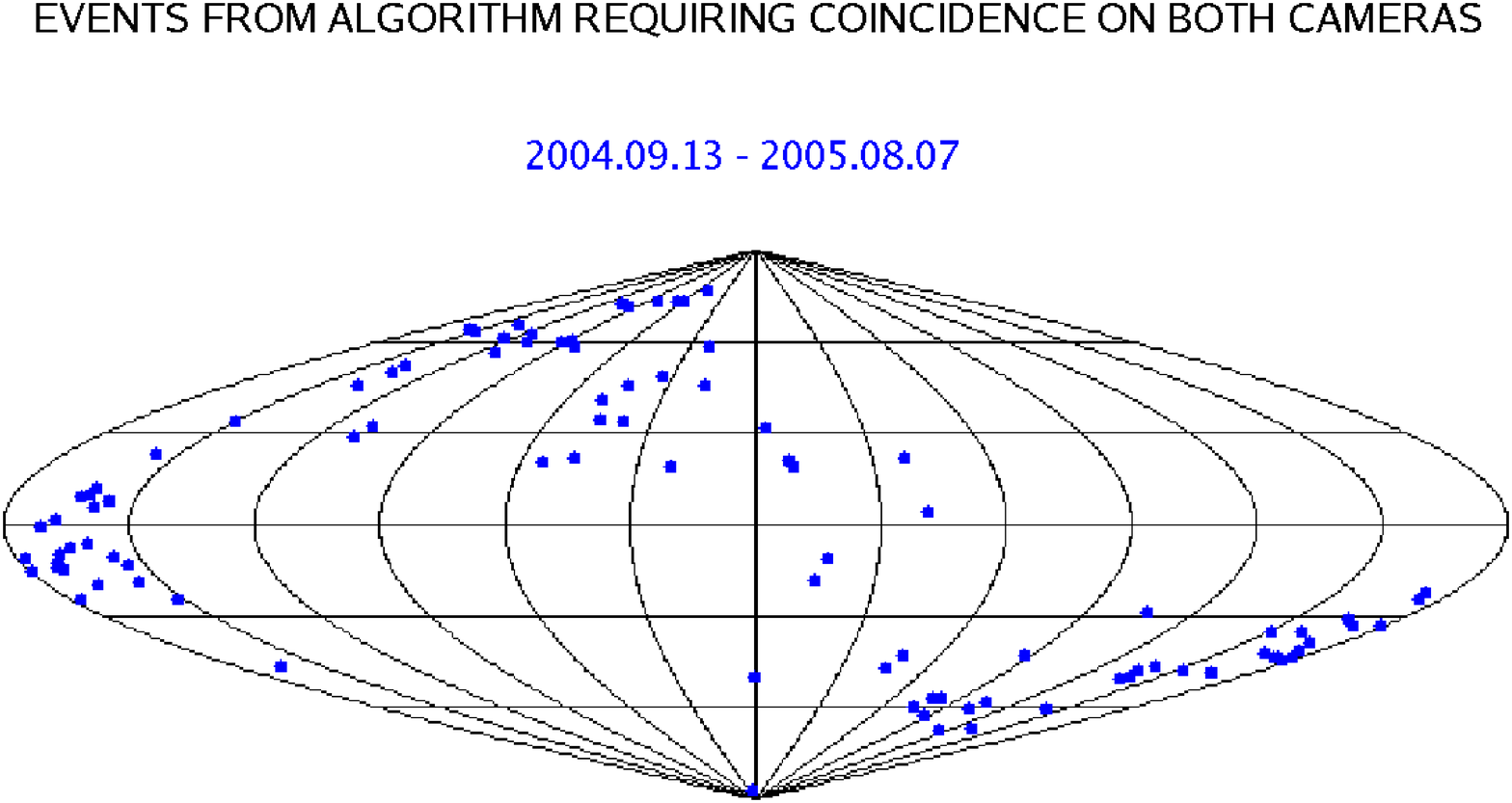}
      \includegraphics[width=2.7in,height=2.7in]{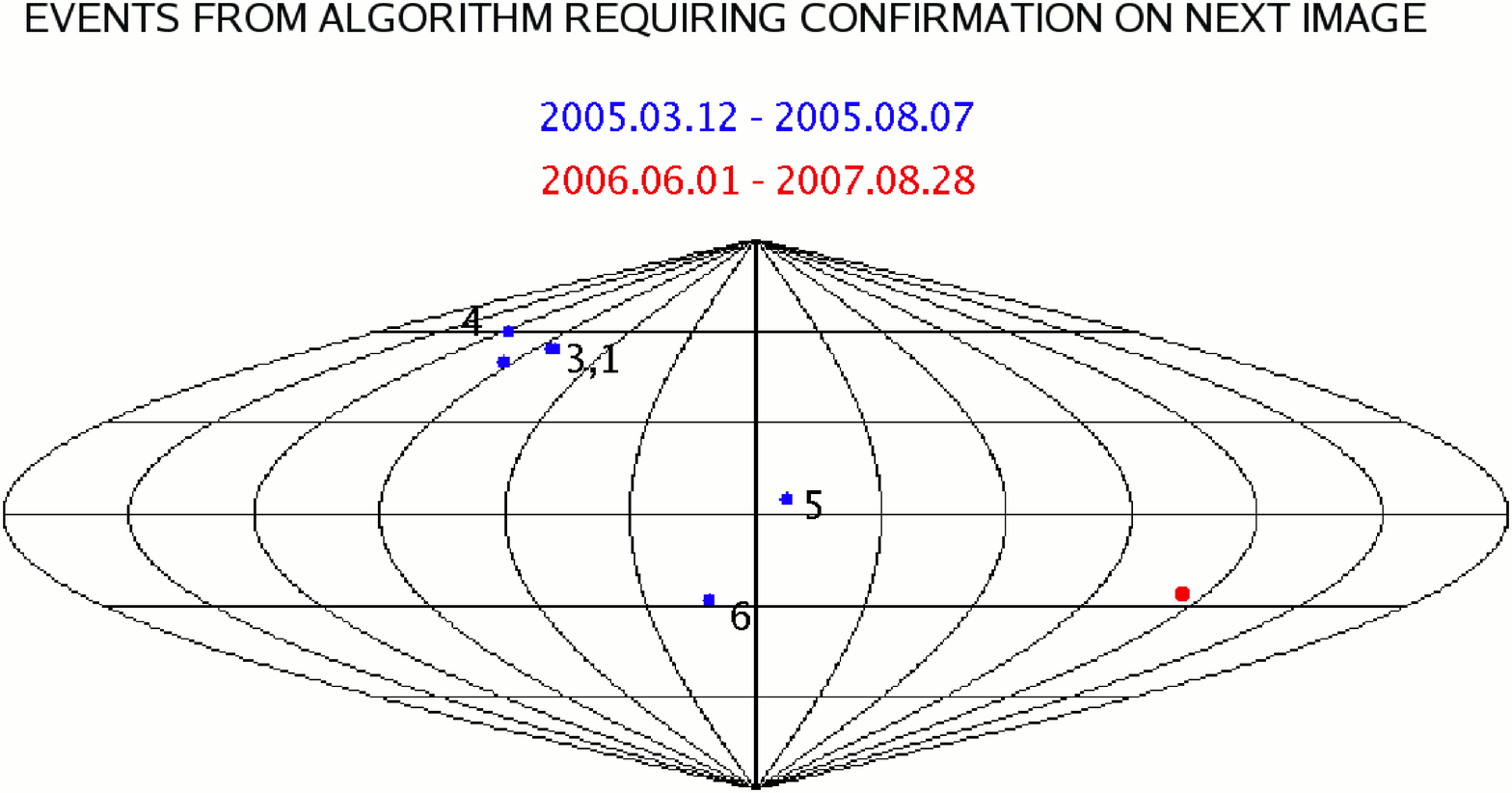}
      \includegraphics[width=2.7in,height=2.7in]{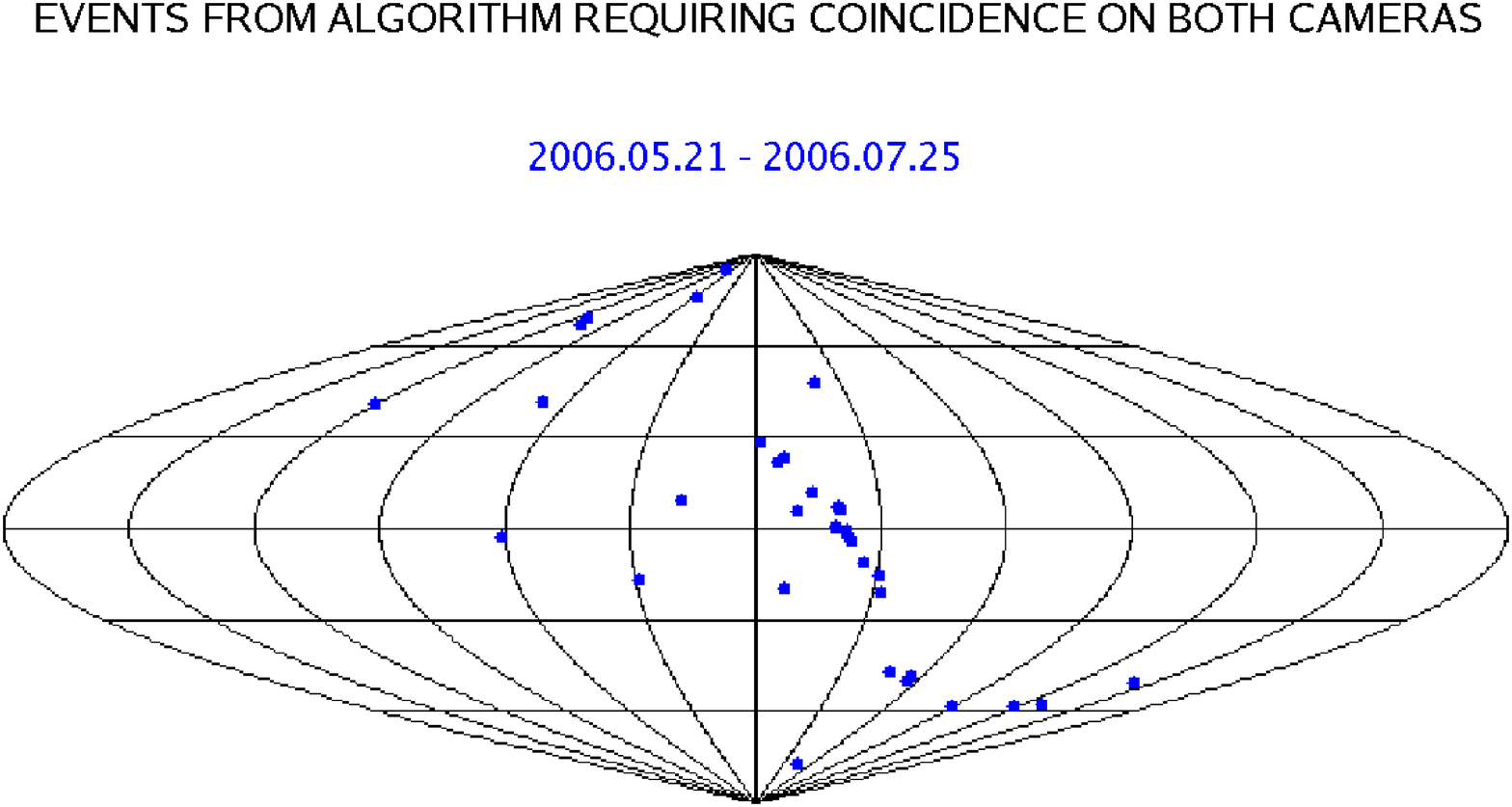}
		\includegraphics[width=2.7in,height=2.7in]{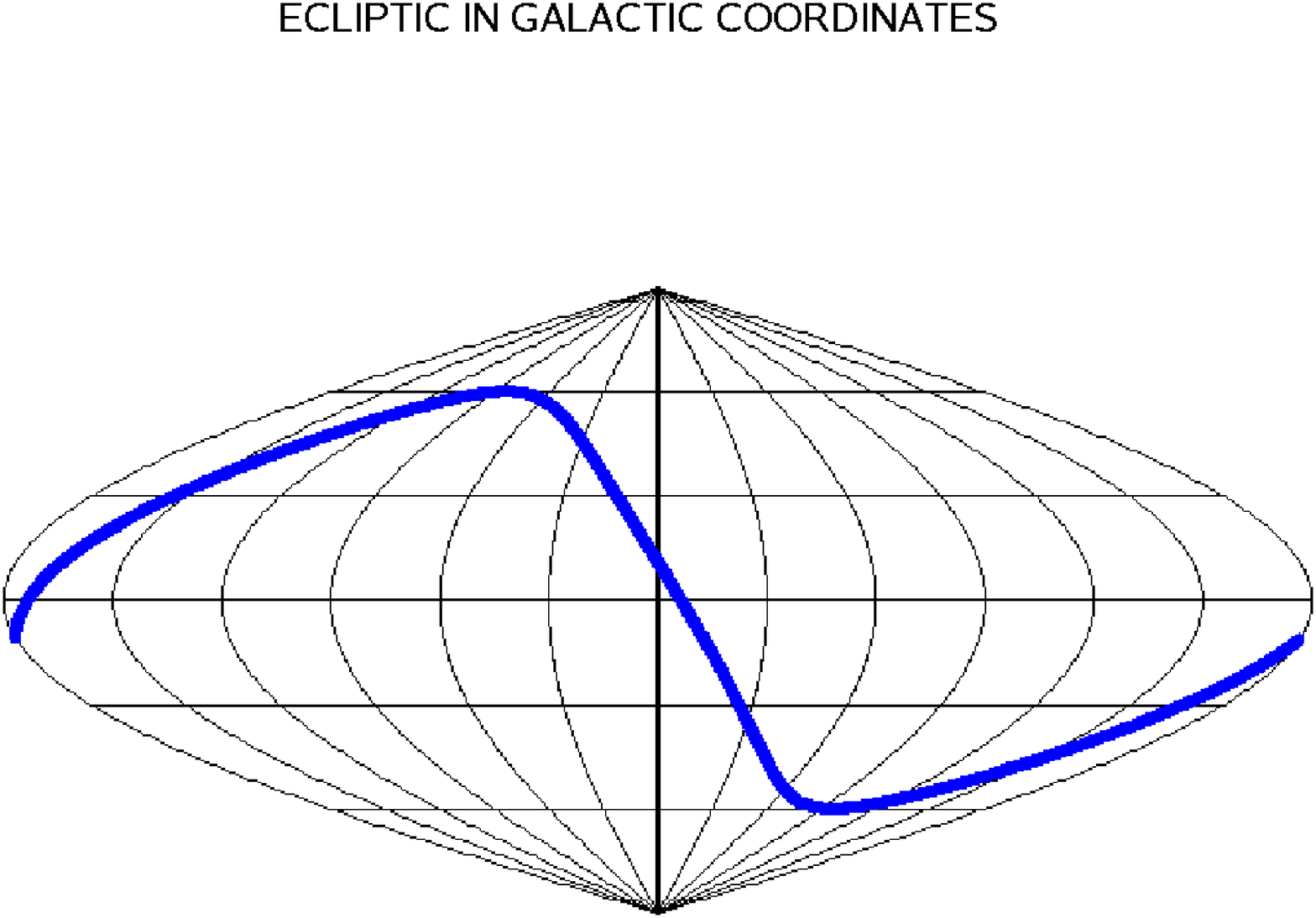}
    \else
      \includegraphics[width=2.7in,height=2.7in]{maps/2004_2005_coinc.eps}
      \includegraphics[width=2.7in,height=2.7in]{maps/conf_next_all.eps}
      \includegraphics[width=2.7in,height=2.7in]{maps/2006_coinc.eps}
		\includegraphics[width=2.7in,height=2.7in]{maps/ecliptic.eps}
    \fi
    \caption{Flashes detected by the "Pi of the Sky" system. The spatial distribution is shown in 
the galactic coordinates. Events from period 2004/2005 are collected near
ecliptic plane due to HETE-2 pointing to anti-solar point} 
    \label{fig_galactic_coord_map}
  \end{center}
\end{figure}

\begin{figure}[!htbp]
  \begin{center}
    \leavevmode
    \ifpdf
		\includegraphics[width=2.7in,height=2.7in]{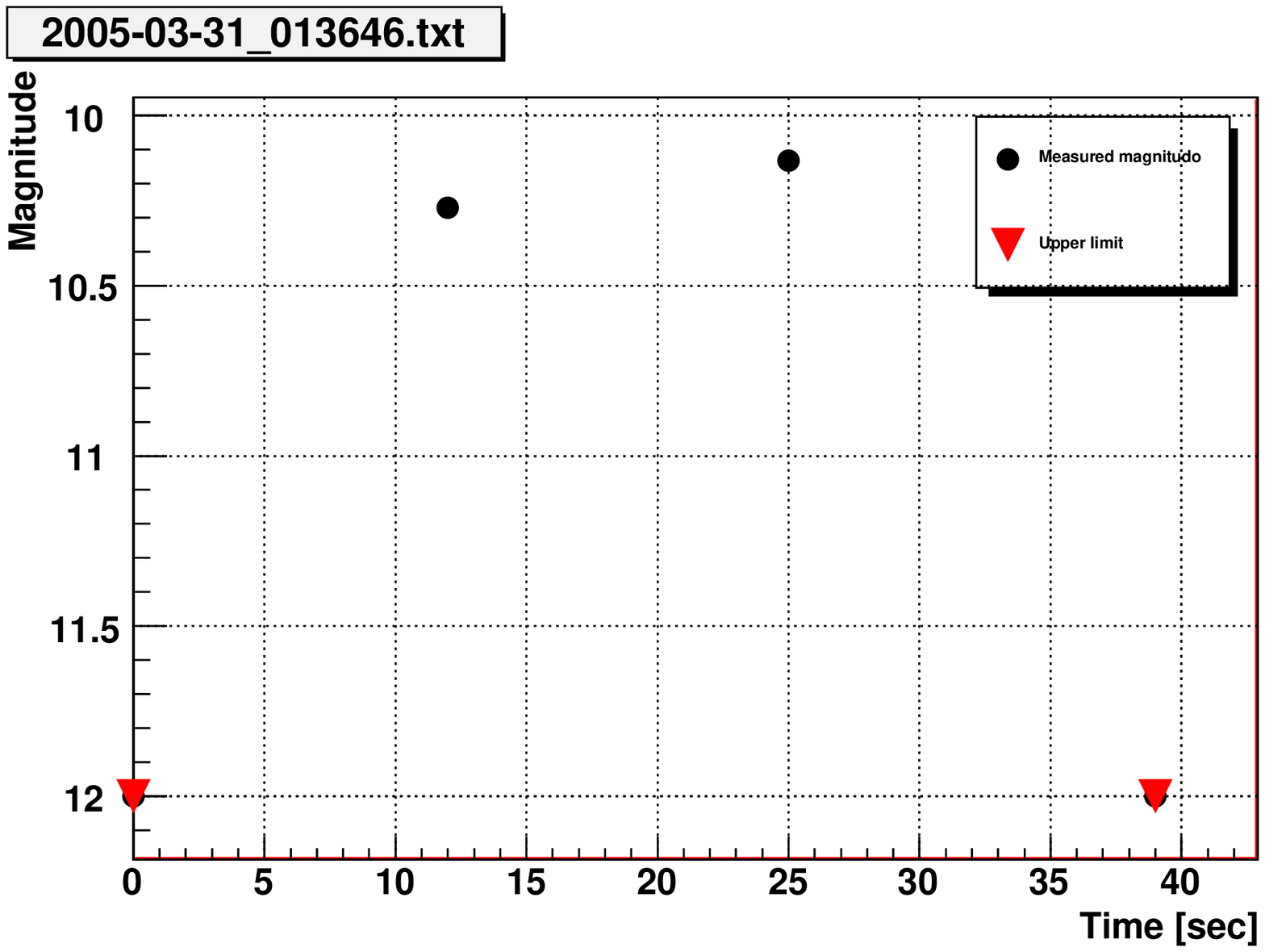}
		\includegraphics[width=2.7in,height=2.7in]{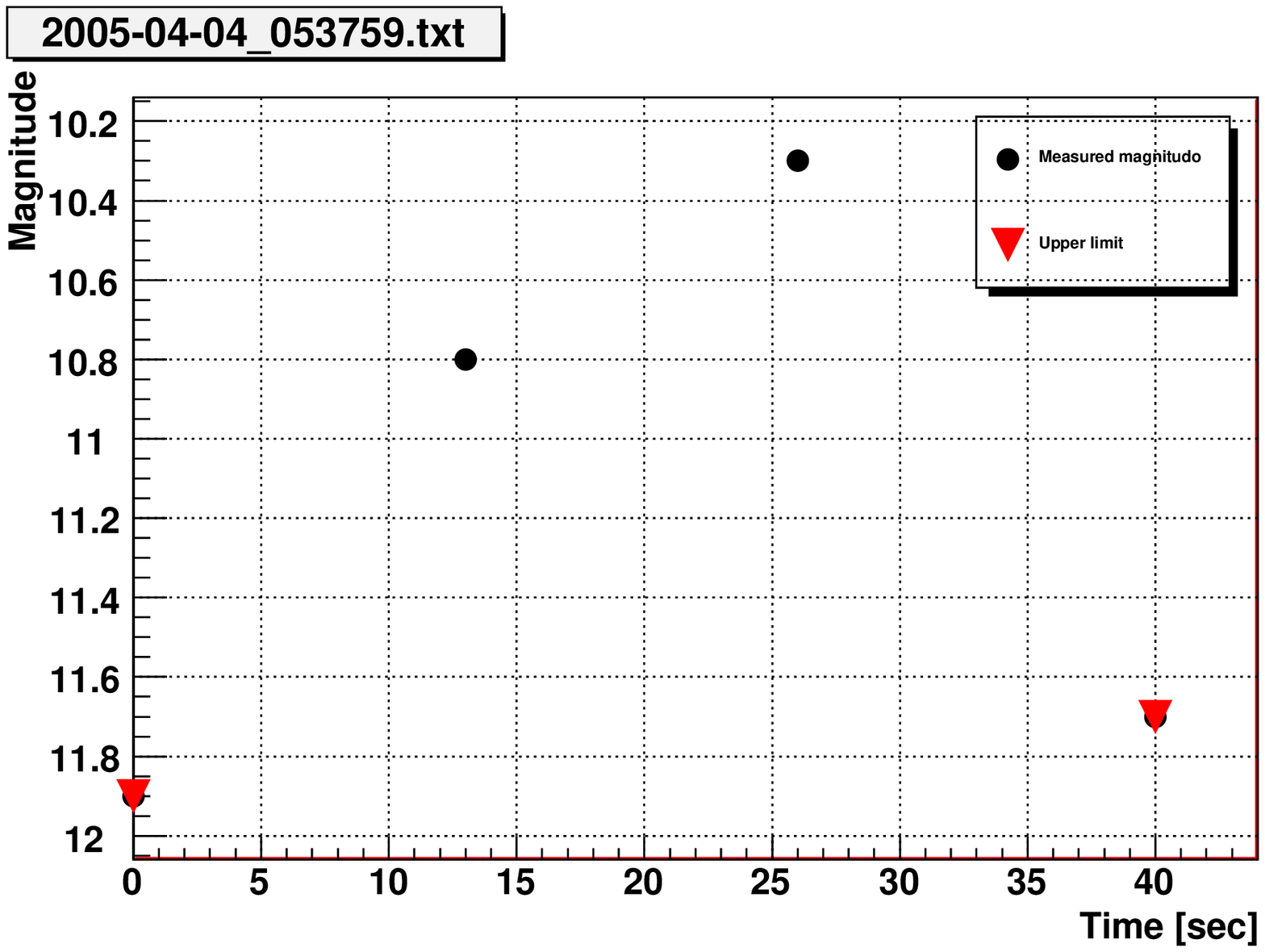}
		\includegraphics[width=2.7in,height=2.7in]{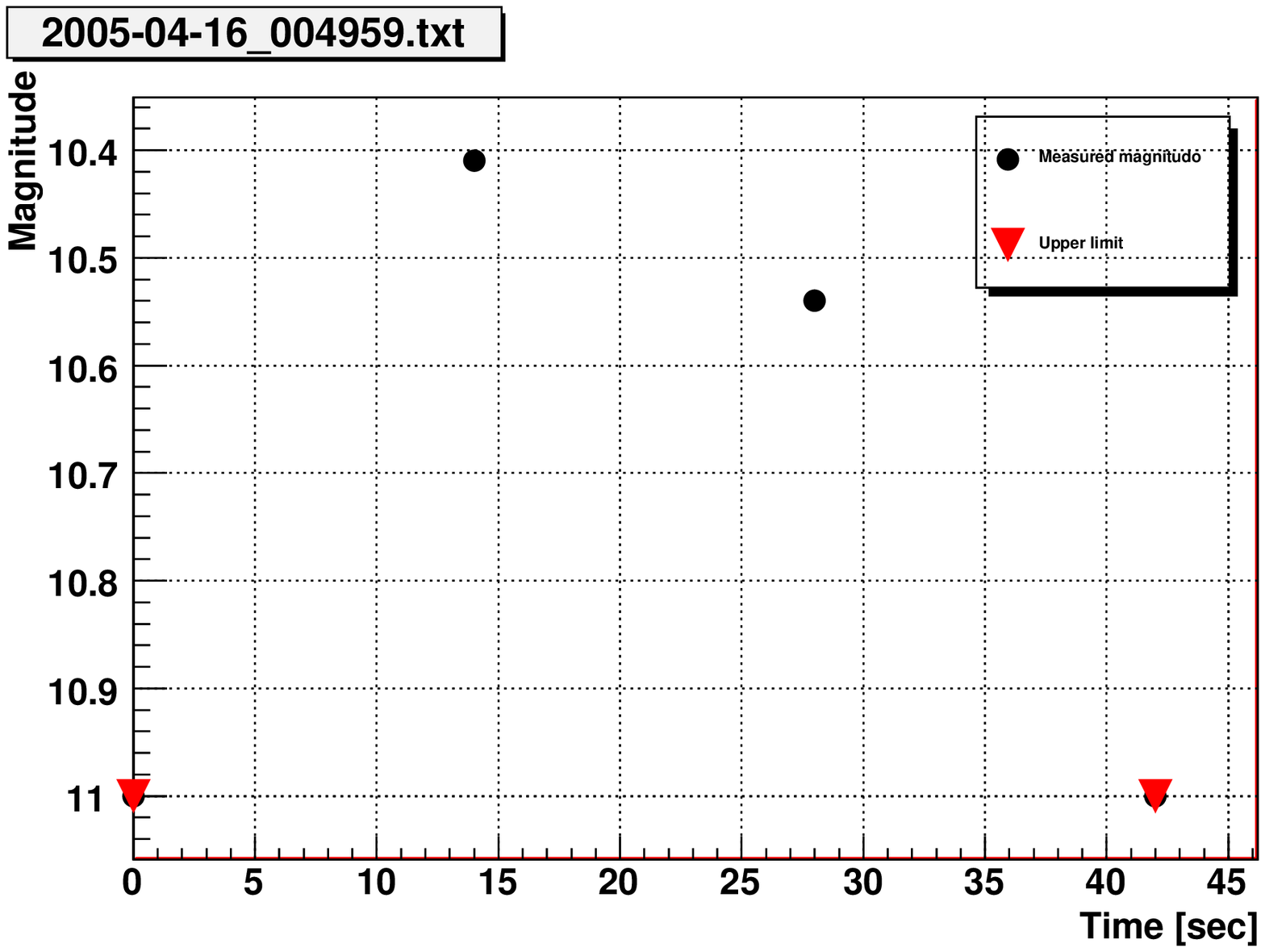}
		\includegraphics[width=2.7in,height=2.7in]{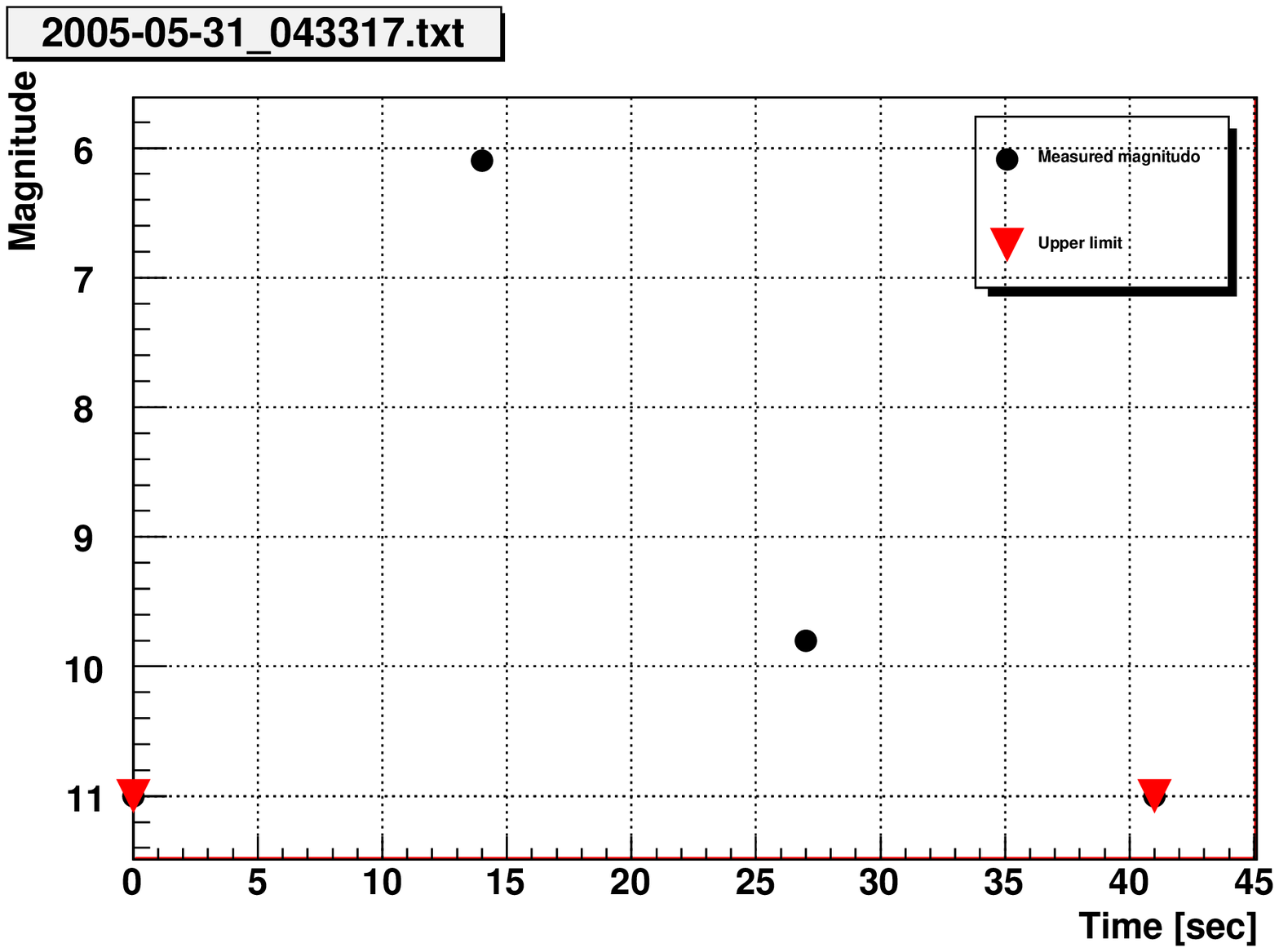}
		\includegraphics[width=2.7in,height=2.7in]{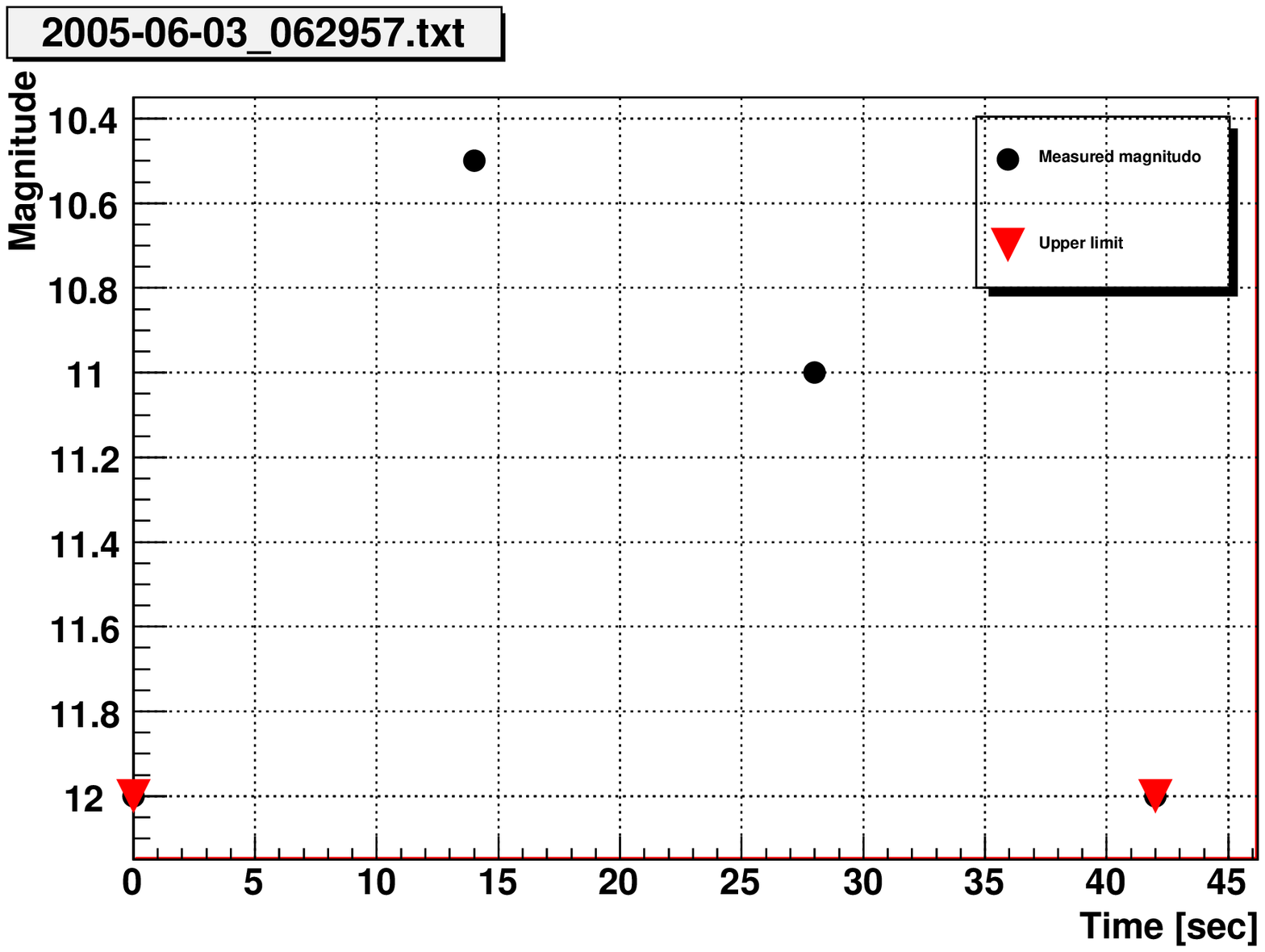}
		\includegraphics[width=2.7in,height=2.7in]{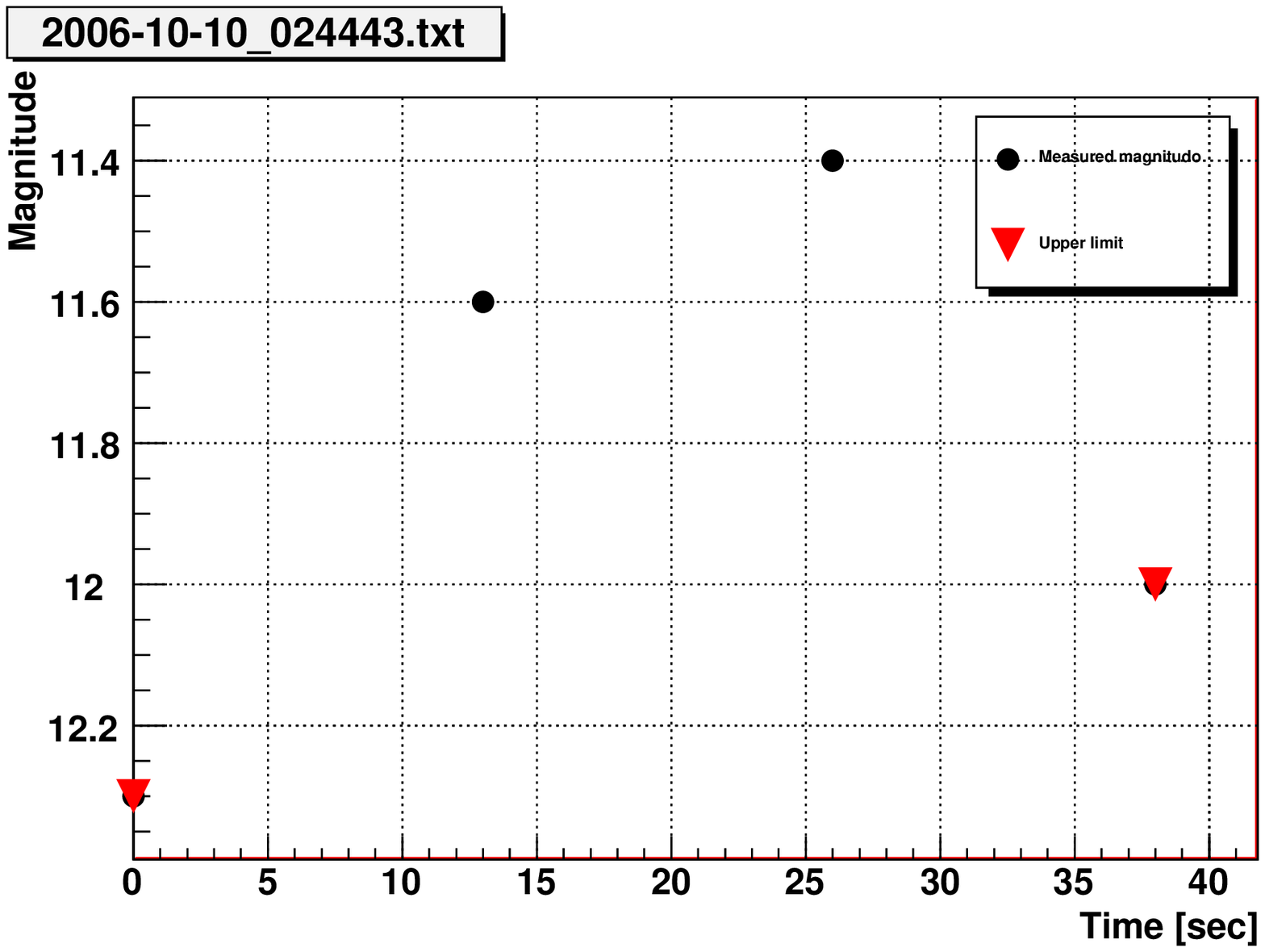}
    \else
		\includegraphics[width=2.7in,height=2.7in]{double_image/LC/2005-03-31_013646.txt.eps}
		\includegraphics[width=2.7in,height=2.7in]{double_image/LC/2005-04-04_053759.txt.eps}
		\includegraphics[width=2.7in,height=2.7in]{double_image/LC/2005-04-16_004959.txt.eps}
		\includegraphics[width=2.7in,height=2.7in]{double_image/LC/2005-05-31_043317.txt.eps}
		\includegraphics[width=2.7in,height=2.7in]{double_image/LC/2005-06-03_062957.txt.eps}
		\includegraphics[width=2.7in,height=2.7in]{double_image/LC/2006-10-10_024443.txt.eps}
    \fi
    \caption{Light curves of flashes visible on at least 2 consecutive 10s exposures}
    \label{fig_double_image_lc}
  \end{center}
\end{figure}

\begin{figure}[!htbp]
  \begin{center}
    \leavevmode
    \ifpdf
      \includegraphics[width=6in]{double_flash_20061010.gif}
    \else
      \includegraphics[width=6in]{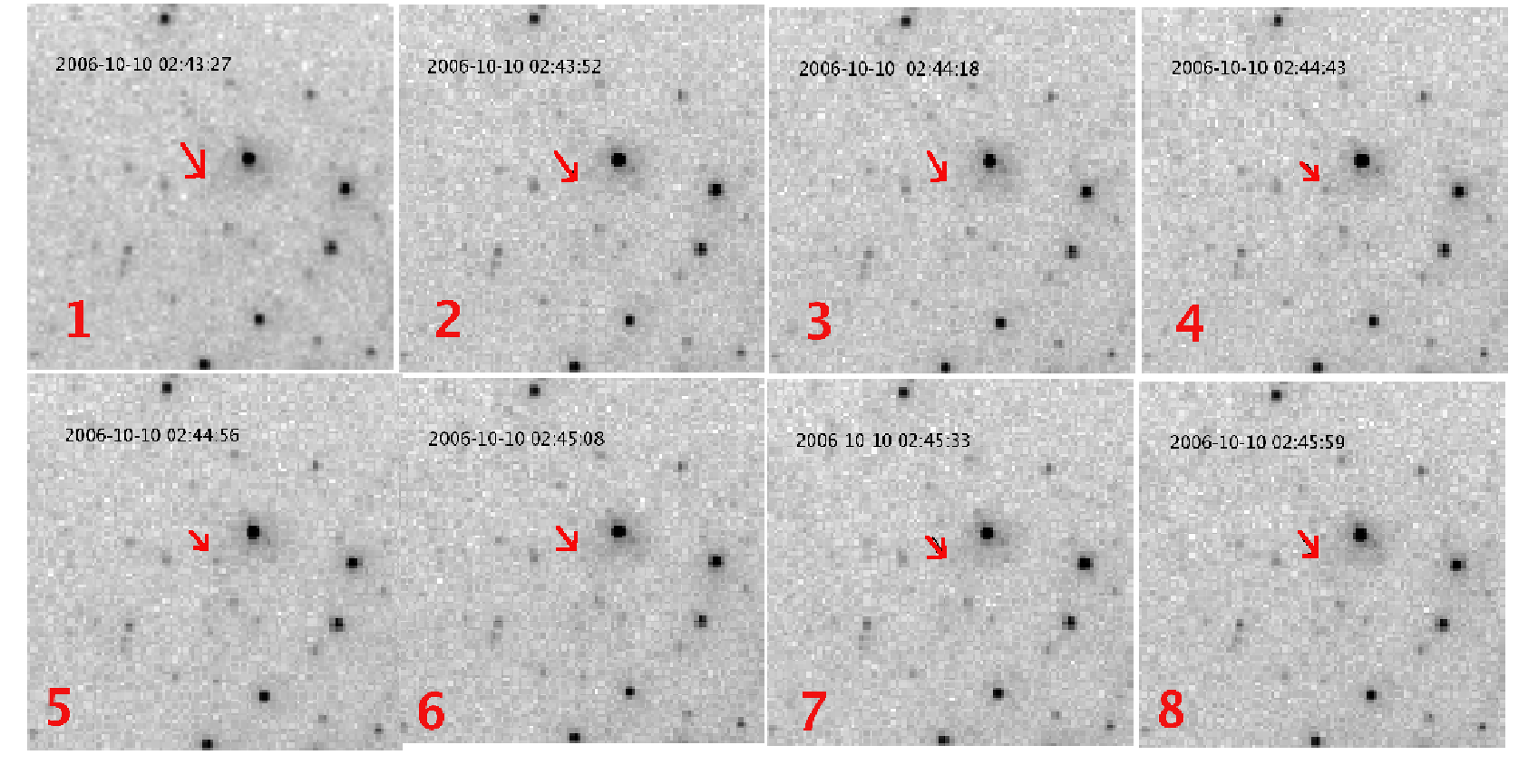}
    \fi
    \caption{Optical flash visible on 2 consecutive images ( 4 and 5 ) on 2006-10-10 02:44:43}
    \label{fig_double_flash_20061010}
  \end{center}
\end{figure}

\begin{figure}[!htbp]
  \begin{center}
    \leavevmode
    \ifpdf
      \includegraphics[width=6in]{cn_leo_even_342_to_356.gif}
    \else
      \includegraphics[width=6in]{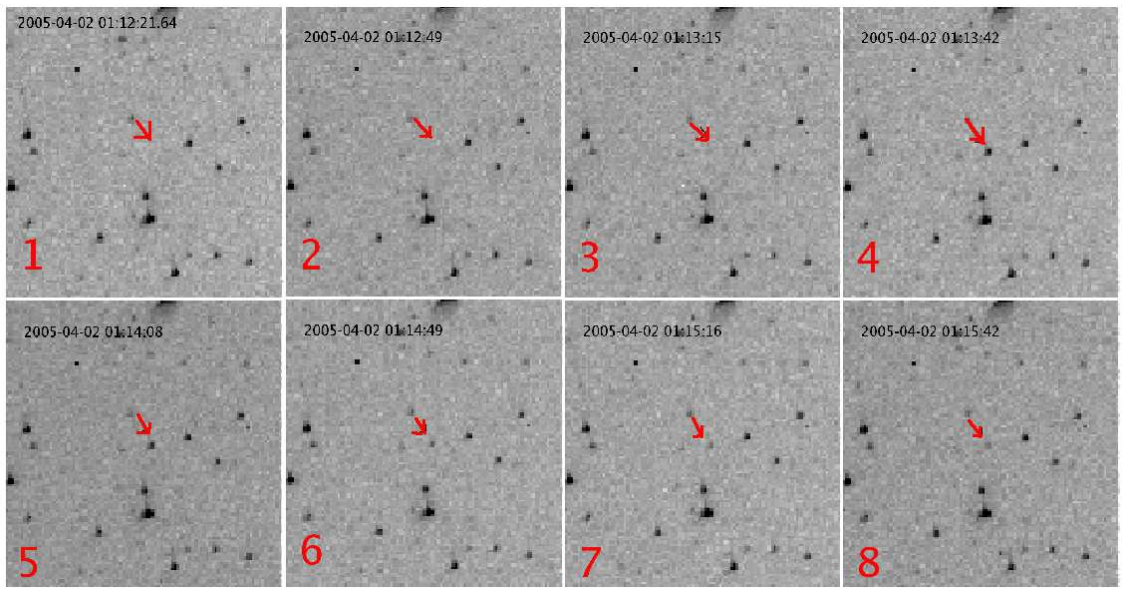}
    \fi
    \caption{Images of CN Leo outburst ( begins at image 4 ) }
    \label{fig_cnleo_images}
  \end{center}
\end{figure}

\begin{figure}[!htbp]
  \begin{center}
    \leavevmode
    \ifpdf
      \includegraphics[width=6in]{cn_leo_050402.gif}
    \else
      \includegraphics[width=6in]{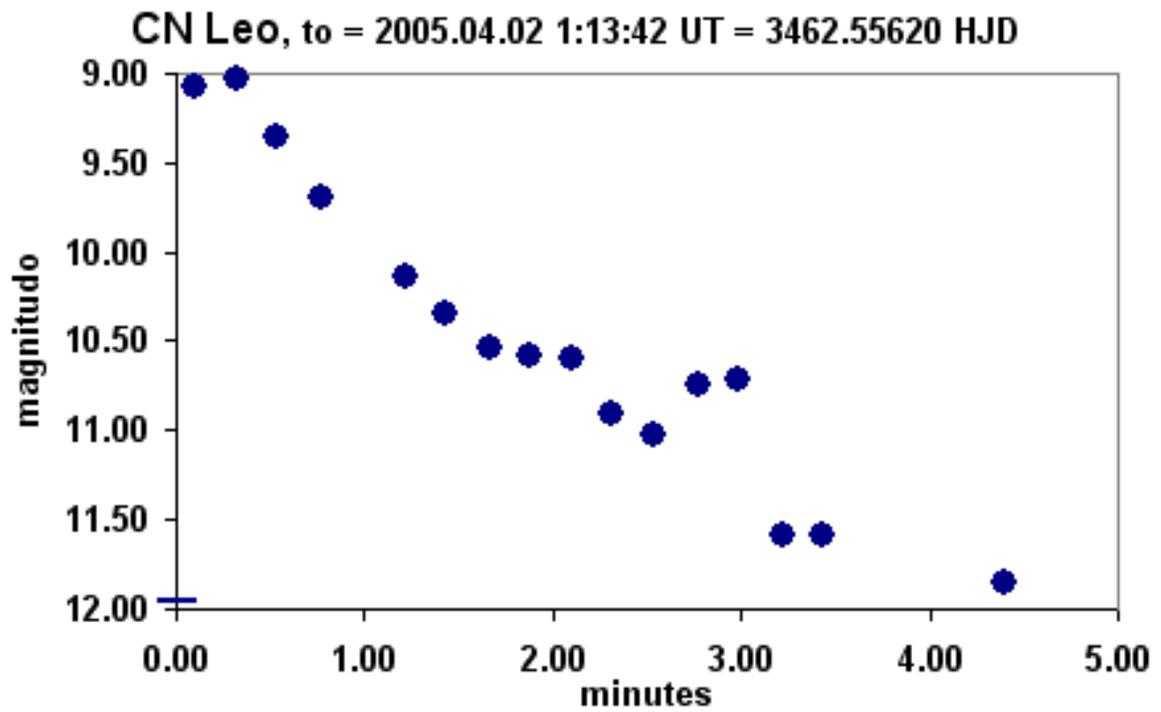}
    \fi
    \caption{Light curve of outburst of flare star CN Leo identified by flash recognition algorithm on 2005.04.02 1:13:40 UT}
    \label{fig_cnleo_lc}
  \end{center}
\end{figure}

In one case the flash was unambiguously identified as known astrophysical object.
This event occurred on 2005.04.02 1:13:40 UT , it is presented in
Figure \ref{fig_cnleo_images}. It was identified as an outburst of the flare star
CN Leo (RA = $10^h56^m29^s$, Dec = $+$7\deg01').
The light curve of this flash is shown in Figure \ref{fig_cnleo_lc}.
The normal brightness of this star, being m=13.5$^m$, increased $\approx100$
times reaching $m_{max}=9^m$. The star was below the limiting magnitude of the
telescope before the explosion and suddenly appeared as a new object.
The signature of the outburst was clearly the flash-like signature which 
allowed for on-line algorithm to identify it. The example of an outburst visible in more then one 
image is shown in Figure \ref{fig_double_flash_20061010}. This outburst 
has not been assigned to any known source.

All other flash events have been analysed. For some of them possible spatial coincidence with known
astrophysical objects was found as shown in the \mbox{Tables \ref{all_singles2006}, \ref{all_singles2005}, \ref{all_singles2004} and
\ref{tab_double_flashes}}. The condition for coincidence was an angular distance of
the flash event to the object $R_{coinc}$<2'. This is not a strict condition and 
coincidence candidates should be treated only as an indication.
No evident coincidence with known GRB event was found. 

\section{Outbursts in 240s timescale}
\label{offline_flashes}

\begin{figure}[!htbp]
  \begin{center}
    \leavevmode
    \ifpdf
      \includegraphics[width=6in]{flare20061128.gif}
    \else
      \includegraphics[width=6in]{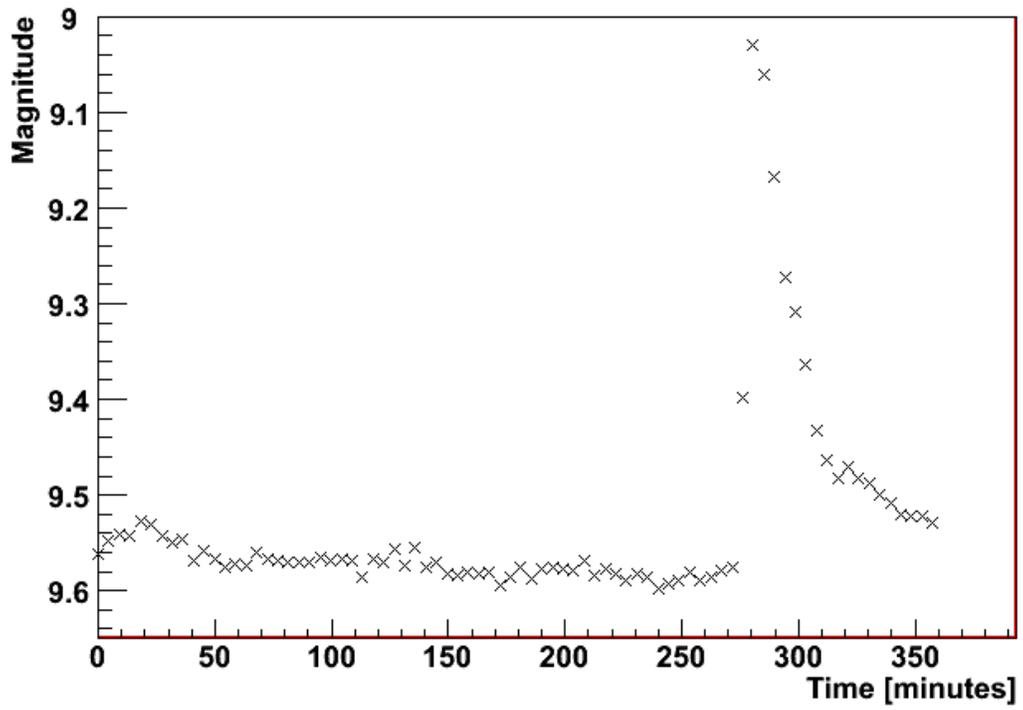}
    \fi
    \caption{Light curve of outburst of flare star GJ 3331A / GJ 3332 in 240s time resolution}
    \label{fig_aver20_gj3331a}
  \end{center}
\end{figure}

\begin{figure}[!htbp]
  \begin{center}
    \leavevmode
    \ifpdf
      \includegraphics[width=6in]{flare20061128_10s.gif}
    \else
      \includegraphics[width=6in]{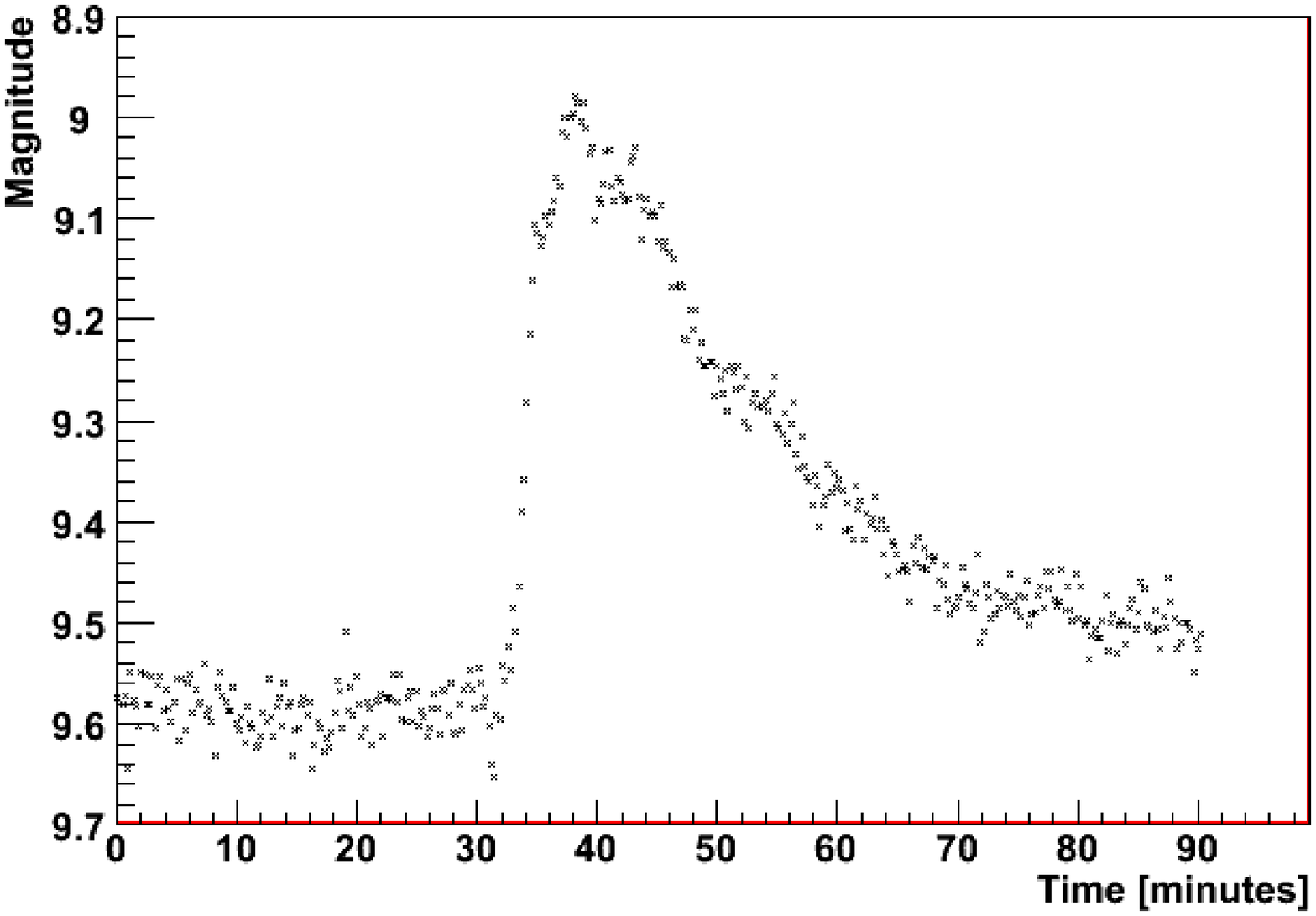}
    \fi
    \caption{Light curve of outburst of flare star GJ 3331A / GJ 3332 in 10s time resolution}
    \label{fig_10s_gj3331a}
  \end{center}
\end{figure}

The off-line algorithms working on 20 averaged images were not implemented 
from the very beginning of run of the prototype. 
A part of the archival data has been analysed and checked, but the amount was too high to inspect all events. 
Since June 2006 these procedure is performed nightly and the results are
inspected after every night. In this period flare identification algorithm
identified a single flare event which occurred on 2006.11.28 06:03 UT and was identified as an outburst of flare
star GJ 3331A / GJ 3332 (RA = $05^h06^m50^s$, Dec = -21\deg35' ).
The light curves of this flare are shown in Figures \ref{fig_aver20_gj3331a} and
\ref{fig_10s_gj3331a}. In this case the star was already present in the "Pi of the Sky" star
catalog and the algorithm found sudden increase of its brightness. The star has risen
by $\Delta$m=$0.55^m$ from m=9.58 to m$_{max}=9.03$. 

\begin{figure}[!htbp]
  \begin{center}
    \leavevmode
    \ifpdf
      \includegraphics[width=4in]{papagena_20070223_small.gif}
    \else
      \includegraphics[width=4in]{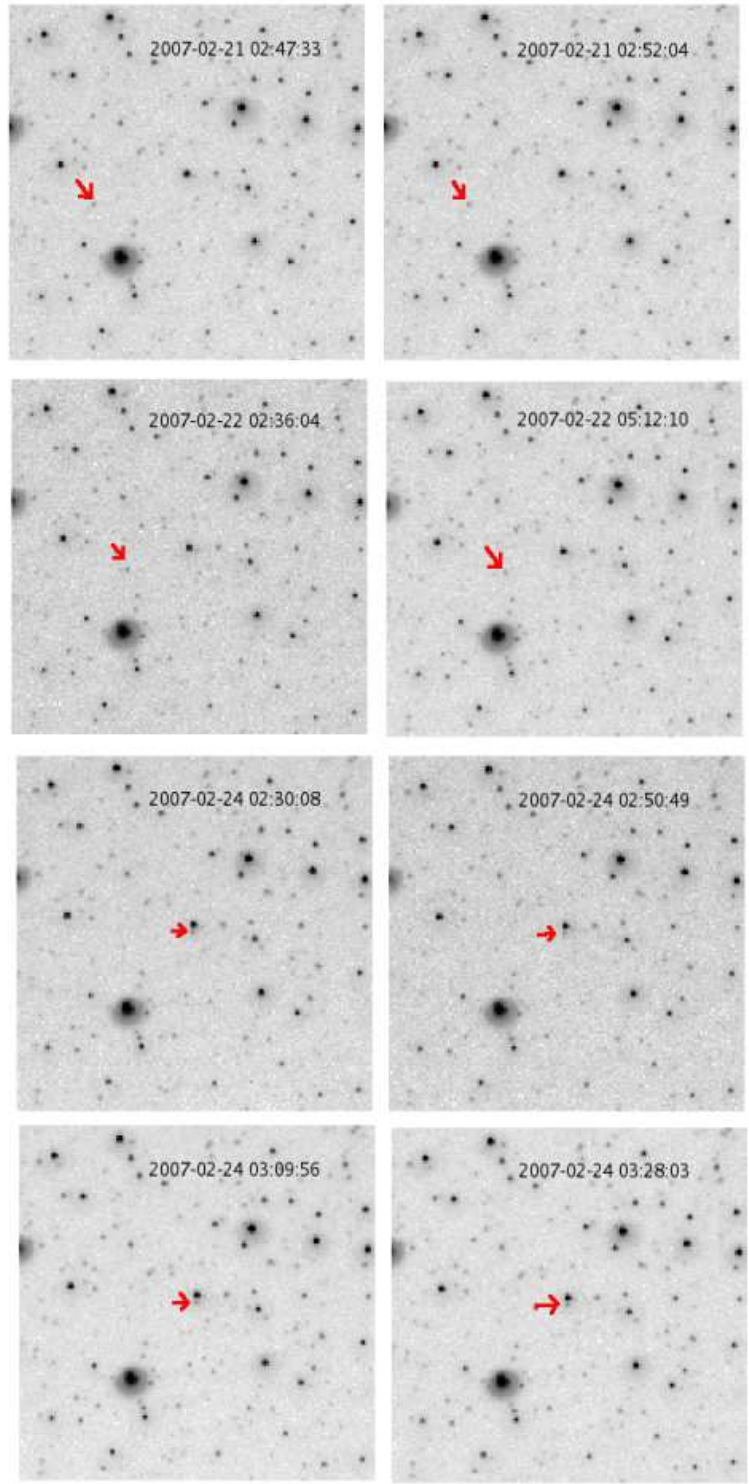}
    \fi
    \caption{Observations of asteroid Papagena in time period from 2007-02-20 to 2007-02-24}
    \label{fig_planetoid_papagena}
  \end{center}
\end{figure}

The off-line algorithm looking for objects with nova-like signature didn't
find any interesting astrophysical event. \\
The only events of astronomical origin detected by this algorithm were background events caused by
known asteroids.  They orbit the Sun and every night can be observed in different
position, thus they are automatically being cataloged as new objects in the star catalog and
found by the algorithm.
The algorithm identified 104 events related to asteroids ( till 2007.10.22 ). They turned out to be a good tool for testing the algorithm performance.
Example of the asteroid detected by the prototype is shown in Figure \ref{fig_planetoid_papagena}.
In the future version of the algorithm known asteroids will be rejected automatically. 

The total amount of data analysed by this algorithm normalized to $4\pi$
coverage is $\approx$ 0.72 days ( till 2007.10.22 ).

\section{GRB observation results}
\label{grb_obs_results}

The main goal of the "Pi of the Sky" project are detections of the optical
counterparts of GRBs. The prototype was not able to cover the whole FOV of
any of the satellites observing in $\gamma$-rays. This strongly limited the chances to observe GRBs 
during $\gamma$ emission. According to our observing strategy ( Sec. \ref{sec_obsstrategy} ) 
in the first period the system followed the FOV of the HETE satellite and since
 June 2006 the SWIFT's FOV was observed.
The Table \ref{tab_grb2004_2005} shows statistics of GRB observations for these two periods. 

\begin{table}[htbp]
\begin{center}
\begin{tiny}
\begin{tabular}{|c|c|c|c|c|c|c|c|c| }
\hline
\textbf{Period} & \begin{minipage}{1.4cm}\textbf{Apparatus Off}\end{minipage} & 
\begin{minipage}{1.4cm}
\textbf{North hemisphere}
\end{minipage}
 & \textbf{Daytime} & 
\begin{minipage}{1.0cm}
\textbf{Below \hspace{6mm} horizon}\end{minipage} & \textbf{Clouds} & \begin{minipage}{1.0cm}\textbf{Outside FOV}\end{minipage} & \begin{minipage}{1.0cm}\textbf{Inside FOV}\end{minipage} & \textbf{Total} \\
\hline
\hline
2004.07.01 - 2005.08.09 & 1 & 18 & 40 & 8 & 4 & 16 & \textbf{2} & 89 \\
\hline
2006.05.20 - 2007.10.22 & 9 & 5 & 70 & 31 & 7 & 19 & \textbf{1} & 142 \\
\hline
\end{tabular}
\end{tiny}
\hfill
\caption{GRB observations statistics in period from 2004.07.01 to 2005.08.09
and since 2006.05.20}
\label{tab_grb2004_2005}
\end{center}
\end{table}


Only 24 GRBs occurred during the night and within the range of
the telescope. All GRB events observed by the prototype are listed in
Table \ref{grb_observed}. In two cases the GRB position was observed before, during and after the GRB. 
The total time of the prototype observations normalized to $4 \pi$ is T$_{total} \approx 2$ days, 
assuming 2-3 GRB occurring every night this means that on average 4-6 GRB
events should occur in the prototype's FOV in the whole period, which is 
in agreement with the observed 3 events.

\begin{table}[htbp]
\begin{center}
\begin{tiny}
\begin{tabular}{|c|c|c|c|c|c| }
\hline
\textbf{GRB} & \textbf{Alert Source} & \textbf{Reaction Time [sec]} & \textbf{Magnitude/Upper Limit} & z & \textbf{GCN Circular} \\
\hline
\hline
GRB070913 & SWIFT & 110 & >12.6$^m$ & -  & - \\
\hline
\textbf{GRB070521} & \textbf{SWIFT} & \textbf{<0} & \textbf{>12.5}\boldmath{$^m$} & \textbf{0.553} & \textbf{GCN6437} \\
\hline
GRB070126 & SWIFT & 180 & >13.0$^m$ ( 1 image ) & - & - \\
\hline
GRB061202 & SWIFT & 170 & >14.3$^m$ ( 10 images ) & - & GCN5891 \\
\hline
GRB060719 & SWIFT & 65 & >12.8$^m$ & - & GCN5346 \\
\hline
GRB060607 & SWIFT & 124 & >13.4$^m$ & 3.082 & GCN5241 \\
\hline
GRB050607 & SWIFT & 60 & >12.5$^m$ & - & GCN3526 \\
\hline
GRB050522 & INTEGRAL & 75 & >11.0$^m$ & - & - \\
\hline
\textbf{GRB050412} & \textbf{SWIFT} & \textbf{<0} & \textbf{>11.5}\boldmath{$^m$} & \textbf{-} & \textbf{GCN3240} \\
\hline
GRB040916 & HETE & 1020 & >13.0$^m$ & - & GCN2725 \\
\hline
\textbf{GRB040825A} & \textbf{HETE} & \textbf{<0} & \textbf{>10.0}\boldmath{$^m$} & \textbf{-} & \textbf{GCN2677} \\
\hline
\end{tabular}
\end{tiny}
\hfill
\caption{Early observations of GRBs performed by the prototype in period 2004-2007. When
it is not explicitly written the limit was determined on 20 co-added images.}
\label{grb_observed}
\end{center}
\end{table}

Early detection of an optical counterpart of a GRB is not an easy task. Among a few
thousands of already detected GRB events only about 25 events have optical
counterparts observed during the first 60 seconds after the GRB trigger ( Tab. \ref{tab_early_60s_ot_detections} ).
A number of early observations ( $T_{obs} \leq$ 60s ) without positive detection is not
much larger, it is around 100 events. There are only 16 
optical observations of GRB position started before and performed during the $\gamma$
emission, they are listed in the Table \ref{tab_bda_limits}. Three of such
observations were performed by the "Pi of the Sky" system and are presented
in this thesis.

\begin{table}[htbp]
\begin{center}
\begin{tabular}{|c|c|c|c|c|}
\hline
\textbf{GRB} & \textbf{Telescope} & \begin{minipage}{3cm}\textbf{Reaction Time [sec]}\end{minipage} & \textbf{Magnitude} & \textbf{GCN Circular} \\
\hline
\hline
990123 & ROTSE & 22 & 11.82 /  8.95 \footnotemark & IAUC7100.1 \\ 
\hline
041219A & RAPTOR & 115 & 19.4 (not sure) & GCN2889 \\
\hline
050319 & ROTSE-III & 27 & 16 & GCN3116 \\
\hline
050401 & ROTSE-III & 33 & 17 & GCN3165 \\
\hline
050502A & ROTSE-III & 23 & 14.3 & GCN3322 \\ 
\hline
050502B & SWIFT/UVOT & 60 & 17.7 & GCN3330 \\
\hline
050801 & ROTSE-III & 22 & 15.0 & GCN3723 \\ 
\hline
051109 & ROTSE-III & 32 & 15.4 & GCN4211 \\ 
\hline
051111 & ROTSE-III & 27 & 13.0 & GCN4247 \\ 
\hline
060110 & RAPTOR & 7 & 16.1 & GCN4474 \\ 
\hline
060111B & TAROT & 28 & 13.8 & GCN4495 \\
\hline
060206 & SWIFT-UVOT & 58 & 16.7 & GCN4684 \\
\hline
060210 & KAIT & 60 & 18.1 & GCN4727 \\
\hline
060418 & PROMPT & 40 & 15.3 & GCN4971 \\
\hline
060526 & WATCHER & 60 & 15 & GCN5165 \\
\hline
060904B & ROTSE-III & 19 & 17.8 & GCN5504 \\ 
\hline
060926 & SWIFT-UVOT & 57 & 19.0 & GCN5625 \\
\hline
060927 & ROTSE-III & 17 & 16.5 & GCN5629 \\ 
\hline
061007 & ROTSE-III & 26 & 13.6 & GCN5706 \\ 
\hline
061025 & ROTSE-III & 45 & 15.6 & GCN5759 \\ 
\hline
061126 & RAPTOR & 21 & 12.3 & GCN5873 \\ 
\hline
070610 & OPTIMA-Burst & 57 & 19 & GCN6492 \\
\hline
070721B & SWIFT-UVOT & 50 & 15.9 & GCN6641 \\
\hline
070808 & ROTSE-III & 29 & 16.2 & GCN6719 \\
\hline
071003 & KAIT & 42 & 12.83 & GCN6844 \\
\hline
\end{tabular}
\hfill
\caption{Positive early ( $T_{obs}$ < 60 sec ) detections of optical counterparts of GRBs}
\label{tab_early_60s_ot_detections}
\end{center}
\end{table}

\footnotetext{Magnitude after 22 seconds and at maximum brightness is shown}

\begin{table}[htbp]
\begin{center}
\begin{tabular}{|c|c|c|c|}
\hline
\textbf{GRB} & \textbf{Telescope} & \textbf{Limit} & \textbf{EXPTIME [sec]} \\
\hline
\hline
020531A & Bootes-1 & 10.5$^m$ before and 8$^m$after & 120 \\
\hline
021201A & Bootes-1 & 10$^m$ & 45 \\
\hline
030115A & Bootes-1 & 10$^m$ & 45 \\
\hline
030226A & Bootes-1 & 11.5$^m$ & 180 \\
\hline
\textbf{040825A} & \textbf{Pi of the Sky} & \textbf{10}\boldmath{$^m$} & \textbf{10} \\
\hline
041211A & Ashra prototype & 11.3$^m$ & ? \\
\hline
050215B & Bootes-2 & 10.0$^m$ & 30 \\ 
\hline
\textbf{050412A} & \textbf{Pi of the Sky} & \textbf{11}\boldmath{$^m$} & \textbf{10} \\
\hline
050504A & Ashra P2/3 & 8$^m$ & 4 \\
\hline
050408A & WIDGET & 10.9$^m$ & 5 x 10 \\
\hline
051211A & Bootes-2 & 10.0$^m$ & 30 \\
\hline
060211A & WIDGET & 10.8$^m$ & 5 ( co-added ) \\
\hline
060413A & WIDGET & 10.0$^m$ & 5 ( co-added ) \\
\hline
\textbf{070521A} & \textbf{Pi of the Sky} & \textbf{12}\boldmath{$^m$} & \textbf{10} \\
\hline 
070616A & WIDGET & 11.8$^m$ & 5 ( co-added ) \\
\hline
070704A & FAVOR & 13.0$^m$ & 0.128 \\
\hline
\end{tabular}
\hfill
\caption{Optical limits derived from images started before the GRB outburst and simultaneous to $\gamma$-ray emission.}
\label{tab_bda_limits}
\end{center}
\end{table}

\section{Interpretation of results}
\label{results_interpretation}

\subsection{Constraints on bright optical flashes related to GRB}
As described in Section \ref{sec_grb_orphaned_ag}, it is expected that a large 
fraction of the GRB events cannot be observed from the Earth in $\gamma$-rays ( \cite{rhoads_1997} , \cite{oag_collim_totani} ). 
These are so called orphan afterglows because in many cases only optical counterparts of 
such events could be observed from the Earth. It is also expected that a part of
GRBs having small Lorentz factor are failed GRBs and do not give significant
signal in $\gamma$-rays ( \cite{oag_collim_huang} , \cite{oag_collim_totani} ). They also could be observed in longer wavelengths.
The number of such events on the whole celestial sphere depends on the model
and it is not well established. 
Recently, a few experiments determined limits on the number of classical ( long time after the GRB event ) 
orphan afterglows on the whole celestial sphere ( \cite{klotz_orphans} , \cite{rau_orphans} ).
However, there are models ( \cite{onaxis_ag_nakar_piran} , \cite{oag_collim_totani} , \cite{oag_collim_huang} )
suggesting that collimation maybe different for different wavelengths. Thus
short optical flashes maybe related to GRBs for which $\gamma$ emission is pointing elsewhere.
The "Pi of the Sky" experiment is a good
tool to observe this kind of events due to its flash recognition algorithm. 
The observed flash events can be considered as candidates for optical flashes related to
GRB events without detected $\gamma$-ray counterpart and used to derive constraints on the number of such events on the whole 
celestial sphere. The number of orphaned optical flashes corrected for the
efficiency of flash identification algorithm can be derived from the
following formula :

\begin{equation}
	N_{oa} = \frac{N_{flashes}}{T_{obs} \cdot \epsilon_{algo} }
\label{eq_num_oa}
\end{equation}

where $T_{obs}$ is the total observing time normalized to $4 \pi$, 
$N_{flashes}$ is the number of optical flashes observed in this period and 
$\epsilon_{algo}$ is the efficiency of the flash recognition algorithm.
The Table \ref{tab_ot_limits} shows the results for a single image flashes and 
flashes visible on more then 1 image of unknown origin. 
The column OT$_{corr}$ ( 5-th ) contains number of flashes corrected for
limited efficiency of flash identification procedure which was determined in
Section \ref{sec_opt_of_algo} ( see equation \ref{eq_aver_single_image_star_eff} ).

\begin{table}[htbp]
\begin{center}
\begin{tiny}
\begin{tabular}{|c|c|c|c|c| }
\hline
\textbf{Time Period} & \textbf{Algorithm} & \textbf{\#OT/day/4$\pi$} & \textbf{\#OT/year/4$\pi$} & \textbf{\#OT$_{corr}$/year/4$\pi$}  \\
\hline
\hline
20040913-20050807 & Coinc ( Carl Zeiss ) & 90/1.68 & 19554 & 55869 \\
\hline
20050312-20050807 & ConfirmOnNext ( Carl Zeiss ) & 5/0.8 & \textbf{2281} & \textbf{6518} \\
\hline
20060512-20060725 & Coinc ( Canon 85mm ) & 35/0.14 & 89336 & 255246 \\
\hline
20060601-20070828 & ConfirmOnNext ( Canon 85mm ) & 1/0.78 & \textbf{471} & \textbf{1346} \\
\hline
20060601-20070828 & Flash algo in 240s timescale & <1/0.69 & <529 & - \\
\hline
\end{tabular}
\end{tiny}
\caption{Upper limits for number of short timescale orphan afterglows determined from the "Pi of the Sky" data}
\label{tab_ot_limits}
\end{center}
\end{table}


The strongest limit comes from the algorithm requiring a presence of a flash
candidate on at least 2 consecutive 10s images. 
The derived number means that there is not more optical flashes brighter
then 12$^m$ then 3.69/day/4$\pi$. This is not very strict limit if we
realize that it is only related to rather bright flashes, but it is comparable 
to the expected number of GRBs on the whole celestial sphere which is
2-3/day/4$\pi$. This number will be used in the next section to derive
constraints on collimation of prompt optical emission of GRBs.

\subsection{Constraints on collimation of optical emission}
\label{ot_jet_limits}

According to presented earlier theoretical predictions, optical signal from GRBs can be
emitted in a larger angle then a $\gamma$ signal. In such a case it should be
possible to observe short optical flashes related to GRBs which do not have
corresponding $\gamma$-ray counterpart visible from the Earth ( \cite{onaxis_ag_nakar_piran} , \cite{oag_collim_totani} , \cite{oag_collim_huang} ).
The constraint on collimation angle of optical emission will be
determined under the assumption that 5 optical flashes observed on at least 2 consecutive
images are related to GRBs which did not have a corresponding $\gamma$ detection.
Number of optical flashes related to GRBs without $\gamma$-ray detection is
related to number of those detected in gamma rays by the following formulae :

\begin{equation}
\frac{N_{all-grb}}{N_{\gamma-visible}} = \left( \frac{ \alpha_{OT} }{ \beta_{\gamma} } \right)^2
\label{eq_angle_ratio}
\end{equation}

\begin{figure}[!htbp]
  \begin{center}
    \leavevmode
    \ifpdf
      \includegraphics[width=4in]{collimation.gif}
    \else
      \includegraphics[width=4in]{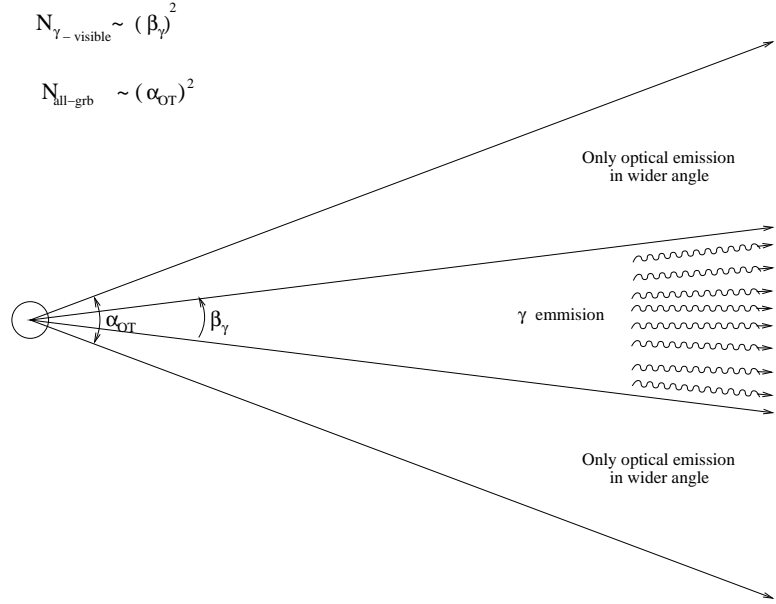}
    \fi
    \caption{A model of an emission mechanism of a GRB with difference in collimation of $\gamma$-ray and optical
emission. There may be a number of events related to GRBs, but invisible in $\gamma$ band from the Earth.}
    \label{fig_collimation}
  \end{center}
\end{figure}

where angles $\alpha_{OT}$ and $\beta_{\gamma}$ are defined in Figure \ref{fig_collimation}
and $N_{all-grb}$ denotes all events related to GRBs which can be observed
from the Earth either in optical or in gamma band.
The events that can be observed by the "Pi of the Sky" system are only those
brighter then 12$^m$. The number of such events among all GRB events can
be estimated from the SWIFT data. The analysis of optical light curves of SWIFT GRBs, 
similar to the one presented in Tab. \ref{tab_predictions},
yields that about a$\approx$8\% of events have optical observations or
brightness extrapolated to time T=0 brighter then 12$^m$, which means :


\begin{equation}
N_{\gamma + OT<12m} = a \cdot N_{\gamma + OT}
\label{eq_bright_ot_ratio}
\end{equation}

Now if we assume that events which do not have $\gamma$-ray counterpart
are not optically brighter then those which have, we can write :

\begin{equation}
N_{No-\gamma + OT<12m} = a_1 \cdot N_{No-\gamma + OT} \leq a \cdot N_{No-\gamma + OT}
\label{eq_no_gamma_with_ot}
\end{equation}

where $N_{No-\gamma + OT}$ are GRBs with no $\gamma$-ray observations,
$N_{No-\gamma + OT<12m}$ are same events, but brighter then 12$^m$ and $a_1$=a is assumed. 
The resulting number of optical flashes $N_{\pi-flash}$ can be used to calculate the upper
limit for a number of orphan afterglows is :\\

\begin{equation}
N_{No-\gamma + OT} = \frac{ N_{No-\gamma + OT<12m} }{ a } \leq \frac{N_{\pi-flash}}{a \cdot \epsilon_{algo} } \approx 45.8 / day 
\end{equation}

where $\epsilon_{algo} \approx$ 0.35 stands for efficiency of a flash
recognition algorithm ( Sec. \ref{sec_opt_of_algo} ) and the strictest obtained limit
 $N_{\pi-flash}$ = 1.28 was used. This means that a number of optical
flashes related to "invisible" GRBs, should not be greater then 21.5 / day.
Thus number of all GRBs visible from the Earth either in
optical or gamma-ray band can be constraint as follows : \\

\begin{equation}
N_{all-grb} \leq N_{No-\gamma + OT} + N_{\gamma-visible} \approx 45.8 + 2.5 = 48.3 / day
\end{equation}

where a number of all GRBs which can be observed from the Earth was assumed to be $N_{\gamma-visible} \approx$ 2.5 .
This gives the constraint on a ratio of the collimation angles :

\begin{equation}
f_{c} = \left( \frac{ \alpha_{OT} }{ \beta_{\gamma} } \right)^2 = \frac{N_{all-grb}}{N_{\gamma-visible}} \leq 48.3 / 2.5 \approx 19.3 
\end{equation}

which gives :

\begin{equation}
\alpha_{OT} \leq 4.4 \cdot \beta_{\gamma}
\end{equation}

This limit means that collimation of optical emission is
not more then $\approx$ 4.4 times larger then collimation of $\gamma$-ray emission.
Collection of more data will improve this limit significantly. It is
possible to compare the obtained result on $f_{c} \approx 19.3$ with the
results obtained in \cite{collim_ratio_guetta} and \cite{collim_ratio_frail}
which are 75$\pm$25 and 500, respectively. However, one must keep in
mind that these results were obtained for classical afterglows, which are
 long duration optical counterparts of GRBs, observed hours/days after the
GRB event.


\subsection{Limits on the optical luminosity of the GRB source}
\label{sec_grb_ot_luminosity}

Using measurements of optical counterparts obtained by other experiments and 
 early brightness limits presented in the previous section, the limitations on the
optical luminosity of the GRB source were determined. They were compared
with the measurements of other optical telescopes. The observed magnitudes 
were corrected for the galactic extinction. The luminosity distance
to the source was calculated in the cosmological model with $\Omega_{\lambda}=0.7$ and $\Omega_{M}=0.3$.
Then luminosity of the source was calculated according to formula :

\begin{equation}
L = \frac{4 \pi D_L^2 f}{1+z} \cdot k
\label{eq_luminosity}
\end{equation} 

where k is cosmological k-correction ( which was not taken into account here
) and f is flux , calculated from magnitude with the following formula :

\begin{equation}
f = f_0 \cdot 10 ^ {-0.4 \cdot m }
\label{eq_mag2flux}
\end{equation}

\begin{figure}[!htbp]
  \begin{center}
    \leavevmode
    \ifpdf
		\includegraphics[width=6in]{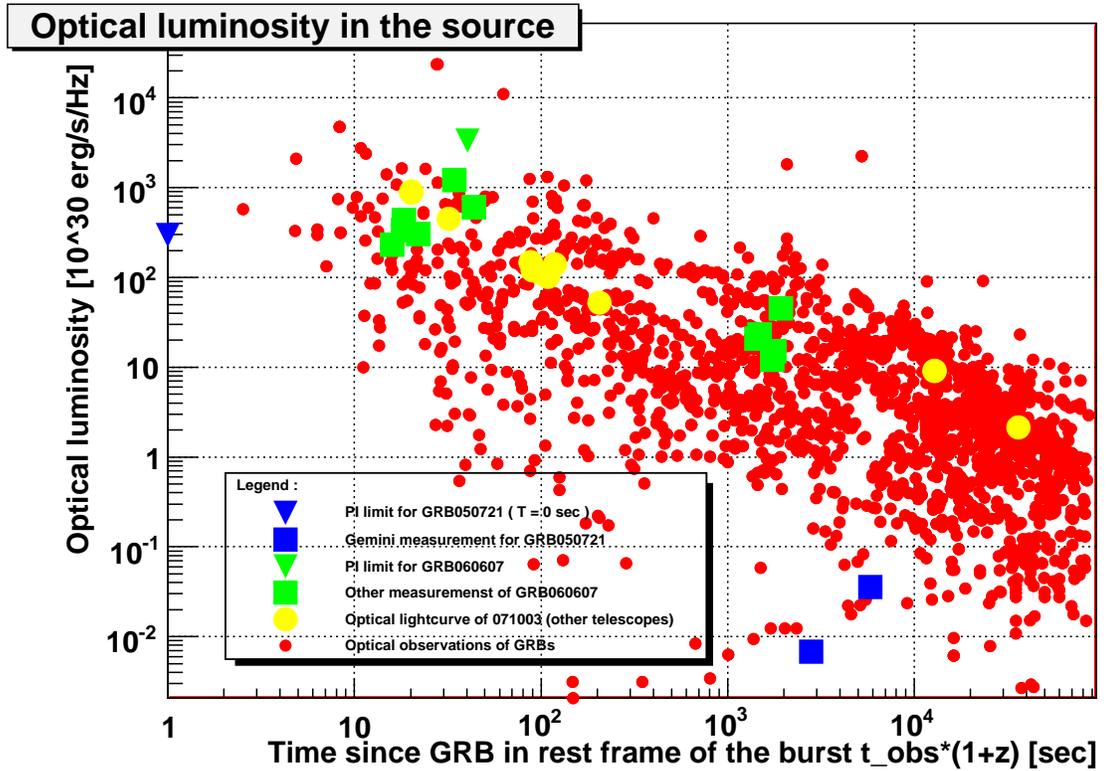}
    \else
		\includegraphics[width=6in]{Lumin/070521_86400sec_all_points.eps}
    \fi
    \caption{Optical luminosity of the source in function of time, up to 24
hours after the burst. The limits obtained from the "Pi of the Sky" measurements and presented in this thesis are shown.}
    \label{fig_lum_vs_time}
  \end{center}
\end{figure}

The luminosity in function of time with the limits determined in this thesis
is shown in Figure \ref{fig_lum_vs_time}

\begin{figure}[!htbp]
  \begin{center}
    \leavevmode
    \ifpdf
		\includegraphics[width=2in,height=2in]{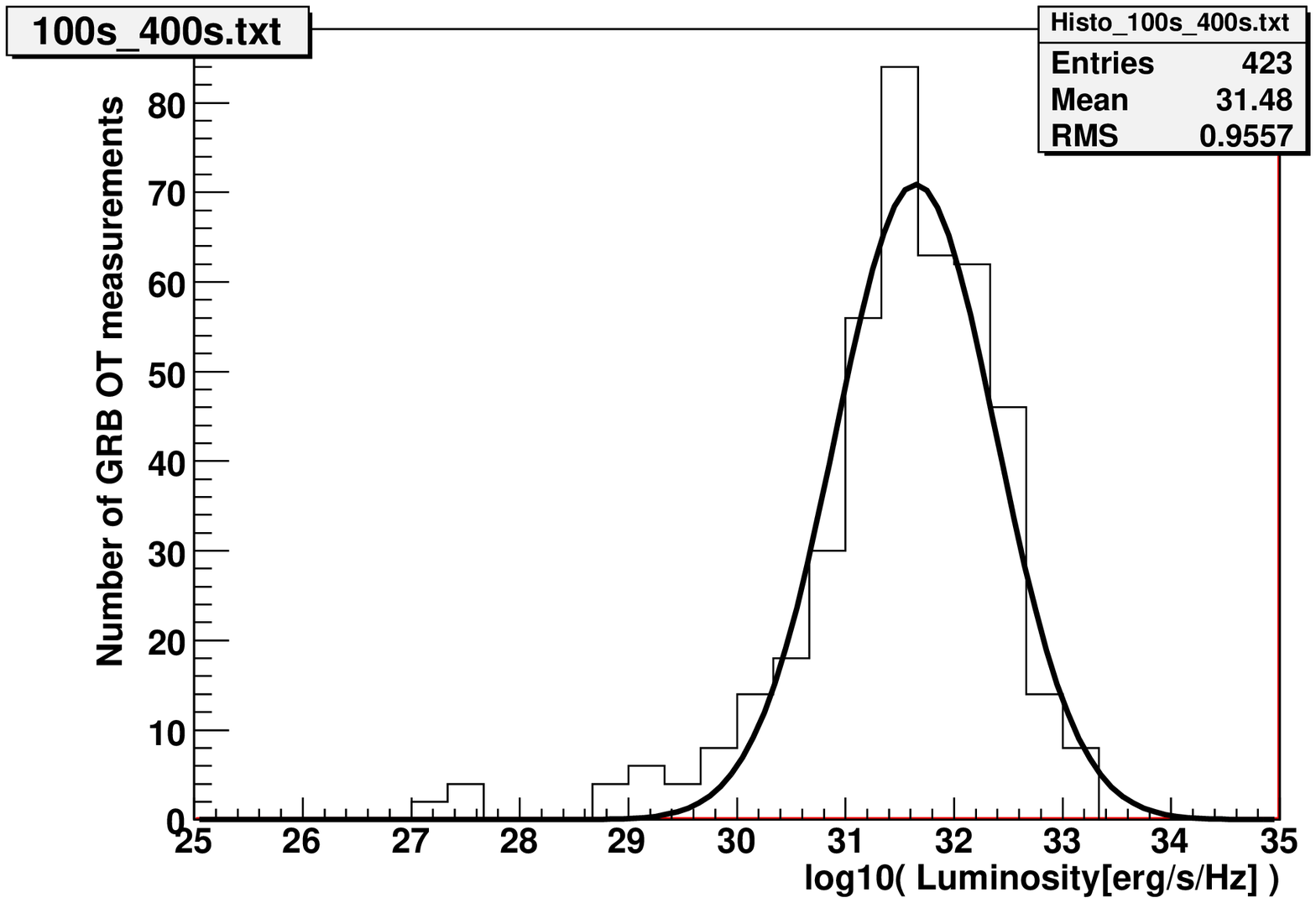}
		\includegraphics[width=2in,height=2in]{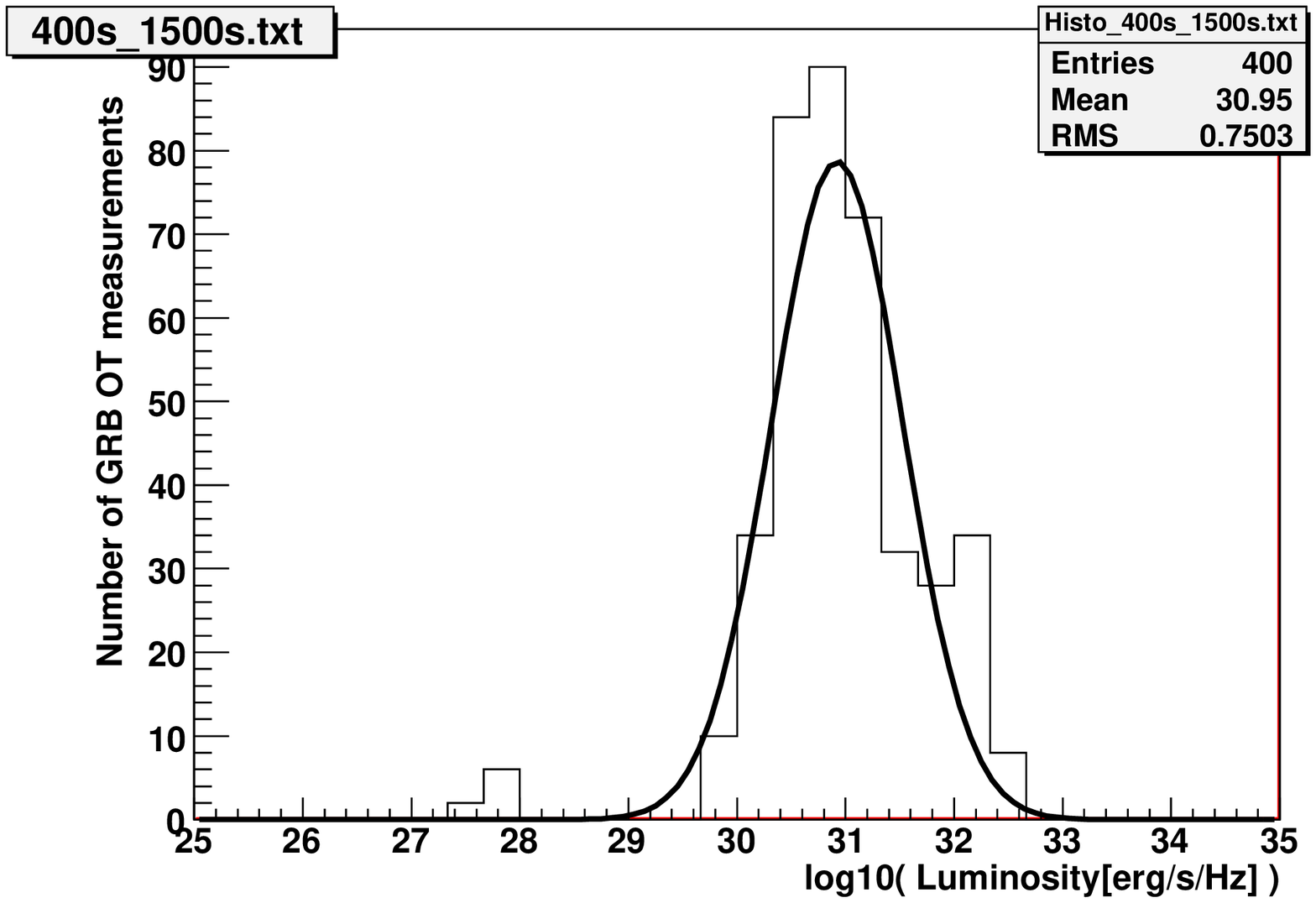}
		\includegraphics[width=2in,height=2in]{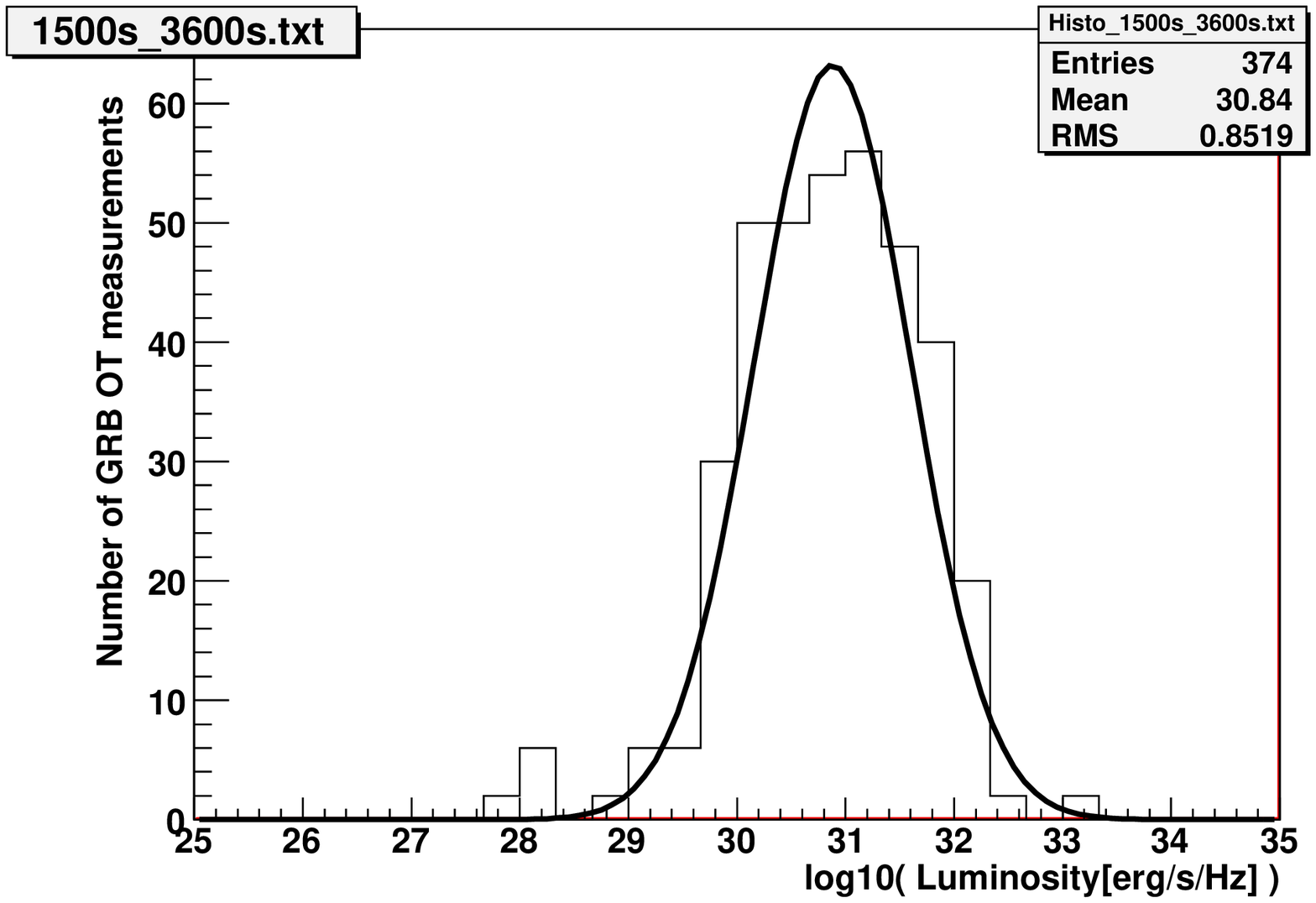}
		\includegraphics[width=2in,height=2in]{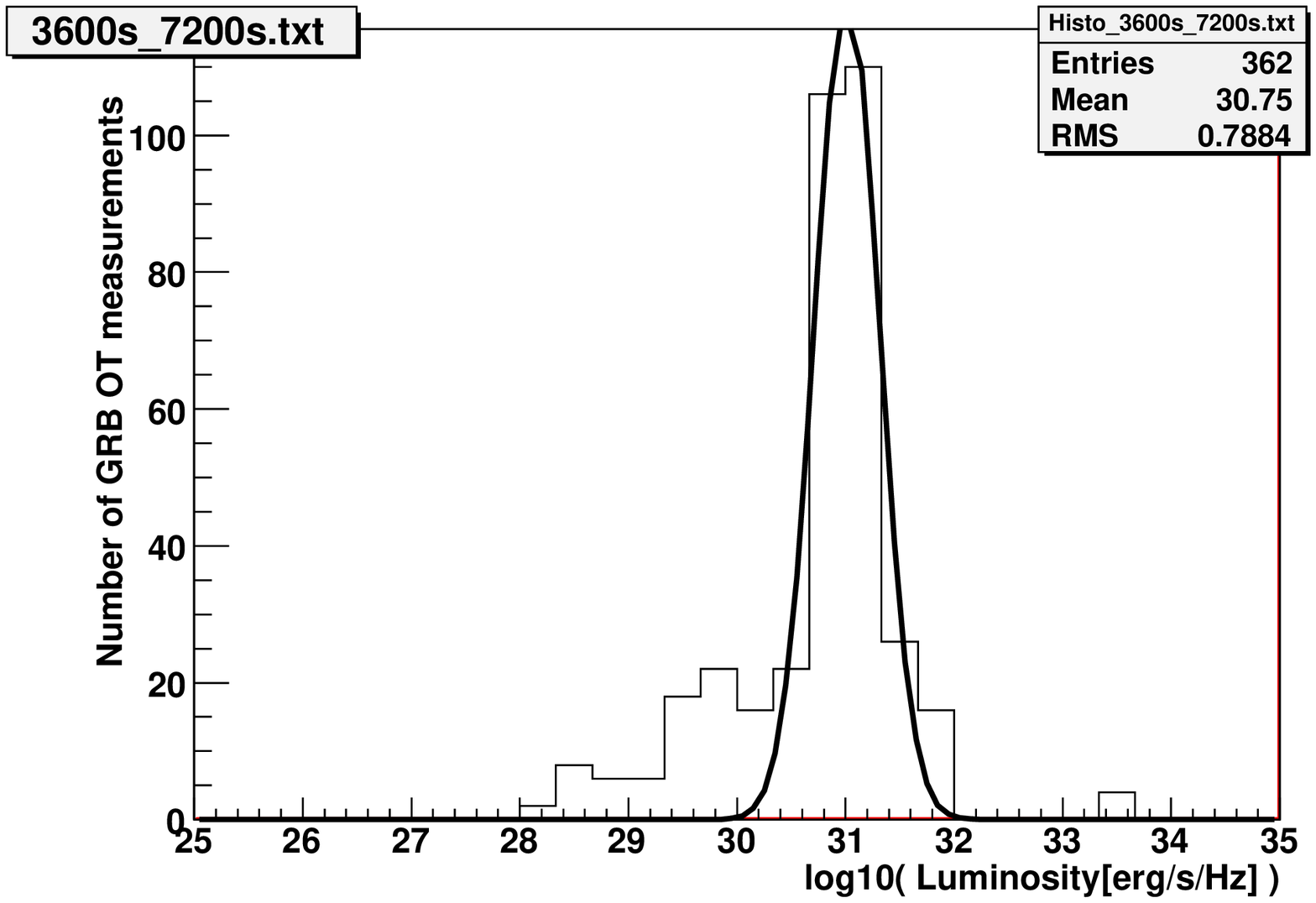}
		\includegraphics[width=2in,height=2in]{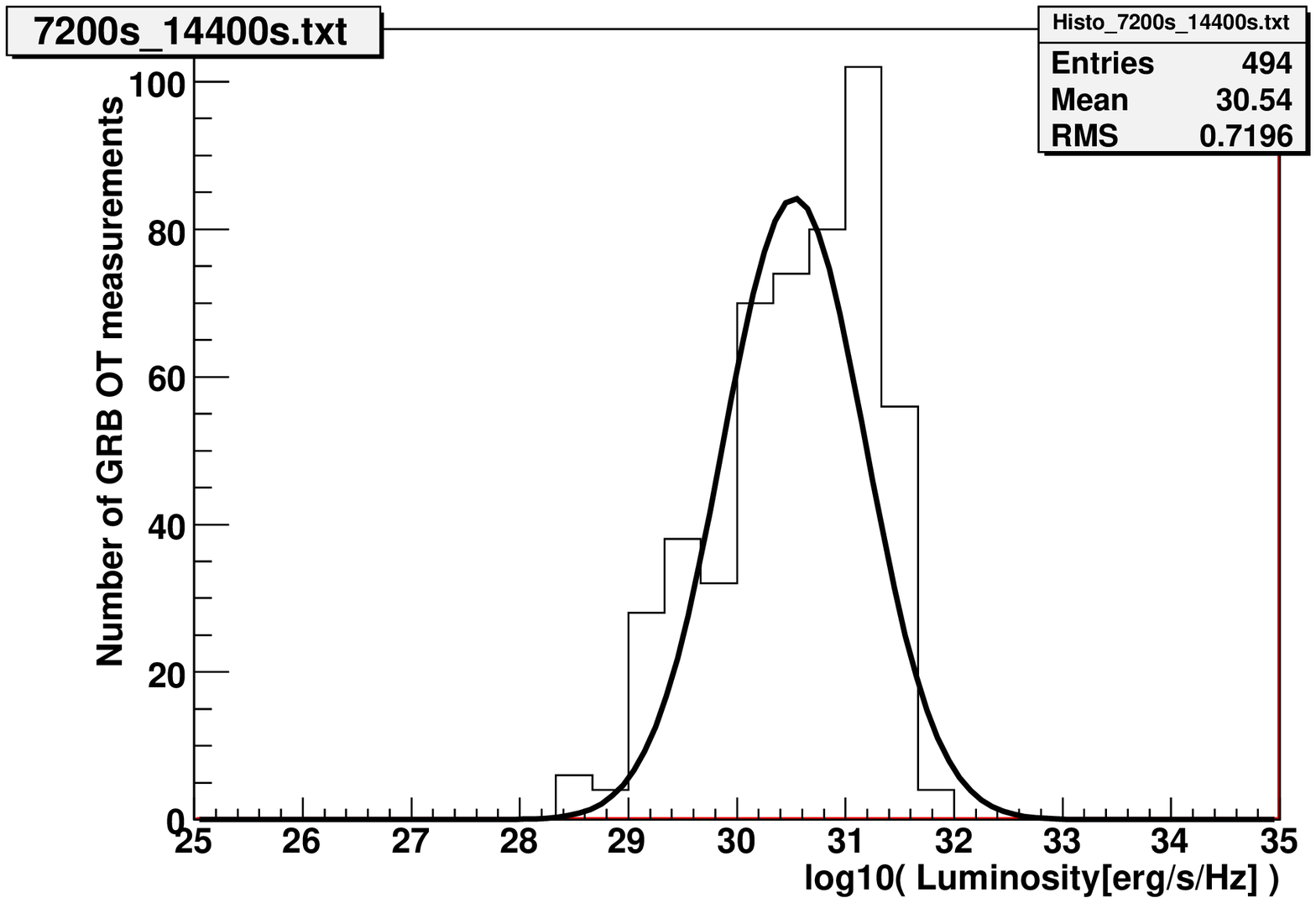}
		\includegraphics[width=2in,height=2in]{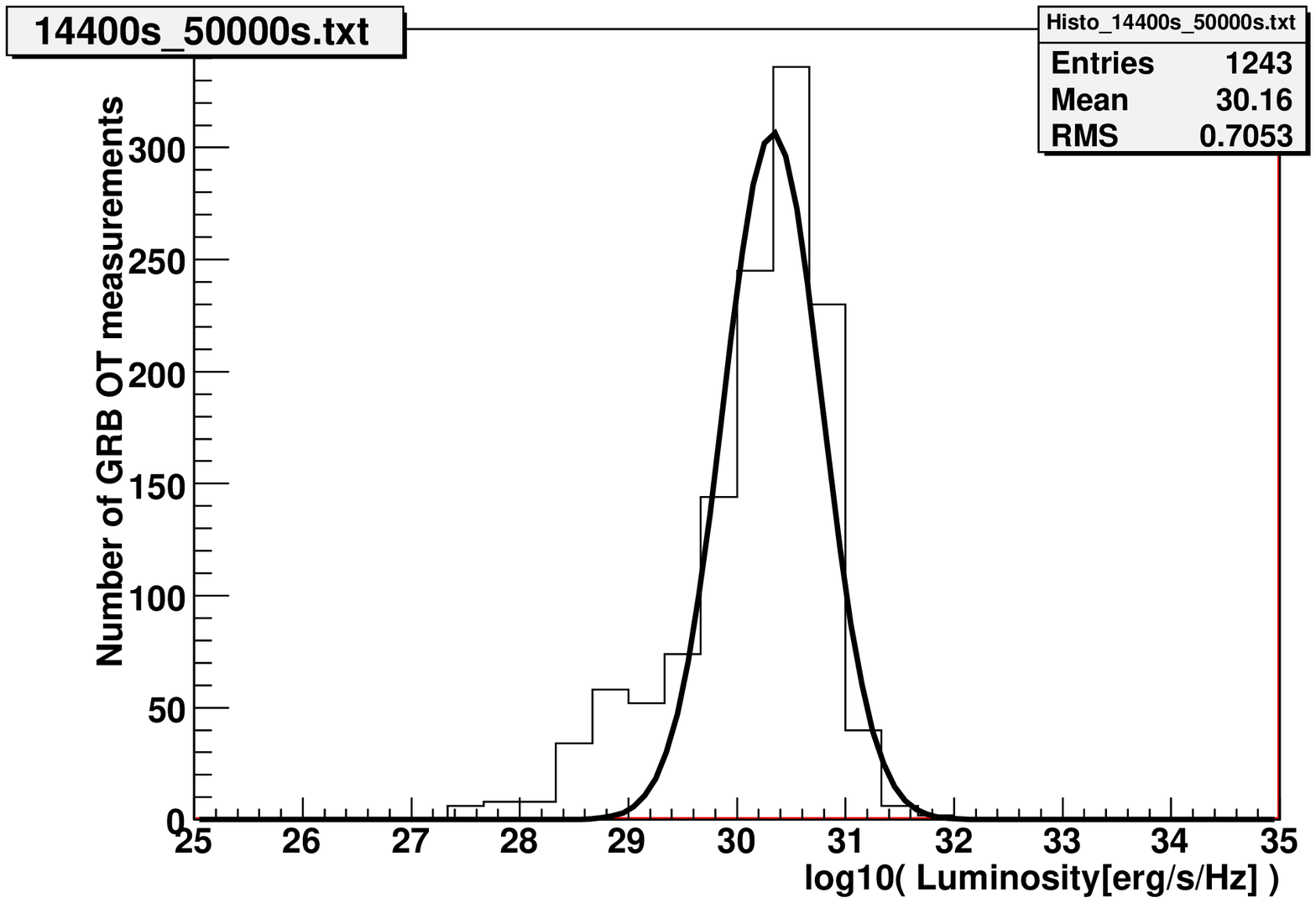}
		\includegraphics[width=2in,height=2in]{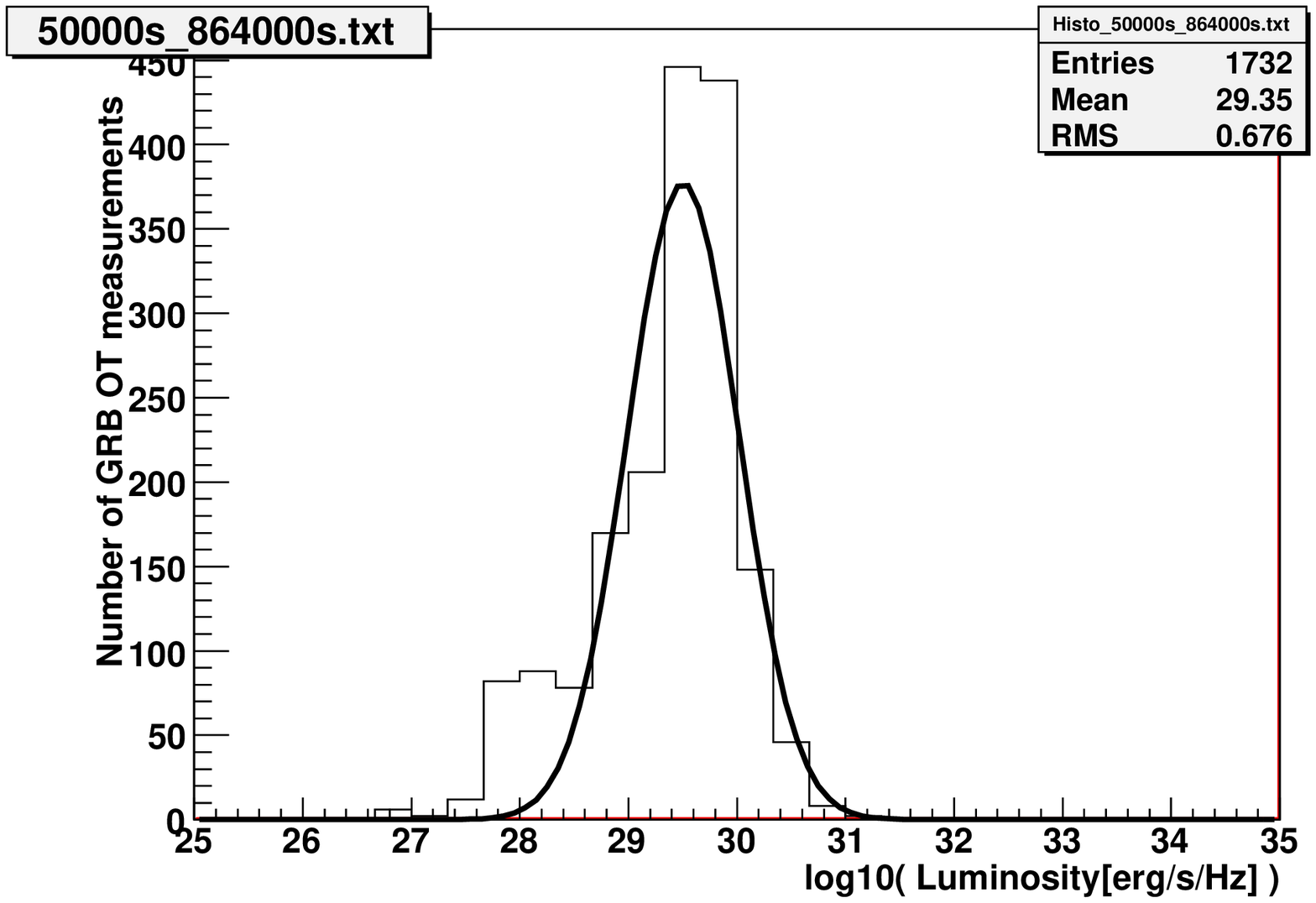}
    \else
		\includegraphics[width=2in,height=2in]{Lumin/fit/100s_400s.txt_histo.eps}
		\includegraphics[width=2in,height=2in]{Lumin/fit/400s_1500s.txt_histo.eps}
		\includegraphics[width=2in,height=2in]{Lumin/fit/1500s_3600s.txt_histo.eps}
		\includegraphics[width=2in,height=2in]{Lumin/fit/3600s_7200s.txt_histo.eps}
		\includegraphics[width=2in,height=2in]{Lumin/fit/7200s_14400s.txt_histo.eps}
		\includegraphics[width=2in,height=2in]{Lumin/fit/14400s_50000s.txt_histo.eps}
		\includegraphics[width=2in,height=2in]{Lumin/fit/50000s_864000s.txt_histo.eps}
    \fi
    \caption{Distributions of luminosity in subranges of time with fitted gauss function}
    \label{fig_gauss_lumin_fits}
  \end{center}
\end{figure}

\begin{figure}[!htbp]
  \begin{center}
    \leavevmode
    \ifpdf
      \includegraphics[width=6in]{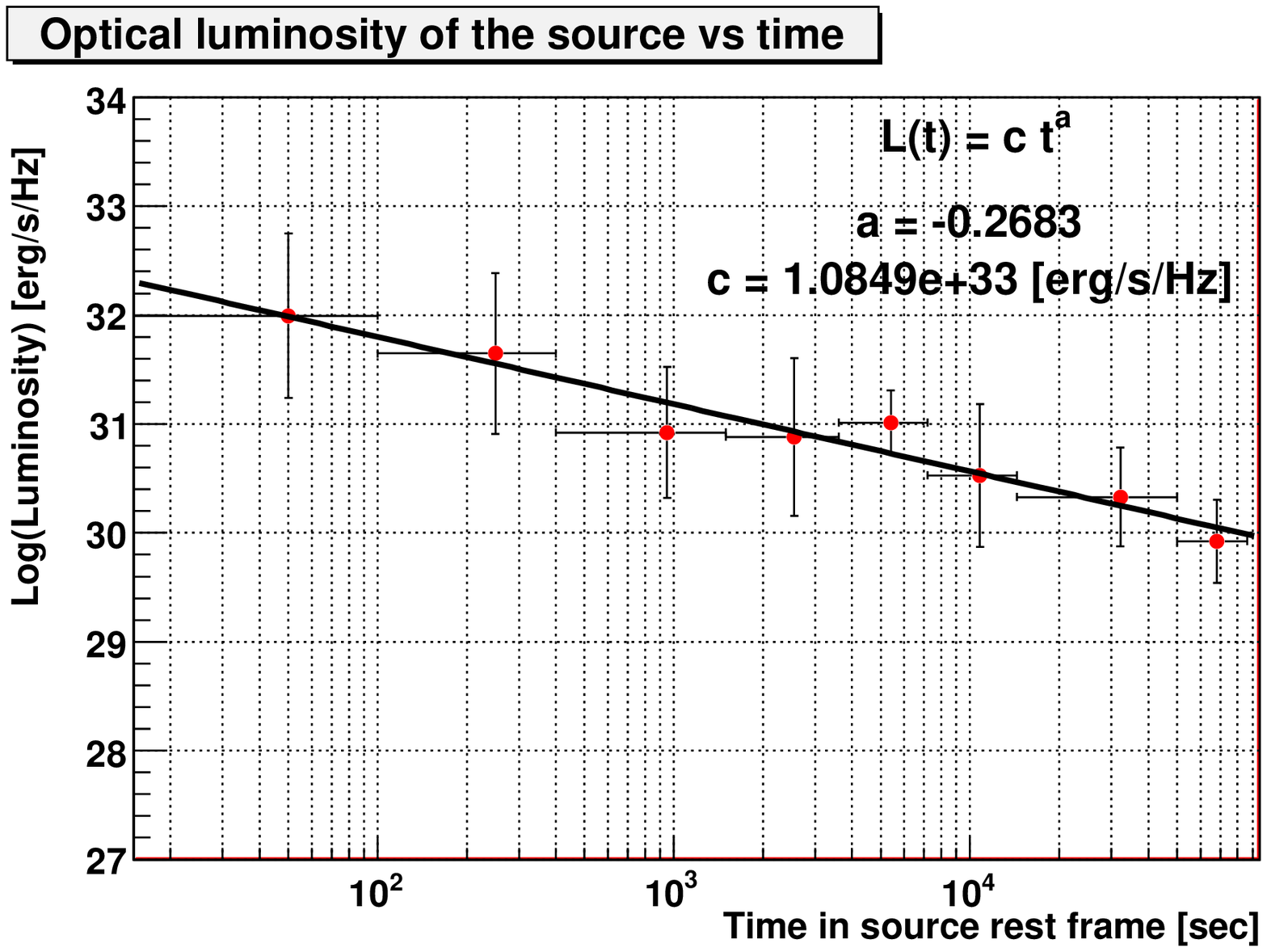}
    \else
      \includegraphics[width=6in]{Lumin/fit/log10L_vs_time.txt.eps}
    \fi
    \caption{Luminosity of the source in function of time since GRB.}
    \label{fig_lumin_vs_time_fit}
  \end{center}
\end{figure}

The measurements has been divided into time bins and Gaussian distribution was
fitted in each time range. These fits are shown in Figure \ref{fig_gauss_lumin_fits}.
The resulting mean values has been plotted against center value of the time.
This dependency with fitted power law function is shown in Figure \ref{fig_lumin_vs_time_fit}.
It is clearly well describing the optical luminosity of the GRB source in
time. The obtained dependency is a quite astonishing result, because it was
obtained from many different GRBs, which indicates that optical counterpart
can have similar properties in many bursts.

The similar analysis was performed in \cite{grb_clustering}. In that case 
a clustering of optical luminosities 12 hours after the burst was suggested,
which is rather not observed in early times analysed in this thesis.



\subsection{Predictions for the full system observations}
\label{sec_full_system_predict}

\begin{figure}[!htbp]
  \begin{center}
    \leavevmode
    \ifpdf
      \includegraphics[width=6in]{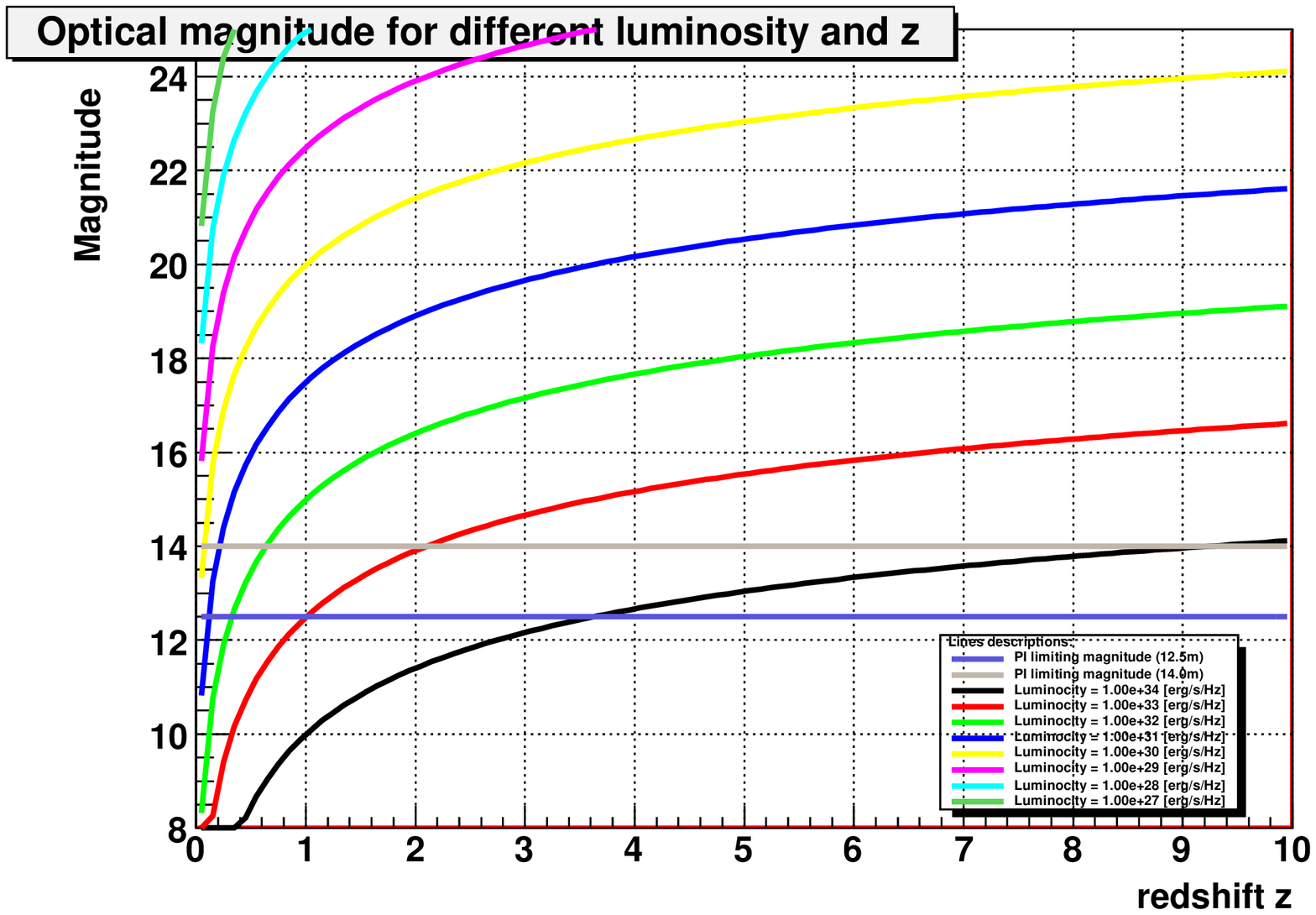}
    \else
      \includegraphics[width=6in]{Lumin/mag_vs_z.eps}
    \fi
    \caption{Optical brightness obtained for different source luminosities in function of redshift}
    \label{fig_m_vs_z}
  \end{center}
\end{figure}

\begin{figure}[!htbp]
  \begin{center}
    \leavevmode
    \ifpdf
      \includegraphics[width=6in]{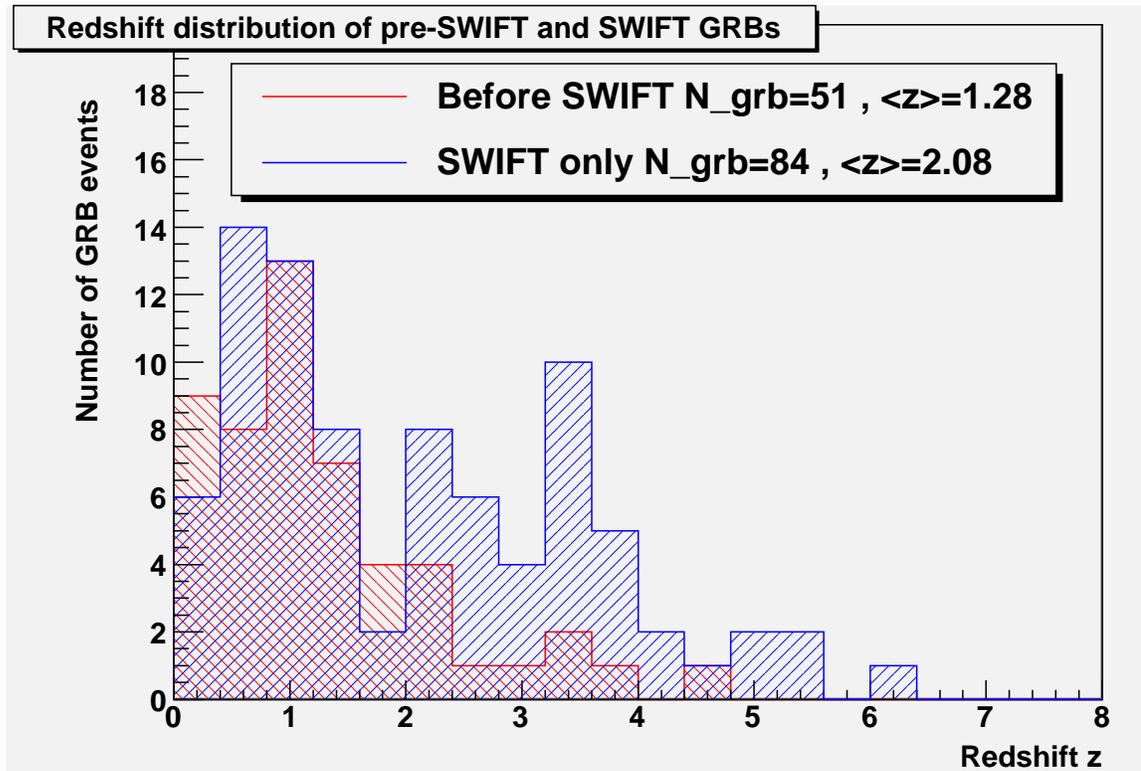}
    \else
      \includegraphics[width=6in]{grb_z/z_distr.eps}
    \fi
    \caption{Distributions of redshifts of pre-SWIFT and SWIFT GRBs}
    \label{fig_z_distr}
  \end{center}
\end{figure}

Assuming luminosities from the range $L_{ot}$=$10^{27} - 10^{34}$ erg/s/Hz
the object brightness was calculated as a function of redshift.
This calculation is shown in Figure \ref{fig_m_vs_z}. The obvious result is
that for redshifts higher then z=3 only extremely bright bursts can be visible 
with the "Pi of the Sky" telescope. However, SWIFT satellite detector works
in 15-350 keV which is rather soft, thus it is more sensitive to more distant GRBs ( with 
high redshift ). The distribution of z for the pre-SWIFT and SWIFT bursts is shown in Figure \ref{fig_z_distr}.
It can clearly be seen that contrary to previous satellites most of the SWIFT bursts are more distant then z=2.
The detector observing in harder bandpass ( like GLAST , 10 MeV - 100 GeV ) would have higher chance
to detect close bursts and thus brighter also in optical band ( Sec. \ref{sec_glast_satellite} ).

\begin{figure}[!htbp]
  \begin{center}
    \leavevmode
    \ifpdf
      \includegraphics[width=6in]{epeak_vs_z.gif}
    \else
      \includegraphics[width=6in]{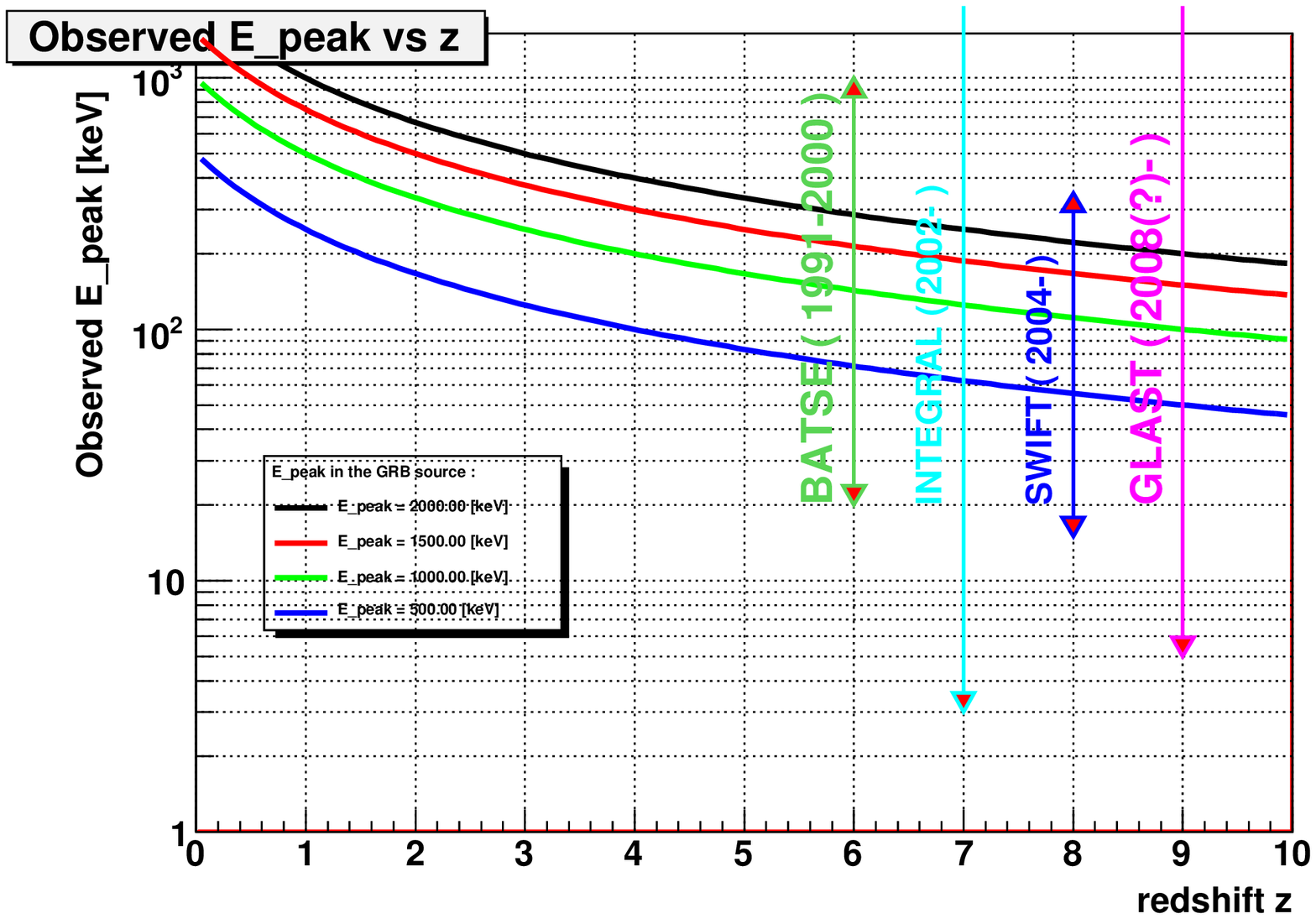}
    \fi
    \caption{Observed $\gamma$-ray energy $E_{peak}^{obs}$ in function of redshift at which GRB occurred. Calculated for $E_{peak}^{source}$ energies of 2.0MeV, 1.5MeV, 1MeV and 500keV. }
    \label{fig_epeak_vs_z}
  \end{center}
\end{figure}

The region above z=5 is only a theoretical prediction and due to neutral
hydrogen present in this region ( not taken into account here ) theoretical lines cannot be treated as realistic shape.
The conclusions is that the brightest optical counterparts can be observed
up to z$\approx$4 and weak bursts can be observed by the "Pi of the Sky"
system only if they occurred relatively close (z<0.2).
This predictions can be compared with 
the acceptance of the GRB detecting satellites presented in Figure \ref{fig_epeak_vs_z}. 
It can clearly be seen that SWIFT satellite has best acceptance for bursts
with z>2 , which strongly limits chances to observe optical counterpart by
the "Pi of the Sky" system. However, GLAST satellite which will be launched
in 2008 will have good acceptance much wider range of gamma energies.

%
%
%

\begin{figure}[!htbp]
  \begin{center}
    \leavevmode
    \ifpdf
		\includegraphics[width=2.7in,height=2.7in]{all_with_uvot.gif}
      \includegraphics[width=2.7in,height=2.7in]{ot_time.gif}
    \else
		\includegraphics[width=2.7in,height=2.7in]{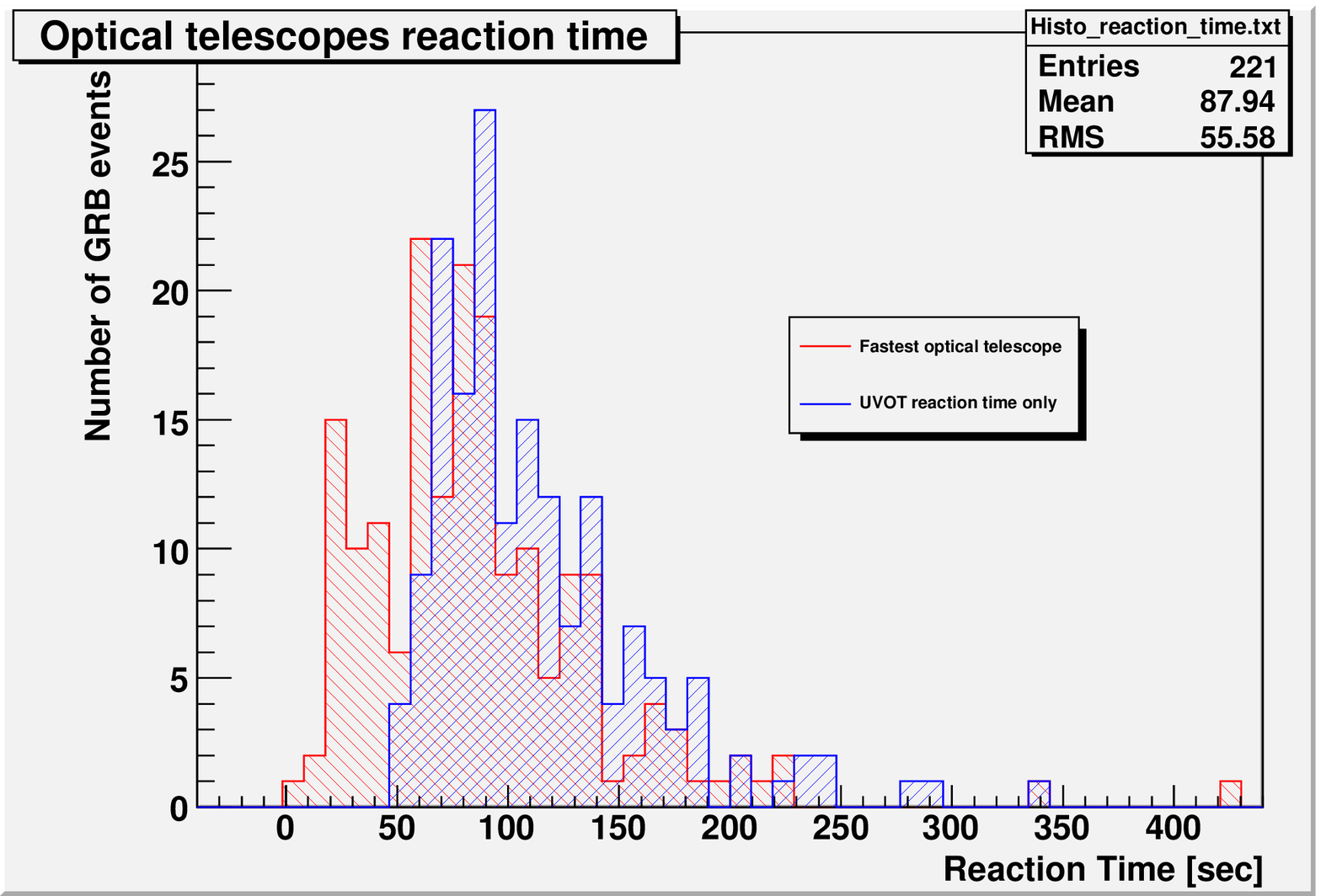}
      \includegraphics[width=2.7in,height=2.7in]{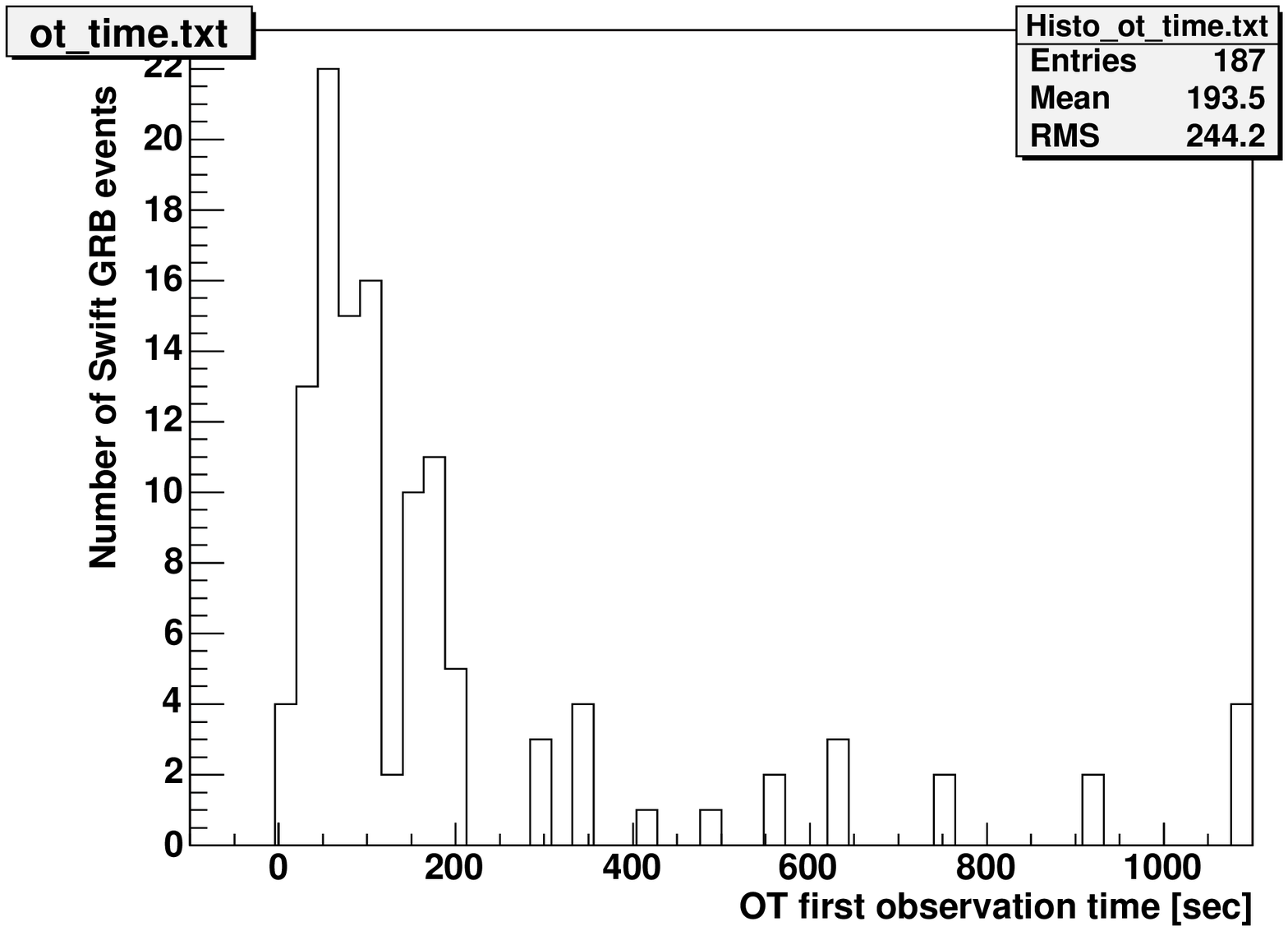}
    \fi
    \caption{Distribution of reaction times (<400 sec) of all optical telescopes
and SWIFT-UVOT shown separately (left plot) and minimal time of positive
optical detection (right plot). Only Swift GRBs were taken into account.}
    \label{fig_distr_mintime}
  \end{center}
\end{figure}

Thanks to the launch of a SWIFT satellite in the fall of 2004 it is possible
to observe optical counterparts of GRBs at very early time. 
Table \ref{tab_swift_grb} shows the number GRBs observed by SWIFT in
subsequent years. On average less then $\approx$50\% of all GRBs have
observed optical counterparts. Figure \ref{fig_distr_mintime} shows the 
distribution of the minimal time of the optical counterparts observations
and reaction time of optical telescopes ( including robotic and SWIFT-UVOT )
The reaction time peaks at $\approx $ 110 seconds, and it is clear that early times
( at T=0 ) are very poorly covered. It is also clear that SWIFT-UVOT
telescope covers only optical data starting from $\gtrsim$ 60 sec, due to
the time needed to slew the spacecraft. This early period can only be
covered by fast robotic telescopes or wide FOV system like "Pi of the Sky"
with negative reaction time.

%
\begin{table}[htbp]
\begin{center}
\begin{tabular}{|c|c|c| }
\hline
 & & \\
\textbf{Period} & \textbf{Number of GRB} & \begin{minipage}{4cm}{\textbf{Number of GRB with observed OT}}\end{minipage} \\ 
 & & \\
\hline
2004.12.17-2004.12.31 & 3 & 1 \\
\hline
2005.01.01-2005.12.31 & 87 & 43 \\
\hline
2006.01.01-2006.12.31 & 105 & 63 \\
\hline
2007.01.01-2007.10.13 & 68  &  38 \\ 
\hline
\end{tabular}
\hfill
\caption{Number of GRBs observed by SWIFT with number of those for which optical counterpart was observed}
\label{tab_swift_grb}
\end{center}
\end{table}


\begin{table}[htbp]
\begin{center}
\begin{tabular}{|c|c|c|c|c| }
\hline
\textbf{GRB} & \begin{minipage}{2.5cm}\textbf{Maximum brightness}\end{minipage} & \textbf{Filter} &  \begin{minipage}{3cm}\textbf{First observation time [s]}\end{minipage} & \textbf{Telescope} \\
\hline
\hline
GRB 071003A & 12.83 & none & 42 & MERGE,KAIT \\
\hline
GRB 061126A & 12.3 & none & 21 & RAPTOR \\
\hline
GRB 061007A & 10.15 & R & 142 & ROTSE,FTS \\
\hline
GRB 060117A & 11.5 &  R & 124 & FRAM \\
\hline
GRB 060111B & 13.0 & none & 27 & TAROT,ROTSE \\
\hline
GRB 051111A & 13.0 & none & 27 & ROTSE \\
\hline
\end{tabular}
\hfill
\caption{The brightest optical counterparts ( called "guaranteed" in the text ) of GRBs observed by SWIFT, which
could be detected by "Pi of the Sky" if it appears within the range of the
telescope. Observed in the time period 2005.01.01-2007.10.20.}
\label{tab_swift_brightest_ot}
\end{center}
\end{table}

The full "Pi of the Sky" system will be a perfect tool for covering this early
period of optical observations. According to the current SWIFT data it is
possible to make conservative estimate on the number of events for which 
the system would see the signal from GRB optical counterpart.
All GRBs detected by SWIFT were analysed and events with observations in the first 15
minutes after the burst were selected. Let us divide them into four classes
 according to "Pi of the Sky" observability : \\

\begin{itemize}
\item \textbf{Guaranteed} : events which have observations brighter then $13^m$ and 
would certainly be detected by the "Pi of the Sky" experiment ( Tab. \ref{tab_swift_brightest_ot} )
\item \textbf{Marginal} : events which have observations brighter then $15^m$ and 
early observation of such events would have great scientific significance
\item \textbf{Extrapolated} :  events which don't have observations brighter then
$15^m$, but light curve of early points extrapolated to time T=0 sec is 
brighter then $15^m$. 
\item \textbf{Beyond Limit} : events with brightest measurements far below limiting magnitude of the "Pi
of the Sky" experiment
\end{itemize}

%
%
\begin{table}[htbp]
\begin{center}
\begin{tabular}{|c|c|c| }
\hline
\textbf{\begin{minipage}{5cm}Observation prediction for "Pi of the Sky"\end{minipage}}  & \textbf{\begin{minipage}{4cm}Number of GRB (2005-2006)\end{minipage}} & \textbf{\%} \\ 
\hline
\hline
Guaranteed & 6 & 2.3 \\
\hline
Marginal & 20 & 7.7 \\
\hline
Extrapolated & 20 & 7.7 \\
\hline
Beyond Limit & 183 & 70.1 \\ 
\hline
No early observations & 32 & 12.3 \\
\hline
\textbf{Total GRB\#} & \textbf{261} & 100.0 \\
\hline
\end{tabular}
\hfill
\caption{Number of GRBs in different prediction groups}
\label{tab_predictions}
\end{center}
\end{table}


Table \ref{tab_predictions} shows the classification results. It can be seen
that 2.3-18\% of all GRBs could be detected if they appear in the FOV of the
"Pi of the Sky" system.
The estimate of the number of events to be observed by the full system is derived from the following formula : \\

\begin{equation}	
\begin{split}
	N_{\Pi-OT / year} = \frac{N^{SWIFT}_{total}}{T^{SWIFT}_{total}} \times f_{Swift-FOV} \times f_{coll} , \\
	f_{Swift-FOV} := f_{night} \times f_{sat fov} , \\
	f_{tot} := f_{night} \times f_{sat fov} \times f_{coll}
\end{split}
\label{eq_ot_per_year}
\end{equation}

where $N^{SWIFT}_{total}$ is the number of events of given type observed by
SWIFT in period $T^{SWIFT}_{total}$, $f_{night}$ is the fraction of time which can be used for
observations at given site, $f_{sat fov}$ is the fraction of night when SWIFT
satellite is above the horizon, $f_{coll}$ is the fraction of nights when
weather was good enough to collect data and the system was functional.
The value $f_{Swift-FOV}$ was calculated for period of one year under assumption
that system collects data only when $h_{SUN} < 10^{\circ}$ and using SWIFT
pointing information from period 2005-2006.
The value of FOV acceptance $f_{sat fov}$ was calculated in conservative way assuming that
it is possible to observe SWIFT's FOV when it is at h>14$^\circ$ above
horizon. However, due to large FOV of SWIFT it is possible to observe part of
its FOV also when it is h<14$^\circ$. The error coming from this should not
exceed few percent. 
The value of $f_{coll} \approx 81\%$ coefficient was estimated for 2006.06.01-2007.10.25
collecting period. If we assume that most of the technical problems will be
solved in the final system and data will not be collected only due to the
bad weather condition this coefficient can be estimated by $f_{weather} \approx$ 91\% ( in Las Campanas Observatory in Chile ).
The values of coefficients for different sites are shown in Table \ref{tab_coeff}.
La Palma site was shown as an example of possible Northern hemisphere
location of the "Pi of the Sky" system.

\begin{table}[htbp]
\begin{center}
\begin{tabular}{|c|c|c|c|c|c| }
\hline
\textbf{Site} & \textbf{Value of} \boldmath{$f_{Swift-FOV}$} & \boldmath{$f_{coll}$} & \boldmath{$f_{weather}$}  & \boldmath{$f_{tot}$} & \boldmath{$f_{tot-max}$} \\
\hline
\hline
LCO & 15\% & 81\% & 91\% & 12\% & 14\% \\
\hline
La Palma & 23\% & 65\% & 75\% \cite{la_palma_weather} & 15\% & 17\% \\
\hline
\end{tabular}
\hfill
\caption{Values of coefficients used to calculate $\pi$ acceptance for SWIFT
GRBs. The value of $f_{tot-max}$ was calculated under the assumption that the
system will not loss any night due to technical problems and $f_{coll} = f_{weather}$ }
\label{tab_coeff}
\end{center}
\end{table}

\begin{table}[htbp]
\begin{center}
\begin{tabular}{|c|c|c|r@{.}l| }
\hline
\textbf{OT Class} & \textbf{Site} & \textbf{\begin{minipage}{3cm}\# of OT in 2005-2007\end{minipage}} & \multicolumn{2}{c|}{\textbf{\begin{minipage}{3cm}Expected \# of observations / year\end{minipage}}} \\
\hline
\hline
Guaranteed & LCO & 6 & 0&28 \\
\hline
Marginal & LCO & 20 & 0&9 \\
\hline
Extrapolated & LCO & 20 & 0&9 \\ 
\hline 
\textbf{Total LCO} & \textbf{LCO} & \textbf{46} & \textbf{2}&\textbf{08} \\
\hline
\hline
Guaranteed & La Palma & 6 & 0&34 \\
\hline
Marginal & La Palma & 20 & 1&1 \\
\hline
Extrapolated & La Palma & 20 & 1&1 \\
\hline
\textbf{Total La Palma} & \textbf{La Palma} & \textbf{46} & \textbf{2}&\textbf{54} \\ 
\hline
\end{tabular}
\hfill
\caption{Expected number of events observed by full $\pi$ system. Values
were obtained under assumption of the failure free system only limited by the
weather conditions.}
\label{tab_ot_num_estimate}
\end{center}
\end{table}

The final estimates for numbers of events of given class are listed in Table \ref{tab_ot_num_estimate}. 
It can be seen that number of sure positive observations is very small.
However, it must be stressed that period at T=0 moment is
almost unknown. The estimation is conservative, GLAST satellite will
 increase the value of $f_{sat fov}$ coefficient. 

There is also an advantage of observing short GRBs during the $\gamma$-ray outburst. The 
numbers shown in Table \ref{tab_ot_num_estimate} are safe estimates
according to what is already observed for long GRBs. They show big chances of observing 
positive signal of great scientific significance in the first 2 years of the
experiment and thus can be considered as optimistic. It is important that in
fact only a few optical counterparts of Swift GRB were observed in T=0 by optical telescope \cite{raptor_041219a},
thus it is not well known what can be expected. The predictions in Table
\ref{tab_ot_num_estimate} are determined according to observations performed 
several dozens of seconds after the burst.


\def\baselinestretch{1}
\chapter{Summary}
\ifpdf
    \graphicspath{{Conclusions/ConclusionsFigs/PNG/}{Conclusions/ConclusionsFigs/PDF/}{Conclusions/ConclusionsFigs/}}
\else
    \graphicspath{{Conclusions/ConclusionsFigs/EPS/}{Conclusions/ConclusionsFigs/}}
\fi

\def\baselinestretch{1.66}

The prototype system is working since July 2004. In this period it
observed FOV $33^o \times 33^o$ during 322 nights. Since June 2006 FOV $20^o \times 20^o$ 
was observed during 276 nights. 
During the entire period satellites detected 231 Gamma Ray
Bursts, 3 events occurred in the field of view of the "Pi of the Sky"
prototype system. In 35 cases the burst occurred outside the FOV and the
position was observed a few minutes after the $\gamma$ detection.
Optical counterpart was not observed in any of the cases. The upper limits
 on the optical brightness have been determined for all 38 events. 

The algorithms for identification of optical flashes and
brightness variations has been designed and tested on the collected data. 
The automatic on-line algorithms have discovered 135 optical flashes visible
in single 10s image on two cameras working in coincidence and 6 flashes visible in at least 2
consecutive images, but only on single camera ( second was not working ).
One of these flashes has been unambiguously identified as the outburst of the
known flare star CN Leo. This confirmed efficiency of the experiment strategy.
The off-line algorithm for identification of brightness increases automatically 
 discovered outburst of known star system GJ3331A/GJ3332. The off-line "nova-algorithm"
finding new objects appearing in the sky during normal observations was well
tested on the asteroids which imitate events of nova-like signature.
The short optical flash visible on at least 2 consecutive images found in the
second period of data collection was used to derive upper limit on the ratio of
optical to $\gamma$-ray collimation of the prompt emission, which is $R \leq 4.4$.
The full system will allow to improve this limit by at least a factor of 16 
( under assumption the system will consist of 16 cameras ). 
Number of predicted positive observations of the optical counterparts of
GRBs is $\approx 2.5$ events/year. This gives very optimistic perspectives
of obtaining meaningful scientific results.




\appendix


\ifpdf
	\graphicspath{{Appendix1/Appendix1Figs/PNG/}{Appendix1/Appendix1Figs/PDF/}{Appendix1/Appendix1Figs/}}
\else
	\graphicspath{{Appendix1/Appendix1Figs/EPS/}{Appendix1/Appendix1Figs/}}
\fi

\chapter{Technical Details on system controll}

\section{Libraries and programs for data aquisition and analysis}
\label{sec_appendix1}

Table below lists libraries which were in big part developed by author and
are components of Data AQuisition and Data Analysis programs. Paths to
source location are relative and should be proceeded with /opt/pi/dev/pisys/daq/src/.

\begin{table}[htbp]
\begin{center}
\begin{tiny}
\begin{tabular}{|c|c|c|c|}
\hline
\textbf{NAME} & \textbf{LIBRARY FILE} & \textbf{SOURCE LOCATON} & \textbf{DESCRIPTION} \\
\hline
baselib & libbaselib.a & cmn/baselib/ & low level common classes for file handling, date/time etc \\
\hline
mathlib & libmathlib.a & cmn/mathlib/ & mathematical functions \\
\hline
log4pi  & liblog4pi.a  & cmn/log4pi/ & interface for logging to sysylog \\
\hline
pidblib & libpidblib.a & ccd/pidblib/ & postgres database interface \\
\hline
asaslib & libasaslib.a & ccd/fitslib/asaslib/ & procedures adopted from ASAS experiment \\
\hline
myfitslib & libmyfitslib.a & ccd/fitslib/myfitslib/ & classes for \texttt{fits} files I/O \\
\hline
cfglib & libcfglib.a & ccd/cfg/ & definitions of default values of paramters \\
\hline
picamdrv & libpicamdrv.a & ccd/ccddriver/picamdrv/ & camera driver library \\
\hline
ccdscript & libccdscript.a & ccd/ccdscripts/ & script generator and pointing classes \\
\hline
ccdsat & libccdsat.a & ccd/ccdsat/  & 
\begin{minipage}{6cm}
interfece to external library for calculating satellites positions \\
\end{minipage} \\
\hline
ccdlib & libccdlib.a & ccd/ccdlib/ & 
\begin{minipage}{6cm}
main data aquistion library, implementing classes for image collection and on-line analysis 
\end{minipage} \\
\hline
ccdinterface & libccdinterface.a & ccd/ccdinterface/ &
\begin{minipage}{6cm}
interface classes to communicate with other parts of the system and external world 
\end{minipage} \\
\hline
ccdastro & libccdastro.a & ccd/ccdastro/ & library for astronomical fomulae \\
\hline
\end{tabular}
\end{tiny}
\caption{Most important libraries developed for data collection and analysis}
\label{tab_daqlibs}
\end{center}
\end{table}

Table below lists most important programs for collecting data and photometry
and astrometry.

\begin{table}[htbp]
\begin{center}
\begin{tiny}
\begin{tabular}{|c|c|c|}
\hline
\textbf{PROGRAM NAME} & \textbf{SOURCE LOCATION} & \textbf{DESCRIPTION} \\
\hline
ccdsingle & ccd/TESTY/ccdsingle/ & program for collection and analysis of images from single camera \\
\hline
ccddouble & ccd/TESTY/ccddouble/ & program for collection and analysis of images from two cameras \\
\hline
test2K2K & ccd/TESTY/test2K2K/ & 
\begin{minipage}{8cm}
program for operating cameras in interactive mode. The usage of the program
is the following :
  test2K2K CAMERA OPTIONS
In order to specify camera, its name k2a, k2b, etc can used or if unknown
the number 0,1,2, etc.
In order to specify IP adress of the camera option -eth=100.100.100.1 must
be provided. For more details see web page : http://hep.fuw.edu.pl/~msok/test2K2K.html
\end{minipage} \\
\hline
piastrometry & ccd/TESTY/piastrometry/ & program for running astrometry \\
\hline
piphoto & ccd/TESTY/piphoto/ & program for fast photometry algorithm \\
\hline
\end{tabular}
\end{tiny}
\caption{Most important programs for data aquisition and analysis. All
programs show short help when launched with option -h}
\label{tab_daqprograms}
\end{center}
\end{table}

\newpage
\section{Description of FITS header keywords}
\label{sec_pi_fits_files}

The table below lists keywords saved to fits header files.

\begin{table}[htbp]
\begin{center}
\begin{tiny}
\begin{tabular}{|c|c| }
\hline
\normalsize\textbf{KEYWORD} & \normalsize\textbf{DESCRIPTION}  \\
\hline
\multicolumn{2}{|c|}{\normalsize\textbf{Standard FITS format keywords}} \\
\hline
SIMPLE & Mandatory, T if conforms to standard, F otherwise \\
BITPIX & Mandatory, number of bits representing data value \\
NAXIS & Mandatory, number of axis in dat array \\
NAXIS1 & Mandatory, number of elements along X axis \\
NAXIS2 & Mandatory, number of elements along Y axis \\
EXTEND & T if FITS may contain extensions \\
BZERO & 
\begin{minipage}{8cm}
Used in case array value are not physcial, but : physical\_value = BZERO + BSCALE * array\_value \\
\end{minipage} \\
BSCALE & 
\begin{minipage}{8cm}
Used in case array value are not physcial, but : physical\_value = BZERO + BSCALE * array\_value \\
\end{minipage} \\
\hline
\multicolumn{2}{|c|}{\normalsize\textbf{Observed object description}} \\
\hline
OBJECT & 
\begin{minipage}{8cm}
Name of object or field observed, project conention is syntax :
H0000+00, where letter is first letter of object name and digits stand for coordinates : HHMM$\pm$DD
\end{minipage} \\
ROTATE & if image is rotated ( 0 / 1 )\\
DRVFLIP & if imaged was fliped by driver on PC side ( 0 / 1 )\\
RA & Right Ascension of image center in hours decimal \\
DEC & Declination of image center in degrees decimal \\
HA & Hour angle of image center in hours decimal \\
AZIM & Azimuth of image center in degrees decimal \\
ALT & Altitude of image center in degrees decimal \\
ZENITH\_D & Zenithal distance of image center in degrees decima \\
OBSMODE & Observation mode - 0 - mount not tracking, 1 - mount tracking \\
AVERAGE & Mean pixel value on image \\
RMS & RMS of pixel value on image \\
\hline
\multicolumn{2}{|c|}{\normalsize\textbf{Observing Site}} \\
\hline
ORIGIN & Name of project \\
SITE & Code of site, LCO - Las Campanas, BRW - Brwinow \\
TELLONG & Geographical longitude [deg] \\
TELLAT & Geographical latitude [deg] \\
TELALT & Altiutude above the see level [m] \\
\hline
\multicolumn{2}{|c|}{\normalsize\textbf{Instrument}} \\
\hline
INSTRUME & Name of instrument \\
CAMERA & Name of camera \\
CAMOPTIC & Camera optic description \\
FILTER & Used filters \\
PIXSCALE & Angular size of pixel [ arcsec ] \\
PIXSIZE & Physcial size of pixel [ $\mu$m ]\\
\hline
\end{tabular}
\end{tiny}
\end{center}
\end{table}

\begin{table}[htbp]
\begin{center}
\begin{tiny}
\begin{tabular}{|c|c| }
\hline
\normalsize\textbf{KEYWORD} & \normalsize\textbf{DESCRIPTION}  \\
\hline
\multicolumn{2}{|c|}{\normalsize\textbf{Exposure id}} \\
\hline
OBSERVER & Observer name \\
DIMAGE & Image number ( during single night )\\
NIMAGE & Obsolate \\
SOFTWARE & Software version \\
BUILD & Software build date and time \\
DRVTYPE & Camera driver type \\
CAMID & Camera ID ( 2-k2a, 3-k2b ) \\
CAMIIDX & Camera internal index \\
FILENAME & FITS file name \\
\hline
\multicolumn{2}{|c|}{\normalsize\textbf{Exposure settings}} \\
\hline
EXPTIME & Require exposure time [sec] \\
REXPTIME & Measured exposure time [sec] ( not very precise ) \\
SHUTTER & values : OPEN / DARK \\
SHUTMODE & Shutter mode OPENED - permanently opened, NORMAL - open/close mode \\
ADCGAIN & Analog to Digital Converter (ADC) gain value \\
ADCBIAS & ADC offset value [mV] \\
ADCGSET & ADC gain setting value \\
LNAGAIN & Low Noise pre-Amplifier (LNA) gain value x8 or x20 ( current value is x8 ) \\
ADCBSET & ADC offset setting value \\
ADCRANGE & 2V or 4V \\
ADCCLAMP & ADC Clamping 2V or 4V \\
ELECGAIN & Gain of camera [e/ADU] \\
COOLING & Cooling enabled / disabled ( YES / NO ) \\
ABINN & Analog binning ( disabled ) \\
SBINN & Software  binning ( disabled ) \\
SPEED & OBSOLATE \\
SPEEDMH & Speed description Vertical / Horizontal - to be corrected (horizontal is missing ) !!! \\
MPP\_BC & Multi-pinned phase or BC mode \\
RO-TIME & Measured data transfer time [sec] ( from camera RAM to PC ) \\
CROTIME & Measured chip readout time [sec] \\
FOCUS & value of current step motors position [steps] \\
HITLENS & Lens hitting ON/OFF \\
SAVEAREA & Area of chip saved in this FITS file [x\_start y\_start x\_end y\_end]\\
USBMODE & USB mode ( 1.1 or 2.0 )\\
FPGAVER & FPGA firmware version date \\
CPRSVER & Cypress firmware version date \\
VERDESC & Camera Firmawre version description \\
RNOISE & readout noise [ADU] - to be verified !!! \\
RELNOISE & readout noise [e] - to be verified !!! \\
\hline
\multicolumn{2}{|c|}{\normalsize\textbf{Exposure environment}} \\
\hline
CHIPTSET & Required chip temperature [Celsius] \\
CHIPTEMP & Measured chip temperature [Celsius] \\
CHIP\_TEM & Measured chip temperature [Celsius] \\
CASTEMP & Measured case temperature [Celsius] \\
AMBTEMP & Measured ambient temperature [Celsius] \\
CAMHUMID & Measured camera humidity \\
AMBHUMID & Measured ambient humidity \\
INTRTEMP & Not implemented \\
AIRMASS & Airmass - not implemented \\
DOME & Dome status : OPENED / CLOSED\\
\hline
\end{tabular}
\end{tiny}
\end{center}
\end{table}

\begin{table}[htbp]
\begin{center}
\begin{tiny}
\begin{tabular}{|c|c| }
\hline
\normalsize\textbf{KEYWORD} & \normalsize\textbf{DESCRIPTION}  \\
\hline
\multicolumn{2}{|c|}{\normalsize\textbf{Exposure date}} \\
\hline
UT-START & UT time of image start \\
DATE-OBS & UT date of image start \\
DATE\_OBS & Local date of image start \\
TIME\_UT & unixtime of image start [sec] \\
UT-END & UT time of image end \\
TIME\_OBS & UT time of image end \\
DATE-END & UT date of image end \\
LOCTIME & local time of image start \\
LOCDATE & local date of image start \\
EPOCH & current epoch \\
EQUINOX & current equinox \\
ST & siderial time [radians]\\ 
JD & Julian date \\
HJD & Heliocentric Julian date \\
\hline
\multicolumn{2}{|c|}{\normalsize\textbf{Astrometry}} \\
\hline
ASTROOK & Astro OK or FAILED ( 1 / 0 ) \\
POSANGLE & Rotation angle [deg] \\
AST\_ORD & Order of transformation equation ( 4 ) \\
ASTUTIME & unixtime of astrometry ( if not performed on every image ) \\ 
AST\_ERR & Error of astrometry [pixels]\\ 
PAR\_X\_0, PAR\_X\_0, ... , PAR\_X\_13 & Coefficients of astrometric transformation \\
PAR\_Y\_0, PAR\_Y\_1, ..., PAR\_Y\_13 & Coefficients of astrometric transformation \\
\hline
\multicolumn{2}{|c|}{\normalsize\textbf{Photometry}} \\
\hline
NSTARS & Number of stars detected on image \\
\hline
\multicolumn{2}{|c|}{\normalsize\textbf{Mount}} \\
\hline
MOUNTRA & Right Ascension as obtained from mount [hours decimal]\\
MOUNTDEC & Declination as obtained from mount [degrees decimal]\\
MOUNTAZ & Azimuth as obtained from mount [degrees decimal]\\
MOUNTTRK & Tracking mode as obtained from mount [0-non tracking, 1-tracking]\\
MOUNTTM & Timestamp of mount information\\
MOUNTHA & Hour angle as obtained from mount [hours decimal]\\
MOUNTALT & Altitude as obtained from mount [degrees decimal]\\
MOUNTDTM & Timestamp of mount information [unixtime - sec]\\
MOUNTMV & If image taken just after mount move ( 0 / 1 )\\
\hline
\end{tabular}
\end{tiny}
\caption{Keywords in FITS files}
\label{tab_cam_fits_header}
\end{center}
\end{table}

\newpage
\section{System log files}
\label{app_log_files}

The most important log files are listed in table below :

\begin{table}[htbp]
\begin{center}
\begin{tiny}
\begin{tabular}{|c|c|}
\hline
\textbf{LOG FILE NAME} & \textbf{DESCRIPTION} \\
\hline
pi.log & log file from all modules written by syslog daemon \\
\hline
run\_daq\_YYYYMMDD.out & standard output from DAQ program ( ccdsingle or ccddouble ) \\
\hline
mount.logfile\_YYYYMMDD\_HHMMSS & mount controll program log file \\
\hline
piman.logfile\_YYYYMMDD\_HHMMSS & piman server log file \\
\hline
gcn\_server.log & log file of listener program receiving and redirecting GCN messages to gcn program \\
\hline
gcn.logfile & log file of gcn program sending alerts to piman server \\
\hline
gcn\_imalive.logfile\_YYYYMMDD & log file containing I'am alive packets from external GCN server \\
\hline
integral\_pointdir.log & INTEGRAL satellite pointing information obtained from GCN server \\
\hline
swift\_pointdir.log & SWIFT satellite pointing information obtained from GCN server \\
\hline
\end{tabular}
\end{tiny}
\caption{Most important log files from $\pi$-system, files are stored in
directory /opt/pi/dev/pisys/log/, unless different path is specified,
YYYYMMDD is date ( for example 20070101 ) and HHMMSS is time ( for example 200112 ) }
\label{tab_log_files}
\end{center}
\end{table}

\newpage
\section{DAQ controll from piman and pishell}
\label{sec_daq_controll_commands}

\begin{table}[htbp]
\begin{center}
\begin{tiny}
\begin{tabular}{|c|c|c|c|}
\hline
\textbf{PISHELL COMMAND} & \textbf{FUNCTION NAME} & \textbf{INPUT} & \textbf{DESCRIPTION} \\
\hline
start\_daq & StartDAQ & - & 
\begin{minipage}{4cm}
starts dark collection and sets daq status to started  
\end{minipage} \\
\hline
- & IsDaqStarted & - & 
\begin{minipage}{4cm}
returns 1 if DAQ is already started, 0 otherwise
\end{minipage} \\
\hline
start\_analysis & StartAnalysis & mount position (RA,DEC) & 
\begin{minipage}{4cm}
sets new position and starts image collection and analysis
\end{minipage} \\
\hline
stop\_analysis & StopAnalysis & - & 
\begin{minipage}{4cm}
stops image collection ( but waits for last image to be finished )
\end{minipage} \\ 
\hline
stop\_analysis\_nowait & StopAnalysisNoWait & - & 
\begin{minipage}{4cm}
stops image collection without waiting for last image
\end{minipage} \\
\hline
- & SendAlert & GCN alert time and position & 
\begin{minipage}{4cm}
informs DAQ about GCN trigger which is being observed
\end{minipage} \\
\hline
- & SetOnTriggerPosition & Alert position (RA,DEC) & 
\begin{minipage}{4cm}
starts image collection after trigger is received
\end{minipage} \\
\hline
fast\_post\_to\_daq & CorrectMountPosition &  mount position (RA,DEC) & corrects mount position \\
\hline
dodarks & DoDarks & N & collects N dark images \\
\hline
take\_npictures & TakeNPictures & N & collects N images \\
\hline
take\_npictures\_synchro & TakeNPicturesSynchro & N & collects N images in synchronize mode \\
\hline
set\_cooling & SetCoolingOnOff & temperature and camera & sets temperature for given camera \\
\hline
change\_param & ChangeParam & param name, value and camera & changes parameter of daq \\
\hline
stat & GetStatus & - & returns status of daq program \\
\hline
set\_fits\_key & SetCustomKey & \texttt{fits} header key name and value & sets value of specified \texttt{fits} header key \\
\hline
\end{tabular}
\end{tiny}
\caption{Functions exported by DAQ}
\label{tab_daq_exp_functions}
\end{center}
\end{table}

\newpage
\section{System controll commands}
\label{sec_system_commands}

Example of night script for automatic system controll : \\

	\begin{figure}[!htbp]
   \begin{center}
	\begin{mylisting}
	\begin{verbatim}
# auto-generated script
# night : 20060605
# camera : Cannon EOS f=85mm
# system start time is : 20060605_180643 local ( 20060605_220643 UT )
# PRIMARY SATELLITE = INTEGRAL
# SECONDARY SATELLITE = HETE
# SUN sets at 1840 LCO time, at (AZ,H)=(110.44,-9.90) [deg]
# SUN rises at 0645 LCO time
# SWIFT at 20060605_180643 local time is at (RA,DEC)=(144.94,15.00)
# HETE info file date : 20060605_072000
# HETE RA=249.05=16h36m12.00s DEC=-63.06
# HETE rises above horizont at 327, sets at 116
#  hete at h_hete >= 30.00 at 1901
# MOON RA=190.54=12h42m10.68s DEC=-5.28 illum = 72.06 %
# MOON will set at 20060606_025143, illum = 73.46 %
# INTEGRAL RA=125.47=08h21m52.32s DEC=-47.51
# INTEGRAL rises above horizont at 737, sets at 36
piman    0 cron_point_hete_off
piman    1 exec_script_synchro(startup.pish)
# 
daq      1825 start_daq
piman    1825 auto_ag_mode_on
piman    1830 manual_mode_off
# Following INTEGRAL at (RA,DEC)=(125.47,-47.51) (az,h)=(46.77,55.89)
# Closest field (RA,DEC)=(128.50,-50.00), (az,h)=(41.23,56.76) => OBJECT=I0834-50
# At 1920 field (RA,DEC)=(128.50,-50.00) of object INTEGRAL is at (az,h)=(45.10,50.74)
# turning OFF cron
piman 1840 manual_mode_on
# internal 1149547243 goto_ra_dec(128.50,-50.00)
piman    1840 cron_point_hete_off
internal 1840 goto_ra_dec_auto_corr(128.50,-50.00,I0834-50)
mount 1840 raw_cmd(autoguide on)
internal 1840 send_pos_to_mount
mount 1840 stat
piman    1845 cron_send_pos_to_mount_on
internal 1850 send_pos_to_mount
mount 1850 stat
# turning ON cron
piman 1850 manual_mode_off
# End of tracking (ra,dec)=(128.50,-50.00) at 2155 , (az,h)=(45.26,26.05) deg
\end{verbatim}
\end{mylisting}
\end{center}
\end{figure}

\begin{figure}[!htbp]
\begin{center}
\leavevmode
\begin{mylisting}
\begin{verbatim}

piman    1900 manual_mode_on
piman    1900 cron_point_hete_off
piman    1900 cron_send_pos_to_mount_off
piman    1900 exec_script_synchro(scan_evening.pish)
piman    1930 manual_mode_off
# Following INTEGRAL at (RA,DEC)=(125.47,-47.51) (az,h)=(49.99,47.46)
# Closest field (RA,DEC)=(128.50,-50.00), (az,h)=(45.79,48.99) => OBJECT=I0834-50
# At 2010 field (RA,DEC)=(128.50,-50.00) of object INTEGRAL is at (az,h)=(47.14,42.62)
# turning OFF cron
piman 1930 manual_mode_on
# internal 1149550254 goto_ra_dec(128.50,-50.00)
piman    1930 cron_point_hete_off
internal 1930 goto_ra_dec_auto_corr(128.50,-50.00,I0834-50)
mount 1930 raw_cmd(autoguide on)
internal 1930 send_pos_to_mount
mount 1930 stat
piman    1935 cron_send_pos_to_mount_on
internal 1940 send_pos_to_mount
mount 1940 stat
# turning ON cron
piman 1940 manual_mode_off

# End of tracking (ra,dec)=(128.50,-50.00) at 2145 , (az,h)=(45.64,27.43) deg
# Following HETE at (RA,DEC)=(249.05,-63.06) (az,h)=(334.65,47.49)
# Closest field (RA,DEC)=(255.00,-60.00), (az,h)=(328.70,46.52) => OBJECT=H1700-60
internal 2140 send_pos_to_mount
# At 2225 field (RA,DEC)=(255.00,-60.00) of object HETE is at (az,h)=(332.55,50.84)
# turning OFF cron
piman 2145 manual_mode_on
# internal 1149558354 goto_ra_dec(255.00,-60.00)
piman    2145 cron_point_hete_off
internal 2145 goto_ra_dec_auto_corr(255.00,-60.00,H1700-60)
mount 2145 raw_cmd(autoguide on)
internal 2145 send_pos_to_mount
mount 2145 stat
piman    2150 cron_send_pos_to_mount_on
internal 2155 send_pos_to_mount
mount 2155 stat
	\end{verbatim}
	\end{mylisting}
  \end{center}
\end{figure}

\begin{figure}[!htbp]
\begin{center}
\leavevmode
\begin{mylisting}
\begin{verbatim}

# turning ON cron
piman 2155 manual_mode_off
# End of tracking (ra,dec)=(255.00,-60.00) at 0630 , (az,h)=(33.97,26.78) deg
# morning
piman    0545 manual_mode_on
piman    0545 cron_point_hete_off
piman    0545 cron_send_pos_to_mount_off
piman    0550 exec_script_synchro(scan_morning.pish)
piman    0615 manual_mode_off
# HETE (RA,DEC)=(249.05,-63.06) at 0615 is at (AZ,H)=(30.32,26.21)
# Waiting for HETE at (RA,DEC)=(252.96,-63.41) (az,h)=(30.32,28.00)
# Closest field (RA,DEC)=(255.00,-60.00), (az,h)=(34.32,28.39) => OBJECT=H1700-60
# At 0655 field (RA,DEC)=(255.00,-60.00) of object HETE is at (az,h)=(33.00,23.52)
# turning OFF cron
piman 0615 manual_mode_on
# internal 1149588948 goto_ra_dec(255.00,-60.00)
piman    0615 cron_point_hete_off
internal 0615 goto_ra_dec_auto_corr(255.00,-60.00,H1700-60)
mount 0615 raw_cmd(autoguide on)
internal 0615 send_pos_to_mount
mount 0615 stat
piman    0620 cron_send_pos_to_mount_on
internal 0625 send_pos_to_mount
mount 0625 stat
# turning ON cron
piman 0625 manual_mode_off
# End of tracking (ra,dec)=(255.00,-60.00) at 0630 , (az,h)=(33.91,26.55) deg
# Do not worry about order, these two commads always go just before shutdown :
piman    0640 cron_point_hete_off
piman    0640 cron_send_pos_to_mount_off
piman    0650 exec_script(shutdown.pish)
	\end{verbatim}
	\end{mylisting}
	\label{lis_pish_old}
   \caption{Example of night pish script}
  \end{center}
\end{figure}

\section{Observation targets}
\label{sec_obs_targets}

The table below lists observation targets in decreasing order of priority :

\begin{table}[htbp]
\begin{center}
\begin{tiny}
\begin{tabular}{|c|c|c|}
\hline
\normalsize{\textbf{OBJECT NAME}} & \normalsize{\textbf{POSITION}} & \normalsize{\textbf{INFO}} \\
\hline
SWIFT Satellite & - & FOV$\approx$2 staradians  \\
INTEGRAL Satellite & - & FOV$\approx$ 15$^\circ$ \\
PKS 2155-304    &  215852-301332.0  & blazar \\
AO 0235+164     &  023838+163659.0  & blazar \\
4C 29.45        &  115931+291444.0  & quasar \\
OI 158          &  073807+174219.0  & quasar \\
OJ 287          &  085448+200629.0  & quasar \\
3C 273          &  122906+020307.0  & quasar \\
OR -017         &  151250-090558.0  & quasar \\
W Com           &  122131+281358.0  & blazar \\
J0210-5055      &  021046-510100.0  & blazar \\
OJ 049          &  083148+042939.0  & blazar \\
GeV J1832-2128  &  183300-213600.0  & blazar \\
OP 151          &  133335+164904.0  & blazar \\
PKS 0537-441    &  053850-440508.0  & blazar \\
Mrk 501         &  165352+394535.0  & blazar \\
Mrk 421         &  110427+381230.0  & blazar \\
QQ Vul          &  200541+223959.0  & polar \\
RR Aqr (D)      &  211843-020812.0  & variable star \\
RR Aqr          &  211501-025344.0  & variable star \\
RR Aqr (C)      &  211917-014609.0  & variable star \\
VV Pup          &  081506-190315.0  & variable star \\
EF Eri          &  031413-223540.0  & polar \\
V834 Cen        &  140907-451717.0  & polar \\
V2214 Oph       &  171201-293732.0  & polar \\
J0458-4635      &  045550-461557.0  & blazar \\
BL Hyi          &  014100-675326.0  & polar \\
MR Ser          &  155247+185626.0  & polar \\
4C 11.69        &  223236+114350.0  & blazar \\
OS 319          &  161341+341247.0  & blazar \\
3C 279          &  125611-054721.0  & blazar \\
V347 Pav        &  184448-741833.0  & polar \\
GQ Mus          &  115202-671220.0  & polar \\
RR Aqr (G)      &  211622-023914.0  & variable star \\
PKS 1229-021    &  123159-022405.0  & blazar \\
\hline
\end{tabular}
\end{tiny}
\caption{List of objects to be automatically followed in order of priority}
\label{tab_pointing_objects}
\end{center}
\end{table}





\ifpdf
	\graphicspath{{Appendix2/Appendix2Figs/PNG/}{Appendix2/Appendix2Figs/PDF/}{Appendix2/Appendix2Figs/}}
\else
	\graphicspath{{Appendix2/Appendix2Figs/EPS/}{Appendix2/Appendix2Figs/}}
\fi

\chapter{Technical Details on Data Analysis}

\section{Parameters of on-line flash recognition algorithm}
\label{sec_appendix2}

\begin{table}[htbp]
\begin{center}
\begin{tabular}{|c|c|}
\hline
\textbf{Parameter Name} & \textbf{Parameter Name in configuration file} \\
\hline 
LaplaceType & CCD\_LAPLACE\_TYPE \\
\hline
$T_n$ & CCD\_NEW\_LAPLACE\_TRESHOLD\_IN\_SIGMA \\ 
\hline
$T_v$ & CCD\_MAX\_LAPLACE\_ON\_OTHER\_IN\_SIGMA \\
$N_{aver}$ & CCD\_AVERAGE\_OF\_PREV\_N \\
\hline
$T_{MinLap}$ & CCD\_MIN\_LAPLACE\_ON\_OTHER \\
\hline
$N_{MaxTv}$ & CCD\_MAX\_NUMBER\_OF\_EVENTS\_AFTER\_TV \\
\hline
enable/disable & CCD\_SKIP\_OVERLAPS \\
$R_{overlap}$ & CCD\_OVERLAP\_REDIAL \\
\hline
$T_{shape}$ & CCD\_CHECK\_EVENT\_SHAPE \\
\hline
$T_{black}$ & CCD\_BLACK\_PIXELS\_RATIO \\
\hline
$T_{hot}$ & CCD\_REJECT\_HOT\_PIXELS\_BY\_AVERAGE\_TRESH \\
\hline
$N_{IfMore}$ & CCD\_SKIP\_IF\_MORE\_THEN \\
$R_{IfMore}$ & CCD\_SKIP\_IF\_MORE\_THEN\_REDIAL \\
\hline
\end{tabular}
\caption{Table translating FLT paramter names into real names used in configuration file}
\label{tab_fltparams_trans}
\end{center}
\end{table}

\begin{table}[htbp]
\begin{center}
\begin{tabular}{|c|c|}
\hline
\textbf{Parameter Name} & \textbf{Parameter Name in configuration file} \\
\hline 
$R_{coinc}$ & CCD\_COIC\_RADIUS\_IN\_SEC \\
\hline
$N_{confirm}$ & CCD\_CONFIRM\_ON\_N\_NEXT\_FRAMES \\
\hline
$R_{sat}$ & CCD\_SAT\_REJ\_RADIUS\_IN\_SEC \\
\hline
$R_{star}$ & CCD\_MATCH\_STAR\_TO\_CAT\_RADIUS\_IN\_ARCSEC \\
$Mag_{max}$ & CCD\_STARCAT\_MAX\_MAG \\
\hline
$N_{track}$ & CCD\_NUM\_BACK\_FRAMES\_FOR\_TRACK \\
$\chi^2$ & CCD\_MAX\_CHI2\_FOR\_POINT\_TO\_MATCH\_LINE \\
$\chi^2$ & CCD\_MAX\_CHI2\_IN\_TRACK \\
\hline
\end{tabular}
\caption{Table translating SLT paramter names into real names used in configuration file}
\label{tab_sltparams_trans}
\end{center}
\end{table}

\begin{table}[htbp]
\begin{center}
\begin{tiny}
\begin{tabular}{|c|c|}
\hline
\textbf{Cut} & \textbf{Parameter Name} \\
\hline
$T_{hough}$ & CCD\_HOUGH\_TRANSFORM\_TRESH \\
$T_{hough_distr}$ & CCD\_HOUGH\_DISTR\_TRESH \\
$T_{hough_height}$ & CCD\_HOUGH\_DISTR\_MAX\_LIMIT \\
\hline
$N_{tstars}$ & CCD\_SLT\_TYPICAL\_STARS\_COUNT \\
$R_{clouds}$ & CCD\_SLT\_REJECT\_FRAME\_IF\_LESS \\
\hline
$L_{diff}$ & CCD\_LAP\_DIFF\_MIN\_RATIO \\
\hline
$T_n^{TLT}$ & CCD\_NEW\_LAPLACE\_TRESHOLD\_IN\_SIGMA \\
\hline
\end{tabular}
\end{tiny}
\caption{Table translating TLT paramter names into real names used in configuration file slt.cfg}
\label{tab_tltparams_trans}
\end{center}
\end{table}



\bibliographystyle{Classes/my}
\bibliography{References/references}
\addcontentsline{toc}{chapter}{References}

\end{document}